\documentclass[10pt,journal,compsoc]{IEEEtran}

\ifCLASSOPTIONcompsoc
  \usepackage[nocompress]{cite}
\else
  \usepackage{cite}
\fi

\ifCLASSINFOpdf

\else

\fi

\usepackage{epsfig}
\usepackage{graphicx}
\usepackage{amsmath}
\usepackage{amssymb}
\usepackage{comment}
\usepackage{multirow}

\usepackage{booktabs}
\usepackage{array, caption, threeparttable}
\usepackage{caption}

\usepackage[table,xcdraw]{xcolor}

\usepackage{algorithm}
\usepackage{algpseudocode}
\usepackage{listings}
\usepackage{color}

\usepackage{mathtools}

\usepackage{todonotes}
\usepackage{gensymb}

\usepackage[breaklinks=true,bookmarks=false]{hyperref}
\usepackage{hyperref}
\hypersetup{hidelinks}
\usepackage{float}
\usepackage{dblfloatfix}
\usepackage{placeins}
\usepackage{subcaption}
\usepackage{longtable}

\begin{document}

\title{Benchmarking Cross-Domain Audio-Visual Deception Detection}

\author{Xiaobao Guo, Zitong Yu,~\IEEEmembership{Senior Member,~IEEE}, Nithish Muthuchamy Selvaraj, Bingquan Shen, \\ Adams Wai-Kin Kong,~\IEEEmembership{Senior Member,~IEEE} and Alex C. Kot,~\IEEEmembership{Life Fellow,~IEEE}

\thanks{This work was supported by National Natural Science Foundation of China (Grant Nos. 62306061 and 62576076), and sponsored by CCF-Tencent Rhino-Bird Open Research Fund. Corresponding author: Zitong Yu.\\
Xiaobao Guo is with the Rapid-Rich Object Search (ROSE) Lab and the College of Computing and Data Science, Nanyang Technological University (NTU), Singapore 639798, Singapore (e-mail: xiaobao001@ntu.edu.sg). \\
Zitong Yu is with School of Computing and Information Technology and Dongguan Key Laboratory for Intelligence and Information Technology, Great Bay University, Dongguan 523000, China.(e-mail: zitong.yu@ieee.org). \\
Nithish Muthuchamy Selvaraj is with the Rapid-Rich Object Search (ROSE) Lab, Nanyang Technological University (NTU), Singapore 639798, Singapore (e-mail: ms.nithish@ntu.edu.sg). \\
Bingquan Shen is with DSO National Laboratories, Singapore 118230, Singapore (e-mail: sbingqua@dso.org.sg). \\
Adams Wai-Kin Kong is with the College of Computing and Data Science, Nanyang Technological University (NTU), Singapore 639798, Singapore (e-mail: adamskong@ntu.edu.sg).\\
Alex C. Kot is with SMBU, Shenzhen 518172, China; VinUniversity, Hanoi 100000, Vietnam; and NTU, Singapore (e-mail: eackot@ntu.edu.sg). }

}

\markboth{IEEE Transactions on Affective Computing}%
{Shell \MakeLowercase{\textit{\emph{et al.}}}: Bare Advanced Demo of IEEEtran.cls for IEEE Computer Society Journals}

\IEEEtitleabstractindextext{%
\begin{abstract}

Automated deception detection is crucial for assisting humans in accurately assessing truthfulness and identifying deceptive behavior. 
Conventional contact-based techniques, like polygraph devices, rely on physiological signals to determine the authenticity of an individual's statements. Nevertheless, recent developments in automated deception detection have demonstrated that multimodal features derived from both audio and video modalities may outperform human observers on publicly available datasets. Despite these promising results, the generalizability of existing audio-visual deception detection approaches across different scenarios remains largely unexplored. To bridge this gap, we present the first cross-domain audio-visual deception detection benchmark, that enables us to assess how well these methods generalize for use in real-world scenarios. We used widely adopted audio and visual features and various architectures to assess single-to-single versus multi-to-single domain generalization performance. To further exploit the impacts using data from multiple source domains for training, we investigate three types of domain sampling strategies, including domain-simultaneous, domain-alternating, and domain-by-domain for multi-to-single domain generalization evaluation. We also propose an algorithm to enhance the generalization performance by maximizing the gradient inner products between modality encoders, named ``MM-IDGM". Furthermore, we proposed the Attention-Mixer fusion method to improve performance, and this benchmark will facilitate future research in audio-visual deception detection.

\end{abstract}

}

\maketitle

\IEEEdisplaynontitleabstractindextext

%
\IEEEpeerreviewmaketitle

\ifCLASSOPTIONcompsoc
\IEEEraisesectionheading{\section{Introduction}\label{sec:introduction}}
\else
\section{Introduction}
\label{sec:introduction}
\fi

\begin{figure}[t]
\centering
\vspace{-1.8em}
\includegraphics[width=0.85\linewidth]{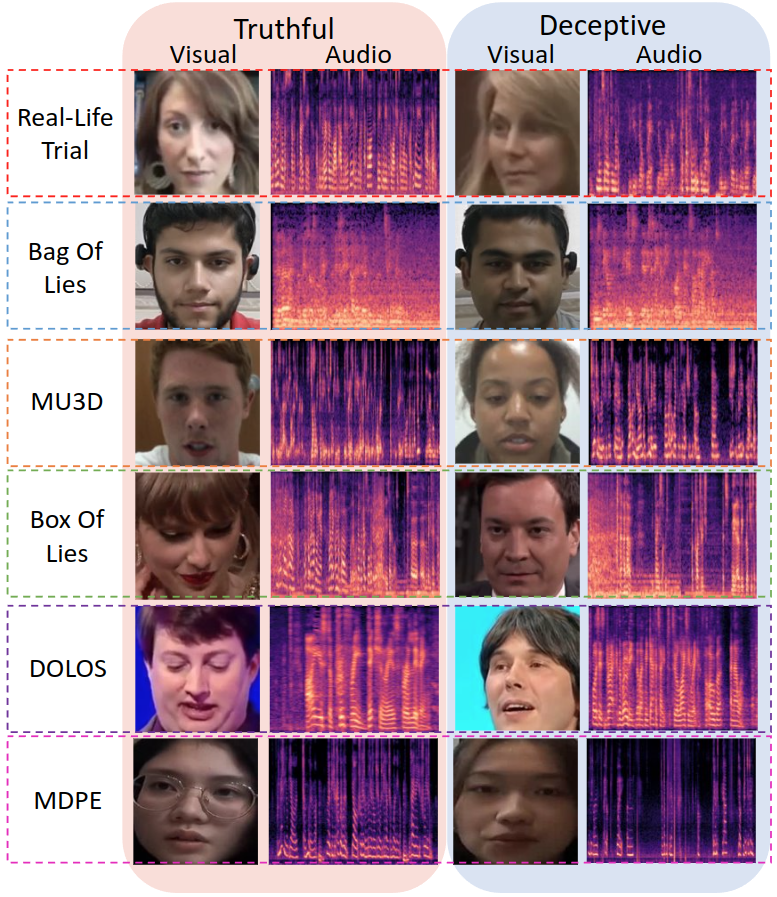}
\vspace{-0.8em}
\caption{Representative samples from six publicly available deception detection datasets: Real-Life Trial~\cite{perez2015deception}, Bag of Lies~\cite{gupta2019bag}, MU3D~\cite{lloyd2019miami}, Box of Lies~\cite{soldner2019box}, DOLOS~\cite{guo2023audio}, and MDPE~\cite{cai2024mdpe}. Each row shows one dataset, with columns grouped by modality (visual or audio) and ground truth (truthful vs. deceptive). The figure is intended to provide a qualitative illustration of domain differences across datasets. The visual samples suggest differences in resolution, illumination, and subjects' ethnicities, while the audio spectrograms reflect variations in pitch structure and background noise. Quantitative dataset statistics are further summarized in Table~\ref{tab:stats}.}

\label{fig:crossvisual}
\end{figure}

\IEEEPARstart{A}{utomated} deception detection is the task of predicting whether a subject is truthful or deceptive from observable behavioral evidence. In this work, audio-visual deception detection refers to a multimodal setting that uses both visual cues, such as facial movements and expressions, and audio cues, such as speech patterns and vocal characteristics, to perform this prediction~\cite{ding2019face,wu2018deception,gogate2017deep, mathur2021unsupervised}. Deception detection has a significant impact on various real-world applications such as law enforcement~\cite{wang2004criminal}, healthcare~\cite{joudaki2014using}, and business~\cite{glancy2011computational}. It has the potential to prevent fraud, improve security measures, and enhance trust and confidence. A reliable deception detection tool can support more accurate decision-makings.

Traditional deception detection is often a contact-based method. It assesses whether someone is telling the truth or not by monitoring physiological responses like skin conductance and heart rate~\cite{synnott2015review, li2024deception,javaid2022eeg}. Experts' behavioral observation and analysis are another technique that evaluates changes in a person's body language, speech patterns, and eye movements~\cite{detect1999lie,nortje2019good}. However, such an assessment can be time-consuming and require significant expertise to perform accurately.

Recently, the development of automated deception detection systems using AI and machine learning techniques has gained significant attention as the traditional methods mentioned above have limitations in terms of reliability, accuracy, and scalability. Various multimodal datasets have been introduced, including real-life trials from court scenes~\cite{perez2015deception}, lab-based setups~\cite{gupta2019bag,lloyd2019miami,cai2024mdpe}, and game show scenarios~\cite{soldner2019box,guo2023audio}. These datasets provide a wide variety of deceptive samples from different domains, enabling researchers to examine the effectiveness of AI models on deception detection. Based on these datasets, progress has been made in deception detection techniques within individual datasets where models are trained and evaluated on the same domain~\cite{gogate2017deep, wu2018deception,karimi2018toward,king2024applications}. Recent studies have utilized rich visual and audio features~\cite{karnati2021lienet, avola2019automatic, yang2021multimodal,karimi2018toward,chebbi2023deception,ding2019face,mathur2020introducing}, such as Mel Spectrogram, emotional states, and facial action units, to enhance the performance of deception detection tasks. 

However, there remains a substantial research gap that needs to be addressed. Specifically, fewer studies have explored the cross-domain issue, despite the presence of significant domain shifts in public deception detection datasets. 
As illustrated in Fig. \ref{fig:crossvisual}, the publicly available datasets used in deception detection exhibit substantial domain shifts across both visual and audio modalities.
Figure \ref{fig:crossvisual} provides a qualitative overview of these differences through representative visual samples and audio spectrograms, while Table~\ref{tab:stats} provides quantitative statistics to further support the dataset gaps. 
The datasets span multiple domains, including game shows (e.g., Box of Lies), courtroom scenarios (Real-Life Trials), and lab settings (e.g., Bag of Lies). The differences include video resolution, illumination, subjects' ethnicities in visuals, signal-to-noise ratios (SNRs), pitch dynamics, and background noise in audio. We quantified these differences by summarizing statistics for visual and audio information (details in Sec. 4.1 and Table~\ref{tab:stats}). These domain discrepancies underline the challenge of generalizing deception detection models beyond the training domain, motivating our benchmark's focus on cross-domain evaluation. Effective methods must be proposed to alleviate the domain shift issue by fusing both audio and visual features in a meaningful way. Addressing these issues can benefit automated deception detection systems in improving generalizability in real-world applications.

To address the issue of cross-domain deception detection, we introduce a new benchmark that evaluates the generalization capacity of AI models using audio and visual features over publicly available datasets\footnote{Protocols and source code are available at \url{https://github.com/Redaimao/cross\_domain\_DD}.}. Unlike previous works that assess models only within a single dataset, our benchmark systematically evaluates cross-domain generalization using widely adopted audio and visual features through two setups: single-to-single domain generalization and multi-to-single domain generalization. Specifically, for the multi-to-single setting, three domain sampling strategies, \emph{i.e.,} domain simultaneous, domain alternating, and domain-by-domain, are implemented for benchmarking. These strategies control how batch data is sampled during training from multiple source datasets, allowing models to learn either domain-invariant or domain-specific patterns. To further enhance performance, we propose Multimodal Inter-Domain Gradient Matching (MM-IDGM) to improve multi-to-single domain generalization performance and an Attention-Mixer fusion method based on MLP-Mixer~\cite{tolstikhin2021mlp}. 
In our experiments, we compared single-to-single and multi-to-single domain generalization, incorporating three sampling strategies, one domain generalization method, and six fusion methods across two modalities. The benchmarking results among different features reveal that the best average fusion accuracy was 56.82\% for single-to-single cross-domain generalization and 58.88\% for multi-to-single cross-domain generalization. Moreover, for MM-IDGM, we achieved the highest average accuracy of 59.02\% based on domain-simultaneous sampling. This benchmarking framework fills a critical gap in the literature and serves as an important tool for evaluating the effectiveness of audio-visual deception detection models in diverse contexts, which will help improve the capabilities of automated deception detection systems in real-world settings. Additionally, we hope our work will inspire further research on multimodal models that address domain shift issues. In summary, our main contributions include:
\begin{itemize}
   \item Introducing a new benchmark for cross-domain audio-visual deception detection: We present a comprehensive benchmark designed to evaluate the model generalization using audio and visual modalities across diverse datasets with domain shifts.
    \item Comparing two evaluation protocols across various architectures: single-to-single and multi-to-single domain generalization. These two protocols enable comprehensive performance comparisons across varying levels of domain discrepancy.
    \item Developing and testing three domain sampling strategies for multi-to-single domain generalization: We propose and evaluate three distinct domain sampling strategies, \emph{i.e.}, domain simultaneous, domain alternating, and domain-by-domain, to assess their effectiveness and offer diverse approaches under cross-domain settings.
    \item Introducing Multimodal Inter-Domain Gradient Matching (MM-IDGM) for domain generalization: We propose MM-IDGM to improve multi-to-single domain generalization by aligning gradient updates between audio and visual modalities, enhancing the model’s ability to transfer knowledge across different domains.
    \item Proposing the Attention-Mixer fusion method for enhanced multimodal performance: We introduce the Attention-Mixer fusion method that combines MLP-Mixer and self-attention layers to better capture both intra- and inter-modal interactions, resulting in improved multimodal integration and cross-domain generalization.
\end{itemize}

In the rest of the paper, Sec.~\ref{sec:relatedwork} provides a review of related psychological studies on cues to deception and multimodal deception detection works. Sec.~\ref{sec:pyramid} introduces our benchmarking approach and fusion method. Sec.~\ref{sec:experiment} provides the cross-domain benchmark results and fusion results. Finally, conclusions and future works are given in Sec.~\ref{sec:conclusion}.

\section{Related work}
\label{sec:relatedwork}

\subsection{Cues to Deception}

The research on using behavioral cues for deception has gradually become active over the past few decades. Psychological researchers have published a large number of works on the analysis of cues to deception~\cite{buller1996interpersonal, hartwig2011lie, vrij2000detecting}. Among the studied behavioral cues, verbal and nonverbal cues were preferred as humans may behave differently between lying and telling the truth. DePaulo \emph{et al.}~\cite{depaulo2003cues} studied and reported experimental results on 158 cues to deception. They revealed that, in general, people who tell lies are less forthcoming and less convincing than those who tell the truth. Liars usually talk about fewer details and make fewer spontaneous corrections. They also sound less involved but more vocally tense. Through the study, the researchers statistically found that liars often press their lips, repeat words, raise their chins, and show less genuine smiles. The results show that some behavioral cues do potentially appear in deception and are even more pronounced when liars are more motivated to cheat.

Levine \emph{et al.}~\cite{levine2014theorizing} reviewed the status quo and provided a new perspective on the theories of deception. They pointed out that lying usually happens when problematic information is involved. It is critical to understand the verbal content in the context. 
Vrij \emph{et al.}~\cite{vrij2012eliciting} realized that interviewers play a vital role in eliciting and enhancing cues to deceit. The authors proposed the ``interviewing to detect deception" technique to open a new path in the deception detection research field. They argued that different psychological states can be exploited by adopting appropriate interview techniques of liars and truth-tellers.
Warren \emph{et al.}~\cite{warren2009detecting} conducted experiments to investigate the relationship between affective facial expressions to deception. The results indicated that leaked emotions with the incongruous intended message can provide useful cues to deception, which supported the nonverbal leakage theory~\cite{ekman1969nonverbal}.

\begin{figure*}[t]
\centering
\includegraphics[width=0.9\linewidth]{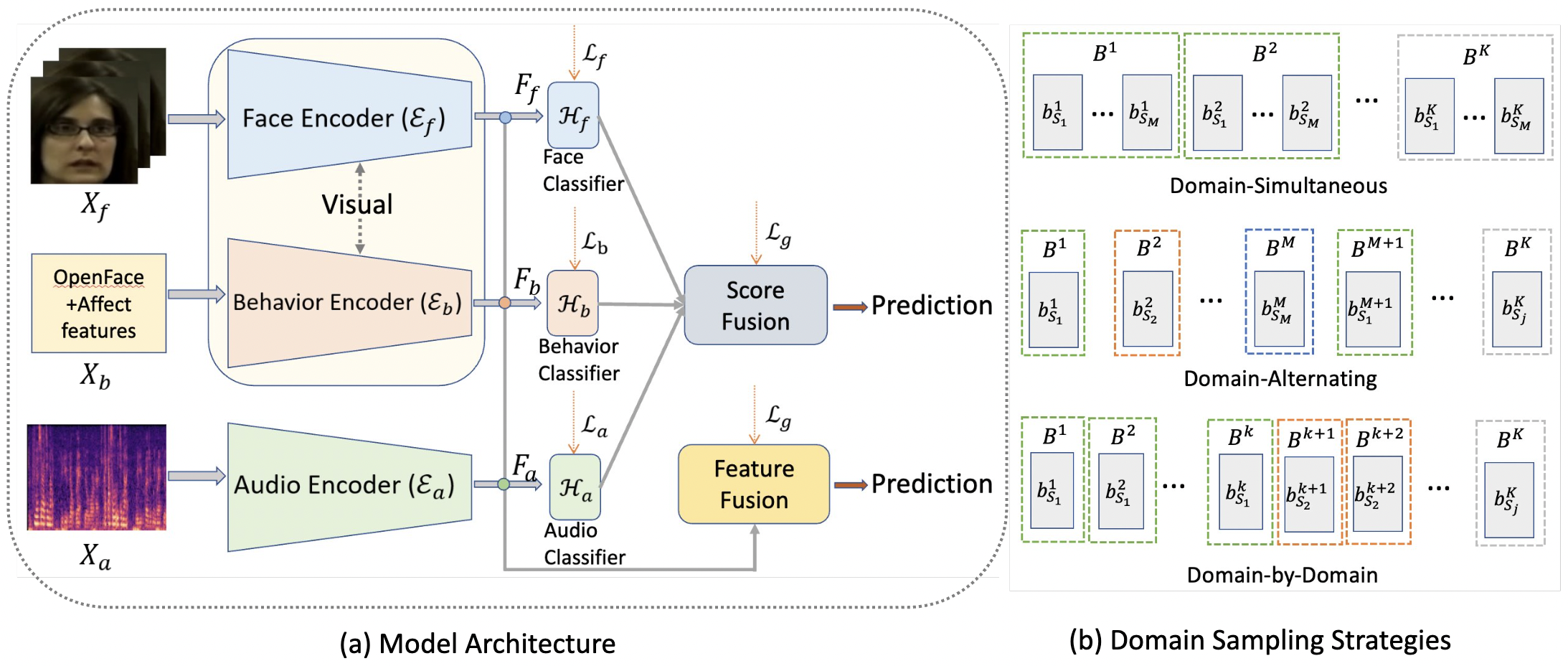}
  \caption{Main network and method. (a) Model architecture. Visual modality includes face and behavior inputs. Audio modality includes Mel Spectrogram input. The features obtained by the respective encoders. The fusion methods include score fusion and feature fusion. (b) Domain sampling strategies. Domain-simultaneous: each batch consists of samples from multiple sources. Domain-alternating: each batch is alternatively sampled from multiple sources. Domain-by-domain: the batches are sampled from one source and then from another. $F_{f}$, $F_{b}$, and $F_{a}$ represent the face, behavior, and audio features, respectively. $L_{f}$, $L_{b}$, $L_{a}$, and $L_{g}$ denote the losses corresponding to the face, behavior, and audio classifiers, as well as the fusion layer.}
\label{fig:mainfig}
\vspace{-1em}
\end{figure*}

\subsection{Multimodal Deception Detection}
Recent advances in deception detection have increasingly incorporated both verbal and non-verbal cues, leveraging multimodal data to enhance detection performance. Often, those methods integrate features from different modalities, such as visual, audio, and textual modalities, and propose fusion strategies to combine them efficiently~\cite{mathur2020introducing, karimi2018toward, karnati2021lienet, avola2019automatic}. In this section, we review and discuss the strengths and limitations of recent audio-visual deception detection methods, including the fusion techniques used and the gaps in addressing cross-domain issues.

\noindent\textbf{Visual Features in Deception Detection.}\quad
Many recent works have utilized facial features extracted from RGB images to perform deception detection~\cite{karimi2018toward, avola2019automatic, ding2019face}. They focus on capturing facial movements and expressions, which are crucial indicators of deception. Many studies have employed Facial Action Units (AUs) as a key feature~\cite{karimi2018toward, avola2019automatic} to analyze subtle facial movements. AUs provide detailed representations of specific facial muscle activations, making them highly useful for identifying deceptive cues. Additionally, gaze movements and facial expressions have been widely adopted as visual features in deception detection models~\cite{wu2018deception, mathur2020introducing, yildirim2023influence,gallardo2021detecting}. By capturing emotional states that may signal deception, these methods can further enhance the detection capabilities.

\noindent\textbf{Audio Features for Deception Detection.}\quad
Along with visual cues, many recent works have incorporated audio features to boost the performance of deception detection systems. Audio signals can provide complementary information, such as changes in speech patterns, tone, and hesitation, which are often associated with deceptive behavior~\cite{gogate2017deep, wu2018deception, karimi2018toward}. For example, Wu \emph{et al.}~\cite{wu2018deception} used MFCC (Mel-frequency Cepstral Coefficients) features, and  Karimi \emph{et al.}~\cite{karimi2018toward}  explored the use of raw audio signals for deception detection, demonstrating the potential of unprocessed data to capture subtle vocal cues indicative of deception.

\noindent\textbf{Multimodal Fusion Techniques.}\quad
The integration of multiple modalities, such as visual, audio, and textual information, has been shown to improve deception detection performance significantly. Most recent works mentioned above have considered multimodal fusion approaches that extract visual, audio, and text information to boost performance~\cite{chebbi2023deception,kumar2021deception}. In addition to visual, audio, and text, Karnati \emph{et al.}~\cite{karnati2021lienet} exploited physiological signals, \emph{i.e.,} EEG representations for deception detection. Fusion methods can be broadly categorized into two types: feature-level fusion and decision-level fusion. Feature fusion combines raw or intermediate features from each modality into a single representation, allowing the model to learn cross-modal dynamics~\cite{karimi2018toward, avola2019automatic, ding2019face, wu2018deception, gogate2017deep}. This method often uses different encoders, linear layers, or fusion modules to blend features and capture interactions between modalities. On the other hand, decision-level fusion merges the outputs or produced logits of individual unimodal models at a later stage. It can reduce computational complexity and enable the system to learn effective marginal representations from each modality~\cite{karnati2021lienet, gogate2017deep}. Both approaches have shown promising results in improving the performance of deception detection systems.

Despite significant progress, there still exists the challenge of cross-domain generalization in multimodal deception detection.
Previous works have mainly focused on optimizing unimodal or fusion methods within a single domain, without considering the domain shift when the system is applied to different environments or populations. For example, models trained on controlled lab data may not generalize well to real-world settings, where the recording conditions and communication styles may introduce significant variability.

In this work, we aim to build a benchmark for cross-domain generalization performance on the widely used audio-visual deception detection datasets, which is crucial for evaluating and improving the robustness of deception detection models in different scenarios. Establishing such a benchmark will provide clearer comparisons, highlight model weaknesses, and guide the development of systems across different domains.

\section{Methodology} \label{sec:pyramid}

The mainstream architecture for audio-visual deception detection usually includes encoders for unimodal feature extraction and/or a fusion module. We follow the widely adopted architecture to build the benchmark on cross-domain audio-visual deception detection in this work. As shown in Fig.~\ref{fig:mainfig}, audio and visual features are extracted from audio and visual encoders. The fusion module is performed based on audio and visual features. The fused feature is input to the classifier for classification. The final prediction is the classifier output logit optimized with the binary cross-entropy loss in Eq.~\ref{eq2}. We build the benchmark for cross-domain generalization performance based on such network architecture with different encoders. We conducted single-to-single and multi-to-single evaluations where three domain sampling strategies included, \emph{i.e.,} domain-simultaneous, domain-alternating, and domain-by-domain. It is worth noting that these domain sampling strategies are implemented in the data loading pipeline rather than in the network architecture itself, and they are introduced to evaluate how different multi-domain training protocols affect cross-domain generalization.

\subsection{Audio and Visual Feature Learning}

To establish a benchmark for cross-domain audio-visual deception detection, we utilize widely adopted audio and visual features along with their respective encoders. Our approach treats audio and visual features as equally important, extracting different types of features simultaneously. Under this unified architecture, different training protocols can then be fairly compared while keeping the backbone and fusion design consistent. As depicted in Fig.~\ref{fig:mainfig}, this network structure offers several advantages: (1) flexibility in network selection: different audio or visual encoders can be effortlessly incorporated and compared in a fair manner, (2) adaptability: the addition or removal of specific modules and/or losses is straightforward. For instance, a fusion module can be inserted before classifiers, and (3) easy performance benchmarking: the system facilitates evaluating performance in various settings, such as score-level fusion and feature-level fusion. In this work, we focus on audio and visual modalities for deception detection. In particular, two kinds of visual features are extracted, \emph{i.e.,} face features from RGB face images and behavior features consist of AUs, affects, etc.

As shown in Fig.~\ref{fig:mainfig}, given a detected RGB face image as input $X_{f}$, the deep features $F_{f}$ could be extracted via face encoder networks $\mathcal{E}_{f}$ (e.g., ResNet18~\cite{He2015Deep}). Similarly, behavior inputs such as the AU and/or affect features $X_{b}$ are encoded by OpenFace~\cite{amos2016openface} or affect model (e.g., EmotionNet~\cite{toisoul2021estimation}) $\mathcal{E}_{b}$ to output behavior features $F_{b}$. Note that we regard both face frames and behavior features as the visual modality but differentiate them in this work as they have different types of information and representations. Given audio input $X_{a}$ (either Mel Spectrogram~\cite{eyben2010opensmile} or waveforms), audio features $F_{a}$ are extracted through audio encoder $\mathcal{E}_{a}$. The corresponding classifiers for face frames ($\mathcal{H}_{f}$), behavior features ($\mathcal{H}_{b}$), and audio features ($\mathcal{H}_{a}$) output the prediction logits $\hat{Y}_f$, $\hat{Y}_b$, and $\hat{Y}_a$, respectively. The fusion layer $\mathcal{G}$ takes $F_{f}$, $F_{b}$, and $F_{a}$ as input. $\mathcal{G}$ is determined by the actual fusion method, e.g., liner layer, transformer layers, MLP, etc. The output logit of $\mathcal{G}$ is denoted by $\hat{Y}_g$. Therefore, the audio and visual learning process can be denoted as follows:
\begin{equation}\small
\label{eq1}
    \begin{split}
        F_{f} &= \mathcal{E}_{f} (X_{f}), \hat{Y}_f = \mathcal{H}_{f} (F_{f}), \\
        F_{b} &= \mathcal{E}_{b} (X_{b}), \hat{Y}_b = \mathcal{H}_{b} (F_{b}), \\
        F_{a} &= \mathcal{E}_{a} (X_{a}), \hat{Y}_a = \mathcal{H}_{a} (F_{a}), \\
        \hat{Y}_g &= \mathcal{G} (F_{f}, F_{b}, F_{a}).
    \end{split}
\end{equation}

\textbf{Loss Function.}\quad  For deception detection ground truth $Y$, where $Y=0$ for truthful and $Y=1$ for deception, the binary cross-entropy loss (BCE) is adopted. The loss for each sample with a certain modality or fused prediction can be denoted as
\begin{equation}
\label{eq2}
    \mathcal{L}_{m} =  -(Ylog(\hat{Y}_{m})+(1-Y)log(1-\hat{Y}_{m})),
\end{equation}
\noindent where $m \in \{f,b,a,g\}$, $\hat{Y}_{m}$ is the corresponding prediction logits. In other words, the BCE loss is calculated separately for each type of modality and/or its fused feature depending on whether a sample has any face frames, visual inputs, or audio inputs. The overall loss function can be described as follows:
\begin{equation}
\label{eq3}
\mathcal{L}_{}=  \frac{1}{N} \sum_{i=1}^N \left ( \sum_{m={f, b, a}} \mathcal{L}_{m, i} + \lambda \mathcal{L}_{g, i} \right ),
\end{equation}
\noindent where $N$ is the number of data samples and $\lambda$ is a trade-off parameter between modality loss and fusion loss. $\lambda$ is set to 0.5 in our experiments. For score fusion, the fusion score is the average of the individual predicted scores for each class, and the cross-entropy loss function is applied to the fused score. For feature fusion, individual features are the inputs for the fusion block.

\subsection{Cross-domain Generalization}

We benchmark the cross-domain generalization on the deception detection task. First, we introduce the notations and definitions in this section. A domain is composed of data that are sampled from a distribution (dataset), which can be denoted as $\mathcal{S} = \{(X; Y)_{i}\}_{i=1}^N\sim P_S$, where $X = (X_f, X_b, X_a)$, $X_f, X_b, X_a$ represent samples of face frames, behavior, and audio modalities, respectively. $Y$ denotes the label, and $P_S$ denotes the joint distribution of the input samples and the output label. In this paper, for simplicity, we follow similar definitions in~\cite{wang2022domain,varanka2023data} to treat each dataset as an individual domain due to their obvious distribution gaps, but more fine-grained intra-domain factors would be explored in future work. For domain generalization, $M$ source domains (training datasets) are given, \emph{i.e.,} $\mathcal{S}_{train} = \{S_j |j=1,\cdots, M\}$, where $\mathcal{S}_j = \{(X; Y)_{i}\}_{i=1}^{N_{j}}\sim P_{S_{j}}$ denotes the $j$-th domain, and $P_{S_{i}} \neq P_{S_{j}}$ for $ 1 \leq i, j \leq M$. $N_{j}$ is the number of total samples in $S_{j}$. The goal of domain generalization is to learn the predictive function $h$ in $M$ source domains to achieve minimum error on an unseen test domain $\mathcal{S}_{test} \sim P_{S_{test}}$, and $P_{S_{test}} \neq P_{S_{j}}$ for $ 1 \leq j \leq M$:
\begin{equation}
\label{eq4}
    min~\mathbb{E}_{(X; Y) \in \mathcal{S}_{test}}\left[\mathcal{L}(h(X), Y)\right],
\end{equation}
\noindent where $X = (X_f, X_b, X_a)$ is the input, $Y$ is the label, $\mathcal{L}$ is the loss function, and $\mathbb{E}$ denotes the expectation.

When $M=1$, it is a \textit{Single-to-single Cross-domain Generalization} task, where the modal is trained on one training dataset and tested on another dataset. When $M\geqslant2$, we propose three strategies to learn from multiple domains for the \textit{Multi-to-single Cross-domain Generalization}. Previous works~\cite{jiang2023domain, gulrajanisearch} have indicated that different sampling strategies for domain generalization may result in different performances. Therefore, in this work, we propose three multi-to-single domain generalization sampling methods and compare their performance. Let $B$ denote one batch of training data with a size of $N_B$. Given multiple training domains $\mathcal{S}_{train} = \{S_j |j=1,\cdots, M\}$, $B$ is a set of training data sampled from $\mathcal{S}_{train}$.

\vspace{0.3em}
\noindent\textbf{Domain-Simultaneous}\quad means to train multiple domains in parallel within each batch of data. In domain simultaneous training, the $k-$th batch of training data is a group of samples from different domains, \emph{i.e.,} $B^k = (b_{S_{1}}^k, \cdots, b_{S_{M}}^k)$, $k \in [1, \cdots K]$, where $b_{S_{j}}^k$ is the batch samples from domain $S_j$ for $j=1,\cdots, M$ and $K$ is the number of batches during training. The total number of $b_{S_{j}}^k$ is $N_B$. As shown in Fig.~\ref{fig:mainfig} (b), each training batch contains smaller batch samples from all the source domains during training. Models are trained to learn from different domains simultaneously by feeding the mixed batch data. This approach helps the model learn domain-invariant features by exposing it to data from multiple domains simultaneously. However, it may cause the model to focus on ``average" features across domains, leading to poor generalization to the unseen domains. Mixing domains ensures broader feature exposure, but can dilute domain-specific nuances that are useful for recognizing patterns in unseen domains.

\vspace{0.3em}
\noindent\textbf{Domain-Alternating}\quad is different from domain simultaneous strategy in terms of batch samples. In domain-alternating, $B^k = b_{S_j}^k$ for $j={k-\lfloor {(k-1)\over{M}} \rfloor \cdot M}$, where $\lfloor{\cdot}\rfloor$ is the flooring operator. The number of $b_{S_{j}}^k$ is $N_B$. Fig.~\ref{fig:mainfig} (b) shows that the consecutive batch samples come from different domains. Domain-alternating method samples from only one domain per training batch, and the domain alternates at each batch. It helps the model specialize for each domain and may lead to better generalization across domains. However, there is a risk of overfitting to individual domains if the diversity of training data is not enough.

\vspace{0.3em}
\noindent\textbf{Domain-by-Domain}\quad aims to train the model by feeding data from source domain data one by one.  $B^k = b_{S_j}^k$ for $\lceil {\sum_{i=0}^{i=j-1}N_{i}\over{N_B}}\rceil \leqslant k \leqslant \lceil {\sum_{i=0}^{i=j}N_{i}\over{N_B}}\rceil$, $N_0 = 0$,  where $\lceil{\cdot}\rceil$ is the ceiling operator. The number of $b_{S_{j}}^k$ is $N_B$. As shown in Fig.~\ref{fig:mainfig}, the batch data samples from one domain after finishing sampling from its previous domains. If the domain shifts are substantial, this strategy may overfit the characteristics of the domain it has just been trained on. Focusing entirely on one domain at a time creates a risk of the model becoming too specialized in that domain, which may hamper its ability to generalize across unseen domains. However, domain-by-domain sampling allows the model to capture domain-specific variations, which might be useful for applications where learning from each domain is necessary before attempting to generalize.

\noindent\textbf{Multimodal Inter-Domain Gradient Matching.}
Beyond the benchmarking on the three different multi-to-single sampling strategies, inspired by Shi et al.~\cite{shigradient}, we propose Multimodal Inter-Domain Gradient Matching (MM-IDGM). IDGM was proposed by Shi et al.~\cite{shigradient} and uses an effective optimization algorithm named ``Fish" to approximate the second-order derivative in IDGM and to reduce computational cost. The main idea of IDGM is to maximize the gradient inner product to align the gradient directions across domains, so as to learn the weights that can produce the closer input-output correspondence. However, considering the multimodal nature of our task, directly optimizing IDGM in multimodal models may not be desirable. Different from IDGM in~\cite{shigradient} which aims to learn the features that are invariant across domains, MM-IDGM prioritizes learning the invariant features across domains for the same modalities by maximizing the gradient inner product between modality encoders. These inner gradient products are then dynamically adjusted based on the unimodal losses from previous steps. The remaining parts of the models are then optimized across domain by IDGM. 
Compared with IDGM, MM-IDGM has three key differences: 1) modality-specific encoder
decomposition, 2) loss-adaptive modality-wise inner updates, and 3) hybrid optimization in
which modality encoders and the remaining shared modules are updated differently. 
The MM-IDGM algorithm is developed based on the ``Fish" algorithm and is presented in Algorithm~\ref{alg:cap}.

\begin{algorithm}
\caption{MM-IDGM for Modalities $m_1$ and $m_2$}\label{alg:cap}
\begin{algorithmic}
\Require $L_{m_1}$, $L_{m_2}$   \Comment{from the previous step}
\While{ iterations $\leq$ maximum iterations}
\State $\tilde{\theta}_{m_1} \gets \theta_{m_1}$, $\tilde{\theta}_{m_2} \gets \theta_{m_2}$, $\tilde{\theta}_r \gets \theta_r$
\For{\texttt{$S_i \in $permute$\{S_1, S_2, \cdots, S_M\}$}}
\State Sample batch $s_i \sim S_i$
\State $g_{m_1,i} = \mathbb{E}_{s_i}[\frac{\partial L(X_{m_1}, Y); \tilde{\theta}_{m_1}}{\partial \tilde{\theta}_{m_1}}]$ \Comment{encoder $m_1$ }
\State $g_{m_2,i} = \mathbb{E}_{s_i}[\frac{\partial L(X_{m_2}, Y); \tilde{\theta}_{m_2}}{\partial \tilde{\theta}_{m_2}}]$ \Comment{encoder $m_2$ }
\State $g_{r,i} = \mathbb{E}_{s_i}[\frac{\partial L(X_{r}, Y); \tilde{\theta}_{r}}{\partial \tilde{\theta}_{r}}]$ \Comment{other }
\State Update $\tilde{\theta}_{m_1} \gets \tilde{\theta}_{m_1} - \alpha (\frac{L_{m_2}}{L_{m_1}+L_{m_2}})g_{m_1,i}$
\State Update $\tilde{\theta}_{m_2} \gets \tilde{\theta}_{m_2} - \alpha (\frac{L_{m_1}}{L_{m_1}+L_{m_2}})g_{m_2,i}$
\State Update $\tilde{\theta}_{r} \gets \tilde{\theta}_{r} - \alpha g_{r,i}$
\EndFor
\State Update $\theta \gets \theta + \epsilon(\tilde{\theta} - \theta)$
\State  Update  $L_{m_1}$, $L_{m_2}$
\EndWhile
\end{algorithmic}
\end{algorithm}

MM-IDGM first performs inner gradient updates on a cloned version of the original model, denoted as $\tilde{\theta}$, focusing on each modality encoder individually. The gradients for each modality encoder are dynamically adjusted based on previous unimodal losses, ensuring a more balanced learning rate across modalities. Finally, the weights of the original model, $\theta$, are updated by applying a weighted difference between the cloned model and the original one.

\subsection{Attention-Mixer Fusion}

Besides investigating cross-domain sampling strategies, inspired by~\cite{tolstikhin2021mlp}, we propose \textit{Attention-Mixer Fusion} to enhance the performance by fusing audio-visual modalities, where the attention mixer layer takes multimodal features as input to produce fused features. In particular, an attention mixer layer is composed of unimodal MLP layers, self-attention layers~\cite{vaswani2017attention}, and crossmodal MLP layers. First, for batch size $N_B$, the input features from different modalities are concatenated and projected to be a tensor $F_{g} \in \mathbb{R}^{N_B \times N_m \times D}$ by a Liner layer, followed by several attention mixer layers, where $N_m$ is the number of input modalities. Specifically, the unimodal MLP layer, the self-attention layer, and the crossmodal MLP layer can be respectively described as
\begin{equation}
\label{eq5}
    U^{*, *, i} = F_{g} ^{*, *, i} + \mathbf{W_2}\ \sigma(\mathbf{W_1}\ LN( F_{g} ^{*, *, i})),\ i = \left[1, D\right],
\end{equation}
\begin{equation}\small
\label{eq6}
  U  = \left [\left (softmax \left( U{\mathbf{W_3} (U\mathbf{W_4})^{T}}\over{\sqrt{D}} \right) U\mathbf{W_5}\right)_{h}\right] \mathbf{W_6},\ h = [1, H],
\end{equation}
\begin{equation}
\label{eq7}
  U^{*, j, *}  = U^{*, j, *} + \mathbf{W_8}\ \sigma(\mathbf{W_7}\ LN( U^{*, j, *})),\ j = \left[1, N_m\right],
\end{equation}
\noindent where $LN(\cdot)$ denotes the Layer Normalization, $\mathbf{W}_{1-8}$ are trainable weights, $H$ is the number of heads in multihead self-attention, and $*$ denotes all the entries in that dimension. Several attention mixer layers are stacked as a deep block, which is set as a hyperparameter in practice. We set it to 6 in our experiment. Finally, the output tensor $U \in \mathbb{R}^{N_B \times N_m \times D}$ is reduced to  $U \in \mathbb{R}^{N_B \times N_m \times 1}$ by obtaining the mean value on the feature dimension. In Eq.~\ref{eq5}, the unimodal MLP layer is conducted along the feature dimension to learn the dynamics in each unimodal feature. Eq.~\ref{eq6} shows the multi-head self-attention operation on the tensor $U$, which further explores the attention between the unimodal features. In Eq.~\ref{eq7}, the crossmodal MLP layer learns the dynamics across the modality dimension from the corresponding feature tokens. In contrast to the MLP-Mixer, which utilizes a simple two-layer MLP structure, our approach introduces a self-attention layer between the two MLP layers. This allows for more efficient global operation from MLP-Mixer and also adaptively weighing token interaction by self-attention. This architectural enhancement allows our model to more effectively capture both intra- and inter-modal interactions, making it especially well-suited for processing multimodal data such as audio and visual inputs.

\section{Experiments}
\label{sec:experiment}

In this part, extensive experiments are conducted to benchmark the cross-domain performances of public deception detection datasets.

\subsection{Databases and Metrics}
\label{sec:dataset}
\begin{table*}[t]
\centering
\caption{Statistics on benchmark datasets, Real-life Trial (R), Bag of Lies (B1), Box of Lies (B2), MU3D (M), DOLOS (D), and MDPE (E).}
\resizebox{0.95\textwidth}{!}{
\begin{tabular}{|c |c |c |c |c |c |
}
\hline
\textbf{Datasets} & \textbf{Avg Resolution} & \textbf{Subject Ethnicity} & \textbf{Demographics} & \textbf{Avg SNR (dB)} & \textbf{Std of SNR (dB)}   \\ \hline 
Real-life Trial (R)     & $\sim$360×640 & mix, unspecified & 58, gender bias & 29.30      &   5.13  \\ \hline
Bag of Lies (B1)        & 720×1280     & 100\% Indian   & 35, gendar bias & 22.63          &   9.31       \\ \hline
Box of Lies (B2)        & $\sim$480×640    & mix, unspecified &  26, (6M / 20F)& 25.59     &   3.25    \\ \hline
MU3D (M)                & 720×1280    & 50\% Black and 50\% White & 80, gender balanced & 37.07          &    4.01    \\ \hline
DOLOS (D)               & 720×1280   &  mix, unspecified &213, gendar bias &  30.72         &    4.97   \\ \hline
MDPE (E)                & 1920×1080   &   100\% Chinese  & 193, gendar bias &   37.17        &    4.28        \\ \hline
\end{tabular}}
\label{tab:stats}
\end{table*}

\textbf{Datasets.}\quad
We benchmarked the cross-domain generalization performance based on four publicly available datasets. We selected six publicly accessible, widely used multimodal deception datasets to construct our cross-domain benchmark. Each dataset brings unique characteristics that are beneficial for evaluating generalization.
 \begin{itemize}     
   \item \textbf{Real Life Trials~\cite{perez2015deception} (R)} is a popular real-world dataset collected from public court trials. Courtroom videos capturing genuine high-stakes deception. It consists of 121 videos including 61 deceptive and 60 truthful video clips. As it is a real-world dataset, the Real Life trial dataset has more noise on both the video and audio. We filtered out some corrupted videos and obtained 108 videos (54 truthful and 54 deceptive) with 58 subjects for our experiments. 
    \item \textbf{Bag of Lies~\cite{gupta2019bag} (B1) }is a multimodal dataset collected from well-controlled lab-based scenarios, where video, audio, EEG, and gaze data are collected. It has 35 subjects, 163 truthful and 162 deceptive video clips. The backgrounds for the videos are relatively clean and it is less noisy.
    \item \textbf{MU3D~\cite{lloyd2019miami} (M)} has 320 video clips and 80 subjects that cover different races and genders. It is also a lab-based dataset that uses the personal description paradigm to stimuli real-world cases. Each participant tells a positive truth, a positive lie, a negative truth, and a negative lie.
    \item \textbf{Box of Lies~\cite{soldner2019box} (B2)} is a deception dataset collected from an online gameshow, which has 25 videos and 26 participants (6 male and 20 female). The full video set contains 29 truthful and 36 deceptive rounds of games. However, the quality of the original Box of Lies dataset is not satisfactory. The visual (the face of the participant) and audio from many clips are not matching due to the frequent changes of viewpoints. To perform a fair comparison, we preprocessed and cleaned the Box of Lies dataset. After preprocessing, 101 video clips were extracted for testing.
   \item \textbf{DOLOS~\cite{guo2023audio} (D)} is the largest game show deception detection dataset, featuring rich deceptive conversations. It consists of 1,675 clips from 213 subjects, including 899 deceptive and 776 truthful samples, with fine-grained audio-visual annotations. The dataset is notable for its higher quality compared to existing alternatives.
   \item \textbf{MDPE~\cite{cai2024mdpe} (E)} contains over 104 hours of deception content with 193 subjects. It supports not only deception detection but also downstream tasks such as personality recognition, emotion analysis. It is a lab-based dataset with monetary incentives to encourage naturalistic deceptive behaviors. Each video has a 5-grade truthful to deceptive labels from the subjects. This enables analysis of subject-level variation in deception behavior.
 \end{itemize}
 
Taken together, these six datasets present diverse domains in terms of subject demographics, recording conditions, modalities, and data quality. This diversity is essential to rigorously evaluate cross-domain generalization for multimodal deception detection. Some of the typical samples from these datasets are shown in Fig.~\ref{fig:crossvisual} and more statistics are provided in Table~\ref{tab:stats}.

\vspace{0.3em}
\noindent\textbf{Evaluation Metrics.}\quad
In this work, we followed the widely adopted metric, binary classification accuracy (\%), for experimental evaluation. The deceptive clips were labeled as 1 and the truthful clips were labeled as 0. 
\begin{table*}[h]
\centering
\caption{The results of intra-domain testing accuracy (\%) on benchmark datasets, Real-life Trial (R), Bag of Lies (B1), MU3D (M), DOLOS (D), and MDPE (E). Abbrev.: V = visual, A = audio, AG = AU+gaze, AGA = AU+gaze+affect, AP = acoustic+prosodic, Mel = mel spectrogram, RN18 = ResNet18. Abbreviations are consistent across all tables.}
\resizebox{0.65\textwidth}{!}{
\begin{tabular}{|c |c |c |c |c |c | c|
}
\hline
\textbf{Input} & \textbf{Model} & \textbf{R to R} & \textbf{B1 to B1} & \textbf{M to M} & \textbf{D to D} & \textbf{E to E}    \\ \hline
V-Gaze        & MLP        &    59.96      &     53.45        &  52.03       & 54.95  &   51.89  \\ \hline
V-AGA     & MLP         & 62.87   & 55.84            &   55.23  &57.18 &     55.38  \\ \hline
V-LM & LSTM & 58.32 & 50.07 & 51.24& 52.03& 51.26 \\ \hline
V-Face        & RN18        & 70.10           & 57.78            &        55.15  & 65.23 &    56.57 \\ \hline
V-Face & RN18+KNN &65.28 &53.68 &  52.04 & 61.36 & 54.92 \\ \hline
V-Face & RN18+SVM & 66.52& 54.21 & 54.36 & 62.17 & 56.08 \\ \hline
V-Face & FFCSN &67.33 & 56.82 & 56.93 & 63.26 & \underline{57.80} \\ \hline
V-Face & DINOv2 & 63.64 & 58.42 & 57.50 & 65.52 & 57.37 \\ \hline
A-AP & MLP  &    66.82        &    50.49       &   55.31    & 58.96  & 50.47    \\ \hline
A-Mel     & RN18        & \textbf{75.78}           & 55.69          &     56.45  & 60.06 &   52.99 \\ \hline
A-Mel & RN18+KNN & 73.24 & 56.38 & 57.22 & 59.07 & 52.30 \\\hline
A-Mel & RN18+SVM & \underline{73.25} & 56.88 & 58.04 & 60.11 & 52.69 \\\hline
& concat       &     59.09     &  54.46          &   56.30 & 70.40 & 55.78  \\
V-Face& KNN        &  60.22    &    54.83      &  \textbf{70.21}    & 69.50 & 56.22  \\
+ V-AGA & SVM       &  61.03     &   55.21       &  \underline{69.96}    & 70.05 & 55.39  \\
 &Atten-Mixer  & 63.05 & 55.45 & 57.59 & 71.55 &  56.18\\
\hline
& concat        &  68.75     &  54.74        &    56.19  & 66.95 &53.39  \\
V-Face& KNN        &  67.33    &     55.32     &   56.62   & 65.98  & 53.16 \\
+ A-Mel& SVM        &  68.45     &  56.08        & 56.36     & 67.45 & 54.03 \\
 & Atten-Mixer& 71.00 & 56.44 & 57.14 & 68.10&54.58 \\
\hline

& concat       & 68.88   &  56.25       &    53.27 & 68.97& 51.39 \\
V-AGA& KNN        &  67.92    &   54.23       &  50.32    & 65.23 &  50.46 \\
+ A-Mel& SVM        &   68.03    &   54.50       &  51.06    & 64.92 & 51.35 \\
 &Atten-Mixer & 69.30& 56.78 & 55.23 & 70.98 & 52.99\\
\hline

&concat       & 69.33         & \underline{58.43}           &  55.24 & \underline{72.70} & 54.98 \\
V-Face& KNN        &  68.55    &   54.25       &  52.28    &  68.48 &  53.19\\
+ V-AGA& SVM        &   68.22    &    54.39      &  52.47    & 68.65 & 54.24  \\
 + A-Mel& Atten-Mixer       & 70.52 & \textbf{59.41} & 56.88  & \textbf{73.85} & \textbf{58.17} \\
\hline
\end{tabular}}
\label{tab:intra}
\end{table*}

\subsection{Implementation Details}
\label{sec:Details}

\textbf{Feature Extraction.}\quad
Several widely-adopted audio and visual features were extracted by different tools. For visual features, OpenFace~\cite{amos2016openface} was used to extract 35-dimensional AUs and 8-dimensional gaze features. Face frames were extracted and aligned by MTCNN~\cite{zhang2016joint}, where we uniformly sampled 64 face frames for each video clip. Affect features were extracted by Emonet~\cite{toisoul2021estimation}, where the feature included 5-class emotions, arousal, and valence. For audio features, Mel Spectrograms and acoustic features were extracted by the OpenSmile toolkit~\cite{eyben2010opensmile}. Raw audio waveforms were also used in our experiments. For compact presentation in Tables~\ref{tab:intra}--\ref{tab:mm_idgm}, abbreviated input labels are used: V = visual, A = audio, LM = facial landmark, AG = AU+gaze, AGA = AU+gaze+affect, Face = face frames, AP = acoustic+prosodic, Mel = mel spectrogram, and Wav = waveform.

\vspace{0.3em}
\noindent\textbf{Protocols.}\quad
Inspired by~\cite{wang2022domain,varanka2023data}, we treated each dataset as a domain. To evaluate the models' cross-domain generalization capacity and alleviate domain information leakage, all the preprocessed data including original training and test data from each dataset was used for either training or testing. Note that the Box of Lies dataset was only used for testing, as many samples were filtered out due to their unsatisfactory quality. Because of shifting camera angles and background audience noise, the speaker’s face and speech may not be captured clearly in the video. The experiments were conducted on the \textit{single-to-single domain} (e.g., R to B1 stands for training on Real-life Trial (R) and testing on Bag of Lies (B1)) and \textit{multi-to-single domain} (e.g., R\&M to B2 stands for training on Real-life Trial (R) and MU3D (M) and testing on Box of Lies (B2)).

 \vspace{0.3em}
\noindent\textbf{Model Selection.}\quad
Different audio and visual features are used for benchmarking, including gaze, AUs, affect, facial landmark (2D, 68 points, using dlib package), acoustic and prosodic features (88-dimensional features extracted from the OpenSmile toolkit), Mel spectrograms, and audio waveforms. We also benchmark and re-implement the methods that are widely adopted by domain-specific deception detection task~\cite{king2024applications}, such as ResNet18, MLP, SVM (Linear)~\cite{karimi2018toward,mathur2020introducing}, KNN~\cite{chebbi2023deception}, FFCSN~\cite{ding2019face}, LSTM~\cite{karimi2018toward}, DINOv2~\cite{Oquab2024DINOv2}, etc. 

Models for audio and visual modalities were selected to fit the data volume. For face frames, we adopted ResNet18~\cite{He2015Deep}, Gate Recurrent Unit (GRU)~\cite{chung2014empirical}, SVM~\cite{karimi2018toward,mathur2020introducing}, and KNN~\cite{chebbi2023deception} models for facial feature extraction and temporal modeling, respectively.
We also use DINOv2 as a pretrained face frames encoder. Specifically, we adopt the ``DINOv2-Vit/S-14" backbone and add a linear classifier for binary prediction. Each face frame isencoded independently, and clip-level prediction is obtained by averaging the frame-level logits. To preserve DINOv2's general visual features while adapting to our task with limited data, we fine-tune only the last transformer block, final normalization layer, and binary classifier (Linear layer), while freezing the rest of the backbone. Two-layer multilayer perception (MLP)~\cite{noriega2005multilayer} or LSTM~\cite{karimi2018toward} models were used for AUs, gaze, and affect feature representation. Specifically, the first MLP Layer initially operates on the dimension that represents the number
of features,  with sizes following the sequence: ``input dimension $\rightarrow$ 64 $\rightarrow$ 128 $\rightarrow$ 1. Afterward, it reshapes the tensor, and the MLP Layer processes on the feature dimension,  where the sizes progress as ``64 $\rightarrow$ 32 $\rightarrow$ 16 $\rightarrow$ 2. The MLP networks have identical structures, except for their first linear layers, as the input dimensions differ. For acoustic features, MLP is used. For the audio-based Mel spectrogram, we used the ResNet18~\cite{He2015Deep}, KNN, and SVM models for time-frequency feature representation and classification. For audio waveforms, the Wave2Vec~\cite{baevski2020wav2vec} model was applied for audio feature extraction. For fair comparison in the fusion experiments, we use the same models for unimodal feature extraction, where the MLP is used for AUs, gaze, and affect features, and ResNet18 is used for Mel spectrograms, and RestNet18+GRU is used for face frames. The extracted unimodal features are then fed into the attention mixer for fusion and classification.

\vspace{0.3em}
\noindent\textbf{Experimental Setting.}\quad Our proposed method was implemented with Pytorch. The ImageNet pretrained models (e.g., ResNet18) for classification were trained on the benchmark datasets using SGD optimizer with the initial learning rate (lr), momentum, and weight decay (wd) were 1e-3, 0.9, and 5e-5, respectively. We trained models with a maximum of 30 epochs and batchsize 32 on a single Nvidia V100 GPU. As for the fusion models (e.g., Atten-Mixer on face frames and Mel Spectrogram), Adam optimizer with initial lr=1e-3 and wd=5e-5 was used. The models were trained with batchsize 16 for a maximum of 30 epochs.

\subsection{Cross-domain Testing with Unimodal Features}
\label{sec:single}

In this subsection, we present the benchmark results of cross-domain testing by investigating unimodal features to evaluate their generalization capacities. 

\begin{table*}[h]
\centering
\caption{The results of single-to-single cross-domain generalization accuracy (\%) on benchmark datasets, Real-life Trial (R), Bag of Lies (B1), Box of Lies (B2), MU3D (M), DOLOS (D), and MDPE (E). Abbrev.: V-Aff= visual (affect), A-Wav = audio (waveform). Abbreviations are consistent across all tables.}
\resizebox{0.98\textwidth}{!}{
\begin{tabular}{|
>{\columncolor[HTML]{FFFFFF}}c |
>{\columncolor[HTML]{FFFFFF}}c |
>{\columncolor[HTML]{FFFFFF}}c |
>{\columncolor[HTML]{FFFFFF}}c |
>{\columncolor[HTML]{FFFFFF}}c |
>{\columncolor[HTML]{FFFFFF}}c |
>{\columncolor[HTML]{FFFFFF}}c |
>{\columncolor[HTML]{FFFFFF}}c |
>{\columncolor[HTML]{FFFFFF}}c |
>{\columncolor[HTML]{FFFFFF}}c |
>{\columncolor[HTML]{FFFFFF}}c |
>{\columncolor[HTML]{FFFFFF}}c |
>{\columncolor[HTML]{FFFFFF}}c |
>{\columncolor[HTML]{FFFFFF}}c |
>{\columncolor[HTML]{FFFFFF}}c |}
\hline
\textbf{Input} & \textbf{Model} & \textbf{R to B1} & \textbf{R to B2} & \textbf{R to M} & \textbf{B1 to R} & \textbf{B1 to B2} & \textbf{B1 to M} & \textbf{M to R} & \textbf{M to B1} & \textbf{M to B2} & \textbf{D to R}  & \textbf{E to R} & \textbf{E to B1} & \textbf{E to B2}  \\ \hline
V-AU                 & LSTM            & 48.11            & -                & -               & \textbf{61.21}   & -                 & -                & -               & -                & -                & -            & - & - & -  \\ \hline
V-Gaze & MLP &47.22 & 59.98 & 54.15 & 57.23 & 55.42 & 49.06 & 45.38 & 50.35 & 52.22 & 53.08 & 50.16 & 51.45 & 53.07 \\\hline
V-AG            & MLP         & 50.77            & \textbf{65.35}   & \textbf{56.87}  & \underline{58.88}   & 58.42             & 50.94            & 46.73           & 51.69            & 53.47            & 54.79      & 51.29 & 52.33 & 54.68    \\ \hline
V-Aff             & MLP         & 50.46            & 58.42            & 50.31           & 50.47            & 51.49             & \underline{52.19}            & \textbf{66.36}           & 51.08            & \textbf{60.40}   & 54.58         & 51.36 & 52.07 & 54.88 \\ \hline
V-AGA     & MLP         & \underline{54.46}   & 59.41            & \underline{54.37}           & 50.47            & 57.43             & \textbf{54.69}   & 60.75           & 51.69            & 55.45            & 55.41 & 51.89 & 52.77 & 55.45 \\ \hline
V-Face        & RN18        & 52.00            & 61.39            & 51.25           & 50.93            & \underline{57.43}             & 50.62            & 57.94           & 51.69            & 57.43            & 54.52          & 51.02 & 53.18 & 55.92 \\ \hline
V-Face        & RN18+GRU    & 53.54            & \underline{63.37}            & 52.81           & 57.41            & 59.41             & 51.56            & 46.73           & 52.92            & 55.45            & 54.80        & 52.03 & 55.06 & 56.32 \\ \hline
V-Face & RN18+KNN &53.13& 63.15 & 51.72 & 56.38 & 59.25 & 51.06 & 47.02 & 53.22 & 56.11 & 55.36 & 51.96 & \textbf{55.88} & 56.15\\\hline
V-Face & RN18+SVM &53.66 & 63.28 & 52.13 & 56.04 & 60.24 & 52.15 & 47.88 & \underline{53.26} & 55.67 & \underline{56.36} & 52.14 & \underline{55.39} & 56.17 \\\hline
V-Face & FFCSN & 54.32 & 62.88 & 51.06 & 59.36 & 53.11 & 48.25 & 52.65 & 52.17 & 54.36 & \textbf{56.85} & \textbf{53.36} & 54.88 & \underline{56.37} \\\hline
V-Face & DINOv2 & \textbf{55.08} & 62.33  & 53.00 & 55.27 & 58.42 & 51.56 & 53.40 & 52.31 & \underline{59.40} & 54.21 & 52.00 & 53.69 & \textbf{56.44}\\\hline
A-AP & MLP &43.06 & 52.18 & 50.36 & 49.44 & 59.56 & 51.46 & 50.17 & 50.45 & 52.04  & 50.38 & 46.98 & 49.77 & 50.03 \\\hline
A-Mel     & RN18        & 46.77            & 53.47            & 52.19           & 50.47            & \textbf{66.34}    & 50.62            & 54.21           & 51.38            & 55.45            & 53.43         & 50.56 & 52.04 & 53.68 \\ \hline
A-Mel & RN18+KNN &46.88 & 54.02 & 51.39 & 50.69 & \underline{62.77} & 49.36 & 51.09 & 51.53 & 52.72 & 53.17 & 50.48 & 51.72 & 53.23 \\\hline
A-Mel & RN18+SVM & 49.98 & 54.02 & 51.36 & 50.45 & 62.17 & 50.33 & 52.04 & 52.48 & 53.66 & 54.25  &51.26 & 52.33 & 54.05\\\hline
A-Wav            & Wave2Vec        & 51.08            & 48.51            & 50.94           & 46.73            & 58.42             & 50.00            & \underline{63.55}  & \textbf{56.31}   & 56.44            & 53.55         & \underline{52.77} & 53.49 & 55.68 \\ \hline
\end{tabular}}

\vspace{1em}

\resizebox{0.98\textwidth}{!}{
\begin{tabular}{|
>{\columncolor[HTML]{FFFFFF}}c |
>{\columncolor[HTML]{FFFFFF}}c |
>{\columncolor[HTML]{FFFFFF}}c |
>{\columncolor[HTML]{FFFFFF}}c |
>{\columncolor[HTML]{FFFFFF}}c |
>{\columncolor[HTML]{FFFFFF}}c |
>{\columncolor[HTML]{FFFFFF}}c |
>{\columncolor[HTML]{FFFFFF}}c |
>{\columncolor[HTML]{FFFFFF}}c |
>{\columncolor[HTML]{FFFFFF}}c |
>{\columncolor[HTML]{FFFFFF}}c |
>{\columncolor[HTML]{FFFFFF}}c |
>{\columncolor[HTML]{FFFFFF}}c |
>{\columncolor[HTML]{FFFFFF}}c |
>{\columncolor[HTML]{FFFFFF}}c |}
\hline
\textbf{Input} & \textbf{Model} & \textbf{R to D} & \textbf{R to E} & \textbf{B1 to D} & \textbf{B1 to E} & \textbf{M to D} & \textbf{M to E} & \textbf{D to B1} & \textbf{D to B2} & \textbf{D to M} & \textbf{D to E}  & \textbf{E to M} & \textbf{E to D} & \textbf{Avg}  \\ \hline
V-AU                 & LSTM            & 49.92            & -                & -               & -   & -                 & -                & -               & -                & -                & -             & - & - & 53.08 \\ \hline
V-Gaze & MLP & 49.98 & 57.46 & 50.21 & 50.55 & 52.02 & 49.87 & \textbf{55.36} & 50.11 & 53.29 & 52.41 & 49.32 & 50.21 & 51.98\\\hline
V-AG            & MLP         & 51.25            & 52.33   & 51.64  & 53.33   & 52.47             & 50.66            & 50.71           & 51.37            & 53.68            & \textbf{55.22 }        & 50.04 & 51.42 & 53.21 \\ \hline
V-Aff             & MLP         & 51.32            & 53.35            & 52.56           & 53.47            & 50.98             & 51.42           & 51.34           & 51.08            & 54.37   & 54.56          & 51.21 & 51.49 & 53.25\\ \hline
V-AGA     & MLP         & 52.38   & 54.24            & 53.19           & 53.98           & 51.32            & 52.26   & 51.20          & 50.25           &  55.08           & 55.02 & 51.26 & 51.49 & 53.84\\ \hline
V-Face        & RN18        & 52.75            & 51.28            & 52.18           & 50.65           & 52.32             & 52.36            & 50.89           & 51.33            &55.42            & 54.62        & 51.25 & 52.08 &  53.30\\ \hline
V-Face        & RN18+GRU    & 52.54            & 52.36            & \textbf{53.88}           & 50.87            & 52.68            & \textbf{53.94}            & 51.34         & 52.36       & \underline{55.69}            &     \textbf{55.22}     & 52.36 & 51.44 &53.84 \\ \hline
V-Face & RN18+KNN & 53.37 & \textbf{62.32} & 51.26 & 56.37 & 59.04 & 50.77 & 48.21 & 51.48 & 54.39 & \underline{55.08} & 51.49 & 52.33 & 54.26\\\hline
V-Face & RN18+SVM &\textbf{54.15} & 61.28 & 50.36 & 55.49 & 59.68 & 50.28 & 49.47 & 50.28 & 53.66 & 54.88 & \underline{52.74} & 54.29 & 54.44 \\\hline
V-Face & FFCSN & \underline{53.87} & \underline{61.46} & 51.42 & \underline{56.23} & 60.13 & 50.35 &50.31 & 53.19 & 53.62 & 54.07 & \textbf{53.28} & \underline{55.03} & \underline{54.50} \\\hline
V-Face & DINOv2 & 53.41 & 52.16 & \underline{53.22} & 54.05 & 58.99 & 51.16 & 54.85 & 52.48 & 54.56 & 54.57 & 51.25 & \textbf{55.61} & \textbf{54.54} \\\hline
A-AP & MLP &45.58 & 51.04 & 51.26 & 50.33 & \textbf{60.82} & 49.92 & 51.83 &50.22 & 53.16 & 50.59& 49.25 & 50.21 & 50.80\\\hline
A-Mel     & RN18        & 50.01            & 52.11          & 50.59          & 50.48           & \underline{60.37}    & 51.29          & 53.08           & 53.86           & 54.36          &  53.68       & 50.26 & 51.42 & 52.88\\ \hline
A-Mel & RN18+KNN &48.82 & 49.98 & 50.21 & 50.28 & 60.08 & 50.49 & 54.11 & 52.38 & 54.59 & 54.32 & 50.38 & 51.27 & 52.24 \\\hline
A-Mel & RN18+SVM & 49.05 & 50.31 & 49.89 & \textbf{60.49} & 56.75 & \underline{52.98} & \underline{55.32} & \textbf{53.98} & \textbf{55.84} & 54.28 & 51.26 & 52.38 & 53.24\\\hline
A-Wav            & Wave2Vec        & 50.21            & 49.51            & 49.94           & 50.43          &     55.83        & 50.82      & 54.26  & \underline{53.65}   & 54.38           & 53.49         & 51.21 & 50.65 & 52.87 \\ \hline
\end{tabular}}

\label{tab:single2single}
\end{table*}

\noindent\textbf{Intra-Domain Testing.}\quad
Before performing cross-domain testing, we show the results of intra-domain testing on Real-life Trial (R), Bag of Lies (B1), MU3D (M), DOLOS (D), and MDPE (E) in Table~\ref{tab:intra}. It is important to note that we did not use the Box of Lies (B2) dataset for training due to concerns about its quality. Specifically, we performed a subject-independent five-fold cross-testing and reported the average accuracies. Among unimodal inputs, mel spectrograms with ResNet18 achieve the highest accuracy (75.78\% on R), highlighting the strength of audio cues, particularly in structured settings like courtroom videos. Visual face frames also perform well with higher average results compared with audio features. However, facial landmarks are less reliable and show consistently low accuracy, which may due to the sparse geometric information and its sensitivity to pose, occlusion, and lighting; thus, we exclude them from the following cross-domain tasks. Multimodal fusion consistently improves performance, with the proposed Atten-Mixer achieving the best results across several datasets (e.g., 73.85\% on D, 58.17\% on E). However, the improvement from fusion is not uniform across all datasets. For example, on Real-life Trial, the best unimodal audio model still outperforms the multimodal models, suggesting that audio cues may already be highly informative in well-structured recording environments. In contrast, the larger gains on DOLOS and MDPE indicate that multimodal fusion is more beneficial when single-modality signals are weaker or more noisy. Performance variation across datasets suggests that data quality, recording conditions, and content type significantly impact model effectiveness.

\vspace{0.3em}
\noindent\textbf{Single-to-Single Domain.}\quad 
Specifically, the models were trained on one dataset (one domain) and tested on the other dataset (another domain). As shown in Tabel~\ref{tab:single2single}, overall, visual-based models exhibit strong generalization compared to audio-only models, particularly those leveraging rich visual descriptors (AU, gaze, affect) or deep networks like RN18+KNN and FFCSN. However, the absolute cross-domain accuracies remain relatively modest, indicating a clear domain gap among different deception datasets. Notably, DINOv2 achieves the highest average accuracy (54.54\%), slightly outperforming FFCSN (54.50\%). Since only the last transformer block of DINOv2 is fine-tuned while most pretrained parameters remain frozen, it suggests that pretrained foundation visual representations provide robust and transferable facial features for cross-domain deception detection while FFCSN shows the effectiveness of visual temporal modeling. The small performance gap among these top-performing models suggests that the choice of robust visual representation is more important than the specific classifier design in this single-to-single transfer setting. Although audio contributes valuable cues, it struggles more with domain shifts compared to visual modalities. Additionally, handcrafted prosodic/acoustic features perform worse overall (e.g., MLP: 50.80\% avg), compared with other audio features. We also observe that cross-domain performance is highly asymmetric. Training on one dataset and testing on another does not necessarily produce the same behavior in the reverse direction. This suggests that some datasets provide more transferable behavioral patterns, while others may contain domain-specific biases caused by recording conditions, protocols, etc.
\begin{table*}[h]
\centering
\caption{The results of multi-to-single cross-domain generalization accuracy (\%) on benchmark datasets, Real-life Trial (R), Bag of Lies (B1), Box of Lies (B2), and MU3D (M), for different generalization strategies.}
\resizebox{0.98\textwidth}{!}{
\begin{tabular}{ccccccccccccc}
\hline
\rowcolor[HTML]{FFFFFF}
\multicolumn{13}{|c|}{\cellcolor[HTML]{EFEFEF}}  \\ 
\multicolumn{13}{|c|}{\cellcolor[HTML]{EFEFEF}\large\textbf{Domain-Simultaneous}}   \\ \hline
\multicolumn{1}{|c|}{\cellcolor[HTML]{FFFFFF}\textbf{Input}} & \multicolumn{1}{c|}{\cellcolor[HTML]{FFFFFF}\textbf{Model}} & \multicolumn{1}{c|}{\cellcolor[HTML]{FFFFFF}\textbf{R\&M to B1}} & \multicolumn{1}{c|}{\cellcolor[HTML]{FFFFFF}\textbf{R\&M to B2}} & \multicolumn{1}{c|}{\cellcolor[HTML]{FFFFFF}\textbf{R\&B1 to B2}} & \multicolumn{1}{c|}{\cellcolor[HTML]{FFFFFF}\textbf{R\&B1 to M}} & \multicolumn{1}{c|}{\cellcolor[HTML]{FFFFFF}\textbf{B1\&M to R}} & \multicolumn{1}{c|}{\cellcolor[HTML]{FFFFFF}\textbf{B1\&M to B2}} & \multicolumn{1}{c|}{\cellcolor[HTML]{FFFFFF}\textbf{\shortstack{R\&B1\&M\\to B2}}} & \multicolumn{1}{c|}{\cellcolor[HTML]{FFFFFF}\textbf{R\&M to D}}  & \multicolumn{1}{c|}{\cellcolor[HTML]{FFFFFF}\textbf{R\&M to E}} & \multicolumn{1}{c|}{\cellcolor[HTML]{FFFFFF}\textbf{R\&B1 to D}} & \multicolumn{1}{c|}{\cellcolor[HTML]{FFFFFF}\textbf{R\&B1 to E}} \\ \hline
\rowcolor[HTML]{FFFFFF} 
\multicolumn{1}{|c|}{\cellcolor[HTML]{FFFFFF}V-Face}        & \multicolumn{1}{c|}{\cellcolor[HTML]{FFFFFF}ResNet18}        & \multicolumn{1}{c|}{\cellcolor[HTML]{FFFFFF}\underline{53.85}}      & \multicolumn{1}{c|}{\cellcolor[HTML]{FFFFFF}49.50}               & \multicolumn{1}{c|}{\cellcolor[HTML]{FFFFFF}49.50}                & \multicolumn{1}{c|}{\cellcolor[HTML]{FFFFFF}50.94}               & \multicolumn{1}{c|}{\cellcolor[HTML]{FFFFFF}44.86}               & \multicolumn{1}{c|}{\cellcolor[HTML]{FFFFFF}60.40}                & \multicolumn{1}{c|}{\cellcolor[HTML]{FFFFFF}44.55}                & \multicolumn{1}{c|}{\cellcolor[HTML]{FFFFFF}52.43}  
  & \multicolumn{1}{c|}{\cellcolor[HTML]{FFFFFF}50.88}
  & \multicolumn{1}{c|}{\cellcolor[HTML]{FFFFFF}\textbf{54.82}}
  & \multicolumn{1}{c|}{\cellcolor[HTML]{FFFFFF}49.92}\\
\end{tabular}}
\label{tab:multi2single}
\end{table*}

\noindent\textbf{Multi-to-Single Domain.}\quad Table~\ref{tab:multi2single} reports the multi-to-single domain generalization results under three sampling strategies: Domain-Simultaneous, Domain-Alternating, and Domain-by-Domain. For each strategy, we evaluate 21 domain generalization tasks using the same set of input features and model configurations. The final column of each strategy block reports the average accuracy across all 21 tasks, providing an overall measure of each feature-model configuration under a given sampling strategy. Although these 21 tasks do not cover all possible combinations among the six datasets, they include a broad range of representative transfer scenarios and are sufficiently comprehensive for evaluation. Overall, the results show that sampling strategy, feature selection, and dataset composition all have a substantial impact on cross-domain deception detection performance. We summarize the main observations below.

(1) Among the three strategies, Domain-by-Domain achieves the best overall average accuracy (average across all the results) of 53.66\%, followed by Domain-Alternating (53.06\%) and Domain-Simultaneous (52.30\%). This suggests that training on individual domains sequentially allows models to better abstract transferable patterns, likely due to reduced complexities between domains during training. In contrast, domain-alternating and domain-simultaneous keep switching between source domains, appearing to induce higher variability and lower overall accuracies. 

(2) Across all strategies, visual-based models generally outperform audio-based models. This trend is reflected in the average results of each strategy block. Under Domain-Simultaneous training, the best average performance is achieved by visual behavioral features, i.e., AU+gaze+affect, with an MLP classifier (54.60\%). Under Domain-Alternating training, the best average performance is obtained by face-frame features with RN18+SVM (54.18\%). Under Domain-by-Domain training, DINOv2 achieves the highest average accuracy (55.53\%), suggesting that pretrained visual foundation representations are especially effective when domains are introduced sequentially.

(3) Regarding feature types, visual and behavioral features are generally more robust than audio features, but their effectiveness varies across sampling strategies. For Domain-Simultaneous training, behavioral cues such as AUs, gaze, and affect are highly competitive, likely because their compact and interpretable representations are easier to learn from mixed-domain data. For Domain-Alternating training, deep face-frame representations perform better, suggesting that frame-level visual features can benefit from repeated exposure to different source domains. For Domain-by-Domain training, face features also perform better, especially using DINOv2. This result suggests that preserving general visual knowledge while lightly adapting to deception-related facial cues can improve cross-domain robustness.

(4) Regarding dataset effects, one observation is that training on clean, lab-collected datasets (such as Bag of Lies and MU3D) tends to generalize better to real-world or noisier datasets (like Real-life Trial or Box of Lies). For example, under domain-simultaneous sampling, training on B1\&M and testing on R yields the best result of 69.16\%. This performance is consistent with the domain-by-domain strategy, where B1\&M to R reaches 67.29\%. Conversely, when training noisy or lower-quality datasets (e.g., R, E), performance often degrades, suggesting that domain complexity and data quality significantly affect generalization.

\begin{figure}[t]
\centering
\includegraphics[width=0.9\linewidth]{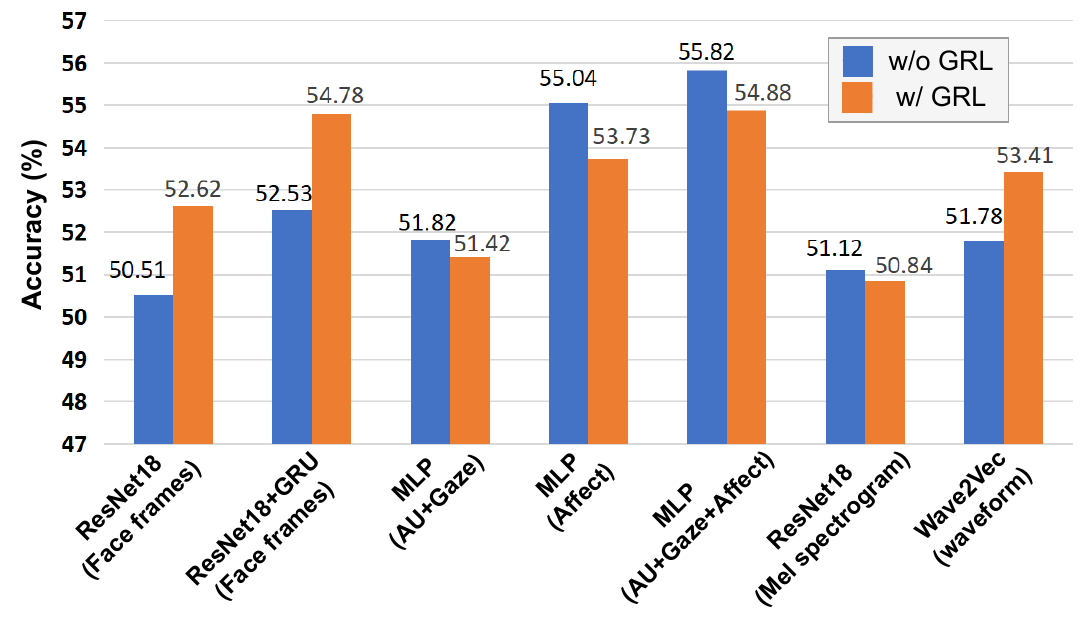}
\vspace{-1.5em}
  \caption{Performance comparisons of Domain-simultaneous training w/ and w/o Gradient Reversal Layer (GRL).
  }
\label{fig:GRL}
\vspace{-1em}
\end{figure}

\begin{table*}[!t]
\centering
\caption{\label{tab:single2singlefusion}Fusion results of single-to-single cross-domain generalization accuracy (\%).}
\resizebox{1.0\textwidth}{!}{
\begin{tabular}{|c|
>{\columncolor[HTML]{FFFFFF}}c |
>{\columncolor[HTML]{FFFFFF}}c |
>{\columncolor[HTML]{FFFFFF}}c |
>{\columncolor[HTML]{FFFFFF}}c |
>{\columncolor[HTML]{FFFFFF}}c |
>{\columncolor[HTML]{FFFFFF}}c |
>{\columncolor[HTML]{FFFFFF}}c |
>{\columncolor[HTML]{FFFFFF}}c |
>{\columncolor[HTML]{FFFFFF}}c |
>{\columncolor[HTML]{FFFFFF}}c |
>{\columncolor[HTML]{FFFFFF}}c |
>{\columncolor[HTML]{FFFFFF}}c |
}
\hline
\cellcolor[HTML]{FFFFFF}{\color[HTML]{000000} \textbf{Input}}   & \cellcolor[HTML]{FFFFFF}{\color[HTML]{000000} \textbf{Type}}   & \cellcolor[HTML]{FFFFFF}{\color[HTML]{000000} \textbf{Method}}   & {\color[HTML]{000000} \textbf{R to B1}}  & {\color[HTML]{000000} \textbf{R to B2}} & {\color[HTML]{000000} \textbf{R to M}}   & {\color[HTML]{000000} \textbf{R to D}}  & {\color[HTML]{000000} \textbf{R to E}}   & {\color[HTML]{000000} \textbf{B1 to R}}    & {\color[HTML]{000000} \textbf{B1 to B2}}  & {\color[HTML]{000000} \textbf{B1 to M}}              & {\color[HTML]{000000} \textbf{B1 to D}}     & {\color[HTML]{000000} \textbf{B1 to E}}   \\ \hline 
\cellcolor[HTML]{ECF4FF}{\color[HTML]{000000} }  & \cellcolor[HTML]{FFFFFF}{\color[HTML]{000000} Score}  & \cellcolor[HTML]{FFFFFF}{\color[HTML]{000000} Avg}  
& {\color[HTML]{000000} 53.23}   & {\color[HTML]{000000} 41.58}   & {\color[HTML]{000000} 51.88}   & {\color[HTML]{000000} 53.85}& {\color[HTML]{000000} 50.16}    & {\color[HTML]{000000}\textbf{64.49}}    & {\color[HTML]{000000} \textbf{62.38}}    & {\color[HTML]{000000} 50.62}    &{\color[HTML]{000000} 52.29} & {\color[HTML]{000000} 50.97}\\ \cline{2-13} 
\cellcolor[HTML]{ECF4FF}{\color[HTML]{000000} }  & \cellcolor[HTML]{FFFFFF}{\color[HTML]{000000} }   & \cellcolor[HTML]{FFFFFF}{\color[HTML]{000000} Concat} & {\color[HTML]{000000} 51.08}  & {\color[HTML]{000000} 55.45}  & {\color[HTML]{000000} 51.88}  & {\color[HTML]{000000} 52.90} & {\color[HTML]{000000} 50.49} & {\color[HTML]{000000} 54.21}   & {\color[HTML]{000000} 58.42}    & {\color[HTML]{000000} 51.25}  & {\color[HTML]{000000} 51.94}&{\color[HTML]{000000}50.08} \\ \cline{3-13} 
\cellcolor[HTML]{ECF4FF}{\color[HTML]{000000} }   & \cellcolor[HTML]{FFFFFF}{\color[HTML]{000000} } & \cellcolor[HTML]{FFFFFF}{\color[HTML]{000000} SE-Concat}  & {\color[HTML]{000000} 53.85}  & {\color[HTML]{000000} \underline{60.40}}  & {\color[HTML]{000000} 51.25}  & {\color[HTML]{000000} 53.07} & {\color[HTML]{000000} 51.29} & {\color[HTML]{000000} 55.14}  & {\color[HTML]{000000} 58.42} & {\color[HTML]{000000} 50.62}   &{\color[HTML]{000000} 55.32} &{\color[HTML]{000000}50.81} \\ \cline{3-13} 
\cellcolor[HTML]{ECF4FF}{\color[HTML]{000000} } & \cellcolor[HTML]{FFFFFF}{\color[HTML]{000000} }  & \cellcolor[HTML]{FFFFFF}{\color[HTML]{000000} Cross-Atten}  & {\color[HTML]{000000} 55.38}  & {\color[HTML]{000000} \textbf{61.39}}  & {\color[HTML]{000000} 52.19}   & {\color[HTML]{000000} 54.11}& {\color[HTML]{000000} \underline{52.35}}  & {\color[HTML]{000000} 51.40} & {\color[HTML]{000000} \underline{60.40}}   & {\color[HTML]{000000} 51.25} & {\color[HTML]{000000} 56.09} & {\color[HTML]{000000}51.05} \\ \cline{3-13} 
\cellcolor[HTML]{ECF4FF}{\color[HTML]{000000} }& \cellcolor[HTML]{FFFFFF}{\color[HTML]{000000} } & \cellcolor[HTML]{FFFFFF}{\color[HTML]{000000} MLP-Mixer} & {\color[HTML]{000000} 55.08} & {\color[HTML]{000000} 48.51} & {\color[HTML]{000000} 53.44}   & {\color[HTML]{000000} 54.02}& {\color[HTML]{000000} 52.18} & {\color[HTML]{000000} 56.07} & {\color[HTML]{000000} 58.42} & {\color[HTML]{000000} \underline{53.75}} & {\color[HTML]{000000}  57.22} &{\color[HTML]{000000} 51.29}\\ \cline{3-13} 
\cellcolor[HTML]{ECF4FF}{\color[HTML]{000000} }& \cellcolor[HTML]{FFFFFF}{\color[HTML]{000000} } & \cellcolor[HTML]{FFFFFF}{\color[HTML]{000000} CLIP-Align} & {\color[HTML]{000000}51.69} & {\color[HTML]{000000} 56.62} & {\color[HTML]{000000} 51.56}   & {\color[HTML]{000000} 54.39}& {\color[HTML]{000000} 51.23} & {\color[HTML]{000000} 57.01} & {\color[HTML]{000000} 57.44} & {\color[HTML]{000000} 52.50} & {\color[HTML]{000000}  51.51} &{\color[HTML]{000000} 52.21}\\ \cline{3-13} 
\multirow{-6}{*}{\cellcolor[HTML]{ECF4FF}{\color[HTML]{000000} \begin{tabular}[c]{@{}c@{}}V-Face\\ +\\ V-AGA\end{tabular}}} & \multirow{-5}{*}{\cellcolor[HTML]{FFFFFF}{\color[HTML]{000000} Feature}} & \cellcolor[HTML]{FFFFFF}{\color[HTML]{000000} \textbf{Atten-Mixer}} & {\color[HTML]{000000} \underline{56.92}}  & {\color[HTML]{000000} 59.41}   & {\color[HTML]{000000} \textbf{57.94}}    & {\color[HTML]{000000} \underline{55.32}}& {\color[HTML]{000000}\textbf{53.22}} & {\color[HTML]{000000} 63.37}  & {\color[HTML]{000000} 53.75}  & {\color[HTML]{000000} \underline{53.75}}       & {\color[HTML]{000000} 57.39} & {\color[HTML]{000000} 51.38} \\ \hline 
\cellcolor[HTML]{FFFFEB}{\color[HTML]{000000} }   & {\color[HTML]{000000} Score}   & {\color[HTML]{000000} Avg}  & \cellcolor[HTML]{FFFFFF}{\color[HTML]{000000} 53.23}   & \cellcolor[HTML]{FFFFFF}{\color[HTML]{000000} 49.50} & \cellcolor[HTML]{FFFFFF}{\color[HTML]{000000} 51.88}   &{\color[HTML]{000000} 53.69} &    {\color[HTML]{000000} 50.81}   & \cellcolor[HTML]{FFFFFF}{\color[HTML]{000000} 50.47} & \cellcolor[HTML]{FFFFFF}{\color[HTML]{000000} 59.41} & \cellcolor[HTML]{FFFFFF}{\color[HTML]{000000} \underline{53.75}} & {\color[HTML]{000000} 52.90} & {\color[HTML]{000000} 50.49}   \\ \cline{2-13} 
\cellcolor[HTML]{FFFFEB}{\color[HTML]{000000} }  & \cellcolor[HTML]{FFFFFF}{\color[HTML]{000000} }   & {\color[HTML]{000000} Concat}               & \cellcolor[HTML]{FFFFFF}{\color[HTML]{000000} 50.77}   & \cellcolor[HTML]{FFFFFF}{\color[HTML]{000000} 53.47} & \cellcolor[HTML]{FFFFFF}{\color[HTML]{000000} 50.47}    &{\color[HTML]{000000} 52.64} &  {\color[HTML]{000000} 50.24}     & \cellcolor[HTML]{FFFFFF}{\color[HTML]{000000} 62.38} & \cellcolor[HTML]{FFFFFF}{\color[HTML]{000000} 51.56} & \cellcolor[HTML]{FFFFFF}{\color[HTML]{000000} 51.88} &{\color[HTML]{000000}  51.69} & {\color[HTML]{000000} 49.35}\\ \cline{3-13} 
\cellcolor[HTML]{FFFFEB}{\color[HTML]{000000} }  & \cellcolor[HTML]{FFFFFF}{\color[HTML]{000000} }   & {\color[HTML]{000000} SE-Concat}                                          & \cellcolor[HTML]{FFFFFF}{\color[HTML]{000000} 50.15}          & \cellcolor[HTML]{FFFFFF}{\color[HTML]{000000} 44.55} & \cellcolor[HTML]{FFFFFF}{\color[HTML]{000000} 51.40}     & {\color[HTML]{000000} 51.86}&   {\color[HTML]{000000} 50.16}  & \cellcolor[HTML]{FFFFFF}{\color[HTML]{000000} 61.39} & \cellcolor[HTML]{FFFFFF}{\color[HTML]{000000} 53.12} & \cellcolor[HTML]{FFFFFF}{\color[HTML]{000000} 52.50} &{\color[HTML]{000000}  54.02} & {\color[HTML]{000000} 50.24}\\ \cline{3-13} 
\cellcolor[HTML]{FFFFEB}{\color[HTML]{000000} }    & \cellcolor[HTML]{FFFFFF}{\color[HTML]{000000} }     & {\color[HTML]{000000} Cross-Atten}                                        & \cellcolor[HTML]{FFFFFF}{\color[HTML]{000000} 54.46}   & \cellcolor[HTML]{FFFFFF}{\color[HTML]{000000} 51.49} & \cellcolor[HTML]{FFFFFF}{\color[HTML]{000000} 55.14}     &{\color[HTML]{000000} 53.93}&   {\color[HTML]{000000} 51.21}  & \cellcolor[HTML]{FFFFFF}{\color[HTML]{000000} 58.42} & \cellcolor[HTML]{FFFFFF}{\color[HTML]{000000} 51.25} & \cellcolor[HTML]{FFFFFF}{\color[HTML]{000000} 52.19} &{\color[HTML]{000000}  54.80} &{\color[HTML]{000000} 50.49} \\ \cline{3-13} 
\cellcolor[HTML]{FFFFEB}{\color[HTML]{000000} }  & \cellcolor[HTML]{FFFFFF}{\color[HTML]{000000} }   & {\color[HTML]{000000} MLP-Mixer}    & \cellcolor[HTML]{FFFFFF}{\color[HTML]{000000} 52.31}          & \cellcolor[HTML]{FFFFFF}{\color[HTML]{000000} 55.45} & \cellcolor[HTML]{FFFFFF}{\color[HTML]{000000} \textbf{57.94}}  &{\color[HTML]{000000} 54.36}& {\color[HTML]{000000} 51.54} & \cellcolor[HTML]{FFFFFF}{\color[HTML]{000000} 63.37} & \cellcolor[HTML]{FFFFFF}{\color[HTML]{000000} 53.12} & \cellcolor[HTML]{FFFFFF}{\color[HTML]{000000} 51.25} &{\color[HTML]{000000} 55.66} &{\color[HTML]{000000} 51.21} \\ \cline{3-13} 
\cellcolor[HTML]{FFFFEB}{\color[HTML]{000000} }& \cellcolor[HTML]{FFFFFF}{\color[HTML]{000000} } & \cellcolor[HTML]{FFFFFF}{\color[HTML]{000000} CLIP-Align} & {\color[HTML]{000000} 53.22} & {\color[HTML]{000000} 55.38} & {\color[HTML]{000000} 52.06}   & {\color[HTML]{000000} 53.56}& {\color[HTML]{000000} 52.11} & {\color[HTML]{000000} 62.43} & {\color[HTML]{000000} 52.41} & {\color[HTML]{000000} 52.37} & {\color[HTML]{000000} 52.98} &{\color[HTML]{000000} 51.36}\\ \cline{3-13} 
\multirow{-6}{*}{\cellcolor[HTML]{FFFFEB}{\color[HTML]{000000} \begin{tabular}[c]{@{}c@{}}V-Face\\ +\\ A-Mel\end{tabular}}}    & \multirow{-5}{*}{\cellcolor[HTML]{FFFFFF}{\color[HTML]{000000} Feature}} & {\color[HTML]{000000} \textbf{Atten-Mixer}}    & \cellcolor[HTML]{FFFFFF}{\color[HTML]{000000} \textbf{57.54}} & \cellcolor[HTML]{FFFFFF}{\color[HTML]{000000} 55.45} & \cellcolor[HTML]{FFFFFF}{\color[HTML]{000000} \underline{56.07}}     &{\color[HTML]{000000} 55.06} &  {\color[HTML]{000000} 51.78}   & \cellcolor[HTML]{FFFFFF}{\color[HTML]{000000} 61.39} & \cellcolor[HTML]{FFFFFF}{\color[HTML]{000000} 50.94} & \cellcolor[HTML]{FFFFFF}{\color[HTML]{000000} 53.12} & {\color[HTML]{000000}  57.13}& {\color[HTML]{000000} \underline{52.35}} \\ \hline 
\cellcolor[HTML]{FFF7F7}{\color[HTML]{000000} }     & {\color[HTML]{000000} Score}   & {\color[HTML]{000000} Avg}  & {\color[HTML]{000000} 49.85}  & {\color[HTML]{000000} 58.42}     & {\color[HTML]{000000} 54.06}      & {\color[HTML]{000000} 52.16}& {\color[HTML]{000000} 50.08}  & {\color[HTML]{000000} 45.79}    & {\color[HTML]{000000} \underline{60.40}}    & {\color[HTML]{000000} 50.62}    &{\color[HTML]{000000}  51.08}&  {\color[HTML]{000000} 50.57}\\ \cline{2-13} 
\cellcolor[HTML]{FFF7F7}{\color[HTML]{000000} }     & \cellcolor[HTML]{FFFFFF}{\color[HTML]{000000} }     & {\color[HTML]{000000} Concat}      & {\color[HTML]{000000} 49.54}   & {\color[HTML]{000000} 53.47}    & {\color[HTML]{000000} 47.66}    & {\color[HTML]{000000} 50.91}&    {\color[HTML]{000000} 50.57}  & {\color[HTML]{000000} 61.39}     & {\color[HTML]{000000} 53.44}   & {\color[HTML]{000000} 51.88}    & {\color[HTML]{000000}  50.39}& {\color[HTML]{000000}50.49} \\ \cline{3-13} 
\cellcolor[HTML]{FFF7F7}{\color[HTML]{000000} }    & \cellcolor[HTML]{FFFFFF}{\color[HTML]{000000} }   & {\color[HTML]{000000} SE-Concat}             & {\color[HTML]{000000} 50.15}  & {\color[HTML]{000000} 48.51}   & {\color[HTML]{000000} 55.14}      & {\color[HTML]{000000} 50.99}&  {\color[HTML]{000000} 50.65}   & {\color[HTML]{000000} 60.40}  & {\color[HTML]{000000} 53.12}    & {\color[HTML]{000000} 50.00} & {\color[HTML]{000000} 52.93}&  {\color[HTML]{000000} 50.24} \\ \cline{3-13} 
\cellcolor[HTML]{FFF7F7}{\color[HTML]{000000} }  & \cellcolor[HTML]{FFFFFF}{\color[HTML]{000000} }  & {\color[HTML]{000000} Cross-Atten}   & {\color[HTML]{000000} 53.23}    & {\color[HTML]{000000} 44.55}  & {\color[HTML]{000000} \textbf{57.94}}       & {\color[HTML]{000000} 51.25}& {\color[HTML]{000000} 50.89} & {\color[HTML]{000000} 55.45}   & {\color[HTML]{000000} 54.69}    & {\color[HTML]{000000} \textbf{54.06}} & {\color[HTML]{000000} 55.14} &  {\color[HTML]{000000} 51.38}\\ \cline{3-13} 
\cellcolor[HTML]{FFF7F7}{\color[HTML]{000000} }         & \cellcolor[HTML]{FFFFFF}{\color[HTML]{000000} }    & {\color[HTML]{000000} MLP-Mixer}  & {\color[HTML]{000000} 49.54}    & {\color[HTML]{000000} 57.43} & {\color[HTML]{000000} 50.47}  & {\color[HTML]{000000} 53.33}&  {\color[HTML]{000000} 51.38} & {\color[HTML]{000000} 63.37}     & {\color[HTML]{000000} 54.06}       & {\color[HTML]{000000} 53.12}   &{\color[HTML]{000000} 56.18} & {\color[HTML]{000000} 51.70}\\ \cline{3-13} 
\cellcolor[HTML]{FFF7F7}{\color[HTML]{000000} }& \cellcolor[HTML]{FFFFFF}{\color[HTML]{000000} } & \cellcolor[HTML]{FFFFFF}{\color[HTML]{000000} CLIP-Align} & {\color[HTML]{000000} 52.00} & {\color[HTML]{000000} 55.42} & {\color[HTML]{000000} 51.25}   & {\color[HTML]{000000} 53.23}& {\color[HTML]{000000} 51.66} & {\color[HTML]{000000} 63.21} & {\color[HTML]{000000} 54.23} & {\color[HTML]{000000} 52.45} & {\color[HTML]{000000}  52.66} &{\color[HTML]{000000} 51.88}\\ \cline{3-13} 
\multirow{-6}{*}{\cellcolor[HTML]{FFF7F7}{\color[HTML]{000000} \begin{tabular}[c]{@{}c@{}}V-AGA\\ +\\ A-Mel\end{tabular}}} & \multirow{-5}{*}{\cellcolor[HTML]{FFFFFF}{\color[HTML]{000000} Feature}} & {\color[HTML]{000000} \textbf{Atten-Mixer}}     & {\color[HTML]{000000} 53.52}  & {\color[HTML]{000000} 54.46}   & {\color[HTML]{000000} 57.94}   & {\color[HTML]{000000} 54.97}&      {\color[HTML]{000000} 51.13}     & {\color[HTML]{000000} 61.39}    & {\color[HTML]{000000} 53.12}    & {\color[HTML]{000000} 51.56}  &{\color[HTML]{000000}  \underline{57.48}}& {\color[HTML]{000000} \textbf{52.43}} \\ \hline 
\cellcolor[HTML]{EEFBED}{\color[HTML]{000000} }  & {\color[HTML]{000000} Score}  & {\color[HTML]{000000} Avg}  & {\color[HTML]{000000} 52.00} & {\color[HTML]{000000} 58.42}   & {\color[HTML]{000000} 51.88}   & {\color[HTML]{000000} 52.43}&  {\color[HTML]{000000} 50.97} & {\color[HTML]{000000} 53.27}  & {\color[HTML]{000000} 58.42}   & {\color[HTML]{000000} 50.94}   & {\color[HTML]{000000} 54.11} & {\color[HTML]{000000} 50.16} \\ \cline{2-13} 
\cellcolor[HTML]{EEFBED}{\color[HTML]{000000} }& \cellcolor[HTML]{FFFFFF}{\color[HTML]{000000} }  & {\color[HTML]{000000} Concat}   & {\color[HTML]{000000} 53.23}  & {\color[HTML]{000000} 58.42}  & {\color[HTML]{000000} 55.14}     & {\color[HTML]{000000} 53.93}& {\color[HTML]{000000} 50.32} & {\color[HTML]{000000} 61.39}   & {\color[HTML]{000000} 52.50} & {\color[HTML]{000000} 52.19} &{\color[HTML]{000000} 55.23} & {\color[HTML]{000000} 50.32} \\ \cline{3-13} 
\cellcolor[HTML]{EEFBED}{\color[HTML]{000000} }  & \cellcolor[HTML]{FFFFFF}{\color[HTML]{000000} }  & {\color[HTML]{000000} SE-Concat}  & {\color[HTML]{000000} 51.38}   & {\color[HTML]{000000} 58.42}   & {\color[HTML]{000000} 51.40}  & {\color[HTML]{000000} 53.59}& {\color[HTML]{000000} 50.24}  & {\color[HTML]{000000} 62.38}   & {\color[HTML]{000000} 52.81}    & {\color[HTML]{000000} 50.94}   &{\color[HTML]{000000} 56.53} & {\color[HTML]{000000} 51.62} \\ \cline{3-13} 
\cellcolor[HTML]{EEFBED}{\color[HTML]{000000} }  & \cellcolor[HTML]{FFFFFF}{\color[HTML]{000000} }  & {\color[HTML]{000000} Cross-Atten}  & {\color[HTML]{000000} 51.08}  & {\color[HTML]{000000} 48.51}  & {\color[HTML]{000000} 55.14}    & {\color[HTML]{000000} 54.11}& {\color[HTML]{000000} 51.38}   & {\color[HTML]{000000} 60.40} & {\color[HTML]{000000} 53.12}   & {\color[HTML]{000000} 53.44}  & {\color[HTML]{000000} 57.22} & {\color[HTML]{000000}50.08} \\ \cline{3-13} 
\cellcolor[HTML]{EEFBED}{\color[HTML]{000000} }     & \cellcolor[HTML]{FFFFFF}{\color[HTML]{000000} }    & {\color[HTML]{000000} MLP-Mixer}  & {\color[HTML]{000000} 55.69}     & {\color[HTML]{000000} 46.53}  & {\color[HTML]{000000} 44.86}  & {\color[HTML]{000000} 54.97}&  {\color[HTML]{000000} 51.05} & {\color[HTML]{000000} 63.37}    & {\color[HTML]{000000} 51.56}    & {\color[HTML]{000000} 50.94}    &{\color[HTML]{000000} 57.04} & {\color[HTML]{000000} 51.21}\\ \cline{3-13} 
\cellcolor[HTML]{EEFBED}{\color[HTML]{000000} }& \cellcolor[HTML]{FFFFFF}{\color[HTML]{000000} } & \cellcolor[HTML]{FFFFFF}{\color[HTML]{000000} CLIP-Align} & {\color[HTML]{000000} 55.30} & {\color[HTML]{000000} 58.53} & {\color[HTML]{000000} 55.06}   & {\color[HTML]{000000} 52.00}& {\color[HTML]{000000} 62.16} & {\color[HTML]{000000} 53.68} & {\color[HTML]{000000} 53.24} & {\color[HTML]{000000} 53.62} & {\color[HTML]{000000}  56.09} &{\color[HTML]{000000} 51.23}\\ \cline{3-13} 
\multirow{-6}{*}{\cellcolor[HTML]{EEFBED}{\color[HTML]{000000} \begin{tabular}[c]{@{}c@{}}V-Face\\ +\\ V-AGA\\ +\\ A-Mel\end{tabular}}} & \multirow{-5}{*}{\cellcolor[HTML]{FFFFFF}{\color[HTML]{000000} Feature}} & {\color[HTML]{000000} \textbf{Atten-Mixer}}   & {\color[HTML]{000000} 55.08}   & {\color[HTML]{000000} \underline{60.40}}  & {\color[HTML]{000000} 57.01}      & {\color[HTML]{000000} \textbf{56.35}}&  {\color[HTML]{000000} 51.86} & {\color[HTML]{000000} \underline{64.36}}  & {\color[HTML]{000000} 53.44}  & {\color[HTML]{000000} 51.25}  &{\color[HTML]{000000} \textbf{57.56}} & {\color[HTML]{000000} 51.62} \\ \hline
\end{tabular}}

\vspace{1em}

\resizebox{1.0\textwidth}{!}{
\begin{tabular}{|c|
>{\columncolor[HTML]{FFFFFF}}c |
>{\columncolor[HTML]{FFFFFF}}c |
>{\columncolor[HTML]{FFFFFF}}c |
>{\columncolor[HTML]{FFFFFF}}c |
>{\columncolor[HTML]{FFFFFF}}c |
>{\columncolor[HTML]{FFFFFF}}c |
>{\columncolor[HTML]{FFFFFF}}c |
>{\columncolor[HTML]{FFFFFF}}c |
>{\columncolor[HTML]{FFFFFF}}c |
>{\columncolor[HTML]{FFFFFF}}c |
>{\columncolor[HTML]{FFFFFF}}c |
>{\columncolor[HTML]{FFFFFF}}c |}
\hline
\cellcolor[HTML]{FFFFFF}{\color[HTML]{000000} \textbf{Input}}   & \cellcolor[HTML]{FFFFFF}{\color[HTML]{000000} \textbf{Type}}  & \cellcolor[HTML]{FFFFFF}{\color[HTML]{000000} \textbf{Method}}     & {\color[HTML]{000000} \textbf{M to R}}  & {\color[HTML]{000000} \textbf{M to B1}} & {\color[HTML]{000000} \textbf{M to B2}}   & {\color[HTML]{000000} \textbf{M to D}}  & {\color[HTML]{000000} \textbf{M to E}}      & {\color[HTML]{000000} \textbf{D to R}} & {\color[HTML]{000000} \textbf{D to B1}} & {\color[HTML]{000000} \textbf{D to B2}} & \cellcolor[HTML]{FFFFFF}\textbf{D to M} & \cellcolor[HTML]{FFFFFF}\textbf{D to E} \\ \hline 
\cellcolor[HTML]{ECF4FF}{\color[HTML]{000000} } & \cellcolor[HTML]{FFFFFF}{\color[HTML]{000000} Score}  & \cellcolor[HTML]{FFFFFF}{\color[HTML]{000000} Avg}  & {\color[HTML]{000000} 57.94} & {\color[HTML]{000000} 53.23}  & 62.38   & {\color[HTML]{000000} 53.41}  & {\color[HTML]{000000} 50.32} & {\color[HTML]{000000} 52.31}  & {\color[HTML]{000000} 51.69}& {\color[HTML]{000000} 52.28}  & 50.46                         & 50.32  \\ \cline{2-13} 
\cellcolor[HTML]{ECF4FF}{\color[HTML]{000000} }  & \cellcolor[HTML]{FFFFFF}{\color[HTML]{000000} }  & \cellcolor[HTML]{FFFFFF}{\color[HTML]{000000} Concat} & {\color[HTML]{000000} 57.01} & {\color[HTML]{000000} 50.15}  & 61.39   & {\color[HTML]{000000} 53.50}  & {\color[HTML]{000000} 51.21}   & {\color[HTML]{000000} 53.84} & {\color[HTML]{000000} 50.46}  & {\color[HTML]{000000} 54.08}  & 50.40   & 50.16  \\ \cline{3-13} 
\cellcolor[HTML]{ECF4FF}{\color[HTML]{000000} }  & \cellcolor[HTML]{FFFFFF}{\color[HTML]{000000} }  & \cellcolor[HTML]{FFFFFF}{\color[HTML]{000000} SE-Concat}  & {\color[HTML]{000000} 56.07} & {\color[HTML]{000000} 51.38} & 65.35   & {\color[HTML]{000000} 54.19} & {\color[HTML]{000000} 50.57} & {\color[HTML]{000000} 54.42} & {\color[HTML]{000000} 51.38}  & {\color[HTML]{000000} 54.48} & 50.37 & 50.77 \\ \cline{3-13} 
\cellcolor[HTML]{ECF4FF}{\color[HTML]{000000} }  & \cellcolor[HTML]{FFFFFF}{\color[HTML]{000000} } & \cellcolor[HTML]{FFFFFF}{\color[HTML]{000000} Cross-Atten} & {\color[HTML]{000000} 55.14}  & {\color[HTML]{000000} 56.31}   & 60.40 & {\color[HTML]{000000} 54.39}  & {\color[HTML]{000000} 51.13}   & {\color[HTML]{000000} 55.23} & {\color[HTML]{000000} 52.62} & {\color[HTML]{000000} 57.43} & 50.86 & 50.97  \\ \cline{3-13} 
\cellcolor[HTML]{ECF4FF}{\color[HTML]{000000} }  & \cellcolor[HTML]{FFFFFF}{\color[HTML]{000000} }  & \cellcolor[HTML]{FFFFFF}{\color[HTML]{000000} MLP-Mixer}& {\color[HTML]{000000} 55.14}  & {\color[HTML]{000000} \textbf{59.08}} & 59.41 & {\color[HTML]{000000} 55.14}  & {\color[HTML]{000000} 51.29}& {\color[HTML]{000000} 55.08}   & {\color[HTML]{000000} 52.92}  & {\color[HTML]{000000} 59.41 }& 50.77 & 51.05 \\ \cline{3-13} 
\cellcolor[HTML]{ECF4FF}{\color[HTML]{000000} }& \cellcolor[HTML]{FFFFFF}{\color[HTML]{000000} } & \cellcolor[HTML]{FFFFFF}{\color[HTML]{000000} CLIP-Align} & {\color[HTML]{000000} 58.02} & {\color[HTML]{000000} 53.09} & {\color[HTML]{000000} 62.33}   & {\color[HTML]{000000} 54.02}& {\color[HTML]{000000} 52.23} & {\color[HTML]{000000} 52.22} & {\color[HTML]{000000} 51.51} & {\color[HTML]{000000} 55.11} & {\color[HTML]{000000}  51.23} &{\color[HTML]{000000} 52.01}\\ \cline{3-13} 
\multirow{-6}{*}{\cellcolor[HTML]{ECF4FF}{\color[HTML]{000000} \begin{tabular}[c]{@{}c@{}}V-Face\\ +\\ V-AGA\end{tabular}}} & \multirow{-5}{*}{\cellcolor[HTML]{FFFFFF}{\color[HTML]{000000} Feature}} & \cellcolor[HTML]{FFFFFF}{\color[HTML]{000000} \textbf{Atten-Mixer}}  & {\color[HTML]{000000} 60.75}  & {\color[HTML]{000000} 56.00}  & 61.39  & {\color[HTML]{000000} 54.97} & {\color[HTML]{000000} 52.02} & {\color[HTML]{000000} 56.92}  & {\color[HTML]{000000} 52.10} & {\color[HTML]{000000} 60.04}  & 51.49 & 51.69  \\ \hline 
\cellcolor[HTML]{FFFFEB}{\color[HTML]{000000} } & {\color[HTML]{000000} Score}  & {\color[HTML]{000000} Avg}  & \cellcolor[HTML]{FFFFFF}{\color[HTML]{000000} 65.42} & \cellcolor[HTML]{FFFFFF}{\color[HTML]{000000} 56.31} & 49.50  & \cellcolor[HTML]{FFFFFF}{\color[HTML]{000000} 52.90} & \cellcolor[HTML]{FFFFFF}{\color[HTML]{000000} 50.49} & \cellcolor[HTML]{FFFFFF}{\color[HTML]{000000} 52.00} & \cellcolor[HTML]{FFFFFF}{\color[HTML]{000000} 51.08} & \cellcolor[HTML]{FFFFFF}{\color[HTML]{000000} 53.47} & 50.24 & 50.18  \\ \cline{2-13} 
\cellcolor[HTML]{FFFFEB}{\color[HTML]{000000} }  & \cellcolor[HTML]{FFFFFF}{\color[HTML]{000000} }   & {\color[HTML]{000000} Concat}  & \cellcolor[HTML]{FFFFFF}{\color[HTML]{000000} 54.21} & \cellcolor[HTML]{FFFFFF}{\color[HTML]{000000} 52.62} & 58.42   & \cellcolor[HTML]{FFFFFF}{\color[HTML]{000000} 51.60} & \cellcolor[HTML]{FFFFFF}{\color[HTML]{000000} 50.81} & \cellcolor[HTML]{FFFFFF}{\color[HTML]{000000} 54.15} & \cellcolor[HTML]{FFFFFF}{\color[HTML]{000000} 50.77} & \cellcolor[HTML]{FFFFFF}{\color[HTML]{000000} 55.45} & 50.47  & 50.52   \\ \cline{3-13} 
\cellcolor[HTML]{FFFFEB}{\color[HTML]{000000} }  & \cellcolor[HTML]{FFFFFF}{\color[HTML]{000000} }   & {\color[HTML]{000000} SE-Concat}  & \cellcolor[HTML]{FFFFFF}{\color[HTML]{000000} 65.42} & \cellcolor[HTML]{FFFFFF}{\color[HTML]{000000} 56.31} & {\color[HTML]{000000} \underline{66.34}}  & \cellcolor[HTML]{FFFFFF}{\color[HTML]{000000} 52.03} & \cellcolor[HTML]{FFFFFF}{\color[HTML]{000000} 51.05} & \cellcolor[HTML]{FFFFFF}{\color[HTML]{000000} 54.46} &  \cellcolor[HTML]{FFFFFF}{\color[HTML]{000000} 50.96} & 56.64 &  50.60 & 50.81    \\ \cline{3-13} 
\cellcolor[HTML]{FFFFEB}{\color[HTML]{000000} }   & \cellcolor[HTML]{FFFFFF}{\color[HTML]{000000} }  & {\color[HTML]{000000} Cross-Atten} & \cellcolor[HTML]{FFFFFF}{\color[HTML]{000000} 63.55} & \cellcolor[HTML]{FFFFFF}{\color[HTML]{000000} 56.62} & {\color[HTML]{000000} \underline{66.34}}   & \cellcolor[HTML]{FFFFFF}{\color[HTML]{000000} 52.38} & \cellcolor[HTML]{FFFFFF}{\color[HTML]{000000} 51.29} & \cellcolor[HTML]{FFFFFF}{\color[HTML]{000000} 54.46} & \cellcolor[HTML]{FFFFFF}{\color[HTML]{000000} 50.50} & \cellcolor[HTML]{FFFFFF}{\color[HTML]{000000} 56.84} & 51.20    & 51.08    \\ \cline{3-13} 
\cellcolor[HTML]{FFFFEB}{\color[HTML]{000000} }   & \cellcolor[HTML]{FFFFFF}{\color[HTML]{000000} }  & {\color[HTML]{000000} MLP-Mixer}  & \cellcolor[HTML]{FFFFFF}{\color[HTML]{000000} 64.49} & \cellcolor[HTML]{FFFFFF}{\color[HTML]{000000} 57.85} & 62.38  & \cellcolor[HTML]{FFFFFF}{\color[HTML]{000000} 52.20} & \cellcolor[HTML]{FFFFFF}{\color[HTML]{000000} 51.05} & \cellcolor[HTML]{FFFFFF}{\color[HTML]{000000} 56.00} & \cellcolor[HTML]{FFFFFF}{\color[HTML]{000000} 50.46} & \cellcolor[HTML]{FFFFFF}{\color[HTML]{000000} 57.43} & 51.08  & 51.38   \\ \cline{3-13}
\cellcolor[HTML]{FFFFEB}{\color[HTML]{000000} }& \cellcolor[HTML]{FFFFFF}{\color[HTML]{000000} } & \cellcolor[HTML]{FFFFFF}{\color[HTML]{000000} CLIP-Align} & {\color[HTML]{000000} 55.15} & {\color[HTML]{000000} 53.52} & {\color[HTML]{000000} 58.21}   & {\color[HTML]{000000} 52.45}& {\color[HTML]{000000} 51.33} & {\color[HTML]{000000} 51.35} & {\color[HTML]{000000} 55.62} & {\color[HTML]{000000} 52.34} & {\color[HTML]{000000}  52.66} &{\color[HTML]{000000} 51.28}\\ \cline{3-13} 
\multirow{-6}{*}{\cellcolor[HTML]{FFFFEB}{\color[HTML]{000000} \begin{tabular}[c]{@{}c@{}}V-Face\\ +\\ A-Mel\end{tabular}}}  & \multirow{-5}{*}{\cellcolor[HTML]{FFFFFF}{\color[HTML]{000000} Feature}} & {\color[HTML]{000000} \textbf{Atten-Mixer}}& \cellcolor[HTML]{FFFFFF}{\color[HTML]{000000} \underline{67.29}} & \cellcolor[HTML]{FFFFFF}{\color[HTML]{000000} 57.85} & 64.36 & \cellcolor[HTML]{FFFFFF}{\color[HTML]{000000} \textbf{55.83}} & \cellcolor[HTML]{FFFFFF}{\color[HTML]{000000} \underline{52.35}} & \cellcolor[HTML]{FFFFFF}{\color[HTML]{000000} 56.31} & \cellcolor[HTML]{FFFFFF}{\color[HTML]{000000} 51.69} & \cellcolor[HTML]{FFFFFF}{\color[HTML]{000000} 58.42} & 53.27  & 52.30 \\ \hline

\cellcolor[HTML]{FFF7F7}{\color[HTML]{000000} }  & {\color[HTML]{000000} Score}    & {\color[HTML]{000000} Avg}    & {\color[HTML]{000000} 49.53}     & {\color[HTML]{000000} 56.00}   & 63.37    & {\color[HTML]{000000} 52.03}   & {\color[HTML]{000000} 51.38}    & {\color[HTML]{000000} 53.38}   & {\color[HTML]{000000} 50.35}  & {\color[HTML]{000000} 53.26} & 51.40   & 50.89  \\ \cline{2-13} 
\cellcolor[HTML]{FFF7F7}{\color[HTML]{000000} }  & \cellcolor[HTML]{FFFFFF}{\color[HTML]{000000} }  & {\color[HTML]{000000} Concat}  & {\color[HTML]{000000} 57.01}  & {\color[HTML]{000000} 50.15}     & 58.42 & {\color[HTML]{000000} 52.29}  & {\color[HTML]{000000} 50.97}  & {\color[HTML]{000000} 56.25}  & {\color[HTML]{000000} 50.15} & {\color[HTML]{000000} 53.47}   & 52.34  & 51.05  \\ \cline{3-13} 
\cellcolor[HTML]{FFF7F7}{\color[HTML]{000000} }  & \cellcolor[HTML]{FFFFFF}{\color[HTML]{000000} }   & {\color[HTML]{000000} SE-Concat}   & {\color[HTML]{000000} 57.01} & {\color[HTML]{000000} 49.85}   & 63.37 & {\color[HTML]{000000} 52.38}   & {\color[HTML]{000000} 50.57} & {\color[HTML]{000000} 56.62}  & {\color[HTML]{000000} 50.77}   & {\color[HTML]{000000} 54.46}  & 53.64  & 51.29   \\ \cline{3-13} 
\cellcolor[HTML]{FFF7F7}{\color[HTML]{000000} }  & \cellcolor[HTML]{FFFFFF}{\color[HTML]{000000} }  & {\color[HTML]{000000} Cross-Atten}   & {\color[HTML]{000000} 63.55}   & {\color[HTML]{000000} 51.69} & 64.36  & {\color[HTML]{000000} 53.07}  & {\color[HTML]{000000} 51.05}  & {\color[HTML]{000000} 56.31}& {\color[HTML]{000000} 51.69}  & {\color[HTML]{000000} 55.35}  & 52.92   & 51.78  \\ \cline{3-13} 
\cellcolor[HTML]{FFF7F7}{\color[HTML]{000000} }  & \cellcolor[HTML]{FFFFFF}{\color[HTML]{000000} }   & {\color[HTML]{000000} MLP-Mixer}  & {\color[HTML]{000000} 59.81}    & {\color[HTML]{000000} 55.08}    & \textbf{69.31}  & {\color[HTML]{000000} 54.36}  & {\color[HTML]{000000} 56.85} & {\color[HTML]{000000} 53.12} & {\color[HTML]{000000} 51.38} & {\color[HTML]{000000} 55.42}   & 52.62  & 52.67  \\ \cline{3-13} 
\cellcolor[HTML]{FFF7F7}{\color[HTML]{000000} }& \cellcolor[HTML]{FFFFFF}{\color[HTML]{000000} } & \cellcolor[HTML]{FFFFFF}{\color[HTML]{000000} CLIP-Align} & {\color[HTML]{000000} 57.88} & {\color[HTML]{000000} 52.14} & {\color[HTML]{000000} 59.21}   & {\color[HTML]{000000} 53.56}& {\color[HTML]{000000} 52.45} & {\color[HTML]{000000} 57.14} & {\color[HTML]{000000} 51.25} & {\color[HTML]{000000} 54.62} & {\color[HTML]{000000}  53.71} &{\color[HTML]{000000} 52.28}\\ \cline{3-13} 
\multirow{-6}{*}{\cellcolor[HTML]{FFF7F7}{\color[HTML]{000000} \begin{tabular}[c]{@{}c@{}}V-AGA\\ +\\ A-Mel\end{tabular}}}   & \multirow{-5}{*}{\cellcolor[HTML]{FFFFFF}{\color[HTML]{000000} Feature}} & {\color[HTML]{000000} \textbf{Atten-Mixer}}  & {\color[HTML]{000000} \textbf{69.16}}      & {\color[HTML]{000000} \underline{58.15}}     & 64.36   & {\color[HTML]{000000} \underline{55.40}}   & {\color[HTML]{000000} 51.46} & {\color[HTML]{000000} 57.23}& {\color[HTML]{000000} 53.85} & {\color[HTML]{000000} \underline{61.39}} & \textbf{54.21}  & \underline{53.32}   \\ \hline
\cellcolor[HTML]{EEFBED}{\color[HTML]{000000} }  & {\color[HTML]{000000} Score}  & {\color[HTML]{000000} Avg}  & {\color[HTML]{000000} 57.01}  & {\color[HTML]{000000} 53.54}  & 61.39   & {\color[HTML]{000000} 50.82}  & {\color[HTML]{000000} 51.38}  & {\color[HTML]{000000} 52.92}  & {\color[HTML]{000000} 52.31}  & {\color[HTML]{000000} 58.22} & 51.60  & 52.04  \\ \cline{2-13} 
\cellcolor[HTML]{EEFBED}{\color[HTML]{000000} }  & \cellcolor[HTML]{FFFFFF}{\color[HTML]{000000} } & {\color[HTML]{000000} Concat}   & {\color[HTML]{000000} 56.07}   & {\color[HTML]{000000} 54.15}    & 60.40   & {\color[HTML]{000000} 51.69}  & {\color[HTML]{000000} 50.16}  & {\color[HTML]{000000} 53.23} & {\color[HTML]{000000} 52.00}    & {\color[HTML]{000000} 58.32}  & 51.40  & 52.00 \\ \cline{3-13} 
\cellcolor[HTML]{EEFBED}{\color[HTML]{000000} }& \cellcolor[HTML]{FFFFFF}{\color[HTML]{000000} }  & {\color[HTML]{000000} SE-Concat} & {\color[HTML]{000000} 59.81}    & {\color[HTML]{000000} 54.77}  & 52.48  & {\color[HTML]{000000} 52.03}  & {\color[HTML]{000000} 51.78}  & {\color[HTML]{000000} 56.95}  & {\color[HTML]{000000} 53.65} & {\color[HTML]{000000} 59.45} & 51.50    & 52.40 \\ \cline{3-13} 
\cellcolor[HTML]{EEFBED}{\color[HTML]{000000} }   & \cellcolor[HTML]{FFFFFF}{\color[HTML]{000000} }   & {\color[HTML]{000000} Cross-Atten}   & {\color[HTML]{000000} 60.75} & {\color[HTML]{000000} 56.31}  & 60.40   & {\color[HTML]{000000} 53.24}   & {\color[HTML]{000000} 51.94}   & {\color[HTML]{000000} 57.33}  & {\color[HTML]{000000} 54.15}  & {\color[HTML]{000000} 59.41}   & 52.30   & 52.34 \\ \cline{3-13} 
\cellcolor[HTML]{EEFBED}{\color[HTML]{000000} }   & \cellcolor[HTML]{FFFFFF}{\color[HTML]{000000} }  & {\color[HTML]{000000} MLP-Mixer}  & {\color[HTML]{000000} 64.49}   & {\color[HTML]{000000} 56.00}  & 60.40  & {\color[HTML]{000000} 54.11}  & {\color[HTML]{000000} 51.54}  & {\color[HTML]{000000} \underline{57.54}} & {\color[HTML]{000000} \underline{54.46}} & {\color[HTML]{000000} 60.42}  & 52.31  & 53.27 \\ \cline{3-13}
\cellcolor[HTML]{EEFBED}{\color[HTML]{000000} }& \cellcolor[HTML]{FFFFFF}{\color[HTML]{000000} } & \cellcolor[HTML]{FFFFFF}{\color[HTML]{000000} CLIP-Align} & {\color[HTML]{000000} 57.32} & {\color[HTML]{000000} 55.24} & {\color[HTML]{000000} 61.35}   & {\color[HTML]{000000} 53.22}& {\color[HTML]{000000} 52.25} & {\color[HTML]{000000} 53.14} & {\color[HTML]{000000} 53.50} & {\color[HTML]{000000} 58.33} & {\color[HTML]{000000}  52.06} &{\color[HTML]{000000} 52.62}\\ \cline{3-13}  
\multirow{-6}{*}{\cellcolor[HTML]{EEFBED}{\color[HTML]{000000} \begin{tabular}[c]{@{}c@{}}V-Face\\ +\\ V-AGA\\ +\\ A-Mel\end{tabular}}} & \multirow{-5}{*}{\cellcolor[HTML]{FFFFFF}{\color[HTML]{000000} Feature}} & {\color[HTML]{000000} \textbf{Atten-Mixer}} & {\color[HTML]{000000} \underline{67.29}}  & {\color[HTML]{000000} 56.00} & 62.38   & {\color[HTML]{000000} 55.32}  & {\color[HTML]{000000} \textbf{52.91}}  & {\color[HTML]{000000} \textbf{58.77}} & {\color[HTML]{000000} \textbf{55.08}}  & {\color[HTML]{000000} 63.37}  & \underline{53.85}  & \textbf{53.54} \\ \hline
\end{tabular}}
\end{table*}

\begin{table*}[t]
\centering
\caption{\label{tab:single2singlefusion_cond} (Continued) Fusion results of single-to-single cross-domain generalization accuracy (\%).}
\resizebox{0.65\textwidth}{!}{
\begin{tabular}{|c|
>{\columncolor[HTML]{FFFFFF}}c |
>{\columncolor[HTML]{FFFFFF}}c |
>{\columncolor[HTML]{FFFFFF}}c |
>{\columncolor[HTML]{FFFFFF}}c |
>{\columncolor[HTML]{FFFFFF}}c |
>{\columncolor[HTML]{FFFFFF}}c |
>{\columncolor[HTML]{FFFFFF}}c |
>{\columncolor[HTML]{FFFFFF}}c |
}
\hline
\cellcolor[HTML]{FFFFFF}{\color[HTML]{000000} \textbf{Input}}   & \cellcolor[HTML]{FFFFFF}{\color[HTML]{000000} \textbf{Type}}  & \cellcolor[HTML]{FFFFFF}{\color[HTML]{000000} \textbf{Method}}     & {\color[HTML]{000000} \textbf{E to R}}  & {\color[HTML]{000000} \textbf{E to B1}} & {\color[HTML]{000000} \textbf{E to B2}}   & {\color[HTML]{000000} \textbf{E to M}}  & {\color[HTML]{000000} \textbf{E to D}}      & {\color[HTML]{000000} \textbf{Avg}}  \\ \hline 
\cellcolor[HTML]{ECF4FF}{\color[HTML]{000000} } & \cellcolor[HTML]{FFFFFF}{\color[HTML]{000000} Score}  & \cellcolor[HTML]{FFFFFF}{\color[HTML]{000000} Avg}  & {\color[HTML]{000000} 49.68} & {\color[HTML]{000000} 51.21}  & 56.46   & {\color[HTML]{000000} 49.53}  & {\color[HTML]{000000} 53.50} & {\color[HTML]{000000} 53.05}  \\ \cline{2-9} 
\cellcolor[HTML]{ECF4FF}{\color[HTML]{000000} }  & \cellcolor[HTML]{FFFFFF}{\color[HTML]{000000} }  & \cellcolor[HTML]{FFFFFF}{\color[HTML]{000000} Concat} & {\color[HTML]{000000} 50.73} & {\color[HTML]{000000} 51.49}  & 56.33   & {\color[HTML]{000000} 50.10}  & {\color[HTML]{000000} 52.38}   & {\color[HTML]{000000} 52.84}  \\ \cline{3-9} 
\cellcolor[HTML]{ECF4FF}{\color[HTML]{000000} }  & \cellcolor[HTML]{FFFFFF}{\color[HTML]{000000} }  & \cellcolor[HTML]{FFFFFF}{\color[HTML]{000000} SE-Concat}  & {\color[HTML]{000000} 50.16} & {\color[HTML]{000000} 52.48} & 57.43   & {\color[HTML]{000000} 50.25} & {\color[HTML]{000000} 53.24} & {\color[HTML]{000000} 53.71} \\ \cline{3-9} 
\cellcolor[HTML]{ECF4FF}{\color[HTML]{000000} }  & \cellcolor[HTML]{FFFFFF}{\color[HTML]{000000} } & \cellcolor[HTML]{FFFFFF}{\color[HTML]{000000} Cross-Atten} & {\color[HTML]{000000} 50.65}  & {\color[HTML]{000000} 53.46}   & 57.83 & {\color[HTML]{000000} 50.47}  & {\color[HTML]{000000} 53.07}   & {\color[HTML]{000000} 54.22} \\ \cline{3-9} 
\cellcolor[HTML]{ECF4FF}{\color[HTML]{000000} }  & \cellcolor[HTML]{FFFFFF}{\color[HTML]{000000} }  & \cellcolor[HTML]{FFFFFF}{\color[HTML]{000000} MLP-Mixer}& {\color[HTML]{000000} 51.38}  & {\color[HTML]{000000} 55.42} & 57.25 & {\color[HTML]{000000} 51.24}  & {\color[HTML]{000000} 54.11}& {\color[HTML]{000000} 54.35}   \\ \cline{3-9} 
\cellcolor[HTML]{ECF4FF}{\color[HTML]{000000} }  & \cellcolor[HTML]{FFFFFF}{\color[HTML]{000000} }  & \cellcolor[HTML]{FFFFFF}{\color[HTML]{000000} CLIP-Align}& {\color[HTML]{000000} 52.23}  & {\color[HTML]{000000} 52.90} & 57.32 & {\color[HTML]{000000} 51.22}  & {\color[HTML]{000000} 52.65}& {\color[HTML]{000000} 53.77}   \\ \cline{3-9} 
\multirow{-6}{*}{\cellcolor[HTML]{ECF4FF}{\color[HTML]{000000} \begin{tabular}[c]{@{}c@{}}V-Face\\ +\\ V-AGA\end{tabular}}} & \multirow{-5}{*}{\cellcolor[HTML]{FFFFFF}{\color[HTML]{000000} Feature}} & \cellcolor[HTML]{FFFFFF}{\color[HTML]{000000} \textbf{Atten-Mixer}}  & {\color[HTML]{000000} \underline{52.27}}  & {\color[HTML]{000000} 56.21}  & 60.09  & {\color[HTML]{000000} 51.45} & {\color[HTML]{000000} 54.54} & {\color[HTML]{000000} 55.78} \\ \hline 
\cellcolor[HTML]{FFFFEB}{\color[HTML]{000000} } & {\color[HTML]{000000} Score}  & {\color[HTML]{000000} Avg}  & \cellcolor[HTML]{FFFFFF}{\color[HTML]{000000} 49.84} & \cellcolor[HTML]{FFFFFF}{\color[HTML]{000000} 52.46} & 57.87  & \cellcolor[HTML]{FFFFFF}{\color[HTML]{000000} 50.03} & \cellcolor[HTML]{FFFFFF}{\color[HTML]{000000} 52.72} & \cellcolor[HTML]{FFFFFF}{\color[HTML]{000000} 52.83}   \\ \cline{2-9} 
\cellcolor[HTML]{FFFFEB}{\color[HTML]{000000} }  & \cellcolor[HTML]{FFFFFF}{\color[HTML]{000000} }   & {\color[HTML]{000000} Concat}  & \cellcolor[HTML]{FFFFFF}{\color[HTML]{000000} 51.21} & \cellcolor[HTML]{FFFFFF}{\color[HTML]{000000} 52.69} & 58.72   & \cellcolor[HTML]{FFFFFF}{\color[HTML]{000000} 50.33} & \cellcolor[HTML]{FFFFFF}{\color[HTML]{000000} 51.34} & \cellcolor[HTML]{FFFFFF}{\color[HTML]{000000} 52.71}   \\ \cline{3-9} 
\cellcolor[HTML]{FFFFEB}{\color[HTML]{000000} }  & \cellcolor[HTML]{FFFFFF}{\color[HTML]{000000} }   & {\color[HTML]{000000} SE-Concat}  & \cellcolor[HTML]{FFFFFF}{\color[HTML]{000000} 50.97} & \cellcolor[HTML]{FFFFFF}{\color[HTML]{000000} 53.62} & 59.41  & \cellcolor[HTML]{FFFFFF}{\color[HTML]{000000} 50.54} & \cellcolor[HTML]{FFFFFF}{\color[HTML]{000000} 52.81} & \cellcolor[HTML]{FFFFFF}{\color[HTML]{000000} 53.65}    \\ \cline{3-9} 
\cellcolor[HTML]{FFFFEB}{\color[HTML]{000000} }   & \cellcolor[HTML]{FFFFFF}{\color[HTML]{000000} }  & {\color[HTML]{000000} Cross-Atten} & \cellcolor[HTML]{FFFFFF}{\color[HTML]{000000} 50.57} & \cellcolor[HTML]{FFFFFF}{\color[HTML]{000000} 54.99} & {\color[HTML]{000000}59.70}  & \cellcolor[HTML]{FFFFFF}{\color[HTML]{000000} 50.67} & \cellcolor[HTML]{FFFFFF}{\color[HTML]{000000} 52.72} & \cellcolor[HTML]{FFFFFF}{\color[HTML]{000000} 54.25}    \\ \cline{3-9} 
\cellcolor[HTML]{FFFFEB}{\color[HTML]{000000} }   & \cellcolor[HTML]{FFFFFF}{\color[HTML]{000000} }  & {\color[HTML]{000000} MLP-Mixer}  & \cellcolor[HTML]{FFFFFF}{\color[HTML]{000000} 51.21} & \cellcolor[HTML]{FFFFFF}{\color[HTML]{000000} 55.08} & 60.08  & \cellcolor[HTML]{FFFFFF}{\color[HTML]{000000} 50.95} & \cellcolor[HTML]{FFFFFF}{\color[HTML]{000000} 52.20} & \cellcolor[HTML]{FFFFFF}{\color[HTML]{000000} 54.80}    \\ \cline{3-9}
\cellcolor[HTML]{FFFFEB}{\color[HTML]{000000} }  & \cellcolor[HTML]{FFFFFF}{\color[HTML]{000000} }  & \cellcolor[HTML]{FFFFFF}{\color[HTML]{000000} CLIP-Align}& {\color[HTML]{000000} 51.53}  & {\color[HTML]{000000} 53.05} & 58.92 & {\color[HTML]{000000} 52.04}  & {\color[HTML]{000000} 52.36}& {\color[HTML]{000000} 53.59}   \\ \cline{3-9}
\multirow{-6}{*}{\cellcolor[HTML]{FFFFEB}{\color[HTML]{000000} \begin{tabular}[c]{@{}c@{}}V-Face\\ +\\ A-Mel\end{tabular}}}  & \multirow{-5}{*}{\cellcolor[HTML]{FFFFFF}{\color[HTML]{000000} Feature}} & {\color[HTML]{000000} \textbf{Atten-Mixer}}& \cellcolor[HTML]{FFFFFF}{\color[HTML]{000000} 51.70} & \cellcolor[HTML]{FFFFFF}{\color[HTML]{000000} 55.45} & 61.39 & \cellcolor[HTML]{FFFFFF}{\color[HTML]{000000} 51.63} & \cellcolor[HTML]{FFFFFF}{\color[HTML]{000000} 54.19} & \cellcolor[HTML]{FFFFFF}{\color[HTML]{000000} 55.79}  \\ \hline

\cellcolor[HTML]{FFF7F7}{\color[HTML]{000000} }  & {\color[HTML]{000000} Score}    & {\color[HTML]{000000} Avg}    & {\color[HTML]{000000} 50.24}     & {\color[HTML]{000000} 54.50}   & 57.43    & {\color[HTML]{000000} 50.65}   & {\color[HTML]{000000} 51.08}    & {\color[HTML]{000000} 52.74}   \\ \cline{2-9} 
\cellcolor[HTML]{FFF7F7}{\color[HTML]{000000} }  & \cellcolor[HTML]{FFFFFF}{\color[HTML]{000000} }  & {\color[HTML]{000000} Concat}  & {\color[HTML]{000000} 50.32}  & {\color[HTML]{000000} 54.46}     & 58.42 & {\color[HTML]{000000} 51.70}  & {\color[HTML]{000000} 51.86}  & {\color[HTML]{000000} 52.74}   \\ \cline{3-9} 
\cellcolor[HTML]{FFF7F7}{\color[HTML]{000000} }  & \cellcolor[HTML]{FFFFFF}{\color[HTML]{000000} }   & {\color[HTML]{000000} SE-Concat}   & {\color[HTML]{000000} 50.81} & {\color[HTML]{000000} 54.66}   & 58.22 & {\color[HTML]{000000} 51.38}   & {\color[HTML]{000000} 51.96} & {\color[HTML]{000000} 53.16}   \\ \cline{3-9} 
\cellcolor[HTML]{FFF7F7}{\color[HTML]{000000} }  & \cellcolor[HTML]{FFFFFF}{\color[HTML]{000000} }  & {\color[HTML]{000000} Cross-Atten}   & {\color[HTML]{000000} 50.97}   & {\color[HTML]{000000} 55.32} & 58.62  & {\color[HTML]{000000} 52.18}  & {\color[HTML]{000000} 53.85}  & {\color[HTML]{000000} 54.05} \\ \cline{3-9} 
\cellcolor[HTML]{FFF7F7}{\color[HTML]{000000} }  & \cellcolor[HTML]{FFFFFF}{\color[HTML]{000000} }   & {\color[HTML]{000000} MLP-Mixer}  & {\color[HTML]{000000} 50.65}    & {\color[HTML]{000000} 55.65}    & 58.77  & {\color[HTML]{000000} 52.10}  & {\color[HTML]{000000} 54.11} & {\color[HTML]{000000} 54.90} \\ \cline{3-9} 
\cellcolor[HTML]{FFF7F7}{\color[HTML]{000000} }  & \cellcolor[HTML]{FFFFFF}{\color[HTML]{000000} }  & \cellcolor[HTML]{FFFFFF}{\color[HTML]{000000} CLIP-Align}& {\color[HTML]{000000} 52.04}  & {\color[HTML]{000000} 54.36} & 58.74 & {\color[HTML]{000000} 52.61}  & {\color[HTML]{000000} 53.03}& {\color[HTML]{000000} 54.12}   \\ \cline{3-9}
\multirow{-6}{*}{\cellcolor[HTML]{FFF7F7}{\color[HTML]{000000} \begin{tabular}[c]{@{}c@{}}V-AGA\\ +\\ A-Mel\end{tabular}}}   & \multirow{-5}{*}{\cellcolor[HTML]{FFFFFF}{\color[HTML]{000000} Feature}} & {\color[HTML]{000000} \textbf{Atten-Mixer}}  & {\color[HTML]{000000} 52.02}      & {\color[HTML]{000000} \textbf{56.44}}     & 60.41   & {\color[HTML]{000000} \underline{53.16}}   & {\color[HTML]{000000} 55.06} & {\color[HTML]{000000} \underline{56.14}}  \\ \hline
\cellcolor[HTML]{EEFBED}{\color[HTML]{000000} }  & {\color[HTML]{000000} Score}  & {\color[HTML]{000000} Avg}  & {\color[HTML]{000000} 49.50}  & {\color[HTML]{000000} 53.23}  & 58.00   & {\color[HTML]{000000} 51.21}  & {\color[HTML]{000000} 52.29}  & {\color[HTML]{000000} 53.52}  \\ \cline{2-9} 
\cellcolor[HTML]{EEFBED}{\color[HTML]{000000} }  & \cellcolor[HTML]{FFFFFF}{\color[HTML]{000000} } & {\color[HTML]{000000} Concat}   & {\color[HTML]{000000} 50.50}   & {\color[HTML]{000000} 53.47}    & 58.92  & {\color[HTML]{000000} 51.13}  & {\color[HTML]{000000} 52.90}  & {\color[HTML]{000000} 53.96} \\ \cline{3-9} 
\cellcolor[HTML]{EEFBED}{\color[HTML]{000000} }& \cellcolor[HTML]{FFFFFF}{\color[HTML]{000000} }  & {\color[HTML]{000000} SE-Concat} & {\color[HTML]{000000} 50.47}    & {\color[HTML]{000000} 54.27}  & 59.11  & {\color[HTML]{000000} 51.29}  & {\color[HTML]{000000} 54.02}  & {\color[HTML]{000000} 54.13}  \\ \cline{3-9} 
\cellcolor[HTML]{EEFBED}{\color[HTML]{000000} }   & \cellcolor[HTML]{FFFFFF}{\color[HTML]{000000} }   & {\color[HTML]{000000} Cross-Atten}   & {\color[HTML]{000000} 50.67} & {\color[HTML]{000000} 55.45}  & 61.29   & {\color[HTML]{000000} 52.02}   & {\color[HTML]{000000} 54.28}   & {\color[HTML]{000000} 54.65}  \\ \cline{3-9} 
\cellcolor[HTML]{EEFBED}{\color[HTML]{000000} }   & \cellcolor[HTML]{FFFFFF}{\color[HTML]{000000} }  & {\color[HTML]{000000} MLP-Mixer}  & {\color[HTML]{000000} 51.49}   & {\color[HTML]{000000} \underline{56.31}}  & 60.40  & {\color[HTML]{000000} 52.27}  & {\color[HTML]{000000} \underline{55.92}}  & {\color[HTML]{000000} 54.73}  \\ \cline{3-9} 
\cellcolor[HTML]{EEFBED}{\color[HTML]{000000} }  & \cellcolor[HTML]{FFFFFF}{\color[HTML]{000000} }  & \cellcolor[HTML]{FFFFFF}{\color[HTML]{000000} CLIP-Align}& {\color[HTML]{000000} 51.09}  & {\color[HTML]{000000} 54.63} & 59.00 & {\color[HTML]{000000} 52.27}  & {\color[HTML]{000000} 53.65}& {\color[HTML]{000000} 54.82}   \\ \cline{3-9}
\multirow{-6}{*}{\cellcolor[HTML]{EEFBED}{\color[HTML]{000000} \begin{tabular}[c]{@{}c@{}}V-Face\\ +\\ V-AGA\\ +\\ A-Mel\end{tabular}}} & \multirow{-5}{*}{\cellcolor[HTML]{FFFFFF}{\color[HTML]{000000} Feature}} & {\color[HTML]{000000} \textbf{Atten-Mixer}} & {\color[HTML]{000000} \textbf{52.48}}  & {\color[HTML]{000000} 56.25} & \textbf{63.37}   & {\color[HTML]{000000} \textbf{54.91}}  & {\color[HTML]{000000}\textbf{56.09} }  & {\color[HTML]{000000} \textbf{56.82}} \\ \hline
\end{tabular}}
\end{table*}

\subsection{Domain-Simultaneous with Gradient Reversal Layer (GRL)}
\label{sec:GRL}

Gradient Reversal Layer (GRL)~\cite{ganin2015unsupervised} has become a foundational component in domain adaptation and domain generalization due to its ability to encourage domain-invariant feature learning. Its effectiveness and versatility have been demonstrated across diverse fields such as face anti-spoofing~\cite{wang2022domain}, sleep stage classification~\cite{jia2021multi}, multi-view 3D object detection~\cite{wang2023towards}, etc.

Following the implementation by Ganin and Lempitsky~\cite{ganin2015unsupervised}, we compared the multi-to-single domain generalization accuracies w/w.o GRL. GRL was proposed to mitigate the domain shift issue by manipulating the training gradients. It worked by acting as an identity transform in forward propagation and multiplying the gradient by a certain negative constant during the backpropagation without having trainable parameters. GRL was inserted between encoders and domain classifiers, which was easy to implement.  As GRL is a widely adopted method for domain generalization, it is investigated to show its effectiveness for the deception detection task. We selected domain-simultaneous as the baseline and added GRL to the original network with the same training setups. The average accuracies were reported in Fig.~\ref{fig:GRL}, where different types of visual and audio features and methods were compared. Training with GRL, the performance of ResNet18 and RN18+GRU models using visual (face frames) features and Wave2Vec model using waveform were enhanced. However, we observed that MLP models using visual (AU/gaze/affect) features and the ResNet18 model using Mel spectrograms degraded in performance. Generally, ResNet18 trained with GRL performed better than MLP for visual modality, and Wave2Vec trained with GRL boosted the performance and surpassed the model trained on the Mel spectrogram.

\subsection{Cross-domain Testing with Multimodal Fusion}
\label{fusion}



\begin{table*}[!t]
\centering
\caption{Fusion results for multi-to-single cross-domain accuracy (\%) using domain-simultaneous sampling.}
\resizebox{1.0\textwidth}{!}{
\begin{tabular}{|c
>{\columncolor[HTML]{FFFFFF}}c 
>{\columncolor[HTML]{FFFFFF}}c 
>{\columncolor[HTML]{FFFFFF}}c 
>{\columncolor[HTML]{FFFFFF}}c 
>{\columncolor[HTML]{FFFFFF}}c 
>{\columncolor[HTML]{FFFFFF}}c 
>{\columncolor[HTML]{FFFFFF}}c 
>{\columncolor[HTML]{FFFFFF}}c
>{\columncolor[HTML]{FFFFFF}}c 
>{\columncolor[HTML]{FFFFFF}}c
>{\columncolor[HTML]{FFFFFF}}c 
>{\columncolor[HTML]{FFFFFF}}c
>{\columncolor[HTML]{FFFFFF}}c|}
\hline
\multicolumn{14}{|c|}{\cellcolor[HTML]{EFEFEF}{\color[HTML]{000000} \textbf{Domain-Simultaneous}}} \\ \hline
\multicolumn{1}{|c|}{\cellcolor[HTML]{FFFFFF}{\color[HTML]{000000} \textbf{Input}}} & \multicolumn{1}{c|}{\cellcolor[HTML]{FFFFFF}{\color[HTML]{000000} \textbf{Type}}}         & \multicolumn{1}{c|}{\cellcolor[HTML]{FFFFFF}{\color[HTML]{000000} \textbf{Method}}}     
& \multicolumn{1}{c|}{\cellcolor[HTML]{FFFFFF}{\color[HTML]{000000} \textbf{R\&M to B1}}} 
& \multicolumn{1}{c|}{\cellcolor[HTML]{FFFFFF}{\color[HTML]{000000} \textbf{R\&M to B2}}} 
& \multicolumn{1}{c|}{\cellcolor[HTML]{FFFFFF}{\color[HTML]{000000} \textbf{R\&M to D}}} 
& \multicolumn{1}{c|}{\cellcolor[HTML]{FFFFFF}{\color[HTML]{000000} \textbf{R\&M to E}}} 
& \multicolumn{1}{c|}{\cellcolor[HTML]{FFFFFF}{\color[HTML]{000000} \textbf{R\&B1 to B2}}} 
& \multicolumn{1}{c|}{\cellcolor[HTML]{FFFFFF}{\color[HTML]{000000} \textbf{R\&B1 to M}}} 
& \multicolumn{1}{c|}{\cellcolor[HTML]{FFFFFF}{\color[HTML]{000000} \textbf{R\&B1 to D}}} 
& \multicolumn{1}{c|}{\cellcolor[HTML]{FFFFFF}{\color[HTML]{000000} \textbf{R\&B1 to E}}} 
& \multicolumn{1}{c|}{\cellcolor[HTML]{FFFFFF}{\color[HTML]{000000} \textbf{B1\&M to R}}} 
& \multicolumn{1}{c|}{\cellcolor[HTML]{FFFFFF}{\color[HTML]{000000} \textbf{B1\&M to B2}}} 
& \multicolumn{1}{c|}{\cellcolor[HTML]{FFFFFF}{\color[HTML]{000000} \textbf{\shortstack{R\&B1\&M\\to B2}}}} \\  \hline
\multicolumn{1}{|c|}{\cellcolor[HTML]{ECF4FF}{\color[HTML]{000000} }}   & \multicolumn{1}{c|}{\cellcolor[HTML]{FFFFFF}{\color[HTML]{000000} Score}}  & \multicolumn{1}{c|}{\cellcolor[HTML]{FFFFFF}{\color[HTML]{000000} Avg}}  & \multicolumn{1}{c|}{\cellcolor[HTML]{FFFFFF}{\color[HTML]{000000} 53.23}}   & \multicolumn{1}{c|}{\cellcolor[HTML]{FFFFFF}{\color[HTML]{000000} 43.56}}  & \multicolumn{1}{c|}{\cellcolor[HTML]{FFFFFF}{\color[HTML]{000000} 50.57}}& \multicolumn{1}{c|}{\cellcolor[HTML]{FFFFFF}{\color[HTML]{000000}50.16}} & \multicolumn{1}{c|}{\cellcolor[HTML]{FFFFFF}{\color[HTML]{000000} 57.43}}   & \multicolumn{1}{c|}{\cellcolor[HTML]{FFFFFF}{\color[HTML]{000000} 50.94}}  & \multicolumn{1}{c|}{\cellcolor[HTML]{FFFFFF}{\color[HTML]{000000}56.44}} & \multicolumn{1}{c|}{\cellcolor[HTML]{FFFFFF}{\color[HTML]{000000} 50.16}} & \multicolumn{1}{c|}{\cellcolor[HTML]{FFFFFF}{\color[HTML]{000000} 51.40}}  & \multicolumn{1}{c|}{\cellcolor[HTML]{FFFFFF}{\color[HTML]{000000} 58.42}}       & \multicolumn{1}{c|}{\cellcolor[HTML]{FFFFFF}{\color[HTML]{000000} 48.51}} \\ \cline{2-14} 
\multicolumn{1}{|c|}{\cellcolor[HTML]{ECF4FF}{\color[HTML]{000000} }}  & \multicolumn{1}{c|}{\cellcolor[HTML]{FFFFFF}{\color[HTML]{000000} }}  & \multicolumn{1}{c|}{\cellcolor[HTML]{FFFFFF}{\color[HTML]{000000} Concat}}       & \multicolumn{1}{c|}{\cellcolor[HTML]{FFFFFF}{\color[HTML]{000000} 52.00}}  & \multicolumn{1}{c|}{\cellcolor[HTML]{FFFFFF}{\color[HTML]{000000} 56.44}}  & \multicolumn{1}{c|}{\cellcolor[HTML]{FFFFFF}{\color[HTML]{000000} 51.29}} & \multicolumn{1}{c|}{\cellcolor[HTML]{FFFFFF}{\color[HTML]{000000} 50.86}}     & \multicolumn{1}{c|}{\cellcolor[HTML]{FFFFFF}{\color[HTML]{000000} 60.40}}  & \multicolumn{1}{c|}{\cellcolor[HTML]{FFFFFF}{\color[HTML]{000000} 51.88}} & \multicolumn{1}{c|}{\cellcolor[HTML]{FFFFFF}{\color[HTML]{000000} 57.04}} & \multicolumn{1}{c|}{\cellcolor[HTML]{FFFFFF}{\color[HTML]{000000} 51.70}} & \multicolumn{1}{c|}{\cellcolor[HTML]{FFFFFF}{\color[HTML]{000000} 51.40}}      & \multicolumn{1}{c|}{\cellcolor[HTML]{FFFFFF}{\color[HTML]{000000} \underline{62.38}}}  & \multicolumn{1}{c|}{\cellcolor[HTML]{FFFFFF}{\color[HTML]{000000} 57.43}} \\ \cline{3-14} 
\multicolumn{1}{|c|}{\cellcolor[HTML]{ECF4FF}{\color[HTML]{000000} }}  & \multicolumn{1}{c|}{\cellcolor[HTML]{FFFFFF}{\color[HTML]{000000} }}  & \multicolumn{1}{c|}{\cellcolor[HTML]{FFFFFF}{\color[HTML]{000000} SE-Concat}}   & \multicolumn{1}{c|}{\cellcolor[HTML]{FFFFFF}{\color[HTML]{000000} \textbf{55.69}}}  & \multicolumn{1}{c|}{\cellcolor[HTML]{FFFFFF}{\color[HTML]{000000} 58.42}} & \multicolumn{1}{c|}{\cellcolor[HTML]{FFFFFF}{\color[HTML]{000000} 52.08}}& \multicolumn{1}{c|}{\cellcolor[HTML]{FFFFFF}{\color[HTML]{000000}52.84}} & \multicolumn{1}{c|}{\cellcolor[HTML]{FFFFFF}{\color[HTML]{000000} 44.55}}       & \multicolumn{1}{c|}{\cellcolor[HTML]{FFFFFF}{\color[HTML]{000000} 52.50}}  &  \multicolumn{1}{c|}{\cellcolor[HTML]{FFFFFF}{\color[HTML]{000000}56.87}}&  \multicolumn{1}{c|}{\cellcolor[HTML]{FFFFFF}{\color[HTML]{000000}51.13}} & \multicolumn{1}{c|}{\cellcolor[HTML]{FFFFFF}{\color[HTML]{000000} 56.07}} & \multicolumn{1}{c|}{\cellcolor[HTML]{FFFFFF}{\color[HTML]{000000} 58.42}}        & \multicolumn{1}{c|}{\cellcolor[HTML]{FFFFFF}{\color[HTML]{000000} 58.42}} \\ \cline{3-14}
\multicolumn{1}{|c|}{\cellcolor[HTML]{ECF4FF}{\color[HTML]{000000} }} & \multicolumn{1}{c|}{\cellcolor[HTML]{FFFFFF}{\color[HTML]{000000} }}  & \multicolumn{1}{c|}{\cellcolor[HTML]{FFFFFF}{\color[HTML]{000000} Cross-Atten}}  & \multicolumn{1}{c|}{\cellcolor[HTML]{FFFFFF}{\color[HTML]{000000} 51.08}}  & \multicolumn{1}{c|}{\cellcolor[HTML]{FFFFFF}{\color[HTML]{000000} 52.48}}  &  \multicolumn{1}{c|}{\cellcolor[HTML]{FFFFFF}{\color[HTML]{000000}52.51}}&  \multicolumn{1}{c|}{\cellcolor[HTML]{FFFFFF}{\color[HTML]{000000}52.24}}     & \multicolumn{1}{c|}{\cellcolor[HTML]{FFFFFF}{\color[HTML]{000000} 55.45}}  & \multicolumn{1}{c|}{\cellcolor[HTML]{FFFFFF}{\color[HTML]{000000} 51.56}} &  \multicolumn{1}{c|}{\cellcolor[HTML]{FFFFFF}{\color[HTML]{000000}57.13}}&  \multicolumn{1}{c|}{\cellcolor[HTML]{FFFFFF}{\color[HTML]{000000}51.94}} & \multicolumn{1}{c|}{\cellcolor[HTML]{FFFFFF}{\color[HTML]{000000} 57.01}}       & \multicolumn{1}{c|}{\cellcolor[HTML]{FFFFFF}{\color[HTML]{000000} 60.40}}  & \multicolumn{1}{c|}{\cellcolor[HTML]{FFFFFF}{\color[HTML]{000000} 57.43}}  \\ \cline{3-14} 
\multicolumn{1}{|c|}{\cellcolor[HTML]{ECF4FF}{\color[HTML]{000000} }} & \multicolumn{1}{c|}{\cellcolor[HTML]{FFFFFF}{\color[HTML]{000000} }} & \multicolumn{1}{c|}{\cellcolor[HTML]{FFFFFF}{\color[HTML]{000000} MLP-Mixer}}    & \multicolumn{1}{c|}{\cellcolor[HTML]{FFFFFF}{\color[HTML]{000000} 53.85}}  & \multicolumn{1}{c|}{\cellcolor[HTML]{FFFFFF}{\color[HTML]{000000} 43.56}}  &  \multicolumn{1}{c|}{\cellcolor[HTML]{FFFFFF}{\color[HTML]{000000} 53.02}}&  \multicolumn{1}{c|}{\cellcolor[HTML]{FFFFFF}{\color[HTML]{000000} 53.32}}     & \multicolumn{1}{c|}{\cellcolor[HTML]{FFFFFF}{\color[HTML]{000000} \underline{62.38}}}  & \multicolumn{1}{c|}{\cellcolor[HTML]{FFFFFF}{\color[HTML]{000000} 52.19}} &  \multicolumn{1}{c|}{\cellcolor[HTML]{FFFFFF}{\color[HTML]{000000}57.74}}&  \multicolumn{1}{c|}{\cellcolor[HTML]{FFFFFF}{\color[HTML]{000000}52.10}} & \multicolumn{1}{c|}{\cellcolor[HTML]{FFFFFF}{\color[HTML]{000000} 56.07}} & \multicolumn{1}{c|}{\cellcolor[HTML]{FFFFFF}{\color[HTML]{000000} 61.39}} & \multicolumn{1}{c|}{\cellcolor[HTML]{FFFFFF}{\color[HTML]{000000} 59.41}}\\ \cline{3-14}
\multicolumn{1}{|c|}{\cellcolor[HTML]{ECF4FF}{\color[HTML]{000000} }} & \multicolumn{1}{c|}{\cellcolor[HTML]{FFFFFF}{\color[HTML]{000000} }} & \multicolumn{1}{c|}{\cellcolor[HTML]{FFFFFF}{\color[HTML]{000000} CLIP-Align}}    & \multicolumn{1}{c|}{\cellcolor[HTML]{FFFFFF}{\color[HTML]{000000} 52.58}}  & \multicolumn{1}{c|}{\cellcolor[HTML]{FFFFFF}{\color[HTML]{000000} 56.92}}  &  \multicolumn{1}{c|}{\cellcolor[HTML]{FFFFFF}{\color[HTML]{000000} 52.24}}&  \multicolumn{1}{c|}{\cellcolor[HTML]{FFFFFF}{\color[HTML]{000000} 51.17}}     & \multicolumn{1}{c|}{\cellcolor[HTML]{FFFFFF}{\color[HTML]{000000} 60.88}}  & \multicolumn{1}{c|}{\cellcolor[HTML]{FFFFFF}{\color[HTML]{000000} 52.46}} &  \multicolumn{1}{c|}{\cellcolor[HTML]{FFFFFF}{\color[HTML]{000000}57.86}}&  \multicolumn{1}{c|}{\cellcolor[HTML]{FFFFFF}{\color[HTML]{000000}52.50}} & \multicolumn{1}{c|}{\cellcolor[HTML]{FFFFFF}{\color[HTML]{000000} 51.68}} & \multicolumn{1}{c|}{\cellcolor[HTML]{FFFFFF}{\color[HTML]{000000} 62.60}} & \multicolumn{1}{c|}{\cellcolor[HTML]{FFFFFF}{\color[HTML]{000000} 58.22}}\\ \cline{3-14}
\multicolumn{1}{|c|}{\multirow{-6}{*}{\cellcolor[HTML]{ECF4FF}{\color[HTML]{000000} \begin{tabular}[c]{@{}c@{}}V-Face\\ +\\ V-AGA\end{tabular}}}}                               & \multicolumn{1}{c|}{\multirow{-5}{*}{\cellcolor[HTML]{FFFFFF}{\color[HTML]{000000} Feature}}} & \multicolumn{1}{c|}{\cellcolor[HTML]{FFFFFF}{\color[HTML]{000000} \textbf{Atten-Mixer}}} & \multicolumn{1}{c|}{\cellcolor[HTML]{FFFFFF}{\color[HTML]{000000} 53.23}}       & \multicolumn{1}{c|}{\cellcolor[HTML]{FFFFFF}{\color[HTML]{000000} 57.43}}&  \multicolumn{1}{c|}{\cellcolor[HTML]{FFFFFF}{\color[HTML]{000000} 53.50}}&  \multicolumn{1}{c|}{\cellcolor[HTML]{FFFFFF}{\color[HTML]{000000}53.40}} & \multicolumn{1}{c|}{\cellcolor[HTML]{FFFFFF}{\color[HTML]{000000} \textbf{63.37}}}  & \multicolumn{1}{c|}{\cellcolor[HTML]{FFFFFF}{\color[HTML]{000000} 52.19}} &  \multicolumn{1}{c|}{\cellcolor[HTML]{FFFFFF}{\color[HTML]{000000}57.39}}&  \multicolumn{1}{c|}{\cellcolor[HTML]{FFFFFF}{\color[HTML]{000000}51.86}} & \multicolumn{1}{c|}{\cellcolor[HTML]{FFFFFF}{\color[HTML]{000000} 57.01}}  & \multicolumn{1}{c|}{\cellcolor[HTML]{FFFFFF}{\color[HTML]{000000} \textbf{63.37}}}  & \multicolumn{1}{c|}{\cellcolor[HTML]{FFFFFF}{\color[HTML]{000000} \underline{62.38}}} \\ \hline
\multicolumn{1}{|c|}{\cellcolor[HTML]{FFFFEB}{\color[HTML]{000000} }} & \multicolumn{1}{c|}{\cellcolor[HTML]{FFFFFF}{\color[HTML]{000000} Score}}  & \multicolumn{1}{c|}{\cellcolor[HTML]{FFFFFF}{\color[HTML]{000000} Avg}}     & \multicolumn{1}{c|}{\cellcolor[HTML]{FFFFFF}{\color[HTML]{000000} 49.54}} & \multicolumn{1}{c|}{\cellcolor[HTML]{FFFFFF}{\color[HTML]{000000} 55.45}}&  \multicolumn{1}{c|}{\cellcolor[HTML]{FFFFFF}{\color[HTML]{000000} 50.73}} &  \multicolumn{1}{c|}{\cellcolor[HTML]{FFFFFF}{\color[HTML]{000000} 50.73}}& \multicolumn{1}{c|}{\cellcolor[HTML]{FFFFFF}{\color[HTML]{000000} 52.48}} & \multicolumn{1}{c|}{\cellcolor[HTML]{FFFFFF}{\color[HTML]{000000} 51.25}} &  \multicolumn{1}{c|}{\cellcolor[HTML]{FFFFFF}{\color[HTML]{000000}57.56}} &  \multicolumn{1}{c|}{\cellcolor[HTML]{FFFFFF}{\color[HTML]{000000}50.81}} & \multicolumn{1}{c|}{\cellcolor[HTML]{FFFFFF}{\color[HTML]{000000} 54.21}} & \multicolumn{1}{c|}{\cellcolor[HTML]{FFFFFF}{\color[HTML]{000000} 59.41}} & \multicolumn{1}{c|}{\cellcolor[HTML]{FFFFFF}{\color[HTML]{000000} 55.45}}  \\ \cline{2-14} 
\multicolumn{1}{|c|}{\cellcolor[HTML]{FFFFEB}{\color[HTML]{000000} }} & \multicolumn{1}{c|}{\cellcolor[HTML]{FFFFFF}{\color[HTML]{000000} }} & \multicolumn{1}{c|}{\cellcolor[HTML]{FFFFFF}{\color[HTML]{000000} Concat}} & \multicolumn{1}{c|}{\cellcolor[HTML]{FFFFFF}{\color[HTML]{000000} 49.54}} & \multicolumn{1}{c|}{\cellcolor[HTML]{FFFFFF}{\color[HTML]{000000} 54.46}} &  \multicolumn{1}{c|}{\cellcolor[HTML]{FFFFFF}{\color[HTML]{000000} 50.97}} &  \multicolumn{1}{c|}{\cellcolor[HTML]{FFFFFF}{\color[HTML]{000000} 50.92}} & \multicolumn{1}{c|}{\cellcolor[HTML]{FFFFFF}{\color[HTML]{000000} 49.50}}  & \multicolumn{1}{c|}{\cellcolor[HTML]{FFFFFF}{\color[HTML]{000000} \underline{53.75}}} &  \multicolumn{1}{c|}{\cellcolor[HTML]{FFFFFF}{\color[HTML]{000000}56.96}}&  \multicolumn{1}{c|}{\cellcolor[HTML]{FFFFFF}{\color[HTML]{000000}51.05}} & \multicolumn{1}{c|}{\cellcolor[HTML]{FFFFFF}{\color[HTML]{000000} 44.86}}  & \multicolumn{1}{c|}{\cellcolor[HTML]{FFFFFF}{\color[HTML]{000000} \textbf{63.37}}}  & \multicolumn{1}{c|}{\cellcolor[HTML]{FFFFFF}{\color[HTML]{000000} 54.46}}  \\ \cline{3-14} 
\multicolumn{1}{|c|}{\cellcolor[HTML]{FFFFEB}{\color[HTML]{000000} }} & \multicolumn{1}{c|}{\cellcolor[HTML]{FFFFFF}{\color[HTML]{000000} }}  & \multicolumn{1}{c|}{\cellcolor[HTML]{FFFFFF}{\color[HTML]{000000} SE-Concat}}   & \multicolumn{1}{c|}{\cellcolor[HTML]{FFFFFF}{\color[HTML]{000000} 52.00}} & \multicolumn{1}{c|}{\cellcolor[HTML]{FFFFFF}{\color[HTML]{000000} 58.42}} &  \multicolumn{1}{c|}{\cellcolor[HTML]{FFFFFF}{\color[HTML]{000000} 51.54}} &  \multicolumn{1}{c|}{\cellcolor[HTML]{FFFFFF}{\color[HTML]{000000} 52.00}} & \multicolumn{1}{c|}{\cellcolor[HTML]{FFFFFF}{\color[HTML]{000000} 42.57}} & \multicolumn{1}{c|}{\cellcolor[HTML]{FFFFFF}{\color[HTML]{000000} 51.56}} &  \multicolumn{1}{c|}{\cellcolor[HTML]{FFFFFF}{\color[HTML]{000000}57.22}} &  \multicolumn{1}{c|}{\cellcolor[HTML]{FFFFFF}{\color[HTML]{000000}51.46}}  & \multicolumn{1}{c|}{\cellcolor[HTML]{FFFFFF}{\color[HTML]{000000} 54.21}} & \multicolumn{1}{c|}{\cellcolor[HTML]{FFFFFF}{\color[HTML]{000000} 58.42}} & \multicolumn{1}{c|}{\cellcolor[HTML]{FFFFFF}{\color[HTML]{000000} \textbf{63.37}}} \\ \cline{3-14} 
\multicolumn{1}{|c|}{\cellcolor[HTML]{FFFFEB}{\color[HTML]{000000} }} & \multicolumn{1}{c|}{\cellcolor[HTML]{FFFFFF}{\color[HTML]{000000} }} & \multicolumn{1}{c|}{\cellcolor[HTML]{FFFFFF}{\color[HTML]{000000} Cross-Atten}}& \multicolumn{1}{c|}{\cellcolor[HTML]{FFFFFF}{\color[HTML]{000000} 50.46}}& \multicolumn{1}{c|}{\cellcolor[HTML]{FFFFFF}{\color[HTML]{000000} 52.48}} &  \multicolumn{1}{c|}{\cellcolor[HTML]{FFFFFF}{\color[HTML]{000000}52.02}}&  \multicolumn{1}{c|}{\cellcolor[HTML]{FFFFFF}{\color[HTML]{000000}53.08}}& \multicolumn{1}{c|}{\cellcolor[HTML]{FFFFFF}{\color[HTML]{000000} 60.40}} & \multicolumn{1}{c|}{\cellcolor[HTML]{FFFFFF}{\color[HTML]{000000} 53.12}}&  \multicolumn{1}{c|}{\cellcolor[HTML]{FFFFFF}{\color[HTML]{000000} 57.30}}&  \multicolumn{1}{c|}{\cellcolor[HTML]{FFFFFF}{\color[HTML]{000000}51.13}} & \multicolumn{1}{c|}{\cellcolor[HTML]{FFFFFF}{\color[HTML]{000000} 52.34}} & \multicolumn{1}{c|}{\cellcolor[HTML]{FFFFFF}{\color[HTML]{000000} 58.42}} & \multicolumn{1}{c|}{\cellcolor[HTML]{FFFFFF}{\color[HTML]{000000} 59.41}} \\ \cline{3-14}
\multicolumn{1}{|c|}{\cellcolor[HTML]{FFFFEB}{\color[HTML]{000000} }} & \multicolumn{1}{c|}{\cellcolor[HTML]{FFFFFF}{\color[HTML]{000000} }} & \multicolumn{1}{c|}{\cellcolor[HTML]{FFFFFF}{\color[HTML]{000000} MLP-Mixer}} & \multicolumn{1}{c|}{\cellcolor[HTML]{FFFFFF}{\color[HTML]{000000} 53.54}} & \multicolumn{1}{c|}{\cellcolor[HTML]{FFFFFF}{\color[HTML]{000000} 57.43}}&  \multicolumn{1}{c|}{\cellcolor[HTML]{FFFFFF}{\color[HTML]{000000}52.10}}&  \multicolumn{1}{c|}{\cellcolor[HTML]{FFFFFF}{\color[HTML]{000000}53.32}} & \multicolumn{1}{c|}{\cellcolor[HTML]{FFFFFF}{\color[HTML]{000000} 52.48}} & \multicolumn{1}{c|}{\cellcolor[HTML]{FFFFFF}{\color[HTML]{000000} 50.31}} &  \multicolumn{1}{c|}{\cellcolor[HTML]{FFFFFF}{\color[HTML]{000000}57.56}}&  \multicolumn{1}{c|}{\cellcolor[HTML]{FFFFFF}{\color[HTML]{000000}52.18}}& \multicolumn{1}{c|}{\cellcolor[HTML]{FFFFFF}{\color[HTML]{000000} 52.34}} & \multicolumn{1}{c|}{\cellcolor[HTML]{FFFFFF}{\color[HTML]{000000} 60.40}} & \multicolumn{1}{c|}{\cellcolor[HTML]{FFFFFF}{\color[HTML]{000000} 47.52}} \\ \cline{3-14}
\multicolumn{1}{|c|}{\cellcolor[HTML]{FFFFEB}{\color[HTML]{000000} }} & \multicolumn{1}{c|}{\cellcolor[HTML]{FFFFFF}{\color[HTML]{000000} }} & \multicolumn{1}{c|}{\cellcolor[HTML]{FFFFFF}{\color[HTML]{000000} CLIP-Align}}    & \multicolumn{1}{c|}{\cellcolor[HTML]{FFFFFF}{\color[HTML]{000000} 50.31}}  & \multicolumn{1}{c|}{\cellcolor[HTML]{FFFFFF}{\color[HTML]{000000} 55.06}}  &  \multicolumn{1}{c|}{\cellcolor[HTML]{FFFFFF}{\color[HTML]{000000} 51.43}}&  \multicolumn{1}{c|}{\cellcolor[HTML]{FFFFFF}{\color[HTML]{000000} 51.27}}     & \multicolumn{1}{c|}{\cellcolor[HTML]{FFFFFF}{\color[HTML]{000000} 50.53}}  & \multicolumn{1}{c|}{\cellcolor[HTML]{FFFFFF}{\color[HTML]{000000} 53.99}} &  \multicolumn{1}{c|}{\cellcolor[HTML]{FFFFFF}{\color[HTML]{000000}57.15}}&  \multicolumn{1}{c|}{\cellcolor[HTML]{FFFFFF}{\color[HTML]{000000}52.42}} & \multicolumn{1}{c|}{\cellcolor[HTML]{FFFFFF}{\color[HTML]{000000} 48.33}} & \multicolumn{1}{c|}{\cellcolor[HTML]{FFFFFF}{\color[HTML]{000000} 62.89}} & \multicolumn{1}{c|}{\cellcolor[HTML]{FFFFFF}{\color[HTML]{000000} 55.68}}\\ \cline{3-14}
\multicolumn{1}{|c|}{\multirow{-6}{*}{\cellcolor[HTML]{FFFFEB}{\color[HTML]{000000} \begin{tabular}[c]{@{}c@{}}V-Face\\ +\\ A-Mel\end{tabular}}}}                               & \multicolumn{1}{c|}{\multirow{-5}{*}{\cellcolor[HTML]{FFFFFF}{\color[HTML]{000000} Feature}}} & \multicolumn{1}{c|}{\cellcolor[HTML]{FFFFFF}{\color[HTML]{000000} \textbf{Atten-Mixer}}} & \multicolumn{1}{c|}{\cellcolor[HTML]{FFFFFF}{\color[HTML]{000000} \textbf{55.69}}} & \multicolumn{1}{c|}{\cellcolor[HTML]{FFFFFF}{\color[HTML]{000000} \textbf{64.36}}}&  \multicolumn{1}{c|}{\cellcolor[HTML]{FFFFFF}{\color[HTML]{000000} 53.97}} &  \multicolumn{1}{c|}{\cellcolor[HTML]{FFFFFF}{\color[HTML]{000000} 53.43}} & \multicolumn{1}{c|}{\cellcolor[HTML]{FFFFFF}{\color[HTML]{000000} 56.44}} & \multicolumn{1}{c|}{\cellcolor[HTML]{FFFFFF}{\color[HTML]{000000} \underline{53.75}}} &  \multicolumn{1}{c|}{\cellcolor[HTML]{FFFFFF}{\color[HTML]{000000}57.82}}&  \multicolumn{1}{c|}{\cellcolor[HTML]{FFFFFF}{\color[HTML]{000000}52.35}}& \multicolumn{1}{c|}{\cellcolor[HTML]{FFFFFF}{\color[HTML]{000000} 58.88}}      & \multicolumn{1}{c|}{\cellcolor[HTML]{FFFFFF}{\color[HTML]{000000} 59.41}} & \multicolumn{1}{c|}{\cellcolor[HTML]{FFFFFF}{\color[HTML]{000000} 58.42}} \\ \hline
\multicolumn{1}{|c|}{\cellcolor[HTML]{FFF7F7}{\color[HTML]{000000} }} & \multicolumn{1}{c|}{\cellcolor[HTML]{FFFFFF}{\color[HTML]{000000} Score}} & \multicolumn{1}{c|}{\cellcolor[HTML]{FFFFFF}{\color[HTML]{000000} Avg}}     & \multicolumn{1}{c|}{\cellcolor[HTML]{FFFFFF}{\color[HTML]{000000} 48.62}} & \multicolumn{1}{c|}{\cellcolor[HTML]{FFFFFF}{\color[HTML]{000000} 58.42}} &  \multicolumn{1}{c|}{\cellcolor[HTML]{FFFFFF}{\color[HTML]{000000}51.21}}&  \multicolumn{1}{c|}{\cellcolor[HTML]{FFFFFF}{\color[HTML]{000000}50.29}}& \multicolumn{1}{c|}{\cellcolor[HTML]{FFFFFF}{\color[HTML]{000000} 56.44}} & \multicolumn{1}{c|}{\cellcolor[HTML]{FFFFFF}{\color[HTML]{000000} 50.94}} &  \multicolumn{1}{c|}{\cellcolor[HTML]{FFFFFF}{\color[HTML]{000000}56.48}}&  \multicolumn{1}{c|}{\cellcolor[HTML]{FFFFFF}{\color[HTML]{000000}51.78}}& \multicolumn{1}{c|}{\cellcolor[HTML]{FFFFFF}{\color[HTML]{000000} 51.40}} & \multicolumn{1}{c|}{\cellcolor[HTML]{FFFFFF}{\color[HTML]{000000} 56.44}} & \multicolumn{1}{c|}{\cellcolor[HTML]{FFFFFF}{\color[HTML]{000000} 53.47}} \\ \cline{2-14}
\multicolumn{1}{|c|}{\cellcolor[HTML]{FFF7F7}{\color[HTML]{000000} }} & \multicolumn{1}{c|}{\cellcolor[HTML]{FFFFFF}{\color[HTML]{000000} }} & \multicolumn{1}{c|}{\cellcolor[HTML]{FFFFFF}{\color[HTML]{000000} Concat}} & \multicolumn{1}{c|}{\cellcolor[HTML]{FFFFFF}{\color[HTML]{000000} 48.31}} & \multicolumn{1}{c|}{\cellcolor[HTML]{FFFFFF}{\color[HTML]{000000} 53.47}}&  \multicolumn{1}{c|}{\cellcolor[HTML]{FFFFFF}{\color[HTML]{000000}51.86}}&  \multicolumn{1}{c|}{\cellcolor[HTML]{FFFFFF}{\color[HTML]{000000}50.44}} & \multicolumn{1}{c|}{\cellcolor[HTML]{FFFFFF}{\color[HTML]{000000} 56.44}} & \multicolumn{1}{c|}{\cellcolor[HTML]{FFFFFF}{\color[HTML]{000000} 50.31}}&  \multicolumn{1}{c|}{\cellcolor[HTML]{FFFFFF}{\color[HTML]{000000}57.13}}&  \multicolumn{1}{c|}{\cellcolor[HTML]{FFFFFF}{\color[HTML]{000000}51.29}} & \multicolumn{1}{c|}{\cellcolor[HTML]{FFFFFF}{\color[HTML]{000000} 56.07}} & \multicolumn{1}{c|}{\cellcolor[HTML]{FFFFFF}{\color[HTML]{000000} \textbf{63.37}}} & \multicolumn{1}{c|}{\cellcolor[HTML]{FFFFFF}{\color[HTML]{000000} 57.43}}  \\ \cline{3-14}
\multicolumn{1}{|c|}{\cellcolor[HTML]{FFF7F7}{\color[HTML]{000000} }}& \multicolumn{1}{c|}{\cellcolor[HTML]{FFFFFF}{\color[HTML]{000000} }} & \multicolumn{1}{c|}{\cellcolor[HTML]{FFFFFF}{\color[HTML]{000000} SE-Concat}}  & \multicolumn{1}{c|}{\cellcolor[HTML]{FFFFFF}{\color[HTML]{000000} 49.85}} & \multicolumn{1}{c|}{\cellcolor[HTML]{FFFFFF}{\color[HTML]{000000} 50.50}}&  \multicolumn{1}{c|}{\cellcolor[HTML]{FFFFFF}{\color[HTML]{000000}53.87}}&  \multicolumn{1}{c|}{\cellcolor[HTML]{FFFFFF}{\color[HTML]{000000}52.32}} & \multicolumn{1}{c|}{\cellcolor[HTML]{FFFFFF}{\color[HTML]{000000} 47.52}} & \multicolumn{1}{c|}{\cellcolor[HTML]{FFFFFF}{\color[HTML]{000000} 52.19}}&  \multicolumn{1}{c|}{\cellcolor[HTML]{FFFFFF}{\color[HTML]{000000}57.42}}&  \multicolumn{1}{c|}{\cellcolor[HTML]{FFFFFF}{\color[HTML]{000000}52.02}} & \multicolumn{1}{c|}{\cellcolor[HTML]{FFFFFF}{\color[HTML]{000000} 49.53}} & \multicolumn{1}{c|}{\cellcolor[HTML]{FFFFFF}{\color[HTML]{000000} 57.43}} & \multicolumn{1}{c|}{\cellcolor[HTML]{FFFFFF}{\color[HTML]{000000} 61.39}}  \\ \cline{3-14} 
\multicolumn{1}{|c|}{\cellcolor[HTML]{FFF7F7}{\color[HTML]{000000} }} & \multicolumn{1}{c|}{\cellcolor[HTML]{FFFFFF}{\color[HTML]{000000} }} & \multicolumn{1}{c|}{\cellcolor[HTML]{FFFFFF}{\color[HTML]{000000} Cross-Atten}} & \multicolumn{1}{c|}{\cellcolor[HTML]{FFFFFF}{\color[HTML]{000000} 51.38}} & \multicolumn{1}{c|}{\cellcolor[HTML]{FFFFFF}{\color[HTML]{000000} \underline{63.37}}}&  \multicolumn{1}{c|}{\cellcolor[HTML]{FFFFFF}{\color[HTML]{000000}54.50}}&  \multicolumn{1}{c|}{\cellcolor[HTML]{FFFFFF}{\color[HTML]{000000}53.24}} & \multicolumn{1}{c|}{\cellcolor[HTML]{FFFFFF}{\color[HTML]{000000} 61.39}} & \multicolumn{1}{c|}{\cellcolor[HTML]{FFFFFF}{\color[HTML]{000000} 51.25}} &  \multicolumn{1}{c|}{\cellcolor[HTML]{FFFFFF}{\color[HTML]{000000} 57.74}}&  \multicolumn{1}{c|}{\cellcolor[HTML]{FFFFFF}{\color[HTML]{000000}52.27}}& \multicolumn{1}{c|}{\cellcolor[HTML]{FFFFFF}{\color[HTML]{000000} 48.60}} & \multicolumn{1}{c|}{\cellcolor[HTML]{FFFFFF}{\color[HTML]{000000} 59.41}} & \multicolumn{1}{c|}{\cellcolor[HTML]{FFFFFF}{\color[HTML]{000000} 58.42}}  \\ \cline{3-14} 
\multicolumn{1}{|c|}{\cellcolor[HTML]{FFF7F7}{\color[HTML]{000000} }} & \multicolumn{1}{c|}{\cellcolor[HTML]{FFFFFF}{\color[HTML]{000000} }} & \multicolumn{1}{c|}{\cellcolor[HTML]{FFFFFF}{\color[HTML]{000000} MLP-Mixer}}  & \multicolumn{1}{c|}{\cellcolor[HTML]{FFFFFF}{\color[HTML]{000000} \textbf{55.69}}} & \multicolumn{1}{c|}{\cellcolor[HTML]{FFFFFF}{\color[HTML]{000000} 59.41}}  &  \multicolumn{1}{c|}{\cellcolor[HTML]{FFFFFF}{\color[HTML]{000000} 55.06}}&  \multicolumn{1}{c|}{\cellcolor[HTML]{FFFFFF}{\color[HTML]{000000}53.24}}    & \multicolumn{1}{c|}{\cellcolor[HTML]{FFFFFF}{\color[HTML]{000000} 57.43}} & \multicolumn{1}{c|}{\cellcolor[HTML]{FFFFFF}{\color[HTML]{000000} \underline{53.75}}} &  \multicolumn{1}{c|}{\cellcolor[HTML]{FFFFFF}{\color[HTML]{000000}57.43}}&  \multicolumn{1}{c|}{\cellcolor[HTML]{FFFFFF}{\color[HTML]{000000}52.99}}& \multicolumn{1}{c|}{\cellcolor[HTML]{FFFFFF}{\color[HTML]{000000} 50.47}} & \multicolumn{1}{c|}{\cellcolor[HTML]{FFFFFF}{\color[HTML]{000000} 61.39}} & \multicolumn{1}{c|}{\cellcolor[HTML]{FFFFFF}{\color[HTML]{000000} 50.50}} \\ \cline{3-14}
\multicolumn{1}{|c|}{\cellcolor[HTML]{FFF7F7}{\color[HTML]{000000} }} & \multicolumn{1}{c|}{\cellcolor[HTML]{FFFFFF}{\color[HTML]{000000} }} & \multicolumn{1}{c|}{\cellcolor[HTML]{FFFFFF}{\color[HTML]{000000} CLIP-Align}}    & \multicolumn{1}{c|}{\cellcolor[HTML]{FFFFFF}{\color[HTML]{000000} 51.36}}  & \multicolumn{1}{c|}{\cellcolor[HTML]{FFFFFF}{\color[HTML]{000000} 54.35}}  &  \multicolumn{1}{c|}{\cellcolor[HTML]{FFFFFF}{\color[HTML]{000000} 52.43}}&  \multicolumn{1}{c|}{\cellcolor[HTML]{FFFFFF}{\color[HTML]{000000} 51.66}}     & \multicolumn{1}{c|}{\cellcolor[HTML]{FFFFFF}{\color[HTML]{000000} 56.87}}  & \multicolumn{1}{c|}{\cellcolor[HTML]{FFFFFF}{\color[HTML]{000000} 51.37}} &  \multicolumn{1}{c|}{\cellcolor[HTML]{FFFFFF}{\color[HTML]{000000}57.86}}&  \multicolumn{1}{c|}{\cellcolor[HTML]{FFFFFF}{\color[HTML]{000000}52.50}} & \multicolumn{1}{c|}{\cellcolor[HTML]{FFFFFF}{\color[HTML]{000000} 56.38}} & \multicolumn{1}{c|}{\cellcolor[HTML]{FFFFFF}{\color[HTML]{000000} 63.00}} & \multicolumn{1}{c|}{\cellcolor[HTML]{FFFFFF}{\color[HTML]{000000} 58.62}}\\ \cline{3-14}
\multicolumn{1}{|c|}{\multirow{-6}{*}{\cellcolor[HTML]{FFF7F7}{\color[HTML]{000000} \begin{tabular}[c]{@{}c@{}}V-AGA\\ +\\ A-Mel\end{tabular}}}} & \multicolumn{1}{c|}{\multirow{-5}{*}{\cellcolor[HTML]{FFFFFF}{\color[HTML]{000000} Feature}}} & \multicolumn{1}{c|}{\cellcolor[HTML]{FFFFFF}{\color[HTML]{000000} \textbf{Atten-Mixer}}} & \multicolumn{1}{c|}{\cellcolor[HTML]{FFFFFF}{\color[HTML]{000000} \underline{55.08}}} & \multicolumn{1}{c|}{\cellcolor[HTML]{FFFFFF}{\color[HTML]{000000} 55.45}}&  \multicolumn{1}{c|}{\cellcolor[HTML]{FFFFFF}{\color[HTML]{000000} \textbf{57.39}}}&  \multicolumn{1}{c|}{\cellcolor[HTML]{FFFFFF}{\color[HTML]{000000}\underline{54.20}}} & \multicolumn{1}{c|}{\cellcolor[HTML]{FFFFFF}{\color[HTML]{000000} 58.42}} & \multicolumn{1}{c|}{\cellcolor[HTML]{FFFFFF}{\color[HTML]{000000} \textbf{54.37}}} &  \multicolumn{1}{c|}{\cellcolor[HTML]{FFFFFF}{\color[HTML]{000000}58.02}}&  \multicolumn{1}{c|}{\cellcolor[HTML]{FFFFFF}{\color[HTML]{000000}52.83}}& \multicolumn{1}{c|}{\cellcolor[HTML]{FFFFFF}{\color[HTML]{000000} 52.34}} & \multicolumn{1}{c|}{\cellcolor[HTML]{FFFFFF}{\color[HTML]{000000} 61.39}} & \multicolumn{1}{c|}{\cellcolor[HTML]{FFFFFF}{\color[HTML]{000000} 60.40}}\\ \hline
\multicolumn{1}{|c|}{\cellcolor[HTML]{EEFBED}{\color[HTML]{000000} }} & \multicolumn{1}{c|}{\cellcolor[HTML]{FFFFFF}{\color[HTML]{000000} Score}} & \multicolumn{1}{c|}{\cellcolor[HTML]{FFFFFF}{\color[HTML]{000000} Avg}}    & \multicolumn{1}{c|}{\cellcolor[HTML]{FFFFFF}{\color[HTML]{000000} 54.77}} & \multicolumn{1}{c|}{\cellcolor[HTML]{FFFFFF}{\color[HTML]{000000} 57.43}} &  \multicolumn{1}{c|}{\cellcolor[HTML]{FFFFFF}{\color[HTML]{000000} 55.50}} &  \multicolumn{1}{c|}{\cellcolor[HTML]{FFFFFF}{\color[HTML]{000000} 50.90}} & \multicolumn{1}{c|}{\cellcolor[HTML]{FFFFFF}{\color[HTML]{000000} 57.43}}      & \multicolumn{1}{c|}{\cellcolor[HTML]{FFFFFF}{\color[HTML]{000000} 52.81}} &  \multicolumn{1}{c|}{\cellcolor[HTML]{FFFFFF}{\color[HTML]{000000} 55.45}} &  \multicolumn{1}{c|}{\cellcolor[HTML]{FFFFFF}{\color[HTML]{000000}51.46}} & \multicolumn{1}{c|}{\cellcolor[HTML]{FFFFFF}{\color[HTML]{000000} \textbf{60.75}}} & \multicolumn{1}{c|}{\cellcolor[HTML]{FFFFFF}{\color[HTML]{000000} 58.42}} & \multicolumn{1}{c|}{\cellcolor[HTML]{FFFFFF}{\color[HTML]{000000} 56.44}}  \\ \cline{2-14}
\multicolumn{1}{|c|}{\cellcolor[HTML]{EEFBED}{\color[HTML]{000000} }}& \multicolumn{1}{c|}{\cellcolor[HTML]{FFFFFF}{\color[HTML]{000000} }} & \multicolumn{1}{c|}{\cellcolor[HTML]{FFFFFF}{\color[HTML]{000000} Concat}} & \multicolumn{1}{c|}{\cellcolor[HTML]{FFFFFF}{\color[HTML]{000000} 54.46}} & \multicolumn{1}{c|}{\cellcolor[HTML]{FFFFFF}{\color[HTML]{000000} 54.46}} &  \multicolumn{1}{c|}{\cellcolor[HTML]{FFFFFF}{\color[HTML]{000000} 55.52}} &  \multicolumn{1}{c|}{\cellcolor[HTML]{FFFFFF}{\color[HTML]{000000} 50.82}} & \multicolumn{1}{c|}{\cellcolor[HTML]{FFFFFF}{\color[HTML]{000000} 47.52}} & \multicolumn{1}{c|}{\cellcolor[HTML]{FFFFFF}{\color[HTML]{000000} 53.12}} &  \multicolumn{1}{c|}{\cellcolor[HTML]{FFFFFF}{\color[HTML]{000000}55.85}} &  \multicolumn{1}{c|}{\cellcolor[HTML]{FFFFFF}{\color[HTML]{000000}51.78}} & \multicolumn{1}{c|}{\cellcolor[HTML]{FFFFFF}{\color[HTML]{000000} 53.27}} & \multicolumn{1}{c|}{\cellcolor[HTML]{FFFFFF}{\color[HTML]{000000} 61.39}} & \multicolumn{1}{c|}{\cellcolor[HTML]{FFFFFF}{\color[HTML]{000000} 56.44}}  \\ \cline{3-14}
\multicolumn{1}{|c|}{\cellcolor[HTML]{EEFBED}{\color[HTML]{000000} }} & \multicolumn{1}{c|}{\cellcolor[HTML]{FFFFFF}{\color[HTML]{000000} }} & \multicolumn{1}{c|}{\cellcolor[HTML]{FFFFFF}{\color[HTML]{000000} SE-Concat}} & \multicolumn{1}{c|}{\cellcolor[HTML]{FFFFFF}{\color[HTML]{000000} 53.23}} & \multicolumn{1}{c|}{\cellcolor[HTML]{FFFFFF}{\color[HTML]{000000} 55.45}} &  \multicolumn{1}{c|}{\cellcolor[HTML]{FFFFFF}{\color[HTML]{000000}56.00}} &  \multicolumn{1}{c|}{\cellcolor[HTML]{FFFFFF}{\color[HTML]{000000}51.60}} & \multicolumn{1}{c|}{\cellcolor[HTML]{FFFFFF}{\color[HTML]{000000} 54.46}} & \multicolumn{1}{c|}{\cellcolor[HTML]{FFFFFF}{\color[HTML]{000000} 52.50}}&  \multicolumn{1}{c|}{\cellcolor[HTML]{FFFFFF}{\color[HTML]{000000}56.33}} &  \multicolumn{1}{c|}{\cellcolor[HTML]{FFFFFF}{\color[HTML]{000000} 51.05}} & \multicolumn{1}{c|}{\cellcolor[HTML]{FFFFFF}{\color[HTML]{000000} \underline{59.81}}} & \multicolumn{1}{c|}{\cellcolor[HTML]{FFFFFF}{\color[HTML]{000000} 56.44}} & \multicolumn{1}{c|}{\cellcolor[HTML]{FFFFFF}{\color[HTML]{000000} 58.42}} \\ \cline{3-14} 
\multicolumn{1}{|c|}{\cellcolor[HTML]{EEFBED}{\color[HTML]{000000} }} & \multicolumn{1}{c|}{\cellcolor[HTML]{FFFFFF}{\color[HTML]{000000} }} & \multicolumn{1}{c|}{\cellcolor[HTML]{FFFFFF}{\color[HTML]{000000} Cross-Atten}} & \multicolumn{1}{c|}{\cellcolor[HTML]{FFFFFF}{\color[HTML]{000000} 48.62}} & \multicolumn{1}{c|}{\cellcolor[HTML]{FFFFFF}{\color[HTML]{000000} 57.43}} &  \multicolumn{1}{c|}{\cellcolor[HTML]{FFFFFF}{\color[HTML]{000000} 56.26}} &  \multicolumn{1}{c|}{\cellcolor[HTML]{FFFFFF}{\color[HTML]{000000} 52.18}} & \multicolumn{1}{c|}{\cellcolor[HTML]{FFFFFF}{\color[HTML]{000000} 50.50}} & \multicolumn{1}{c|}{\cellcolor[HTML]{FFFFFF}{\color[HTML]{000000} 53.12}} &  \multicolumn{1}{c|}{\cellcolor[HTML]{FFFFFF}{\color[HTML]{000000}57.46}} &  \multicolumn{1}{c|}{\cellcolor[HTML]{FFFFFF}{\color[HTML]{000000} 51.62}} & \multicolumn{1}{c|}{\cellcolor[HTML]{FFFFFF}{\color[HTML]{000000} 48.60}} & \multicolumn{1}{c|}{\cellcolor[HTML]{FFFFFF}{\color[HTML]{000000} 58.42}} & \multicolumn{1}{c|}{\cellcolor[HTML]{FFFFFF}{\color[HTML]{000000} 56.44}}   \\ \cline{3-14} 
\multicolumn{1}{|c|}{\cellcolor[HTML]{EEFBED}{\color[HTML]{000000} }} & \multicolumn{1}{c|}{\cellcolor[HTML]{FFFFFF}{\color[HTML]{000000} }} & \multicolumn{1}{c|}{\cellcolor[HTML]{FFFFFF}{\color[HTML]{000000} MLP-Mixer}} & \multicolumn{1}{c|}{\cellcolor[HTML]{FFFFFF}{\color[HTML]{000000} 48.92}} & \multicolumn{1}{c|}{\cellcolor[HTML]{FFFFFF}{\color[HTML]{000000} 48.51}}&  \multicolumn{1}{c|}{\cellcolor[HTML]{FFFFFF}{\color[HTML]{000000} 56.61}} &  \multicolumn{1}{c|}{\cellcolor[HTML]{FFFFFF}{\color[HTML]{000000}53.68}} & \multicolumn{1}{c|}{\cellcolor[HTML]{FFFFFF}{\color[HTML]{000000} 59.41}} & \multicolumn{1}{c|}{\cellcolor[HTML]{FFFFFF}{\color[HTML]{000000} 52.81}} &  \multicolumn{1}{c|}{\cellcolor[HTML]{FFFFFF}{\color[HTML]{000000}\underline{58.40}}}&  \multicolumn{1}{c|}{\cellcolor[HTML]{FFFFFF}{\color[HTML]{000000}52.21}}& \multicolumn{1}{c|}{\cellcolor[HTML]{FFFFFF}{\color[HTML]{000000} 54.21}} & \multicolumn{1}{c|}{\cellcolor[HTML]{FFFFFF}{\color[HTML]{000000} 57.43}} & \multicolumn{1}{c|}{\cellcolor[HTML]{FFFFFF}{\color[HTML]{000000} 51.49}}  \\ \cline{3-14} 
\multicolumn{1}{|c|}{\cellcolor[HTML]{EEFBED}{\color[HTML]{000000} }} & \multicolumn{1}{c|}{\cellcolor[HTML]{FFFFFF}{\color[HTML]{000000} }} & \multicolumn{1}{c|}{\cellcolor[HTML]{FFFFFF}{\color[HTML]{000000} CLIP-Align}}    & \multicolumn{1}{c|}{\cellcolor[HTML]{FFFFFF}{\color[HTML]{000000} 55.29}}  & \multicolumn{1}{c|}{\cellcolor[HTML]{FFFFFF}{\color[HTML]{000000} 54.83}}  &  \multicolumn{1}{c|}{\cellcolor[HTML]{FFFFFF}{\color[HTML]{000000} 56.63}}&  \multicolumn{1}{c|}{\cellcolor[HTML]{FFFFFF}{\color[HTML]{000000} 51.46}}     & \multicolumn{1}{c|}{\cellcolor[HTML]{FFFFFF}{\color[HTML]{000000} 47.74}}  & \multicolumn{1}{c|}{\cellcolor[HTML]{FFFFFF}{\color[HTML]{000000} 54.07}} &  \multicolumn{1}{c|}{\cellcolor[HTML]{FFFFFF}{\color[HTML]{000000}56.33}}&  \multicolumn{1}{c|}{\cellcolor[HTML]{FFFFFF}{\color[HTML]{000000}\textbf{52.97}}} & \multicolumn{1}{c|}{\cellcolor[HTML]{FFFFFF}{\color[HTML]{000000} 53.98}} & \multicolumn{1}{c|}{\cellcolor[HTML]{FFFFFF}{\color[HTML]{000000} 61.73}} & \multicolumn{1}{c|}{\cellcolor[HTML]{FFFFFF}{\color[HTML]{000000} 57.47}}\\ \cline{3-14}
\multicolumn{1}{|c|}{\multirow{-6}{*}{\cellcolor[HTML]{EEFBED}{\color[HTML]{000000} \begin{tabular}[c]{@{}c@{}}V-Face\\ +\\ V-AGA\\ +\\ A-Mel\end{tabular}}}} & \multicolumn{1}{c|}{\multirow{-5}{*}{\cellcolor[HTML]{FFFFFF}{\color[HTML]{000000} Feature}}} & \multicolumn{1}{c|}{\cellcolor[HTML]{FFFFFF}{\color[HTML]{000000} \textbf{Atten-Mixer}}} & \multicolumn{1}{c|}{\cellcolor[HTML]{FFFFFF}{\color[HTML]{000000} 52.31}}       & \multicolumn{1}{c|}{\cellcolor[HTML]{FFFFFF}{\color[HTML]{000000} 59.41}} &  \multicolumn{1}{c|}{\cellcolor[HTML]{FFFFFF}{\color[HTML]{000000}\underline{57.30}}}&  \multicolumn{1}{c|}{\cellcolor[HTML]{FFFFFF}{\color[HTML]{000000}\textbf{54.57}}}& \multicolumn{1}{c|}{\cellcolor[HTML]{FFFFFF}{\color[HTML]{000000} 58.42}} & \multicolumn{1}{c|}{\cellcolor[HTML]{FFFFFF}{\color[HTML]{000000} 52.50}} &  \multicolumn{1}{c|}{\cellcolor[HTML]{FFFFFF}{\color[HTML]{000000} \textbf{59.41}}}&  \multicolumn{1}{c|}{\cellcolor[HTML]{FFFFFF}{\color[HTML]{000000} \underline{52.84}}}& \multicolumn{1}{c|}{\cellcolor[HTML]{FFFFFF}{\color[HTML]{000000} 58.88}} & \multicolumn{1}{c|}{\cellcolor[HTML]{FFFFFF}{\color[HTML]{000000} 61.39}} & \multicolumn{1}{c|}{\cellcolor[HTML]{FFFFFF}{\color[HTML]{000000} 60.40}} \\ \hline

\multicolumn{1}{|c|}{\cellcolor[HTML]{FFFFFF}{\color[HTML]{000000} \textbf{Input}}} & \multicolumn{1}{c|}{\cellcolor[HTML]{FFFFFF}{\color[HTML]{000000} \textbf{Type}}}         & \multicolumn{1}{c|}{\cellcolor[HTML]{FFFFFF}{\color[HTML]{000000} \textbf{Method}}}     & \multicolumn{1}{c|}{\cellcolor[HTML]{FFFFFF}{\color[HTML]{000000} \textbf{D\&E to B1}}} & \multicolumn{1}{c|}{\cellcolor[HTML]{FFFFFF}{\color[HTML]{000000} \textbf{D\&E to B2}}} & \multicolumn{1}{c|}{\cellcolor[HTML]{FFFFFF}{\color[HTML]{000000} \textbf{D\&E to M}}} & \multicolumn{1}{c|}{\cellcolor[HTML]{FFFFFF}{\color[HTML]{000000} \textbf{D\&E to R}}} & \multicolumn{1}{c|}{\cellcolor[HTML]{FFFFFF}{\color[HTML]{000000} \textbf{\shortstack{R\&B1\&M\\to D}}}} & \multicolumn{1}{c|}{\cellcolor[HTML]{FFFFFF}{\color[HTML]{000000} \textbf{\shortstack{R\&B1\&M\\to E}}}} & \multicolumn{1}{c|}{\cellcolor[HTML]{FFFFFF}{\color[HTML]{000000} \textbf{\shortstack{R\&D\&E\\to B1}}}} & {\color[HTML]{000000} \textbf{\shortstack{R\&D\&E\\to M}}}         & {\color[HTML]{000000} \textbf{\shortstack{R\&M\&D\&E\\to B2}}}  & {\color[HTML]{000000} \textbf{\shortstack{R\&B1\&M\&\\D\&E to B2}}}     & {\color[HTML]{000000} \textbf{Avg}}       \\  \hline
\multicolumn{1}{|c|}{\cellcolor[HTML]{ECF4FF}{\color[HTML]{000000} }}   & \multicolumn{1}{c|}{\cellcolor[HTML]{FFFFFF}{\color[HTML]{000000} Score}}  & \multicolumn{1}{c|}{\cellcolor[HTML]{FFFFFF}{\color[HTML]{000000} Avg}}     & \multicolumn{1}{c|}{\cellcolor[HTML]{FFFFFF}{\color[HTML]{000000} 50.66}}   & \multicolumn{1}{c|}{\cellcolor[HTML]{FFFFFF}{\color[HTML]{000000} 53.56}}   & \multicolumn{1}{c|}{\cellcolor[HTML]{FFFFFF}{\color[HTML]{000000} 52.50}}   & \multicolumn{1}{c|}{\cellcolor[HTML]{FFFFFF}{\color[HTML]{000000} 52.34}}  & \multicolumn{1}{c|}{\cellcolor[HTML]{FFFFFF}{\color[HTML]{000000} 49.96}}  & \multicolumn{1}{c|}{\cellcolor[HTML]{FFFFFF}{\color[HTML]{000000} 50.47}}       & \multicolumn{1}{c|}{\cellcolor[HTML]{FFFFFF}{\color[HTML]{000000} 52.29}}  & \multicolumn{1}{c|}{\cellcolor[HTML]{FFFFFF}{\color[HTML]{000000}50.44}} &\multicolumn{1}{c|}{\cellcolor[HTML]{FFFFFF}{\color[HTML]{000000} 53.24}}&\multicolumn{1}{c|}{\cellcolor[HTML]{FFFFFF}{\color[HTML]{000000}51.19}}& \multicolumn{1}{c|}{\cellcolor[HTML]{FFFFFF}{\color[HTML]{000000}51.18}}\\ \cline{2-14} 
\multicolumn{1}{|c|}{\cellcolor[HTML]{ECF4FF}{\color[HTML]{000000} }}  & \multicolumn{1}{c|}{\cellcolor[HTML]{FFFFFF}{\color[HTML]{000000} }}  & \multicolumn{1}{c|}{\cellcolor[HTML]{FFFFFF}{\color[HTML]{000000} Concat}}       & \multicolumn{1}{c|}{\cellcolor[HTML]{FFFFFF}{\color[HTML]{000000} 51.38}}  & \multicolumn{1}{c|}{\cellcolor[HTML]{FFFFFF}{\color[HTML]{000000} 54.46}}       & \multicolumn{1}{c|}{\cellcolor[HTML]{FFFFFF}{\color[HTML]{000000} 50.31}}  & \multicolumn{1}{c|}{\cellcolor[HTML]{FFFFFF}{\color[HTML]{000000} 52.31}}  & \multicolumn{1}{c|}{\cellcolor[HTML]{FFFFFF}{\color[HTML]{000000} 53.33}}      & \multicolumn{1}{c|}{\cellcolor[HTML]{FFFFFF}{\color[HTML]{000000} 52.18}}  & \multicolumn{1}{c|}{\cellcolor[HTML]{FFFFFF}{\color[HTML]{000000} 53.40}}  & \multicolumn{1}{c|}{\cellcolor[HTML]{FFFFFF}{\color[HTML]{000000} 51.25}} &\multicolumn{1}{c|}{\cellcolor[HTML]{FFFFFF}{\color[HTML]{000000} 53.24}} &\multicolumn{1}{c|}{\cellcolor[HTML]{FFFFFF}{\color[HTML]{000000} 51.19}} &\multicolumn{1}{c|}{\cellcolor[HTML]{FFFFFF}{\color[HTML]{000000} 51.18}} \\ \cline{3-14} 
\multicolumn{1}{|c|}{\cellcolor[HTML]{ECF4FF}{\color[HTML]{000000} }}  & \multicolumn{1}{c|}{\cellcolor[HTML]{FFFFFF}{\color[HTML]{000000} }}  & \multicolumn{1}{c|}{\cellcolor[HTML]{FFFFFF}{\color[HTML]{000000} SE-Concat}}   & \multicolumn{1}{c|}{\cellcolor[HTML]{FFFFFF}{\color[HTML]{000000} 51.49}}  & \multicolumn{1}{c|}{\cellcolor[HTML]{FFFFFF}{\color[HTML]{000000} 54.86}}  & \multicolumn{1}{c|}{\cellcolor[HTML]{FFFFFF}{\color[HTML]{000000} 51.85}}       & \multicolumn{1}{c|}{\cellcolor[HTML]{FFFFFF}{\color[HTML]{000000} 52.20}}  & \multicolumn{1}{c|}{\cellcolor[HTML]{FFFFFF}{\color[HTML]{000000} 55.49}} & \multicolumn{1}{c|}{\cellcolor[HTML]{FFFFFF}{\color[HTML]{000000} 53.07}}        & \multicolumn{1}{c|}{\cellcolor[HTML]{FFFFFF}{\color[HTML]{000000} 55.15}}  & \multicolumn{1}{c|}{\cellcolor[HTML]{FFFFFF}{\color[HTML]{000000} 51.88}} &\multicolumn{1}{c|}{\cellcolor[HTML]{FFFFFF}{\color[HTML]{000000} 54.87}} &\multicolumn{1}{c|}{\cellcolor[HTML]{FFFFFF}{\color[HTML]{000000} 54.81}} & \multicolumn{1}{c|}{\cellcolor[HTML]{FFFFFF}{\color[HTML]{000000} 53.94}}\\ \cline{3-14}
\multicolumn{1}{|c|}{\cellcolor[HTML]{ECF4FF}{\color[HTML]{000000} }} & \multicolumn{1}{c|}{\cellcolor[HTML]{FFFFFF}{\color[HTML]{000000} }}  & \multicolumn{1}{c|}{\cellcolor[HTML]{FFFFFF}{\color[HTML]{000000} Cross-Atten}}  & \multicolumn{1}{c|}{\cellcolor[HTML]{FFFFFF}{\color[HTML]{000000} 52.00}}  & \multicolumn{1}{c|}{\cellcolor[HTML]{FFFFFF}{\color[HTML]{000000} 55.45}}       & \multicolumn{1}{c|}{\cellcolor[HTML]{FFFFFF}{\color[HTML]{000000} 52.06}}  & \multicolumn{1}{c|}{\cellcolor[HTML]{FFFFFF}{\color[HTML]{000000} 53.20}}  & \multicolumn{1}{c|}{\cellcolor[HTML]{FFFFFF}{\color[HTML]{000000} 56.70}}       & \multicolumn{1}{c|}{\cellcolor[HTML]{FFFFFF}{\color[HTML]{000000} 54.96}}  & \multicolumn{1}{c|}{\cellcolor[HTML]{FFFFFF}{\color[HTML]{000000} 55.45}}  & \multicolumn{1}{c|}{\cellcolor[HTML]{FFFFFF}{\color[HTML]{000000} 52.31}} &\multicolumn{1}{c|}{\cellcolor[HTML]{FFFFFF}{\color[HTML]{000000}55.31}}&\multicolumn{1}{c|}{\cellcolor[HTML]{FFFFFF}{\color[HTML]{000000}56.56}}&\multicolumn{1}{c|}{\cellcolor[HTML]{FFFFFF}{\color[HTML]{000000} 54.44}} \\ \cline{3-14} 
\multicolumn{1}{|c|}{\cellcolor[HTML]{ECF4FF}{\color[HTML]{000000} }} & \multicolumn{1}{c|}{\cellcolor[HTML]{FFFFFF}{\color[HTML]{000000} }} & \multicolumn{1}{c|}{\cellcolor[HTML]{FFFFFF}{\color[HTML]{000000} MLP-Mixer}}    & \multicolumn{1}{c|}{\cellcolor[HTML]{FFFFFF}{\color[HTML]{000000} 52.48}}  & \multicolumn{1}{c|}{\cellcolor[HTML]{FFFFFF}{\color[HTML]{000000} 56.42}}       & \multicolumn{1}{c|}{\cellcolor[HTML]{FFFFFF}{\color[HTML]{000000} 52.94}}  & \multicolumn{1}{c|}{\cellcolor[HTML]{FFFFFF}{\color[HTML]{000000} 53.27}}  & \multicolumn{1}{c|}{\cellcolor[HTML]{FFFFFF}{\color[HTML]{000000} 56.96}} & \multicolumn{1}{c|}{\cellcolor[HTML]{FFFFFF}{\color[HTML]{000000} 56.44}} & \multicolumn{1}{c|}{\cellcolor[HTML]{FFFFFF}{\color[HTML]{000000} 56.44}}& \multicolumn{1}{c|}{\cellcolor[HTML]{FFFFFF}{\color[HTML]{000000} 53.13}} &\multicolumn{1}{c|}{\cellcolor[HTML]{FFFFFF}{\color[HTML]{000000}57.24}}&\multicolumn{1}{c|}{\cellcolor[HTML]{FFFFFF}{\color[HTML]{000000}57.98}}&\multicolumn{1}{c|}{\cellcolor[HTML]{FFFFFF}{\color[HTML]{000000}55.16}} \\ \cline{3-14}
\multicolumn{1}{|c|}{\cellcolor[HTML]{ECF4FF}{\color[HTML]{000000} }} & \multicolumn{1}{c|}{\cellcolor[HTML]{FFFFFF}{\color[HTML]{000000} }} & \multicolumn{1}{c|}{\cellcolor[HTML]{FFFFFF}{\color[HTML]{000000} CLIP-Align}}    & \multicolumn{1}{c|}{\cellcolor[HTML]{FFFFFF}{\color[HTML]{000000} 52.12}}  & \multicolumn{1}{c|}{\cellcolor[HTML]{FFFFFF}{\color[HTML]{000000} 55.23}}       & \multicolumn{1}{c|}{\cellcolor[HTML]{FFFFFF}{\color[HTML]{000000} 51.05}}  & \multicolumn{1}{c|}{\cellcolor[HTML]{FFFFFF}{\color[HTML]{000000} 53.18}}  & \multicolumn{1}{c|}{\cellcolor[HTML]{FFFFFF}{\color[HTML]{000000} 54.02}} & \multicolumn{1}{c|}{\cellcolor[HTML]{FFFFFF}{\color[HTML]{000000} 52.91}} & \multicolumn{1}{c|}{\cellcolor[HTML]{FFFFFF}{\color[HTML]{000000} 54.31}}& \multicolumn{1}{c|}{\cellcolor[HTML]{FFFFFF}{\color[HTML]{000000} 51.92}} &\multicolumn{1}{c|}{\cellcolor[HTML]{FFFFFF}{\color[HTML]{000000}54.06}}&\multicolumn{1}{c|}{\cellcolor[HTML]{FFFFFF}{\color[HTML]{000000}51.77}}&\multicolumn{1}{c|}{\cellcolor[HTML]{FFFFFF}{\color[HTML]{000000}54.26}} \\ \cline{3-14}
\multicolumn{1}{|c|}{\multirow{-6}{*}{\cellcolor[HTML]{ECF4FF}{\color[HTML]{000000} \begin{tabular}[c]{@{}c@{}}V-Face\\ +\\ V-AGA\end{tabular}}}}                               & \multicolumn{1}{c|}{\multirow{-5}{*}{\cellcolor[HTML]{FFFFFF}{\color[HTML]{000000} Feature}}} & \multicolumn{1}{c|}{\cellcolor[HTML]{FFFFFF}{\color[HTML]{000000} \textbf{Atten-Mixer}}} & \multicolumn{1}{c|}{\cellcolor[HTML]{FFFFFF}{\color[HTML]{000000} 53.47}}       & \multicolumn{1}{c|}{\cellcolor[HTML]{FFFFFF}{\color[HTML]{000000} 59.01}} & \multicolumn{1}{c|}{\cellcolor[HTML]{FFFFFF}{\color[HTML]{000000} 53.62}}  & \multicolumn{1}{c|}{\cellcolor[HTML]{FFFFFF}{\color[HTML]{000000} 54.21}}  & \multicolumn{1}{c|}{\cellcolor[HTML]{FFFFFF}{\color[HTML]{000000} 58.43}}  & \multicolumn{1}{c|}{\cellcolor[HTML]{FFFFFF}{\color[HTML]{000000} 57.10}}  & \multicolumn{1}{c|}{\cellcolor[HTML]{FFFFFF}{\color[HTML]{000000} 57.54}}  & \multicolumn{1}{c|}{\cellcolor[HTML]{FFFFFF}{\color[HTML]{000000} 53.75}} &\multicolumn{1}{c|}{\cellcolor[HTML]{FFFFFF}{\color[HTML]{000000} 60.40}} &\multicolumn{1}{c|}{\cellcolor[HTML]{FFFFFF}{\color[HTML]{000000} \textbf{60.99}}} &\multicolumn{1}{c|}{\cellcolor[HTML]{FFFFFF}{\color[HTML]{000000}56.84}}\\ \hline
\multicolumn{1}{|c|}{\cellcolor[HTML]{FFFFEB}{\color[HTML]{000000} }} & \multicolumn{1}{c|}{\cellcolor[HTML]{FFFFFF}{\color[HTML]{000000} Score}}  & \multicolumn{1}{c|}{\cellcolor[HTML]{FFFFFF}{\color[HTML]{000000} Avg}}     & \multicolumn{1}{c|}{\cellcolor[HTML]{FFFFFF}{\color[HTML]{000000} 50.50}} & \multicolumn{1}{c|}{\cellcolor[HTML]{FFFFFF}{\color[HTML]{000000} 53.47}} & \multicolumn{1}{c|}{\cellcolor[HTML]{FFFFFF}{\color[HTML]{000000} 51.25}}      & \multicolumn{1}{c|}{\cellcolor[HTML]{FFFFFF}{\color[HTML]{000000} 52.64}} & \multicolumn{1}{c|}{\cellcolor[HTML]{FFFFFF}{\color[HTML]{000000} 54.80}} & \multicolumn{1}{c|}{\cellcolor[HTML]{FFFFFF}{\color[HTML]{000000} 50.21}} & \multicolumn{1}{c|}{\cellcolor[HTML]{FFFFFF}{\color[HTML]{000000} 52.47}}  & \multicolumn{1}{c|}{\cellcolor[HTML]{FFFFFF}{\color[HTML]{000000} 50.62}} &\multicolumn{1}{c|}{\cellcolor[HTML]{FFFFFF}{\color[HTML]{000000} 53.94}} &\multicolumn{1}{c|}{\cellcolor[HTML]{FFFFFF}{\color[HTML]{000000} 55.00}} & \multicolumn{1}{c|}{\cellcolor[HTML]{FFFFFF}{\color[HTML]{000000}52.98}} \\ \cline{2-14} 
\multicolumn{1}{|c|}{\cellcolor[HTML]{FFFFEB}{\color[HTML]{000000} }} & \multicolumn{1}{c|}{\cellcolor[HTML]{FFFFFF}{\color[HTML]{000000} }} & \multicolumn{1}{c|}{\cellcolor[HTML]{FFFFFF}{\color[HTML]{000000} Concat}} & \multicolumn{1}{c|}{\cellcolor[HTML]{FFFFFF}{\color[HTML]{000000} 52.34}} & \multicolumn{1}{c|}{\cellcolor[HTML]{FFFFFF}{\color[HTML]{000000} 53.87}}  & \multicolumn{1}{c|}{\cellcolor[HTML]{FFFFFF}{\color[HTML]{000000} 50.62}}      & \multicolumn{1}{c|}{\cellcolor[HTML]{FFFFFF}{\color[HTML]{000000} 52.98}}  & \multicolumn{1}{c|}{\cellcolor[HTML]{FFFFFF}{\color[HTML]{000000} 56.78}}  & \multicolumn{1}{c|}{\cellcolor[HTML]{FFFFFF}{\color[HTML]{000000} 53.48}}  & \multicolumn{1}{c|}{\cellcolor[HTML]{FFFFFF}{\color[HTML]{000000} 54.26}}  & \multicolumn{1}{c|}{\cellcolor[HTML]{FFFFFF}{\color[HTML]{000000} 51.55}} &\multicolumn{1}{c|}{\cellcolor[HTML]{FFFFFF}{\color[HTML]{000000}53.69}}&\multicolumn{1}{c|}{\cellcolor[HTML]{FFFFFF}{\color[HTML]{000000}56.16}}&\multicolumn{1}{c|}{\cellcolor[HTML]{FFFFFF}{\color[HTML]{000000}53.21}} \\ \cline{3-14} 
\multicolumn{1}{|c|}{\cellcolor[HTML]{FFFFEB}{\color[HTML]{000000} }} & \multicolumn{1}{c|}{\cellcolor[HTML]{FFFFFF}{\color[HTML]{000000} }}  & \multicolumn{1}{c|}{\cellcolor[HTML]{FFFFFF}{\color[HTML]{000000} SE-Concat}}   & \multicolumn{1}{c|}{\cellcolor[HTML]{FFFFFF}{\color[HTML]{000000} 52.10}} & \multicolumn{1}{c|}{\cellcolor[HTML]{FFFFFF}{\color[HTML]{000000} 55.15}} & \multicolumn{1}{c|}{\cellcolor[HTML]{FFFFFF}{\color[HTML]{000000} 51.38}} & \multicolumn{1}{c|}{\cellcolor[HTML]{FFFFFF}{\color[HTML]{000000} 53.85}} & \multicolumn{1}{c|}{\cellcolor[HTML]{FFFFFF}{\color[HTML]{000000} 58.08}} & \multicolumn{1}{c|}{\cellcolor[HTML]{FFFFFF}{\color[HTML]{000000} 55.75}} & \multicolumn{1}{c|}{\cellcolor[HTML]{FFFFFF}{\color[HTML]{000000} 54.35}} & \multicolumn{1}{c|}{\cellcolor[HTML]{FFFFFF}{\color[HTML]{000000} 53.25}} &\multicolumn{1}{c|}{\cellcolor[HTML]{FFFFFF}{\color[HTML]{000000}54.06}}&\multicolumn{1}{c|}{\cellcolor[HTML]{FFFFFF}{\color[HTML]{000000}56.26}}&\multicolumn{1}{c|}{\cellcolor[HTML]{FFFFFF}{\color[HTML]{000000}54.14}} \\ \cline{3-14} 
\multicolumn{1}{|c|}{\cellcolor[HTML]{FFFFEB}{\color[HTML]{000000} }} & \multicolumn{1}{c|}{\cellcolor[HTML]{FFFFFF}{\color[HTML]{000000} }} & \multicolumn{1}{c|}{\cellcolor[HTML]{FFFFFF}{\color[HTML]{000000} Cross-Atten}}& \multicolumn{1}{c|}{\cellcolor[HTML]{FFFFFF}{\color[HTML]{000000} 53.84}}& \multicolumn{1}{c|}{\cellcolor[HTML]{FFFFFF}{\color[HTML]{000000} 56.44}} & \multicolumn{1}{c|}{\cellcolor[HTML]{FFFFFF}{\color[HTML]{000000} 52.94}} & \multicolumn{1}{c|}{\cellcolor[HTML]{FFFFFF}{\color[HTML]{000000} 54.61}} & \multicolumn{1}{c|}{\cellcolor[HTML]{FFFFFF}{\color[HTML]{000000} 58.17}} & \multicolumn{1}{c|}{\cellcolor[HTML]{FFFFFF}{\color[HTML]{000000} 57.39}} & \multicolumn{1}{c|}{\cellcolor[HTML]{FFFFFF}{\color[HTML]{000000} 55.20}} & \multicolumn{1}{c|}{\cellcolor[HTML]{FFFFFF}{\color[HTML]{000000} 53.69}} &\multicolumn{1}{c|}{\cellcolor[HTML]{FFFFFF}{\color[HTML]{000000}55.13}}&\multicolumn{1}{c|}{\cellcolor[HTML]{FFFFFF}{\color[HTML]{000000}57.15}}&\multicolumn{1}{c|}{\cellcolor[HTML]{FFFFFF}{\color[HTML]{000000}54.99}} \\ \cline{3-14}
\multicolumn{1}{|c|}{\cellcolor[HTML]{FFFFEB}{\color[HTML]{000000} }} & \multicolumn{1}{c|}{\cellcolor[HTML]{FFFFFF}{\color[HTML]{000000} }} & \multicolumn{1}{c|}{\cellcolor[HTML]{FFFFFF}{\color[HTML]{000000} MLP-Mixer}} & \multicolumn{1}{c|}{\cellcolor[HTML]{FFFFFF}{\color[HTML]{000000} 53.18}} & \multicolumn{1}{c|}{\cellcolor[HTML]{FFFFFF}{\color[HTML]{000000} 56.74}} & \multicolumn{1}{c|}{\cellcolor[HTML]{FFFFFF}{\color[HTML]{000000} 53.38}} & \multicolumn{1}{c|}{\cellcolor[HTML]{FFFFFF}{\color[HTML]{000000} 54.35}} & \multicolumn{1}{c|}{\cellcolor[HTML]{FFFFFF}{\color[HTML]{000000} 58.30}} & \multicolumn{1}{c|}{\cellcolor[HTML]{FFFFFF}{\color[HTML]{000000} 58.42}} & \multicolumn{1}{c|}{\cellcolor[HTML]{FFFFFF}{\color[HTML]{000000} 57.37}} & \multicolumn{1}{c|}{\cellcolor[HTML]{FFFFFF}{\color[HTML]{000000} 54.06}} &\multicolumn{1}{c|}{\cellcolor[HTML]{FFFFFF}{\color[HTML]{000000}59.94}}&\multicolumn{1}{c|}{\cellcolor[HTML]{FFFFFF}{\color[HTML]{000000}58.22}}& \multicolumn{1}{c|}{\cellcolor[HTML]{FFFFFF}{\color[HTML]{000000}54.91}}\\ \cline{3-14}
\multicolumn{1}{|c|}{\cellcolor[HTML]{FFFFEB}{\color[HTML]{000000} }} & \multicolumn{1}{c|}{\cellcolor[HTML]{FFFFFF}{\color[HTML]{000000} }} & \multicolumn{1}{c|}{\cellcolor[HTML]{FFFFFF}{\color[HTML]{000000} CLIP-Align}}    & \multicolumn{1}{c|}{\cellcolor[HTML]{FFFFFF}{\color[HTML]{000000} 53.21}}  & \multicolumn{1}{c|}{\cellcolor[HTML]{FFFFFF}{\color[HTML]{000000} 54.62}}       & \multicolumn{1}{c|}{\cellcolor[HTML]{FFFFFF}{\color[HTML]{000000} 51.30}}  & \multicolumn{1}{c|}{\cellcolor[HTML]{FFFFFF}{\color[HTML]{000000} 53.85}}  & \multicolumn{1}{c|}{\cellcolor[HTML]{FFFFFF}{\color[HTML]{000000} 57.61}} & \multicolumn{1}{c|}{\cellcolor[HTML]{FFFFFF}{\color[HTML]{000000} 54.22}} & \multicolumn{1}{c|}{\cellcolor[HTML]{FFFFFF}{\color[HTML]{000000} 54.94}}& \multicolumn{1}{c|}{\cellcolor[HTML]{FFFFFF}{\color[HTML]{000000} 52.43}} &\multicolumn{1}{c|}{\cellcolor[HTML]{FFFFFF}{\color[HTML]{000000}54.51}}&\multicolumn{1}{c|}{\cellcolor[HTML]{FFFFFF}{\color[HTML]{000000}56.89}}&\multicolumn{1}{c|}{\cellcolor[HTML]{FFFFFF}{\color[HTML]{000000}53.94}} \\ \cline{3-14}
\multicolumn{1}{|c|}{\multirow{-6}{*}{\cellcolor[HTML]{FFFFEB}{\color[HTML]{000000} \begin{tabular}[c]{@{}c@{}}V-Face\\ +\\ A-Mel\end{tabular}}}}                               & \multicolumn{1}{c|}{\multirow{-5}{*}{\cellcolor[HTML]{FFFFFF}{\color[HTML]{000000} Feature}}} & \multicolumn{1}{c|}{\cellcolor[HTML]{FFFFFF}{\color[HTML]{000000} \textbf{Atten-Mixer}}} & \multicolumn{1}{c|}{\cellcolor[HTML]{FFFFFF}{\color[HTML]{000000} 54.24}} & \multicolumn{1}{c|}{\cellcolor[HTML]{FFFFFF}{\color[HTML]{000000} 59.41}} & \multicolumn{1}{c|}{\cellcolor[HTML]{FFFFFF}{\color[HTML]{000000} 53.50}} & \multicolumn{1}{c|}{\cellcolor[HTML]{FFFFFF}{\color[HTML]{000000} 55.14}} & \multicolumn{1}{c|}{\cellcolor[HTML]{FFFFFF}{\color[HTML]{000000} 58.53}}      & \multicolumn{1}{c|}{\cellcolor[HTML]{FFFFFF}{\color[HTML]{000000} \underline{58.65}}} & \multicolumn{1}{c|}{\cellcolor[HTML]{FFFFFF}{\color[HTML]{000000} 58.92}}  & \multicolumn{1}{c|}{\cellcolor[HTML]{FFFFFF}{\color[HTML]{000000} \textbf{54.88}}} &\multicolumn{1}{c|}{\cellcolor[HTML]{FFFFFF}{\color[HTML]{000000} \underline{62.38}}}&\multicolumn{1}{c|}{\cellcolor[HTML]{FFFFFF}{\color[HTML]{000000} 59.35}}&\multicolumn{1}{c|}{\cellcolor[HTML]{FFFFFF}{\color[HTML]{000000} \underline{57.12}}} \\ \hline
\multicolumn{1}{|c|}{\cellcolor[HTML]{FFF7F7}{\color[HTML]{000000} }} & \multicolumn{1}{c|}{\cellcolor[HTML]{FFFFFF}{\color[HTML]{000000} Score}} & \multicolumn{1}{c|}{\cellcolor[HTML]{FFFFFF}{\color[HTML]{000000} Avg}}     & \multicolumn{1}{c|}{\cellcolor[HTML]{FFFFFF}{\color[HTML]{000000} 52.16}} & \multicolumn{1}{c|}{\cellcolor[HTML]{FFFFFF}{\color[HTML]{000000} 56.44}} & \multicolumn{1}{c|}{\cellcolor[HTML]{FFFFFF}{\color[HTML]{000000} 51.81}} & \multicolumn{1}{c|}{\cellcolor[HTML]{FFFFFF}{\color[HTML]{000000} 53.50}} & \multicolumn{1}{c|}{\cellcolor[HTML]{FFFFFF}{\color[HTML]{000000} 52.72}} & \multicolumn{1}{c|}{\cellcolor[HTML]{FFFFFF}{\color[HTML]{000000} 50.14}} & \multicolumn{1}{c|}{\cellcolor[HTML]{FFFFFF}{\color[HTML]{000000} 57.00}} & \multicolumn{1}{c|}{\cellcolor[HTML]{FFFFFF}{\color[HTML]{000000} 50.94}} &\multicolumn{1}{c|}{\cellcolor[HTML]{FFFFFF}{\color[HTML]{000000} 53.62}} &\multicolumn{1}{c|}{\cellcolor[HTML]{FFFFFF}{\color[HTML]{000000}54.98}}&\multicolumn{1}{c|}{\cellcolor[HTML]{FFFFFF}{\color[HTML]{000000} 53.28}} \\ \cline{2-14}
\multicolumn{1}{|c|}{\cellcolor[HTML]{FFF7F7}{\color[HTML]{000000} }} & \multicolumn{1}{c|}{\cellcolor[HTML]{FFFFFF}{\color[HTML]{000000} }} & \multicolumn{1}{c|}{\cellcolor[HTML]{FFFFFF}{\color[HTML]{000000} Concat}} & \multicolumn{1}{c|}{\cellcolor[HTML]{FFFFFF}{\color[HTML]{000000} 52.08}} & \multicolumn{1}{c|}{\cellcolor[HTML]{FFFFFF}{\color[HTML]{000000} 56.04}} & \multicolumn{1}{c|}{\cellcolor[HTML]{FFFFFF}{\color[HTML]{000000} 51.20}} & \multicolumn{1}{c|}{\cellcolor[HTML]{FFFFFF}{\color[HTML]{000000} 53.44}} & \multicolumn{1}{c|}{\cellcolor[HTML]{FFFFFF}{\color[HTML]{000000} 56.87}} & \multicolumn{1}{c|}{\cellcolor[HTML]{FFFFFF}{\color[HTML]{000000} 50.77}} & \multicolumn{1}{c|}{\cellcolor[HTML]{FFFFFF}{\color[HTML]{000000} 57.50}} & \multicolumn{1}{c|}{\cellcolor[HTML]{FFFFFF}{\color[HTML]{000000} 51.57}} &\multicolumn{1}{c|}{\cellcolor[HTML]{FFFFFF}{\color[HTML]{000000} 54.69}} &\multicolumn{1}{c|}{\cellcolor[HTML]{FFFFFF}{\color[HTML]{000000} 57.39}} & \multicolumn{1}{c|}{\cellcolor[HTML]{FFFFFF}{\color[HTML]{000000} 54.18}} \\ \cline{3-14}
\multicolumn{1}{|c|}{\cellcolor[HTML]{FFF7F7}{\color[HTML]{000000} }}& \multicolumn{1}{c|}{\cellcolor[HTML]{FFFFFF}{\color[HTML]{000000} }} & \multicolumn{1}{c|}{\cellcolor[HTML]{FFFFFF}{\color[HTML]{000000} SE-Concat}}  & \multicolumn{1}{c|}{\cellcolor[HTML]{FFFFFF}{\color[HTML]{000000} 53.21}} & \multicolumn{1}{c|}{\cellcolor[HTML]{FFFFFF}{\color[HTML]{000000} 57.27}} & \multicolumn{1}{c|}{\cellcolor[HTML]{FFFFFF}{\color[HTML]{000000} 52.75}} & \multicolumn{1}{c|}{\cellcolor[HTML]{FFFFFF}{\color[HTML]{000000} 54.36}} & \multicolumn{1}{c|}{\cellcolor[HTML]{FFFFFF}{\color[HTML]{000000} 57.13}} & \multicolumn{1}{c|}{\cellcolor[HTML]{FFFFFF}{\color[HTML]{000000} 56.09}} & \multicolumn{1}{c|}{\cellcolor[HTML]{FFFFFF}{\color[HTML]{000000} 58.51}} & \multicolumn{1}{c|}{\cellcolor[HTML]{FFFFFF}{\color[HTML]{000000} 52.69}} &\multicolumn{1}{c|}{\cellcolor[HTML]{FFFFFF}{\color[HTML]{000000} 55.75}} &\multicolumn{1}{c|}{\cellcolor[HTML]{FFFFFF}{\color[HTML]{000000} 58.22}} &\multicolumn{1}{c|}{\cellcolor[HTML]{FFFFFF}{\color[HTML]{000000} 54.29}} \\ \cline{3-14} 
\multicolumn{1}{|c|}{\cellcolor[HTML]{FFF7F7}{\color[HTML]{000000} }} & \multicolumn{1}{c|}{\cellcolor[HTML]{FFFFFF}{\color[HTML]{000000} }} & \multicolumn{1}{c|}{\cellcolor[HTML]{FFFFFF}{\color[HTML]{000000} Cross-Atten}} & \multicolumn{1}{c|}{\cellcolor[HTML]{FFFFFF}{\color[HTML]{000000} 53.85}} & \multicolumn{1}{c|}{\cellcolor[HTML]{FFFFFF}{\color[HTML]{000000} 59.64}} & \multicolumn{1}{c|}{\cellcolor[HTML]{FFFFFF}{\color[HTML]{000000} 52.38}} & \multicolumn{1}{c|}{\cellcolor[HTML]{FFFFFF}{\color[HTML]{000000} 54.45}} & \multicolumn{1}{c|}{\cellcolor[HTML]{FFFFFF}{\color[HTML]{000000} 57.74}} & \multicolumn{1}{c|}{\cellcolor[HTML]{FFFFFF}{\color[HTML]{000000} 56.44}} & \multicolumn{1}{c|}{\cellcolor[HTML]{FFFFFF}{\color[HTML]{000000} 58.49}} & \multicolumn{1}{c|}{\cellcolor[HTML]{FFFFFF}{\color[HTML]{000000} 53.38}} &\multicolumn{1}{c|}{\cellcolor[HTML]{FFFFFF}{\color[HTML]{000000}57.25}}&\multicolumn{1}{c|}{\cellcolor[HTML]{FFFFFF}{\color[HTML]{000000}59.61}}&\multicolumn{1}{c|}{\cellcolor[HTML]{FFFFFF}{\color[HTML]{000000}55.93}} \\ \cline{3-14} 
\multicolumn{1}{|c|}{\cellcolor[HTML]{FFF7F7}{\color[HTML]{000000} }} & \multicolumn{1}{c|}{\cellcolor[HTML]{FFFFFF}{\color[HTML]{000000} }} & \multicolumn{1}{c|}{\cellcolor[HTML]{FFFFFF}{\color[HTML]{000000} MLP-Mixer}}  & \multicolumn{1}{c|}{\cellcolor[HTML]{FFFFFF}{\color[HTML]{000000} 54.14}} & \multicolumn{1}{c|}{\cellcolor[HTML]{FFFFFF}{\color[HTML]{000000} 60.40}}      & \multicolumn{1}{c|}{\cellcolor[HTML]{FFFFFF}{\color[HTML]{000000} 53.56}} & \multicolumn{1}{c|}{\cellcolor[HTML]{FFFFFF}{\color[HTML]{000000} 55.64}} & \multicolumn{1}{c|}{\cellcolor[HTML]{FFFFFF}{\color[HTML]{000000} 58.25}} & \multicolumn{1}{c|}{\cellcolor[HTML]{FFFFFF}{\color[HTML]{000000} 57.03}} & \multicolumn{1}{c|}{\cellcolor[HTML]{FFFFFF}{\color[HTML]{000000} 59.41}} & \multicolumn{1}{c|}{\cellcolor[HTML]{FFFFFF}{\color[HTML]{000000} 54.50}} &\multicolumn{1}{c|}{\cellcolor[HTML]{FFFFFF}{\color[HTML]{000000} 59.38}}&\multicolumn{1}{c|}{\cellcolor[HTML]{FFFFFF}{\color[HTML]{000000}59.04}}&\multicolumn{1}{c|}{\cellcolor[HTML]{FFFFFF}{\color[HTML]{000000}56.13}}\\ \cline{3-14}
\multicolumn{1}{|c|}{\cellcolor[HTML]{FFF7F7}{\color[HTML]{000000} }} & \multicolumn{1}{c|}{\cellcolor[HTML]{FFFFFF}{\color[HTML]{000000} }} & \multicolumn{1}{c|}{\cellcolor[HTML]{FFFFFF}{\color[HTML]{000000} CLIP-Align}}    & \multicolumn{1}{c|}{\cellcolor[HTML]{FFFFFF}{\color[HTML]{000000} 52.81}}  & \multicolumn{1}{c|}{\cellcolor[HTML]{FFFFFF}{\color[HTML]{000000} 56.32}}       & \multicolumn{1}{c|}{\cellcolor[HTML]{FFFFFF}{\color[HTML]{000000} 52.24}}  & \multicolumn{1}{c|}{\cellcolor[HTML]{FFFFFF}{\color[HTML]{000000} 54.05}}  & \multicolumn{1}{c|}{\cellcolor[HTML]{FFFFFF}{\color[HTML]{000000} 57.22}} & \multicolumn{1}{c|}{\cellcolor[HTML]{FFFFFF}{\color[HTML]{000000} 51.74}} & \multicolumn{1}{c|}{\cellcolor[HTML]{FFFFFF}{\color[HTML]{000000} 57.92}}& \multicolumn{1}{c|}{\cellcolor[HTML]{FFFFFF}{\color[HTML]{000000} 52.45}} &\multicolumn{1}{c|}{\cellcolor[HTML]{FFFFFF}{\color[HTML]{000000}54.85}}&\multicolumn{1}{c|}{\cellcolor[HTML]{FFFFFF}{\color[HTML]{000000}58.48}}&\multicolumn{1}{c|}{\cellcolor[HTML]{FFFFFF}{\color[HTML]{000000}54.98}} \\ \cline{3-14}
\multicolumn{1}{|c|}{\multirow{-6}{*}{\cellcolor[HTML]{FFF7F7}{\color[HTML]{000000} \begin{tabular}[c]{@{}c@{}}V-AGA\\ +\\ A-Mel\end{tabular}}}} & \multicolumn{1}{c|}{\multirow{-5}{*}{\cellcolor[HTML]{FFFFFF}{\color[HTML]{000000} Feature}}} & \multicolumn{1}{c|}{\cellcolor[HTML]{FFFFFF}{\color[HTML]{000000} \textbf{Atten-Mixer}}} & \multicolumn{1}{c|}{\cellcolor[HTML]{FFFFFF}{\color[HTML]{000000} 55.11}} & \multicolumn{1}{c|}{\cellcolor[HTML]{FFFFFF}{\color[HTML]{000000} \underline{62.38}}} & \multicolumn{1}{c|}{\cellcolor[HTML]{FFFFFF}{\color[HTML]{000000} \underline{54.02}}} & \multicolumn{1}{c|}{\cellcolor[HTML]{FFFFFF}{\color[HTML]{000000} 56.28}} & \multicolumn{1}{c|}{\cellcolor[HTML]{FFFFFF}{\color[HTML]{000000} \underline{58.48}}} & \multicolumn{1}{c|}{\cellcolor[HTML]{FFFFFF}{\color[HTML]{000000} 57.65}} & \multicolumn{1}{c|}{\cellcolor[HTML]{FFFFFF}{\color[HTML]{000000} \underline{60.19}}}& \multicolumn{1}{c|}{\cellcolor[HTML]{FFFFFF}{\color[HTML]{000000} 54.64}}  &\multicolumn{1}{c|}{\cellcolor[HTML]{FFFFFF}{\color[HTML]{000000} 60.50}}&\multicolumn{1}{c|}{\cellcolor[HTML]{FFFFFF}{\color[HTML]{000000} 60.13}}&\multicolumn{1}{c|}{\cellcolor[HTML]{FFFFFF}{\color[HTML]{000000} 57.11}} \\ \hline
\multicolumn{1}{|c|}{\cellcolor[HTML]{EEFBED}{\color[HTML]{000000} }} & \multicolumn{1}{c|}{\cellcolor[HTML]{FFFFFF}{\color[HTML]{000000} Score}} & \multicolumn{1}{c|}{\cellcolor[HTML]{FFFFFF}{\color[HTML]{000000} Avg}}    & \multicolumn{1}{c|}{\cellcolor[HTML]{FFFFFF}{\color[HTML]{000000} 53.85}} & \multicolumn{1}{c|}{\cellcolor[HTML]{FFFFFF}{\color[HTML]{000000} 54.46}} & \multicolumn{1}{c|}{\cellcolor[HTML]{FFFFFF}{\color[HTML]{000000} 52.69}}      & \multicolumn{1}{c|}{\cellcolor[HTML]{FFFFFF}{\color[HTML]{000000} 53.21}} & \multicolumn{1}{c|}{\cellcolor[HTML]{FFFFFF}{\color[HTML]{000000} 56.27}} & \multicolumn{1}{c|}{\cellcolor[HTML]{FFFFFF}{\color[HTML]{000000} 50.84}} & \multicolumn{1}{c|}{\cellcolor[HTML]{FFFFFF}{\color[HTML]{000000} 54.13}} & \multicolumn{1}{c|}{\cellcolor[HTML]{FFFFFF}{\color[HTML]{000000} 50.93}} &\multicolumn{1}{c|}{\cellcolor[HTML]{FFFFFF}{\color[HTML]{000000} 53.13}} &\multicolumn{1}{c|}{\cellcolor[HTML]{FFFFFF}{\color[HTML]{000000}57.04}} &\multicolumn{1}{c|}{\cellcolor[HTML]{FFFFFF}{\color[HTML]{000000}54.66}} \\ \cline{2-14}
\multicolumn{1}{|c|}{\cellcolor[HTML]{EEFBED}{\color[HTML]{000000} }}& \multicolumn{1}{c|}{\cellcolor[HTML]{FFFFFF}{\color[HTML]{000000} }} & \multicolumn{1}{c|}{\cellcolor[HTML]{FFFFFF}{\color[HTML]{000000} Concat}} & \multicolumn{1}{c|}{\cellcolor[HTML]{FFFFFF}{\color[HTML]{000000} 55.42}} & \multicolumn{1}{c|}{\cellcolor[HTML]{FFFFFF}{\color[HTML]{000000} 57.43}} & \multicolumn{1}{c|}{\cellcolor[HTML]{FFFFFF}{\color[HTML]{000000} 52.62}} & \multicolumn{1}{c|}{\cellcolor[HTML]{FFFFFF}{\color[HTML]{000000} 54.97}} & \multicolumn{1}{c|}{\cellcolor[HTML]{FFFFFF}{\color[HTML]{000000} 56.61}} & \multicolumn{1}{c|}{\cellcolor[HTML]{FFFFFF}{\color[HTML]{000000} 53.33}} & \multicolumn{1}{c|}{\cellcolor[HTML]{FFFFFF}{\color[HTML]{000000} 54.87}} & \multicolumn{1}{c|}{\cellcolor[HTML]{FFFFFF}{\color[HTML]{000000} 51.50}} &\multicolumn{1}{c|}{\cellcolor[HTML]{FFFFFF}{\color[HTML]{000000} 54.06}}&\multicolumn{1}{c|}{\cellcolor[HTML]{FFFFFF}{\color[HTML]{000000}57.54}}&\multicolumn{1}{c|}{\cellcolor[HTML]{FFFFFF}{\color[HTML]{000000}54.43}} \\ \cline{3-14}
\multicolumn{1}{|c|}{\cellcolor[HTML]{EEFBED}{\color[HTML]{000000} }} & \multicolumn{1}{c|}{\cellcolor[HTML]{FFFFFF}{\color[HTML]{000000} }} & \multicolumn{1}{c|}{\cellcolor[HTML]{FFFFFF}{\color[HTML]{000000} SE-Concat}} & \multicolumn{1}{c|}{\cellcolor[HTML]{FFFFFF}{\color[HTML]{000000} 56.44}} & \multicolumn{1}{c|}{\cellcolor[HTML]{FFFFFF}{\color[HTML]{000000} 58.13}} & \multicolumn{1}{c|}{\cellcolor[HTML]{FFFFFF}{\color[HTML]{000000} 52.90}} & \multicolumn{1}{c|}{\cellcolor[HTML]{FFFFFF}{\color[HTML]{000000} 56.07}} & \multicolumn{1}{c|}{\cellcolor[HTML]{FFFFFF}{\color[HTML]{000000} 56.88}} & \multicolumn{1}{c|}{\cellcolor[HTML]{FFFFFF}{\color[HTML]{000000} 56.18}} & \multicolumn{1}{c|}{\cellcolor[HTML]{FFFFFF}{\color[HTML]{000000} 55.02}} &\multicolumn{1}{c|}{\cellcolor[HTML]{FFFFFF}{\color[HTML]{000000} 53.16}} &\multicolumn{1}{c|}{\cellcolor[HTML]{FFFFFF}{\color[HTML]{000000} 54.44}}&\multicolumn{1}{c|}{\cellcolor[HTML]{FFFFFF}{\color[HTML]{000000}57.22}}&\multicolumn{1}{c|}{\cellcolor[HTML]{FFFFFF}{\color[HTML]{000000}55.32}} \\ \cline{3-14} 
\multicolumn{1}{|c|}{\cellcolor[HTML]{EEFBED}{\color[HTML]{000000} }} & \multicolumn{1}{c|}{\cellcolor[HTML]{FFFFFF}{\color[HTML]{000000} }} & \multicolumn{1}{c|}{\cellcolor[HTML]{FFFFFF}{\color[HTML]{000000} Cross-Atten}} & \multicolumn{1}{c|}{\cellcolor[HTML]{FFFFFF}{\color[HTML]{000000} \underline{57.43}}} & \multicolumn{1}{c|}{\cellcolor[HTML]{FFFFFF}{\color[HTML]{000000} 59.41}} & \multicolumn{1}{c|}{\cellcolor[HTML]{FFFFFF}{\color[HTML]{000000} 53.20}} & \multicolumn{1}{c|}{\cellcolor[HTML]{FFFFFF}{\color[HTML]{000000} 56.21}} & \multicolumn{1}{c|}{\cellcolor[HTML]{FFFFFF}{\color[HTML]{000000} 58.22}} & \multicolumn{1}{c|}{\cellcolor[HTML]{FFFFFF}{\color[HTML]{000000} 57.22}} & \multicolumn{1}{c|}{\cellcolor[HTML]{FFFFFF}{\color[HTML]{000000} 57.98}} & \multicolumn{1}{c|}{\cellcolor[HTML]{FFFFFF}{\color[HTML]{000000} 53.29}} &\multicolumn{1}{c|}{\cellcolor[HTML]{FFFFFF}{\color[HTML]{000000} 56.93}}&\multicolumn{1}{c|}{\cellcolor[HTML]{FFFFFF}{\color[HTML]{000000}58.13}}&\multicolumn{1}{c|}{\cellcolor[HTML]{FFFFFF}{\color[HTML]{000000}55.17}}  \\ \cline{3-14} 
\multicolumn{1}{|c|}{\cellcolor[HTML]{EEFBED}{\color[HTML]{000000} }} & \multicolumn{1}{c|}{\cellcolor[HTML]{FFFFFF}{\color[HTML]{000000} }} & \multicolumn{1}{c|}{\cellcolor[HTML]{FFFFFF}{\color[HTML]{000000} MLP-Mixer}} & \multicolumn{1}{c|}{\cellcolor[HTML]{FFFFFF}{\color[HTML]{000000} 56.46}} & \multicolumn{1}{c|}{\cellcolor[HTML]{FFFFFF}{\color[HTML]{000000} 60.34}} & \multicolumn{1}{c|}{\cellcolor[HTML]{FFFFFF}{\color[HTML]{000000} 54.00}} & \multicolumn{1}{c|}{\cellcolor[HTML]{FFFFFF}{\color[HTML]{000000} \underline{57.07}}} & \multicolumn{1}{c|}{\cellcolor[HTML]{FFFFFF}{\color[HTML]{000000} 58.42}} & \multicolumn{1}{c|}{\cellcolor[HTML]{FFFFFF}{\color[HTML]{000000} 58.08}} & \multicolumn{1}{c|}{\cellcolor[HTML]{FFFFFF}{\color[HTML]{000000} 59.63}} & \multicolumn{1}{c|}{\cellcolor[HTML]{FFFFFF}{\color[HTML]{000000}54.31}} &\multicolumn{1}{c|}{\cellcolor[HTML]{FFFFFF}{\color[HTML]{000000}59.75}}&\multicolumn{1}{c|}{\cellcolor[HTML]{FFFFFF}{\color[HTML]{000000}59.44}}&\multicolumn{1}{c|}{\cellcolor[HTML]{FFFFFF}{\color[HTML]{000000}55.77}} \\ \cline{3-14} 
\multicolumn{1}{|c|}{\cellcolor[HTML]{EEFBED}{\color[HTML]{000000} }} & \multicolumn{1}{c|}{\cellcolor[HTML]{FFFFFF}{\color[HTML]{000000} }} & \multicolumn{1}{c|}{\cellcolor[HTML]{FFFFFF}{\color[HTML]{000000} CLIP-Align}}    & \multicolumn{1}{c|}{\cellcolor[HTML]{FFFFFF}{\color[HTML]{000000} 56.18}}  & \multicolumn{1}{c|}{\cellcolor[HTML]{FFFFFF}{\color[HTML]{000000} 58.21}}       & \multicolumn{1}{c|}{\cellcolor[HTML]{FFFFFF}{\color[HTML]{000000} 53.47}}  & \multicolumn{1}{c|}{\cellcolor[HTML]{FFFFFF}{\color[HTML]{000000} 55.39}}  & \multicolumn{1}{c|}{\cellcolor[HTML]{FFFFFF}{\color[HTML]{000000} 57.54}} & \multicolumn{1}{c|}{\cellcolor[HTML]{FFFFFF}{\color[HTML]{000000} 53.98}} & \multicolumn{1}{c|}{\cellcolor[HTML]{FFFFFF}{\color[HTML]{000000} 55.76}}& \multicolumn{1}{c|}{\cellcolor[HTML]{FFFFFF}{\color[HTML]{000000} 52.18}} &\multicolumn{1}{c|}{\cellcolor[HTML]{FFFFFF}{\color[HTML]{000000}54.74}}&\multicolumn{1}{c|}{\cellcolor[HTML]{FFFFFF}{\color[HTML]{000000}58.35}}&\multicolumn{1}{c|}{\cellcolor[HTML]{FFFFFF}{\color[HTML]{000000}55.16}} \\ \cline{3-14}
\multicolumn{1}{|c|}{\multirow{-6}{*}{\cellcolor[HTML]{EEFBED}{\color[HTML]{000000} \begin{tabular}[c]{@{}c@{}}V-Face\\ +\\ V-AGA\\ +\\ A-Mel\end{tabular}}}} & \multicolumn{1}{c|}{\multirow{-5}{*}{\cellcolor[HTML]{FFFFFF}{\color[HTML]{000000} Feature}}} & \multicolumn{1}{c|}{\cellcolor[HTML]{FFFFFF}{\color[HTML]{000000} \textbf{Atten-Mixer}}} & \multicolumn{1}{c|}{\cellcolor[HTML]{FFFFFF}{\color[HTML]{000000} \textbf{57.48}}}       & \multicolumn{1}{c|}{\cellcolor[HTML]{FFFFFF}{\color[HTML]{000000} \textbf{63.17}}} & \multicolumn{1}{c|}{\cellcolor[HTML]{FFFFFF}{\color[HTML]{000000}\textbf{54.35}}} & \multicolumn{1}{c|}{\cellcolor[HTML]{FFFFFF}{\color[HTML]{000000}\textbf{58.88}}} & \multicolumn{1}{c|}{\cellcolor[HTML]{FFFFFF}{\color[HTML]{000000} \textbf{59.15}}} & \multicolumn{1}{c|}{\cellcolor[HTML]{FFFFFF}{\color[HTML]{000000}\textbf{58.96}}} & \multicolumn{1}{c|}{\cellcolor[HTML]{FFFFFF}{\color[HTML]{000000} \textbf{60.42}}} & \multicolumn{1}{c|}{\cellcolor[HTML]{FFFFFF}{\color[HTML]{000000} 54.56}}  &\multicolumn{1}{c|}{\cellcolor[HTML]{FFFFFF}{\color[HTML]{000000} \textbf{62.58}}}&\multicolumn{1}{c|}{\cellcolor[HTML]{FFFFFF}{\color[HTML]{000000} \underline{60.42}}}&\multicolumn{1}{c|}{\cellcolor[HTML]{FFFFFF}{\color[HTML]{000000} \textbf{57.97}}}  \\ \hline
\end{tabular}}
\label{tab:multi2single_fusion_simul}
\vspace{1em}
\end{table*}


\begin{table*}[!t]
\centering
\caption{Fusion results for multi-to-single cross-domain accuracy (\%) using domain-alternating sampling.}
\resizebox{1.0\textwidth}{!}{
\begin{tabular}{|c
>{\columncolor[HTML]{FFFFFF}}c 
>{\columncolor[HTML]{FFFFFF}}c 
>{\columncolor[HTML]{FFFFFF}}c 
>{\columncolor[HTML]{FFFFFF}}c 
>{\columncolor[HTML]{FFFFFF}}c 
>{\columncolor[HTML]{FFFFFF}}c 
>{\columncolor[HTML]{FFFFFF}}c 
>{\columncolor[HTML]{FFFFFF}}c
>{\columncolor[HTML]{FFFFFF}}c 
>{\columncolor[HTML]{FFFFFF}}c
>{\columncolor[HTML]{FFFFFF}}c 
>{\columncolor[HTML]{FFFFFF}}c
>{\columncolor[HTML]{FFFFFF}}c|}
\hline
\multicolumn{14}{|c|}{\cellcolor[HTML]{EFEFEF}{\color[HTML]{000000} \textbf{Domain-Alternating}}} \\ \hline
\multicolumn{1}{|c|}{\cellcolor[HTML]{FFFFFF}{\color[HTML]{000000} \textbf{Input}}} & \multicolumn{1}{c|}{\cellcolor[HTML]{FFFFFF}{\color[HTML]{000000} \textbf{Type}}}  & \multicolumn{1}{c|}{\cellcolor[HTML]{FFFFFF}{\color[HTML]{000000} \textbf{Method}}}   & \multicolumn{1}{c|}{\cellcolor[HTML]{FFFFFF}{\color[HTML]{000000} \textbf{R\&M to B1}}} 
& \multicolumn{1}{c|}{\cellcolor[HTML]{FFFFFF}{\color[HTML]{000000} \textbf{R\&M to B2}}} 
& \multicolumn{1}{c|}{\cellcolor[HTML]{FFFFFF}{\color[HTML]{000000} \textbf{R\&M to D}}} 
& \multicolumn{1}{c|}{\cellcolor[HTML]{FFFFFF}{\color[HTML]{000000} \textbf{R\&M to E}}} 
& \multicolumn{1}{c|}{\cellcolor[HTML]{FFFFFF}{\color[HTML]{000000} \textbf{R\&B1 to B2}}} 
& \multicolumn{1}{c|}{\cellcolor[HTML]{FFFFFF}{\color[HTML]{000000} \textbf{R\&B1 to M}}} 
& \multicolumn{1}{c|}{\cellcolor[HTML]{FFFFFF}{\color[HTML]{000000} \textbf{R\&B1 to D}}} 
& \multicolumn{1}{c|}{\cellcolor[HTML]{FFFFFF}{\color[HTML]{000000} \textbf{R\&B1 to E}}} 
& \multicolumn{1}{c|}{\cellcolor[HTML]{FFFFFF}{\color[HTML]{000000} \textbf{B1\&M to R}}} 
& \multicolumn{1}{c|}{\cellcolor[HTML]{FFFFFF}{\color[HTML]{000000} \textbf{B1\&M to B2}}} 
& \multicolumn{1}{c|}{\cellcolor[HTML]{FFFFFF}{\color[HTML]{000000} \textbf{\shortstack{R\&B1\&M\\to B2}}}} \\  \hline
\multicolumn{1}{|c|}{\cellcolor[HTML]{ECF4FF}{\color[HTML]{000000} }}  & \multicolumn{1}{c|}{\cellcolor[HTML]{FFFFFF}{\color[HTML]{000000} Score}}  & \multicolumn{1}{c|}{\cellcolor[HTML]{FFFFFF}{\color[HTML]{000000} Avg}}     & \multicolumn{1}{c|}{\cellcolor[HTML]{FFFFFF}{\color[HTML]{000000} 55.38}}  & \multicolumn{1}{c|}{\cellcolor[HTML]{FFFFFF}{\color[HTML]{000000} 57.43}} &\multicolumn{1}{c|}{\cellcolor[HTML]{FFFFFF}{\color[HTML]{000000}  50.67  }} &\multicolumn{1}{c|}{\cellcolor[HTML]{FFFFFF}{\color[HTML]{000000}   50.32 }}& \multicolumn{1}{c|}{\cellcolor[HTML]{FFFFFF}{\color[HTML]{000000} 58.42}} & \multicolumn{1}{c|}{\cellcolor[HTML]{FFFFFF}{\color[HTML]{000000} 50.94}} &\multicolumn{1}{c|}{\cellcolor[HTML]{FFFFFF}{\color[HTML]{000000}  55.32  }} &\multicolumn{1}{c|}{\cellcolor[HTML]{FFFFFF}{\color[HTML]{000000}   50.22 }}& \multicolumn{1}{c|}{\cellcolor[HTML]{FFFFFF}{\color[HTML]{000000} 57.94}} & \multicolumn{1}{c|}{\cellcolor[HTML]{FFFFFF}{\color[HTML]{000000} 50.50}} & \multicolumn{1}{c|}{\cellcolor[HTML]{FFFFFF}{\color[HTML]{000000} 61.39}} \\ \cline{2-14} 
\multicolumn{1}{|c|}{\cellcolor[HTML]{ECF4FF}{\color[HTML]{000000} }} & \multicolumn{1}{c|}{\cellcolor[HTML]{FFFFFF}{\color[HTML]{000000} }}  & \multicolumn{1}{c|}{\cellcolor[HTML]{FFFFFF}{\color[HTML]{000000} Concat}}  & \multicolumn{1}{c|}{\cellcolor[HTML]{FFFFFF}{\color[HTML]{000000} 55.08}} & \multicolumn{1}{c|}{\cellcolor[HTML]{FFFFFF}{\color[HTML]{000000} 58.42}} &\multicolumn{1}{c|}{\cellcolor[HTML]{FFFFFF}{\color[HTML]{000000}  50.29  }} &\multicolumn{1}{c|}{\cellcolor[HTML]{FFFFFF}{\color[HTML]{000000}   50.54 }}   & \multicolumn{1}{c|}{\cellcolor[HTML]{FFFFFF}{\color[HTML]{000000} \underline{63.37}}}  & \multicolumn{1}{c|}{\cellcolor[HTML]{FFFFFF}{\color[HTML]{000000} 51.56}}   &\multicolumn{1}{c|}{\cellcolor[HTML]{FFFFFF}{\color[HTML]{000000}  56.98  }} &\multicolumn{1}{c|}{\cellcolor[HTML]{FFFFFF}{\color[HTML]{000000}   52.73 }}   & \multicolumn{1}{c|}{\cellcolor[HTML]{FFFFFF}{\color[HTML]{000000} 62.62}}  & \multicolumn{1}{c|}{\cellcolor[HTML]{FFFFFF}{\color[HTML]{000000} 62.38}}  & \multicolumn{1}{c|}{\cellcolor[HTML]{FFFFFF}{\color[HTML]{000000} 53.47}} \\ \cline{3-14} 
\multicolumn{1}{|c|}{\cellcolor[HTML]{ECF4FF}{\color[HTML]{000000} }}  & \multicolumn{1}{c|}{\cellcolor[HTML]{FFFFFF}{\color[HTML]{000000} }}  & \multicolumn{1}{c|}{\cellcolor[HTML]{FFFFFF}SE-Concat} & \multicolumn{1}{c|}{\cellcolor[HTML]{FFFFFF}50.46} & \multicolumn{1}{c|}{\cellcolor[HTML]{FFFFFF}49.50} &\multicolumn{1}{c|}{\cellcolor[HTML]{FFFFFF}{\color[HTML]{000000}  51.18  }} &\multicolumn{1}{c|}{\cellcolor[HTML]{FFFFFF}{\color[HTML]{000000}   52.78 }} & \multicolumn{1}{c|}{\cellcolor[HTML]{FFFFFF}\underline{63.37}}  & \multicolumn{1}{c|}{\cellcolor[HTML]{FFFFFF}52.19}  &\multicolumn{1}{c|}{\cellcolor[HTML]{FFFFFF}{\color[HTML]{000000}  56.80  }} &\multicolumn{1}{c|}{\cellcolor[HTML]{FFFFFF}{\color[HTML]{000000}   52.05 }}& \multicolumn{1}{c|}{\cellcolor[HTML]{FFFFFF}57.01} & \multicolumn{1}{c|}{\cellcolor[HTML]{FFFFFF}\textbf{65.35}}  & \multicolumn{1}{c|}{\cellcolor[HTML]{FFFFFF}56.44}  \\ \cline{3-14} 
\multicolumn{1}{|c|}{\cellcolor[HTML]{ECF4FF}{\color[HTML]{000000} }} & \multicolumn{1}{c|}{\cellcolor[HTML]{FFFFFF}{\color[HTML]{000000} }} & \multicolumn{1}{c|}{\cellcolor[HTML]{FFFFFF}Cross-Atten}  & \multicolumn{1}{c|}{\cellcolor[HTML]{FFFFFF}55.08} & \multicolumn{1}{c|}{\cellcolor[HTML]{FFFFFF}59.41}&\multicolumn{1}{c|}{\cellcolor[HTML]{FFFFFF}{\color[HTML]{000000}  52.41  }} &\multicolumn{1}{c|}{\cellcolor[HTML]{FFFFFF}{\color[HTML]{000000}   53.01 }} & \multicolumn{1}{c|}{\cellcolor[HTML]{FFFFFF}60.40} & \multicolumn{1}{c|}{\cellcolor[HTML]{FFFFFF}52.50}  &\multicolumn{1}{c|}{\cellcolor[HTML]{FFFFFF}{\color[HTML]{000000}  57.71  }} &\multicolumn{1}{c|}{\cellcolor[HTML]{FFFFFF}{\color[HTML]{000000}   52.13 }}& \multicolumn{1}{c|}{\cellcolor[HTML]{FFFFFF}48.60} & \multicolumn{1}{c|}{\cellcolor[HTML]{FFFFFF}58.42}  & \multicolumn{1}{c|}{\cellcolor[HTML]{FFFFFF}61.39}   \\ \cline{3-14} 
\multicolumn{1}{|c|}{\cellcolor[HTML]{ECF4FF}{\color[HTML]{000000} }}  & \multicolumn{1}{c|}{\cellcolor[HTML]{FFFFFF}{\color[HTML]{000000} }}& \multicolumn{1}{c|}{\cellcolor[HTML]{FFFFFF}MLP-Mixer}  & \multicolumn{1}{c|}{\cellcolor[HTML]{FFFFFF}\textbf{57.54}}  & \multicolumn{1}{c|}{\cellcolor[HTML]{FFFFFF}63.37} &\multicolumn{1}{c|}{\cellcolor[HTML]{FFFFFF}{\color[HTML]{000000}  53.23  }} &\multicolumn{1}{c|}{\cellcolor[HTML]{FFFFFF}{\color[HTML]{000000}   53.22 }}  & \multicolumn{1}{c|}{\cellcolor[HTML]{FFFFFF}\underline{63.37}}  & \multicolumn{1}{c|}{\cellcolor[HTML]{FFFFFF}50.94}     &\multicolumn{1}{c|}{\cellcolor[HTML]{FFFFFF}{\color[HTML]{000000}  57.82  }} &\multicolumn{1}{c|}{\cellcolor[HTML]{FFFFFF}{\color[HTML]{000000}   52.24 }} & \multicolumn{1}{c|}{\cellcolor[HTML]{FFFFFF}\underline{65.42}}   & \multicolumn{1}{c|}{\cellcolor[HTML]{FFFFFF}52.48}    & \multicolumn{1}{c|}{\cellcolor[HTML]{FFFFFF}56.44}   \\ \cline{3-14}
\multicolumn{1}{|c|}{\cellcolor[HTML]{ECF4FF}{\color[HTML]{000000} }} & \multicolumn{1}{c|}{\cellcolor[HTML]{FFFFFF}{\color[HTML]{000000} }} & \multicolumn{1}{c|}{\cellcolor[HTML]{FFFFFF}{\color[HTML]{000000} CLIP-Align}}    & \multicolumn{1}{c|}{\cellcolor[HTML]{FFFFFF}{\color[HTML]{000000}55.45}}  & \multicolumn{1}{c|}{\cellcolor[HTML]{FFFFFF}{\color[HTML]{000000}59.34}}  &  \multicolumn{1}{c|}{\cellcolor[HTML]{FFFFFF}{\color[HTML]{000000}50.47}}&  \multicolumn{1}{c|}{\cellcolor[HTML]{FFFFFF}{\color[HTML]{000000}51.58}}     & \multicolumn{1}{c|}{\cellcolor[HTML]{FFFFFF}{\color[HTML]{000000}64.03}}  & \multicolumn{1}{c|}{\cellcolor[HTML]{FFFFFF}{\color[HTML]{000000}52.05}} &  \multicolumn{1}{c|}{\cellcolor[HTML]{FFFFFF}{\color[HTML]{000000}58.06}}&  \multicolumn{1}{c|}{\cellcolor[HTML]{FFFFFF}{\color[HTML]{000000}52.98}} & \multicolumn{1}{c|}{\cellcolor[HTML]{FFFFFF}{\color[HTML]{000000}63.35}} & \multicolumn{1}{c|}{\cellcolor[HTML]{FFFFFF}{\color[HTML]{000000}\underline{62.94}}} & \multicolumn{1}{c|}{\cellcolor[HTML]{FFFFFF}{\color[HTML]{000000}54.28}}\\ \cline{3-14} 
\multicolumn{1}{|c|}{\multirow{-6}{*}{\cellcolor[HTML]{ECF4FF}{\color[HTML]{000000} \begin{tabular}[c]{@{}c@{}}V-Face\\ +\\ V-AGA\end{tabular}}}}   & \multicolumn{1}{c|}{\multirow{-5}{*}{\cellcolor[HTML]{FFFFFF}{\color[HTML]{000000} Feature}}} & \multicolumn{1}{c|}{\cellcolor[HTML]{FFFFFF}\textbf{Atten-Mixer}}  & \multicolumn{1}{c|}{\cellcolor[HTML]{FFFFFF}\underline{56.62}}     & \multicolumn{1}{c|}{\cellcolor[HTML]{FFFFFF}61.39} &\multicolumn{1}{c|}{\cellcolor[HTML]{FFFFFF}{\color[HTML]{000000}  53.60  }} &\multicolumn{1}{c|}{\cellcolor[HTML]{FFFFFF}{\color[HTML]{000000}   53.50 }}  & \multicolumn{1}{c|}{\cellcolor[HTML]{FFFFFF}\underline{63.37}}  & \multicolumn{1}{c|}{\cellcolor[HTML]{FFFFFF}51.56}  &\multicolumn{1}{c|}{\cellcolor[HTML]{FFFFFF}{\color[HTML]{000000}  57.30  }} &\multicolumn{1}{c|}{\cellcolor[HTML]{FFFFFF}{\color[HTML]{000000}   52.65 }}& \multicolumn{1}{c|}{\cellcolor[HTML]{FFFFFF}58.88}     & \multicolumn{1}{c|}{\cellcolor[HTML]{FFFFFF}62.38}   & \multicolumn{1}{c|}{\cellcolor[HTML]{FFFFFF}58.42}   \\ \hline
\multicolumn{1}{|c|}{\cellcolor[HTML]{FFFFEB}}  & \multicolumn{1}{c|}{\cellcolor[HTML]{FFFFFF}Score}     & \multicolumn{1}{c|}{\cellcolor[HTML]{FFFFFF}Avg}  & \multicolumn{1}{c|}{\cellcolor[HTML]{FFFFFF}50.15}  & \multicolumn{1}{c|}{\cellcolor[HTML]{FFFFFF}58.42} &\multicolumn{1}{c|}{\cellcolor[HTML]{FFFFFF}{\color[HTML]{000000}  50.65  }} &\multicolumn{1}{c|}{\cellcolor[HTML]{FFFFFF}{\color[HTML]{000000}   50.87 }} & \multicolumn{1}{c|}{\cellcolor[HTML]{FFFFFF}60.40}    & \multicolumn{1}{c|}{\cellcolor[HTML]{FFFFFF}50.62} &\multicolumn{1}{c|}{\cellcolor[HTML]{FFFFFF}{\color[HTML]{000000}  57.88  }} &\multicolumn{1}{c|}{\cellcolor[HTML]{FFFFFF}{\color[HTML]{000000}   50.06 }} & \multicolumn{1}{c|}{\cellcolor[HTML]{FFFFFF}51.40}    & \multicolumn{1}{c|}{\cellcolor[HTML]{FFFFFF}60.40} & \multicolumn{1}{c|}{\cellcolor[HTML]{FFFFFF}58.42}      \\ \cline{2-14} 
\multicolumn{1}{|c|}{\cellcolor[HTML]{FFFFEB}}  & \multicolumn{1}{c|}{\cellcolor[HTML]{FFFFFF}} & \multicolumn{1}{c|}{\cellcolor[HTML]{FFFFFF}Concat}   & \multicolumn{1}{c|}{\cellcolor[HTML]{FFFFFF}53.23}     & \multicolumn{1}{c|}{\cellcolor[HTML]{FFFFFF}62.38}&\multicolumn{1}{c|}{\cellcolor[HTML]{FFFFFF}{\color[HTML]{000000}  51.11  }} &\multicolumn{1}{c|}{\cellcolor[HTML]{FFFFFF}{\color[HTML]{000000}   50.95 }}& \multicolumn{1}{c|}{\cellcolor[HTML]{FFFFFF}54.46}  & \multicolumn{1}{c|}{\cellcolor[HTML]{FFFFFF}51.25}&\multicolumn{1}{c|}{\cellcolor[HTML]{FFFFFF}{\color[HTML]{000000}  57.04  }} &\multicolumn{1}{c|}{\cellcolor[HTML]{FFFFFF}{\color[HTML]{000000}   51.24 }} & \multicolumn{1}{c|}{\cellcolor[HTML]{FFFFFF}59.81}  & \multicolumn{1}{c|}{\cellcolor[HTML]{FFFFFF}54.46}  & \multicolumn{1}{c|}{\cellcolor[HTML]{FFFFFF}54.46}  \\ \cline{3-14} 
\multicolumn{1}{|c|}{\cellcolor[HTML]{FFFFEB}}  & \multicolumn{1}{c|}{\cellcolor[HTML]{FFFFFF}}   & \multicolumn{1}{c|}{\cellcolor[HTML]{FFFFFF}SE-Concat}  & \multicolumn{1}{c|}{\cellcolor[HTML]{FFFFFF}50.15}  & \multicolumn{1}{c|}{\cellcolor[HTML]{FFFFFF}52.48} &\multicolumn{1}{c|}{\cellcolor[HTML]{FFFFFF}{\color[HTML]{000000}  51.66  }} &\multicolumn{1}{c|}{\cellcolor[HTML]{FFFFFF}{\color[HTML]{000000}   52.55 }} & \multicolumn{1}{c|}{\cellcolor[HTML]{FFFFFF}\textbf{64.36}}  & \multicolumn{1}{c|}{\cellcolor[HTML]{FFFFFF}50.31} &\multicolumn{1}{c|}{\cellcolor[HTML]{FFFFFF}{\color[HTML]{000000}  57.73  }} &\multicolumn{1}{c|}{\cellcolor[HTML]{FFFFFF}{\color[HTML]{000000}   51.26 }} & \multicolumn{1}{c|}{\cellcolor[HTML]{FFFFFF}59.81} & \multicolumn{1}{c|}{\cellcolor[HTML]{FFFFFF}55.45}  & \multicolumn{1}{c|}{\cellcolor[HTML]{FFFFFF}50.50}   \\ \cline{3-14} 
\multicolumn{1}{|c|}{\cellcolor[HTML]{FFFFEB}}  & \multicolumn{1}{c|}{\cellcolor[HTML]{FFFFFF}} & \multicolumn{1}{c|}{\cellcolor[HTML]{FFFFFF}Cross-Atten}   & \multicolumn{1}{c|}{\cellcolor[HTML]{FFFFFF}52.92}  & \multicolumn{1}{c|}{\cellcolor[HTML]{FFFFFF}57.43}  &\multicolumn{1}{c|}{\cellcolor[HTML]{FFFFFF}{\color[HTML]{000000}  52.13  }} &\multicolumn{1}{c|}{\cellcolor[HTML]{FFFFFF}{\color[HTML]{000000}   53.89 }} & \multicolumn{1}{c|}{\cellcolor[HTML]{FFFFFF}49.50}   & \multicolumn{1}{c|}{\cellcolor[HTML]{FFFFFF}\underline{52.81}} &\multicolumn{1}{c|}{\cellcolor[HTML]{FFFFFF}{\color[HTML]{000000}  57.82  }} &\multicolumn{1}{c|}{\cellcolor[HTML]{FFFFFF}{\color[HTML]{000000}   52.08 }} & \multicolumn{1}{c|}{\cellcolor[HTML]{FFFFFF}65.42}   & \multicolumn{1}{c|}{\cellcolor[HTML]{FFFFFF}53.47}  & \multicolumn{1}{c|}{\cellcolor[HTML]{FFFFFF}66.34}  \\ \cline{3-14} 
\multicolumn{1}{|c|}{\cellcolor[HTML]{FFFFEB}}  & \multicolumn{1}{c|}{\cellcolor[HTML]{FFFFFF}}  & \multicolumn{1}{c|}{\cellcolor[HTML]{FFFFFF}MLP-Mixer}  & \multicolumn{1}{c|}{\cellcolor[HTML]{FFFFFF}51.69}   & \multicolumn{1}{c|}{\cellcolor[HTML]{FFFFFF}58.42} &\multicolumn{1}{c|}{\cellcolor[HTML]{FFFFFF}{\color[HTML]{000000}  52.15  }} &\multicolumn{1}{c|}{\cellcolor[HTML]{FFFFFF}{\color[HTML]{000000}   53.99 }} & \multicolumn{1}{c|}{\cellcolor[HTML]{FFFFFF}57.43}   & \multicolumn{1}{c|}{\cellcolor[HTML]{FFFFFF}51.25} &\multicolumn{1}{c|}{\cellcolor[HTML]{FFFFFF}{\color[HTML]{000000}  57.68  }} &\multicolumn{1}{c|}{\cellcolor[HTML]{FFFFFF}{\color[HTML]{000000}   52.11 }} & \multicolumn{1}{c|}{\cellcolor[HTML]{FFFFFF}62.62}    & \multicolumn{1}{c|}{\cellcolor[HTML]{FFFFFF}58.42}  & \multicolumn{1}{c|}{\cellcolor[HTML]{FFFFFF}56.44}     \\ \cline{3-14} 
\multicolumn{1}{|c|}{\cellcolor[HTML]{FFFFEB}{\color[HTML]{000000} }} & \multicolumn{1}{c|}{\cellcolor[HTML]{FFFFFF}{\color[HTML]{000000} }} & \multicolumn{1}{c|}{\cellcolor[HTML]{FFFFFF}{\color[HTML]{000000} CLIP-Align}}    & \multicolumn{1}{c|}{\cellcolor[HTML]{FFFFFF}{\color[HTML]{000000}53.09}}  & \multicolumn{1}{c|}{\cellcolor[HTML]{FFFFFF}{\color[HTML]{000000}63.41}}  &  \multicolumn{1}{c|}{\cellcolor[HTML]{FFFFFF}{\color[HTML]{000000}51.53}}&  \multicolumn{1}{c|}{\cellcolor[HTML]{FFFFFF}{\color[HTML]{000000}51.83}}     & \multicolumn{1}{c|}{\cellcolor[HTML]{FFFFFF}{\color[HTML]{000000}54.62}}  & \multicolumn{1}{c|}{\cellcolor[HTML]{FFFFFF}{\color[HTML]{000000}51.99}} &  \multicolumn{1}{c|}{\cellcolor[HTML]{FFFFFF}{\color[HTML]{000000}58.13}}&  \multicolumn{1}{c|}{\cellcolor[HTML]{FFFFFF}{\color[HTML]{000000}51.55}} & \multicolumn{1}{c|}{\cellcolor[HTML]{FFFFFF}{\color[HTML]{000000}60.37}} & \multicolumn{1}{c|}{\cellcolor[HTML]{FFFFFF}{\color[HTML]{000000}55.43}} & \multicolumn{1}{c|}{\cellcolor[HTML]{FFFFFF}{\color[HTML]{000000}54.71}}\\ \cline{3-14}
\multicolumn{1}{|c|}{\multirow{-6}{*}{\cellcolor[HTML]{FFFFEB}\begin{tabular}[c]{@{}c@{}}V-Face\\ +\\ A-Mel\end{tabular}}}   & \multicolumn{1}{c|}{\multirow{-5}{*}{\cellcolor[HTML]{FFFFFF}Feature}}  & \multicolumn{1}{c|}{\cellcolor[HTML]{FFFFFF}\textbf{Atten-Mixer}} & \multicolumn{1}{c|}{\cellcolor[HTML]{FFFFFF}53.23}  & \multicolumn{1}{c|}{\cellcolor[HTML]{FFFFFF}63.37}    &\multicolumn{1}{c|}{\cellcolor[HTML]{FFFFFF}{\color[HTML]{000000}  54.01  }} &\multicolumn{1}{c|}{\cellcolor[HTML]{FFFFFF}{\color[HTML]{000000}   53.53 }}   & \multicolumn{1}{c|}{\cellcolor[HTML]{FFFFFF}56.44}  & \multicolumn{1}{c|}{\cellcolor[HTML]{FFFFFF}51.25}   &\multicolumn{1}{c|}{\cellcolor[HTML]{FFFFFF}{\color[HTML]{000000}  57.85  }} &\multicolumn{1}{c|}{\cellcolor[HTML]{FFFFFF}{\color[HTML]{000000}   52.32 }}  & \multicolumn{1}{c|}{\cellcolor[HTML]{FFFFFF}63.55}   & \multicolumn{1}{c|}{\cellcolor[HTML]{FFFFFF}59.41}  & \multicolumn{1}{c|}{\cellcolor[HTML]{FFFFFF}56.44}   \\ \hline
\multicolumn{1}{|c|}{\cellcolor[HTML]{FFF7F7}}  & \multicolumn{1}{c|}{\cellcolor[HTML]{FFFFFF}Score}     & \multicolumn{1}{c|}{\cellcolor[HTML]{FFFFFF}Avg}  & \multicolumn{1}{c|}{\cellcolor[HTML]{FFFFFF}51.38}  & \multicolumn{1}{c|}{\cellcolor[HTML]{FFFFFF}56.44} &\multicolumn{1}{c|}{\cellcolor[HTML]{FFFFFF}{\color[HTML]{000000}  51.20  }} &\multicolumn{1}{c|}{\cellcolor[HTML]{FFFFFF}{\color[HTML]{000000}   50.38 }}& \multicolumn{1}{c|}{\cellcolor[HTML]{FFFFFF}54.46}     & \multicolumn{1}{c|}{\cellcolor[HTML]{FFFFFF}50.00} &\multicolumn{1}{c|}{\cellcolor[HTML]{FFFFFF}{\color[HTML]{000000}  56.05  }} &\multicolumn{1}{c|}{\cellcolor[HTML]{FFFFFF}{\color[HTML]{000000}   51.74 }}& \multicolumn{1}{c|}{\cellcolor[HTML]{FFFFFF}64.49}     & \multicolumn{1}{c|}{\cellcolor[HTML]{FFFFFF}61.39}   & \multicolumn{1}{c|}{\cellcolor[HTML]{FFFFFF}49.50}  \\ \cline{2-14}
\multicolumn{1}{|c|}{\cellcolor[HTML]{FFF7F7}} & \multicolumn{1}{c|}{\cellcolor[HTML]{FFFFFF}}  & \multicolumn{1}{c|}{\cellcolor[HTML]{FFFFFF}Concat}  & \multicolumn{1}{c|}{\cellcolor[HTML]{FFFFFF}49.85}   & \multicolumn{1}{c|}{\cellcolor[HTML]{FFFFFF}62.38} &\multicolumn{1}{c|}{\cellcolor[HTML]{FFFFFF}{\color[HTML]{000000}  51.00  }} &\multicolumn{1}{c|}{\cellcolor[HTML]{FFFFFF}{\color[HTML]{000000}   50.48 }} & \multicolumn{1}{c|}{\cellcolor[HTML]{FFFFFF}58.42}   & \multicolumn{1}{c|}{\cellcolor[HTML]{FFFFFF}51.88} &\multicolumn{1}{c|}{\cellcolor[HTML]{FFFFFF}{\color[HTML]{000000}  57.23  }} &\multicolumn{1}{c|}{\cellcolor[HTML]{FFFFFF}{\color[HTML]{000000}   51.86 }} & \multicolumn{1}{c|}{\cellcolor[HTML]{FFFFFF}61.68}     & \multicolumn{1}{c|}{\cellcolor[HTML]{FFFFFF}55.45}    & \multicolumn{1}{c|}{\cellcolor[HTML]{FFFFFF}51.49}   \\ \cline{3-14} 
\multicolumn{1}{|c|}{\cellcolor[HTML]{FFF7F7}}   & \multicolumn{1}{c|}{\cellcolor[HTML]{FFFFFF}}   & \multicolumn{1}{c|}{\cellcolor[HTML]{FFFFFF}SE-Concat}  & \multicolumn{1}{c|}{\cellcolor[HTML]{FFFFFF}50.46}    & \multicolumn{1}{c|}{\cellcolor[HTML]{FFFFFF}62.38}  &\multicolumn{1}{c|}{\cellcolor[HTML]{FFFFFF}{\color[HTML]{000000}  53.24  }} &\multicolumn{1}{c|}{\cellcolor[HTML]{FFFFFF}{\color[HTML]{000000}   53.84 }}  & \multicolumn{1}{c|}{\cellcolor[HTML]{FFFFFF}58.42}   & \multicolumn{1}{c|}{\cellcolor[HTML]{FFFFFF}50.00}  &\multicolumn{1}{c|}{\cellcolor[HTML]{FFFFFF}{\color[HTML]{000000}  57.72  }} &\multicolumn{1}{c|}{\cellcolor[HTML]{FFFFFF}{\color[HTML]{000000}   52.34 }} & \multicolumn{1}{c|}{\cellcolor[HTML]{FFFFFF}57.94}   & \multicolumn{1}{c|}{\cellcolor[HTML]{FFFFFF}59.41}   & \multicolumn{1}{c|}{\cellcolor[HTML]{FFFFFF}\textbf{65.35}}  \\ \cline{3-14} 
\multicolumn{1}{|c|}{\cellcolor[HTML]{FFF7F7}}  & \multicolumn{1}{c|}{\cellcolor[HTML]{FFFFFF}}    & \multicolumn{1}{c|}{\cellcolor[HTML]{FFFFFF}Cross-Atten}  & \multicolumn{1}{c|}{\cellcolor[HTML]{FFFFFF}49.85}  & \multicolumn{1}{c|}{\cellcolor[HTML]{FFFFFF}62.38} &\multicolumn{1}{c|}{\cellcolor[HTML]{FFFFFF}{\color[HTML]{000000}  54.51  }} &\multicolumn{1}{c|}{\cellcolor[HTML]{FFFFFF}{\color[HTML]{000000}   53.92 }} & \multicolumn{1}{c|}{\cellcolor[HTML]{FFFFFF}50.50}  & \multicolumn{1}{c|}{\cellcolor[HTML]{FFFFFF}\textbf{53.44}} &\multicolumn{1}{c|}{\cellcolor[HTML]{FFFFFF}{\color[HTML]{000000}  57.81  }} &\multicolumn{1}{c|}{\cellcolor[HTML]{FFFFFF}{\color[HTML]{000000}   52.88 }} & \multicolumn{1}{c|}{\cellcolor[HTML]{FFFFFF}58.88}  & \multicolumn{1}{c|}{\cellcolor[HTML]{FFFFFF}56.44} & \multicolumn{1}{c|}{\cellcolor[HTML]{FFFFFF}56.44}   \\ \cline{3-14} 
\multicolumn{1}{|c|}{\cellcolor[HTML]{FFF7F7}}  & \multicolumn{1}{c|}{\cellcolor[HTML]{FFFFFF}}  & \multicolumn{1}{c|}{\cellcolor[HTML]{FFFFFF}MLP-Mixer}  & \multicolumn{1}{c|}{\cellcolor[HTML]{FFFFFF}50.15}  & \multicolumn{1}{c|}{\cellcolor[HTML]{FFFFFF}55.45} &\multicolumn{1}{c|}{\cellcolor[HTML]{FFFFFF}{\color[HTML]{000000}  54.59  }} &\multicolumn{1}{c|}{\cellcolor[HTML]{FFFFFF}{\color[HTML]{000000}   53.49 }}& \multicolumn{1}{c|}{\cellcolor[HTML]{FFFFFF}58.42}     & \multicolumn{1}{c|}{\cellcolor[HTML]{FFFFFF}50.31} &\multicolumn{1}{c|}{\cellcolor[HTML]{FFFFFF}{\color[HTML]{000000}  58.03  }} &\multicolumn{1}{c|}{\cellcolor[HTML]{FFFFFF}{\color[HTML]{000000}   53.01 }}& \multicolumn{1}{c|}{\cellcolor[HTML]{FFFFFF}58.88}     & \multicolumn{1}{c|}{\cellcolor[HTML]{FFFFFF}61.39} & \multicolumn{1}{c|}{\cellcolor[HTML]{FFFFFF}57.43}  \\ \cline{3-14} 
\multicolumn{1}{|c|}{\cellcolor[HTML]{FFF7F7}{\color[HTML]{000000} }} & \multicolumn{1}{c|}{\cellcolor[HTML]{FFFFFF}{\color[HTML]{000000} }} & \multicolumn{1}{c|}{\cellcolor[HTML]{FFFFFF}{\color[HTML]{000000} CLIP-Align}}    & \multicolumn{1}{c|}{\cellcolor[HTML]{FFFFFF}{\color[HTML]{000000}50.47}}  & \multicolumn{1}{c|}{\cellcolor[HTML]{FFFFFF}{\color[HTML]{000000}63.29}}  &  \multicolumn{1}{c|}{\cellcolor[HTML]{FFFFFF}{\color[HTML]{000000}51.36}}&  \multicolumn{1}{c|}{\cellcolor[HTML]{FFFFFF}{\color[HTML]{000000}51.27}}     & \multicolumn{1}{c|}{\cellcolor[HTML]{FFFFFF}{\color[HTML]{000000}59.18}}  & \multicolumn{1}{c|}{\cellcolor[HTML]{FFFFFF}{\color[HTML]{000000}52.41}} &  \multicolumn{1}{c|}{\cellcolor[HTML]{FFFFFF}{\color[HTML]{000000}57.97}}&  \multicolumn{1}{c|}{\cellcolor[HTML]{FFFFFF}{\color[HTML]{000000}52.32}} & \multicolumn{1}{c|}{\cellcolor[HTML]{FFFFFF}{\color[HTML]{000000}62.69}} & \multicolumn{1}{c|}{\cellcolor[HTML]{FFFFFF}{\color[HTML]{000000}56.14}} & \multicolumn{1}{c|}{\cellcolor[HTML]{FFFFFF}{\color[HTML]{000000}52.30}}\\ \cline{3-14}
\multicolumn{1}{|c|}{\multirow{-6}{*}{\cellcolor[HTML]{FFF7F7}\begin{tabular}[c]{@{}c@{}}V-AGA\\ +\\ A-Mel\end{tabular}}}  & \multicolumn{1}{c|}{\multirow{-5}{*}{\cellcolor[HTML]{FFFFFF}Feature}}  & \multicolumn{1}{c|}{\cellcolor[HTML]{FFFFFF}\textbf{Atten-Mixer}}  & \multicolumn{1}{c|}{\cellcolor[HTML]{FFFFFF}50.15}  & \multicolumn{1}{c|}{\cellcolor[HTML]{FFFFFF}59.41} &\multicolumn{1}{c|}{\cellcolor[HTML]{FFFFFF}{\color[HTML]{000000}  54.99  }} &\multicolumn{1}{c|}{\cellcolor[HTML]{FFFFFF}{\color[HTML]{000000}   \underline{54.31} }} & \multicolumn{1}{c|}{\cellcolor[HTML]{FFFFFF}50.50} & \multicolumn{1}{c|}{\cellcolor[HTML]{FFFFFF}52.19} &\multicolumn{1}{c|}{\cellcolor[HTML]{FFFFFF}{\color[HTML]{000000}  \underline{58.25}  }} &\multicolumn{1}{c|}{\cellcolor[HTML]{FFFFFF}{\color[HTML]{000000}   53.06 }} & \multicolumn{1}{c|}{\cellcolor[HTML]{FFFFFF}58.88}  & \multicolumn{1}{c|}{\cellcolor[HTML]{FFFFFF}60.40}  & \multicolumn{1}{c|}{\cellcolor[HTML]{FFFFFF}61.39}  \\ \hline
\multicolumn{1}{|c|}{\cellcolor[HTML]{EEFBED}}  & \multicolumn{1}{c|}{\cellcolor[HTML]{FFFFFF}Score}    & \multicolumn{1}{c|}{\cellcolor[HTML]{FFFFFF}Avg} & \multicolumn{1}{c|}{\cellcolor[HTML]{FFFFFF}56.00}    & \multicolumn{1}{c|}{\cellcolor[HTML]{FFFFFF}52.48}  &\multicolumn{1}{c|}{\cellcolor[HTML]{FFFFFF}{\color[HTML]{000000}  55.22  }} &\multicolumn{1}{c|}{\cellcolor[HTML]{FFFFFF}{\color[HTML]{000000}   50.92 }}& \multicolumn{1}{c|}{\cellcolor[HTML]{FFFFFF}59.41}    & \multicolumn{1}{c|}{\cellcolor[HTML]{FFFFFF}51.88} &\multicolumn{1}{c|}{\cellcolor[HTML]{FFFFFF}{\color[HTML]{000000}  55.06  }} &\multicolumn{1}{c|}{\cellcolor[HTML]{FFFFFF}{\color[HTML]{000000}   50.16 }} & \multicolumn{1}{c|}{\cellcolor[HTML]{FFFFFF}51.40}    & \multicolumn{1}{c|}{\cellcolor[HTML]{FFFFFF}60.40}& \multicolumn{1}{c|}{\cellcolor[HTML]{FFFFFF}56.44}  \\ \cline{2-14} 
\multicolumn{1}{|c|}{\cellcolor[HTML]{EEFBED}}   & \multicolumn{1}{c|}{\cellcolor[HTML]{FFFFFF}} & \multicolumn{1}{c|}{\cellcolor[HTML]{FFFFFF}Concat}  & \multicolumn{1}{c|}{\cellcolor[HTML]{FFFFFF}51.69}      & \multicolumn{1}{c|}{\cellcolor[HTML]{FFFFFF}58.42} &\multicolumn{1}{c|}{\cellcolor[HTML]{FFFFFF}{\color[HTML]{000000}  55.38  }} &\multicolumn{1}{c|}{\cellcolor[HTML]{FFFFFF}{\color[HTML]{000000}   51.09 }} & \multicolumn{1}{c|}{\cellcolor[HTML]{FFFFFF}58.42}   & \multicolumn{1}{c|}{\cellcolor[HTML]{FFFFFF}52.81}&\multicolumn{1}{c|}{\cellcolor[HTML]{FFFFFF}{\color[HTML]{000000}  55.32  }} &\multicolumn{1}{c|}{\cellcolor[HTML]{FFFFFF}{\color[HTML]{000000}   51.25 }}  & \multicolumn{1}{c|}{\cellcolor[HTML]{FFFFFF}57.01}     & \multicolumn{1}{c|}{\cellcolor[HTML]{FFFFFF}61.39}  & \multicolumn{1}{c|}{\cellcolor[HTML]{FFFFFF}56.44}     \\ \cline{3-14}
\multicolumn{1}{|c|}{\cellcolor[HTML]{EEFBED}}  & \multicolumn{1}{c|}{\cellcolor[HTML]{FFFFFF}} & \multicolumn{1}{c|}{\cellcolor[HTML]{FFFFFF}{\color[HTML]{000000} SE-Concat}}  & \multicolumn{1}{c|}{\cellcolor[HTML]{FFFFFF}{\color[HTML]{000000} 56.31}}   & \multicolumn{1}{c|}{\cellcolor[HTML]{FFFFFF}{\color[HTML]{000000} 59.41}}&\multicolumn{1}{c|}{\cellcolor[HTML]{FFFFFF}{\color[HTML]{000000}  56.22  }} &\multicolumn{1}{c|}{\cellcolor[HTML]{FFFFFF}{\color[HTML]{000000}   51.28 }} & \multicolumn{1}{c|}{\cellcolor[HTML]{FFFFFF}{\color[HTML]{000000} 58.42}}      & \multicolumn{1}{c|}{\cellcolor[HTML]{FFFFFF}{\color[HTML]{000000} 50.31}} &\multicolumn{1}{c|}{\cellcolor[HTML]{FFFFFF}{\color[HTML]{000000}  56.38  }} &\multicolumn{1}{c|}{\cellcolor[HTML]{FFFFFF}{\color[HTML]{000000}   51.99 }} & \multicolumn{1}{c|}{\cellcolor[HTML]{FFFFFF}{\color[HTML]{000000} 53.27}}  & \multicolumn{1}{c|}{\cellcolor[HTML]{FFFFFF}{\color[HTML]{000000} 61.39}}  & \multicolumn{1}{c|}{\cellcolor[HTML]{FFFFFF}{\color[HTML]{000000} 55.45}}     \\ \cline{3-14}
\multicolumn{1}{|c|}{\cellcolor[HTML]{EEFBED}}  & \multicolumn{1}{c|}{\cellcolor[HTML]{FFFFFF}} & \multicolumn{1}{c|}{\cellcolor[HTML]{FFFFFF}{\color[HTML]{000000} Cross-Atten}}  & \multicolumn{1}{c|}{\cellcolor[HTML]{FFFFFF}{\color[HTML]{000000} 49.54}}  & \multicolumn{1}{c|}{\cellcolor[HTML]{FFFFFF}{\color[HTML]{000000} \textbf{67.33}}} &\multicolumn{1}{c|}{\cellcolor[HTML]{FFFFFF}{\color[HTML]{000000}  56.43  }} &\multicolumn{1}{c|}{\cellcolor[HTML]{FFFFFF}{\color[HTML]{000000}   52.39 }} & \multicolumn{1}{c|}{\cellcolor[HTML]{FFFFFF}{\color[HTML]{000000} 54.46}}  & \multicolumn{1}{c|}{\cellcolor[HTML]{FFFFFF}{\color[HTML]{000000} 51.56}} &\multicolumn{1}{c|}{\cellcolor[HTML]{FFFFFF}{\color[HTML]{000000}  57.06  }} &\multicolumn{1}{c|}{\cellcolor[HTML]{FFFFFF}{\color[HTML]{000000}   52.89 }} & \multicolumn{1}{c|}{\cellcolor[HTML]{FFFFFF}{\color[HTML]{000000} \textbf{66.36}}} & \multicolumn{1}{c|}{\cellcolor[HTML]{FFFFFF}{\color[HTML]{000000} 52.48}} & \multicolumn{1}{c|}{\cellcolor[HTML]{FFFFFF}{\color[HTML]{000000} 60.40}}   \\ \cline{3-14}
\multicolumn{1}{|c|}{\cellcolor[HTML]{EEFBED}}  & \multicolumn{1}{c|}{\cellcolor[HTML]{FFFFFF}}  & \multicolumn{1}{c|}{\cellcolor[HTML]{FFFFFF}MLP-Mixer}   & \multicolumn{1}{c|}{\cellcolor[HTML]{FFFFFF}50.15}  & \multicolumn{1}{c|}{\cellcolor[HTML]{FFFFFF}63.37} &\multicolumn{1}{c|}{\cellcolor[HTML]{FFFFFF}{\color[HTML]{000000}  \underline{57.00}  }} &\multicolumn{1}{c|}{\cellcolor[HTML]{FFFFFF}{\color[HTML]{000000}   53.33 }} & \multicolumn{1}{c|}{\cellcolor[HTML]{FFFFFF}60.40}    & \multicolumn{1}{c|}{\cellcolor[HTML]{FFFFFF}51.56}  &\multicolumn{1}{c|}{\cellcolor[HTML]{FFFFFF}{\color[HTML]{000000}  58.03  }} &\multicolumn{1}{c|}{\cellcolor[HTML]{FFFFFF}{\color[HTML]{000000}   \underline{53.09} }} & \multicolumn{1}{c|}{\cellcolor[HTML]{FFFFFF}62.62}  & \multicolumn{1}{c|}{\cellcolor[HTML]{FFFFFF}61.39}  & \multicolumn{1}{c|}{\cellcolor[HTML]{FFFFFF}61.39}     \\ \cline{3-14}
\multicolumn{1}{|c|}{\cellcolor[HTML]{EEFBED}{\color[HTML]{000000} }} & \multicolumn{1}{c|}{\cellcolor[HTML]{FFFFFF}{\color[HTML]{000000} }} & \multicolumn{1}{c|}{\cellcolor[HTML]{FFFFFF}{\color[HTML]{000000} CLIP-Align}}    & \multicolumn{1}{c|}{\cellcolor[HTML]{FFFFFF}{\color[HTML]{000000}52.32}}  & \multicolumn{1}{c|}{\cellcolor[HTML]{FFFFFF}{\color[HTML]{000000}58.60}}  &  \multicolumn{1}{c|}{\cellcolor[HTML]{FFFFFF}{\color[HTML]{000000}56.42}}&  \multicolumn{1}{c|}{\cellcolor[HTML]{FFFFFF}{\color[HTML]{000000}51.81}}     & \multicolumn{1}{c|}{\cellcolor[HTML]{FFFFFF}{\color[HTML]{000000}58.77}}  & \multicolumn{1}{c|}{\cellcolor[HTML]{FFFFFF}{\color[HTML]{000000}53.72}} &  \multicolumn{1}{c|}{\cellcolor[HTML]{FFFFFF}{\color[HTML]{000000}55.79}}&  \multicolumn{1}{c|}{\cellcolor[HTML]{FFFFFF}{\color[HTML]{000000}52.33}} & \multicolumn{1}{c|}{\cellcolor[HTML]{FFFFFF}{\color[HTML]{000000}57.27}} & \multicolumn{1}{c|}{\cellcolor[HTML]{FFFFFF}{\color[HTML]{000000}61.98}} & \multicolumn{1}{c|}{\cellcolor[HTML]{FFFFFF}{\color[HTML]{000000}57.26}}\\ \cline{3-14}
\multicolumn{1}{|c|}{\multirow{-6}{*}{\cellcolor[HTML]{EEFBED}\begin{tabular}[c]{@{}c@{}}V-Face\\ +\\ V-AGA\\ +\\ A-Mel\end{tabular}}}   & \multicolumn{1}{c|}{\multirow{-5}{*}{\cellcolor[HTML]{FFFFFF}Feature}}   & \multicolumn{1}{c|}{\cellcolor[HTML]{FFFFFF}\textbf{Atten-Mixer}}   & \multicolumn{1}{c|}{\cellcolor[HTML]{FFFFFF}51.69}  & \multicolumn{1}{c|}{\cellcolor[HTML]{FFFFFF}\underline{66.34}} &\multicolumn{1}{c|}{\cellcolor[HTML]{FFFFFF}{\color[HTML]{000000}  \textbf{57.38}  }} &\multicolumn{1}{c|}{\cellcolor[HTML]{FFFFFF}{\color[HTML]{000000}   \textbf{55.02} }} & \multicolumn{1}{c|}{\cellcolor[HTML]{FFFFFF}60.40}   & \multicolumn{1}{c|}{\cellcolor[HTML]{FFFFFF}51.88} &\multicolumn{1}{c|}{\cellcolor[HTML]{FFFFFF}{\color[HTML]{000000}  \textbf{59.82}  }} &\multicolumn{1}{c|}{\cellcolor[HTML]{FFFFFF}{\color[HTML]{000000}   \textbf{53.86} }} & \multicolumn{1}{c|}{\cellcolor[HTML]{FFFFFF}59.81}  & \multicolumn{1}{c|}{\cellcolor[HTML]{FFFFFF}61.39} & \multicolumn{1}{c|}{\cellcolor[HTML]{FFFFFF}\underline{62.38}} \\ \hline

\multicolumn{1}{|c|}{\cellcolor[HTML]{FFFFFF}{\color[HTML]{000000} \textbf{Input}}} & \multicolumn{1}{c|}{\cellcolor[HTML]{FFFFFF}{\color[HTML]{000000} \textbf{Type}}}         & \multicolumn{1}{c|}{\cellcolor[HTML]{FFFFFF}{\color[HTML]{000000} \textbf{Method}}}     & \multicolumn{1}{c|}{\cellcolor[HTML]{FFFFFF}{\color[HTML]{000000} \textbf{D\&E to B1}}} & \multicolumn{1}{c|}{\cellcolor[HTML]{FFFFFF}{\color[HTML]{000000} \textbf{D\&E to B2}}} & \multicolumn{1}{c|}{\cellcolor[HTML]{FFFFFF}{\color[HTML]{000000} \textbf{D\&E to M}}} & \multicolumn{1}{c|}{\cellcolor[HTML]{FFFFFF}{\color[HTML]{000000} \textbf{D\&E to R}}} & \multicolumn{1}{c|}{\cellcolor[HTML]{FFFFFF}{\color[HTML]{000000} \textbf{\shortstack{R\&B1\&M\\to D}}}} & \multicolumn{1}{c|}{\cellcolor[HTML]{FFFFFF}{\color[HTML]{000000} \textbf{\shortstack{R\&B1\&M\\to E}}}} & \multicolumn{1}{c|}{\cellcolor[HTML]{FFFFFF}{\color[HTML]{000000} \textbf{\shortstack{R\&D\&E\\to B1}}}} & {\color[HTML]{000000} \textbf{\shortstack{R\&D\&E\\to M}}}         & {\color[HTML]{000000} \textbf{\shortstack{R\&M\&D\&E\\to B2}}}  & {\color[HTML]{000000} \textbf{\shortstack{R\&B1\&M\&\\D\&E to B2}}}     & {\color[HTML]{000000} \textbf{Avg}}       \\  \hline
\multicolumn{1}{|c|}{\cellcolor[HTML]{ECF4FF}{\color[HTML]{000000} }}   & \multicolumn{1}{c|}{\cellcolor[HTML]{FFFFFF}{\color[HTML]{000000} Score}}  & \multicolumn{1}{c|}{\cellcolor[HTML]{FFFFFF}{\color[HTML]{000000} Avg}}     & \multicolumn{1}{c|}{\cellcolor[HTML]{FFFFFF}{\color[HTML]{000000} 50.65}}   & \multicolumn{1}{c|}{\cellcolor[HTML]{FFFFFF}{\color[HTML]{000000} 53.45}}   & \multicolumn{1}{c|}{\cellcolor[HTML]{FFFFFF}{\color[HTML]{000000} 52.35}} &\multicolumn{1}{c|}{\cellcolor[HTML]{FFFFFF}{\color[HTML]{000000} 51.16}}  & \multicolumn{1}{c|}{\cellcolor[HTML]{FFFFFF}{\color[HTML]{000000} 48.96}}  & \multicolumn{1}{c|}{\cellcolor[HTML]{FFFFFF}{\color[HTML]{000000} 50.67}}  & \multicolumn{1}{c|}{\cellcolor[HTML]{FFFFFF}{\color[HTML]{000000} 52.31}}       & \multicolumn{1}{c|}{\cellcolor[HTML]{FFFFFF}{\color[HTML]{000000} 50.35}}  & \multicolumn{1}{c|}{\cellcolor[HTML]{FFFFFF}{\color[HTML]{000000}  54.34}} &\multicolumn{1}{c|}{\cellcolor[HTML]{FFFFFF}{\color[HTML]{000000} 52.34}}&\multicolumn{1}{c|}{\cellcolor[HTML]{FFFFFF}{\color[HTML]{000000} 53.10}} \\ \cline{2-14} 
\multicolumn{1}{|c|}{\cellcolor[HTML]{ECF4FF}{\color[HTML]{000000} }}  & \multicolumn{1}{c|}{\cellcolor[HTML]{FFFFFF}{\color[HTML]{000000} }}  & \multicolumn{1}{c|}{\cellcolor[HTML]{FFFFFF}{\color[HTML]{000000} Concat}}       & \multicolumn{1}{c|}{\cellcolor[HTML]{FFFFFF}{\color[HTML]{000000} 51.42}}  & \multicolumn{1}{c|}{\cellcolor[HTML]{FFFFFF}{\color[HTML]{000000} 54.44}}       & \multicolumn{1}{c|}{\cellcolor[HTML]{FFFFFF}{\color[HTML]{000000} 52.33}}  & \multicolumn{1}{c|}{\cellcolor[HTML]{FFFFFF}{\color[HTML]{000000} 51.34}}  & \multicolumn{1}{c|}{\cellcolor[HTML]{FFFFFF}{\color[HTML]{000000} 54.32}}      & \multicolumn{1}{c|}{\cellcolor[HTML]{FFFFFF}{\color[HTML]{000000} 52.31}}  & \multicolumn{1}{c|}{\cellcolor[HTML]{FFFFFF}{\color[HTML]{000000} 53.31}}  & \multicolumn{1}{c|}{\cellcolor[HTML]{FFFFFF}{\color[HTML]{000000} 51.29}} &\multicolumn{1}{c|}{\cellcolor[HTML]{FFFFFF}{\color[HTML]{000000} 54.11}}&\multicolumn{1}{c|}{\cellcolor[HTML]{FFFFFF}{\color[HTML]{000000} 54.39}}&\multicolumn{1}{c|}{\cellcolor[HTML]{FFFFFF}{\color[HTML]{000000} 54.60}}\\ \cline{3-14} 
\multicolumn{1}{|c|}{\cellcolor[HTML]{ECF4FF}{\color[HTML]{000000} }}  & \multicolumn{1}{c|}{\cellcolor[HTML]{FFFFFF}{\color[HTML]{000000} }}  & \multicolumn{1}{c|}{\cellcolor[HTML]{FFFFFF}{\color[HTML]{000000} SE-Concat}}   & \multicolumn{1}{c|}{\cellcolor[HTML]{FFFFFF}{\color[HTML]{000000} 51.48}}  & \multicolumn{1}{c|}{\cellcolor[HTML]{FFFFFF}{\color[HTML]{000000} 54.38}}  & \multicolumn{1}{c|}{\cellcolor[HTML]{FFFFFF}{\color[HTML]{000000} 52.06}}       & \multicolumn{1}{c|}{\cellcolor[HTML]{FFFFFF}{\color[HTML]{000000} 52.45}}  & \multicolumn{1}{c|}{\cellcolor[HTML]{FFFFFF}{\color[HTML]{000000} 56.03}} & \multicolumn{1}{c|}{\cellcolor[HTML]{FFFFFF}{\color[HTML]{000000} 53.28}}        & \multicolumn{1}{c|}{\cellcolor[HTML]{FFFFFF}{\color[HTML]{000000} 54.29}}  & \multicolumn{1}{c|}{\cellcolor[HTML]{FFFFFF}{\color[HTML]{000000} 51.85}} &\multicolumn{1}{c|}{\cellcolor[HTML]{FFFFFF}{\color[HTML]{000000} 54.65}}&\multicolumn{1}{c|}{\cellcolor[HTML]{FFFFFF}{\color[HTML]{000000} 54.82}} &\multicolumn{1}{c|}{\cellcolor[HTML]{FFFFFF}{\color[HTML]{000000} 54.40}}\\ \cline{3-14}
\multicolumn{1}{|c|}{\cellcolor[HTML]{ECF4FF}{\color[HTML]{000000} }} & \multicolumn{1}{c|}{\cellcolor[HTML]{FFFFFF}{\color[HTML]{000000} }}  & \multicolumn{1}{c|}{\cellcolor[HTML]{FFFFFF}{\color[HTML]{000000} Cross-Atten}}  & \multicolumn{1}{c|}{\cellcolor[HTML]{FFFFFF}{\color[HTML]{000000} 52.34}}  & \multicolumn{1}{c|}{\cellcolor[HTML]{FFFFFF}{\color[HTML]{000000} 55.26}}       & \multicolumn{1}{c|}{\cellcolor[HTML]{FFFFFF}{\color[HTML]{000000} 53.10}}  & \multicolumn{1}{c|}{\cellcolor[HTML]{FFFFFF}{\color[HTML]{000000} 53.52}}  & \multicolumn{1}{c|}{\cellcolor[HTML]{FFFFFF}{\color[HTML]{000000} 56.13}}       & \multicolumn{1}{c|}{\cellcolor[HTML]{FFFFFF}{\color[HTML]{000000} 54.65}}  & \multicolumn{1}{c|}{\cellcolor[HTML]{FFFFFF}{\color[HTML]{000000} 54.99}}  & \multicolumn{1}{c|}{\cellcolor[HTML]{FFFFFF}{\color[HTML]{000000} 53.05}} &\multicolumn{1}{c|}{\cellcolor[HTML]{FFFFFF}{\color[HTML]{000000} 55.33}}&\multicolumn{1}{c|}{\cellcolor[HTML]{FFFFFF}{\color[HTML]{000000} 55.69}}&\multicolumn{1}{c|}{\cellcolor[HTML]{FFFFFF}{\color[HTML]{000000} 55.01}} \\ \cline{3-14} 
\multicolumn{1}{|c|}{\cellcolor[HTML]{ECF4FF}{\color[HTML]{000000} }} & \multicolumn{1}{c|}{\cellcolor[HTML]{FFFFFF}{\color[HTML]{000000} }} & \multicolumn{1}{c|}{\cellcolor[HTML]{FFFFFF}{\color[HTML]{000000} MLP-Mixer}}    & \multicolumn{1}{c|}{\cellcolor[HTML]{FFFFFF}{\color[HTML]{000000} 52.46}}  & \multicolumn{1}{c|}{\cellcolor[HTML]{FFFFFF}{\color[HTML]{000000} 55.39}}       & \multicolumn{1}{c|}{\cellcolor[HTML]{FFFFFF}{\color[HTML]{000000} 53.15}}  & \multicolumn{1}{c|}{\cellcolor[HTML]{FFFFFF}{\color[HTML]{000000} 53.65}}  & \multicolumn{1}{c|}{\cellcolor[HTML]{FFFFFF}{\color[HTML]{000000} 56.35}} & \multicolumn{1}{c|}{\cellcolor[HTML]{FFFFFF}{\color[HTML]{000000} 56.33}} & \multicolumn{1}{c|}{\cellcolor[HTML]{FFFFFF}{\color[HTML]{000000} 56.05}}& \multicolumn{1}{c|}{\cellcolor[HTML]{FFFFFF}{\color[HTML]{000000} 53.38}} &\multicolumn{1}{c|}{\cellcolor[HTML]{FFFFFF}{\color[HTML]{000000} 56.89}}&\multicolumn{1}{c|}{\cellcolor[HTML]{FFFFFF}{\color[HTML]{000000} 57.82}}&\multicolumn{1}{c|}{\cellcolor[HTML]{FFFFFF}{\color[HTML]{000000} 56.07}} \\ \cline{3-14}
\multicolumn{1}{|c|}{\cellcolor[HTML]{ECF4FF}{\color[HTML]{000000} }} & \multicolumn{1}{c|}{\cellcolor[HTML]{FFFFFF}{\color[HTML]{000000} }} & \multicolumn{1}{c|}{\cellcolor[HTML]{FFFFFF}{\color[HTML]{000000} CLIP-Align}}    & \multicolumn{1}{c|}{\cellcolor[HTML]{FFFFFF}{\color[HTML]{000000}52.14}}  & \multicolumn{1}{c|}{\cellcolor[HTML]{FFFFFF}{\color[HTML]{000000}54.61}}       & \multicolumn{1}{c|}{\cellcolor[HTML]{FFFFFF}{\color[HTML]{000000}53.34}}  & \multicolumn{1}{c|}{\cellcolor[HTML]{FFFFFF}{\color[HTML]{000000}51.81}}  & \multicolumn{1}{c|}{\cellcolor[HTML]{FFFFFF}{\color[HTML]{000000}55.23}} & \multicolumn{1}{c|}{\cellcolor[HTML]{FFFFFF}{\color[HTML]{000000}52.66}} & \multicolumn{1}{c|}{\cellcolor[HTML]{FFFFFF}{\color[HTML]{000000}53.95}}& \multicolumn{1}{c|}{\cellcolor[HTML]{FFFFFF}{\color[HTML]{000000}52.37}} &\multicolumn{1}{c|}{\cellcolor[HTML]{FFFFFF}{\color[HTML]{000000}54.38}}&\multicolumn{1}{c|}{\cellcolor[HTML]{FFFFFF}{\color[HTML]{000000}55.18}}&\multicolumn{1}{c|}{\cellcolor[HTML]{FFFFFF}{\color[HTML]{000000}55.24}} \\ \cline{3-14}
\multicolumn{1}{|c|}{\multirow{-6}{*}{\cellcolor[HTML]{ECF4FF}{\color[HTML]{000000} \begin{tabular}[c]{@{}c@{}}V-Face\\ +\\ V-AGA\end{tabular}}}}                               & \multicolumn{1}{c|}{\multirow{-5}{*}{\cellcolor[HTML]{FFFFFF}{\color[HTML]{000000} Feature}}} & \multicolumn{1}{c|}{\cellcolor[HTML]{FFFFFF}{\color[HTML]{000000} \textbf{Atten-Mixer}}} & \multicolumn{1}{c|}{\cellcolor[HTML]{FFFFFF}{\color[HTML]{000000} 53.89}}       & \multicolumn{1}{c|}{\cellcolor[HTML]{FFFFFF}{\color[HTML]{000000} 57.02}} & \multicolumn{1}{c|}{\cellcolor[HTML]{FFFFFF}{\color[HTML]{000000} 53.45}}  & \multicolumn{1}{c|}{\cellcolor[HTML]{FFFFFF}{\color[HTML]{000000} 54.62}}  & \multicolumn{1}{c|}{\cellcolor[HTML]{FFFFFF}{\color[HTML]{000000} 58.16}}  & \multicolumn{1}{c|}{\cellcolor[HTML]{FFFFFF}{\color[HTML]{000000} 57.01}}  & \multicolumn{1}{c|}{\cellcolor[HTML]{FFFFFF}{\color[HTML]{000000} 58.24}}  & \multicolumn{1}{c|}{\cellcolor[HTML]{FFFFFF}{\color[HTML]{000000} 54.02}} &\multicolumn{1}{c|}{\cellcolor[HTML]{FFFFFF}{\color[HTML]{000000} 59.89}}&\multicolumn{1}{c|}{\cellcolor[HTML]{FFFFFF}{\color[HTML]{000000} \textbf{59.85}}} &\multicolumn{1}{c|}{\cellcolor[HTML]{FFFFFF}{\color[HTML]{000000} 56.94}}\\ \hline
\multicolumn{1}{|c|}{\cellcolor[HTML]{FFFFEB}{\color[HTML]{000000} }} & \multicolumn{1}{c|}{\cellcolor[HTML]{FFFFFF}{\color[HTML]{000000} Score}}  & \multicolumn{1}{c|}{\cellcolor[HTML]{FFFFFF}{\color[HTML]{000000} Avg}}     & \multicolumn{1}{c|}{\cellcolor[HTML]{FFFFFF}{\color[HTML]{000000} 50.53}} & \multicolumn{1}{c|}{\cellcolor[HTML]{FFFFFF}{\color[HTML]{000000} 52.88}} & \multicolumn{1}{c|}{\cellcolor[HTML]{FFFFFF}{\color[HTML]{000000} 51.38}}      & \multicolumn{1}{c|}{\cellcolor[HTML]{FFFFFF}{\color[HTML]{000000} 52.63}} & \multicolumn{1}{c|}{\cellcolor[HTML]{FFFFFF}{\color[HTML]{000000} 54.28}} & \multicolumn{1}{c|}{\cellcolor[HTML]{FFFFFF}{\color[HTML]{000000} 51.04}} & \multicolumn{1}{c|}{\cellcolor[HTML]{FFFFFF}{\color[HTML]{000000} 52.85}}  & \multicolumn{1}{c|}{\cellcolor[HTML]{FFFFFF}{\color[HTML]{000000} 51.25}} &\multicolumn{1}{c|}{\cellcolor[HTML]{FFFFFF}{\color[HTML]{000000} 53.22}} &\multicolumn{1}{c|}{\cellcolor[HTML]{FFFFFF}{\color[HTML]{000000} 55.68}} &\multicolumn{1}{c|}{\cellcolor[HTML]{FFFFFF}{\color[HTML]{000000}  53.57 }} \\ \cline{2-14} 
\multicolumn{1}{|c|}{\cellcolor[HTML]{FFFFEB}{\color[HTML]{000000} }} & \multicolumn{1}{c|}{\cellcolor[HTML]{FFFFFF}{\color[HTML]{000000} }} & \multicolumn{1}{c|}{\cellcolor[HTML]{FFFFFF}{\color[HTML]{000000} Concat}} & \multicolumn{1}{c|}{\cellcolor[HTML]{FFFFFF}{\color[HTML]{000000} 51.45}} & \multicolumn{1}{c|}{\cellcolor[HTML]{FFFFFF}{\color[HTML]{000000} 53.66}}  & \multicolumn{1}{c|}{\cellcolor[HTML]{FFFFFF}{\color[HTML]{000000} 52.36}}      & \multicolumn{1}{c|}{\cellcolor[HTML]{FFFFFF}{\color[HTML]{000000} 52.89}}  & \multicolumn{1}{c|}{\cellcolor[HTML]{FFFFFF}{\color[HTML]{000000} 56.35}}  & \multicolumn{1}{c|}{\cellcolor[HTML]{FFFFFF}{\color[HTML]{000000} 53.28}}  & \multicolumn{1}{c|}{\cellcolor[HTML]{FFFFFF}{\color[HTML]{000000} 54.39}}  & \multicolumn{1}{c|}{\cellcolor[HTML]{FFFFFF}{\color[HTML]{000000} 52.22}} &\multicolumn{1}{c|}{\cellcolor[HTML]{FFFFFF}{\color[HTML]{000000} 53.31}} &\multicolumn{1}{c|}{\cellcolor[HTML]{FFFFFF}{\color[HTML]{000000} 56.83}} &\multicolumn{1}{c|}{\cellcolor[HTML]{FFFFFF}{\color[HTML]{000000}54.15}} \\ \cline{3-14} 
\multicolumn{1}{|c|}{\cellcolor[HTML]{FFFFEB}{\color[HTML]{000000} }} & \multicolumn{1}{c|}{\cellcolor[HTML]{FFFFFF}{\color[HTML]{000000} }}  & \multicolumn{1}{c|}{\cellcolor[HTML]{FFFFFF}{\color[HTML]{000000} SE-Concat}}   & \multicolumn{1}{c|}{\cellcolor[HTML]{FFFFFF}{\color[HTML]{000000} 52.11}} & \multicolumn{1}{c|}{\cellcolor[HTML]{FFFFFF}{\color[HTML]{000000} 55.02}} & \multicolumn{1}{c|}{\cellcolor[HTML]{FFFFFF}{\color[HTML]{000000} 52.39}} & \multicolumn{1}{c|}{\cellcolor[HTML]{FFFFFF}{\color[HTML]{000000} 53.08}} & \multicolumn{1}{c|}{\cellcolor[HTML]{FFFFFF}{\color[HTML]{000000} 58.04}} & \multicolumn{1}{c|}{\cellcolor[HTML]{FFFFFF}{\color[HTML]{000000} 54.99}} & \multicolumn{1}{c|}{\cellcolor[HTML]{FFFFFF}{\color[HTML]{000000} 54.78}} & \multicolumn{1}{c|}{\cellcolor[HTML]{FFFFFF}{\color[HTML]{000000} 53.48}} &\multicolumn{1}{c|}{\cellcolor[HTML]{FFFFFF}{\color[HTML]{000000} 54.09}}& \multicolumn{1}{c|}{\cellcolor[HTML]{FFFFFF}{\color[HTML]{000000} 56.89}} &\multicolumn{1}{c|}{\cellcolor[HTML]{FFFFFF}{\color[HTML]{000000} 54.34}} \\ \cline{3-14} 
\multicolumn{1}{|c|}{\cellcolor[HTML]{FFFFEB}{\color[HTML]{000000} }} & \multicolumn{1}{c|}{\cellcolor[HTML]{FFFFFF}{\color[HTML]{000000} }} & \multicolumn{1}{c|}{\cellcolor[HTML]{FFFFFF}{\color[HTML]{000000} Cross-Atten}}& \multicolumn{1}{c|}{\cellcolor[HTML]{FFFFFF}{\color[HTML]{000000} 52.98}}& \multicolumn{1}{c|}{\cellcolor[HTML]{FFFFFF}{\color[HTML]{000000} 56.36}} & \multicolumn{1}{c|}{\cellcolor[HTML]{FFFFFF}{\color[HTML]{000000} 53.24}} & \multicolumn{1}{c|}{\cellcolor[HTML]{FFFFFF}{\color[HTML]{000000} 53.96}} & \multicolumn{1}{c|}{\cellcolor[HTML]{FFFFFF}{\color[HTML]{000000} 58.26}} & \multicolumn{1}{c|}{\cellcolor[HTML]{FFFFFF}{\color[HTML]{000000} 56.84}} & \multicolumn{1}{c|}{\cellcolor[HTML]{FFFFFF}{\color[HTML]{000000} 55.76}} & \multicolumn{1}{c|}{\cellcolor[HTML]{FFFFFF}{\color[HTML]{000000} 53.87}} &\multicolumn{1}{c|}{\cellcolor[HTML]{FFFFFF}{\color[HTML]{000000} 55.68}} &\multicolumn{1}{c|}{\cellcolor[HTML]{FFFFFF}{\color[HTML]{000000} 57.63}} &\multicolumn{1}{c|}{\cellcolor[HTML]{FFFFFF}{\color[HTML]{000000} 55.62}} \\ \cline{3-14}
\multicolumn{1}{|c|}{\cellcolor[HTML]{FFFFEB}{\color[HTML]{000000} }} & \multicolumn{1}{c|}{\cellcolor[HTML]{FFFFFF}{\color[HTML]{000000} }} & \multicolumn{1}{c|}{\cellcolor[HTML]{FFFFFF}{\color[HTML]{000000} MLP-Mixer}} & \multicolumn{1}{c|}{\cellcolor[HTML]{FFFFFF}{\color[HTML]{000000} 53.42}} & \multicolumn{1}{c|}{\cellcolor[HTML]{FFFFFF}{\color[HTML]{000000} 56.79}} & \multicolumn{1}{c|}{\cellcolor[HTML]{FFFFFF}{\color[HTML]{000000} 53.33}} & \multicolumn{1}{c|}{\cellcolor[HTML]{FFFFFF}{\color[HTML]{000000} 54.21}} & \multicolumn{1}{c|}{\cellcolor[HTML]{FFFFFF}{\color[HTML]{000000} 58.45}} & \multicolumn{1}{c|}{\cellcolor[HTML]{FFFFFF}{\color[HTML]{000000} 57.05}} & \multicolumn{1}{c|}{\cellcolor[HTML]{FFFFFF}{\color[HTML]{000000} 57.35}} & \multicolumn{1}{c|}{\cellcolor[HTML]{FFFFFF}{\color[HTML]{000000} 54.33}} &\multicolumn{1}{c|}{\cellcolor[HTML]{FFFFFF}{\color[HTML]{000000} 59.88}} &\multicolumn{1}{c|}{\cellcolor[HTML]{FFFFFF}{\color[HTML]{000000} 58.47}} &\multicolumn{1}{c|}{\cellcolor[HTML]{FFFFFF}{\color[HTML]{000000} 55.98}} \\ \cline{3-14}
\multicolumn{1}{|c|}{\cellcolor[HTML]{FFFFEB}{\color[HTML]{000000} }} & \multicolumn{1}{c|}{\cellcolor[HTML]{FFFFFF}{\color[HTML]{000000} }} & \multicolumn{1}{c|}{\cellcolor[HTML]{FFFFFF}{\color[HTML]{000000} CLIP-Align}}    & \multicolumn{1}{c|}{\cellcolor[HTML]{FFFFFF}{\color[HTML]{000000}51.21}}  & \multicolumn{1}{c|}{\cellcolor[HTML]{FFFFFF}{\color[HTML]{000000}53.82}}       & \multicolumn{1}{c|}{\cellcolor[HTML]{FFFFFF}{\color[HTML]{000000}53.38}}  & \multicolumn{1}{c|}{\cellcolor[HTML]{FFFFFF}{\color[HTML]{000000}53.45}}  & \multicolumn{1}{c|}{\cellcolor[HTML]{FFFFFF}{\color[HTML]{000000}56.76}} & \multicolumn{1}{c|}{\cellcolor[HTML]{FFFFFF}{\color[HTML]{000000}54.25}} & \multicolumn{1}{c|}{\cellcolor[HTML]{FFFFFF}{\color[HTML]{000000}54.60}}& \multicolumn{1}{c|}{\cellcolor[HTML]{FFFFFF}{\color[HTML]{000000}53.04}} &\multicolumn{1}{c|}{\cellcolor[HTML]{FFFFFF}{\color[HTML]{000000}54.41}}&\multicolumn{1}{c|}{\cellcolor[HTML]{FFFFFF}{\color[HTML]{000000}57.63}}&\multicolumn{1}{c|}{\cellcolor[HTML]{FFFFFF}{\color[HTML]{000000}54.72}} \\ \cline{3-14}
\multicolumn{1}{|c|}{\multirow{-6}{*}{\cellcolor[HTML]{FFFFEB}{\color[HTML]{000000} \begin{tabular}[c]{@{}c@{}}V-Face\\ +\\ A-Mel\end{tabular}}}}                               & \multicolumn{1}{c|}{\multirow{-5}{*}{\cellcolor[HTML]{FFFFFF}{\color[HTML]{000000} Feature}}} & \multicolumn{1}{c|}{\cellcolor[HTML]{FFFFFF}{\color[HTML]{000000} \textbf{Atten-Mixer}}} & \multicolumn{1}{c|}{\cellcolor[HTML]{FFFFFF}{\color[HTML]{000000} 54.64}} & \multicolumn{1}{c|}{\cellcolor[HTML]{FFFFFF}{\color[HTML]{000000} 58.98}} & \multicolumn{1}{c|}{\cellcolor[HTML]{FFFFFF}{\color[HTML]{000000} 53.06}} & \multicolumn{1}{c|}{\cellcolor[HTML]{FFFFFF}{\color[HTML]{000000} 54.76}} & \multicolumn{1}{c|}{\cellcolor[HTML]{FFFFFF}{\color[HTML]{000000} \underline{58.95}}}      & \multicolumn{1}{c|}{\cellcolor[HTML]{FFFFFF}{\color[HTML]{000000} 58.28}} & \multicolumn{1}{c|}{\cellcolor[HTML]{FFFFFF}{\color[HTML]{000000} 58.92}}  & \multicolumn{1}{c|}{\cellcolor[HTML]{FFFFFF}{\color[HTML]{000000} \underline{55.21}}} &\multicolumn{1}{c|}{\cellcolor[HTML]{FFFFFF}{\color[HTML]{000000} \textbf{62.05}}} &\multicolumn{1}{c|}{\cellcolor[HTML]{FFFFFF}{\color[HTML]{000000} 59.86}} &\multicolumn{1}{c|}{\cellcolor[HTML]{FFFFFF}{\color[HTML]{000000} 56.96}} \\ \hline
\multicolumn{1}{|c|}{\cellcolor[HTML]{FFF7F7}{\color[HTML]{000000} }} & \multicolumn{1}{c|}{\cellcolor[HTML]{FFFFFF}{\color[HTML]{000000} Score}} & \multicolumn{1}{c|}{\cellcolor[HTML]{FFFFFF}{\color[HTML]{000000} Avg}}     & \multicolumn{1}{c|}{\cellcolor[HTML]{FFFFFF}{\color[HTML]{000000} 52.17}} & \multicolumn{1}{c|}{\cellcolor[HTML]{FFFFFF}{\color[HTML]{000000} 56.42}} & \multicolumn{1}{c|}{\cellcolor[HTML]{FFFFFF}{\color[HTML]{000000} 51.28}} & \multicolumn{1}{c|}{\cellcolor[HTML]{FFFFFF}{\color[HTML]{000000} 52.36}} & \multicolumn{1}{c|}{\cellcolor[HTML]{FFFFFF}{\color[HTML]{000000} 53.15}} & \multicolumn{1}{c|}{\cellcolor[HTML]{FFFFFF}{\color[HTML]{000000} 50.58}} & \multicolumn{1}{c|}{\cellcolor[HTML]{FFFFFF}{\color[HTML]{000000} 57.08}} & \multicolumn{1}{c|}{\cellcolor[HTML]{FFFFFF}{\color[HTML]{000000} 51.43}} &\multicolumn{1}{c|}{\cellcolor[HTML]{FFFFFF}{\color[HTML]{000000} 53.34}} &\multicolumn{1}{c|}{\cellcolor[HTML]{FFFFFF}{\color[HTML]{000000} 54.27}}& \multicolumn{1}{c|}{\cellcolor[HTML]{FFFFFF}{\color[HTML]{000000} 53.79}} \\ \cline{2-14}
\multicolumn{1}{|c|}{\cellcolor[HTML]{FFF7F7}{\color[HTML]{000000} }} & \multicolumn{1}{c|}{\cellcolor[HTML]{FFFFFF}{\color[HTML]{000000} }} & \multicolumn{1}{c|}{\cellcolor[HTML]{FFFFFF}{\color[HTML]{000000} Concat}} & \multicolumn{1}{c|}{\cellcolor[HTML]{FFFFFF}{\color[HTML]{000000} 52.89}} & \multicolumn{1}{c|}{\cellcolor[HTML]{FFFFFF}{\color[HTML]{000000} 56.63}} & \multicolumn{1}{c|}{\cellcolor[HTML]{FFFFFF}{\color[HTML]{000000} 51.64}} & \multicolumn{1}{c|}{\cellcolor[HTML]{FFFFFF}{\color[HTML]{000000} 53.34}} & \multicolumn{1}{c|}{\cellcolor[HTML]{FFFFFF}{\color[HTML]{000000} 56.24}} & \multicolumn{1}{c|}{\cellcolor[HTML]{FFFFFF}{\color[HTML]{000000} 50.87}} & \multicolumn{1}{c|}{\cellcolor[HTML]{FFFFFF}{\color[HTML]{000000} 57.06}} & \multicolumn{1}{c|}{\cellcolor[HTML]{FFFFFF}{\color[HTML]{000000} 52.11}} &\multicolumn{1}{c|}{\cellcolor[HTML]{FFFFFF}{\color[HTML]{000000} 54.99}} &\multicolumn{1}{c|}{\cellcolor[HTML]{FFFFFF}{\color[HTML]{000000} 57.25}} &\multicolumn{1}{c|}{\cellcolor[HTML]{FFFFFF}{\color[HTML]{000000} 54.50}} \\ \cline{3-14}
\multicolumn{1}{|c|}{\cellcolor[HTML]{FFF7F7}{\color[HTML]{000000} }}& \multicolumn{1}{c|}{\cellcolor[HTML]{FFFFFF}{\color[HTML]{000000} }} & \multicolumn{1}{c|}{\cellcolor[HTML]{FFFFFF}{\color[HTML]{000000} SE-Concat}}  & \multicolumn{1}{c|}{\cellcolor[HTML]{FFFFFF}{\color[HTML]{000000} 53.42}} & \multicolumn{1}{c|}{\cellcolor[HTML]{FFFFFF}{\color[HTML]{000000} 57.06}} & \multicolumn{1}{c|}{\cellcolor[HTML]{FFFFFF}{\color[HTML]{000000} 52.03}} & \multicolumn{1}{c|}{\cellcolor[HTML]{FFFFFF}{\color[HTML]{000000} 54.68}} & \multicolumn{1}{c|}{\cellcolor[HTML]{FFFFFF}{\color[HTML]{000000} 57.03}} & \multicolumn{1}{c|}{\cellcolor[HTML]{FFFFFF}{\color[HTML]{000000} 56.92}} & \multicolumn{1}{c|}{\cellcolor[HTML]{FFFFFF}{\color[HTML]{000000} 58.45}} & \multicolumn{1}{c|}{\cellcolor[HTML]{FFFFFF}{\color[HTML]{000000} 52.98}} &\multicolumn{1}{c|}{\cellcolor[HTML]{FFFFFF}{\color[HTML]{000000} 55.56}} &\multicolumn{1}{c|}{\cellcolor[HTML]{FFFFFF}{\color[HTML]{000000} 58.62}} &\multicolumn{1}{c|}{\cellcolor[HTML]{FFFFFF}{\color[HTML]{000000} 56.07}} \\ \cline{3-14} 
\multicolumn{1}{|c|}{\cellcolor[HTML]{FFF7F7}{\color[HTML]{000000} }} & \multicolumn{1}{c|}{\cellcolor[HTML]{FFFFFF}{\color[HTML]{000000} }} & \multicolumn{1}{c|}{\cellcolor[HTML]{FFFFFF}{\color[HTML]{000000} Cross-Atten}} & \multicolumn{1}{c|}{\cellcolor[HTML]{FFFFFF}{\color[HTML]{000000} 53.22}} & \multicolumn{1}{c|}{\cellcolor[HTML]{FFFFFF}{\color[HTML]{000000} 58.99}} & \multicolumn{1}{c|}{\cellcolor[HTML]{FFFFFF}{\color[HTML]{000000} 52.65}} & \multicolumn{1}{c|}{\cellcolor[HTML]{FFFFFF}{\color[HTML]{000000} 54.92}} & \multicolumn{1}{c|}{\cellcolor[HTML]{FFFFFF}{\color[HTML]{000000} 57.05}} & \multicolumn{1}{c|}{\cellcolor[HTML]{FFFFFF}{\color[HTML]{000000} 56.83}} & \multicolumn{1}{c|}{\cellcolor[HTML]{FFFFFF}{\color[HTML]{000000} 58.69}} & \multicolumn{1}{c|}{\cellcolor[HTML]{FFFFFF}{\color[HTML]{000000} 53.89}} &\multicolumn{1}{c|}{\cellcolor[HTML]{FFFFFF}{\color[HTML]{000000} 57.84}} &\multicolumn{1}{c|}{\cellcolor[HTML]{FFFFFF}{\color[HTML]{000000} 59.37}} &\multicolumn{1}{c|}{\cellcolor[HTML]{FFFFFF}{\color[HTML]{000000} 55.74}} \\ \cline{3-14} 
\multicolumn{1}{|c|}{\cellcolor[HTML]{FFF7F7}{\color[HTML]{000000} }} & \multicolumn{1}{c|}{\cellcolor[HTML]{FFFFFF}{\color[HTML]{000000} }} & \multicolumn{1}{c|}{\cellcolor[HTML]{FFFFFF}{\color[HTML]{000000} MLP-Mixer}}  & \multicolumn{1}{c|}{\cellcolor[HTML]{FFFFFF}{\color[HTML]{000000} 54.65}} & \multicolumn{1}{c|}{\cellcolor[HTML]{FFFFFF}{\color[HTML]{000000} 59.78}} & \multicolumn{1}{c|}{\cellcolor[HTML]{FFFFFF}{\color[HTML]{000000} 53.08}} & \multicolumn{1}{c|}{\cellcolor[HTML]{FFFFFF}{\color[HTML]{000000} 55.01}} & \multicolumn{1}{c|}{\cellcolor[HTML]{FFFFFF}{\color[HTML]{000000} 58.34}} & \multicolumn{1}{c|}{\cellcolor[HTML]{FFFFFF}{\color[HTML]{000000} 57.36}} & \multicolumn{1}{c|}{\cellcolor[HTML]{FFFFFF}{\color[HTML]{000000} 59.05}} & \multicolumn{1}{c|}{\cellcolor[HTML]{FFFFFF}{\color[HTML]{000000} 54.54}} & \multicolumn{1}{c|}{\cellcolor[HTML]{FFFFFF}{\color[HTML]{000000} 59.08}} &\multicolumn{1}{c|}{\cellcolor[HTML]{FFFFFF}{\color[HTML]{000000} 59.38}} & \multicolumn{1}{c|}{\cellcolor[HTML]{FFFFFF}{\color[HTML]{000000} 56.26}} \\ \cline{3-14}
\multicolumn{1}{|c|}{\cellcolor[HTML]{FFF7F7}{\color[HTML]{000000} }} & \multicolumn{1}{c|}{\cellcolor[HTML]{FFFFFF}{\color[HTML]{000000} }} & \multicolumn{1}{c|}{\cellcolor[HTML]{FFFFFF}{\color[HTML]{000000} CLIP-Align}}    & \multicolumn{1}{c|}{\cellcolor[HTML]{FFFFFF}{\color[HTML]{000000}53.42}}  & \multicolumn{1}{c|}{\cellcolor[HTML]{FFFFFF}{\color[HTML]{000000}57.31}}       & \multicolumn{1}{c|}{\cellcolor[HTML]{FFFFFF}{\color[HTML]{000000}52.49}}  & \multicolumn{1}{c|}{\cellcolor[HTML]{FFFFFF}{\color[HTML]{000000}54.12}}  & \multicolumn{1}{c|}{\cellcolor[HTML]{FFFFFF}{\color[HTML]{000000}56.98}} & \multicolumn{1}{c|}{\cellcolor[HTML]{FFFFFF}{\color[HTML]{000000}51.56}} & \multicolumn{1}{c|}{\cellcolor[HTML]{FFFFFF}{\color[HTML]{000000}57.72}}& \multicolumn{1}{c|}{\cellcolor[HTML]{FFFFFF}{\color[HTML]{000000}52.84}} &\multicolumn{1}{c|}{\cellcolor[HTML]{FFFFFF}{\color[HTML]{000000}55.68}}&\multicolumn{1}{c|}{\cellcolor[HTML]{FFFFFF}{\color[HTML]{000000}58.09}}&\multicolumn{1}{c|}{\cellcolor[HTML]{FFFFFF}{\color[HTML]{000000}55.22}} \\ \cline{3-14}
\multicolumn{1}{|c|}{\multirow{-6}{*}{\cellcolor[HTML]{FFF7F7}{\color[HTML]{000000} \begin{tabular}[c]{@{}c@{}}V-AGA\\ +\\ A-Mel\end{tabular}}}} & \multicolumn{1}{c|}{\multirow{-5}{*}{\cellcolor[HTML]{FFFFFF}{\color[HTML]{000000} Feature}}} & \multicolumn{1}{c|}{\cellcolor[HTML]{FFFFFF}{\color[HTML]{000000} \textbf{Atten-Mixer}}} & \multicolumn{1}{c|}{\cellcolor[HTML]{FFFFFF}{\color[HTML]{000000} 54.38}} & \multicolumn{1}{c|}{\cellcolor[HTML]{FFFFFF}{\color[HTML]{000000} \underline{61.49}}} & \multicolumn{1}{c|}{\cellcolor[HTML]{FFFFFF}{\color[HTML]{000000} 54.56}} & \multicolumn{1}{c|}{\cellcolor[HTML]{FFFFFF}{\color[HTML]{000000} 56.38}} & \multicolumn{1}{c|}{\cellcolor[HTML]{FFFFFF}{\color[HTML]{000000} 58.65}} & \multicolumn{1}{c|}{\cellcolor[HTML]{FFFFFF}{\color[HTML]{000000} 57.82}} & \multicolumn{1}{c|}{\cellcolor[HTML]{FFFFFF}{\color[HTML]{000000} \underline{60.18}}}& \multicolumn{1}{c|}{\cellcolor[HTML]{FFFFFF}{\color[HTML]{000000} \textbf{55.38}}}  &\multicolumn{1}{c|}{\cellcolor[HTML]{FFFFFF}{\color[HTML]{000000} \underline{61.05}}}&\multicolumn{1}{c|}{\cellcolor[HTML]{FFFFFF}{\color[HTML]{000000} \underline{61.20}}}&\multicolumn{1}{c|}{\cellcolor[HTML]{FFFFFF}{\color[HTML]{000000} 56.89}} \\ \hline
\multicolumn{1}{|c|}{\cellcolor[HTML]{EEFBED}{\color[HTML]{000000} }} & \multicolumn{1}{c|}{\cellcolor[HTML]{FFFFFF}{\color[HTML]{000000} Score}} & \multicolumn{1}{c|}{\cellcolor[HTML]{FFFFFF}{\color[HTML]{000000} Avg}}    & \multicolumn{1}{c|}{\cellcolor[HTML]{FFFFFF}{\color[HTML]{000000} 52.85}} & \multicolumn{1}{c|}{\cellcolor[HTML]{FFFFFF}{\color[HTML]{000000} 56.84}} & \multicolumn{1}{c|}{\cellcolor[HTML]{FFFFFF}{\color[HTML]{000000} 52.41}}      & \multicolumn{1}{c|}{\cellcolor[HTML]{FFFFFF}{\color[HTML]{000000} 53.04}} & \multicolumn{1}{c|}{\cellcolor[HTML]{FFFFFF}{\color[HTML]{000000} 56.48}} & \multicolumn{1}{c|}{\cellcolor[HTML]{FFFFFF}{\color[HTML]{000000} 51.22}} & \multicolumn{1}{c|}{\cellcolor[HTML]{FFFFFF}{\color[HTML]{000000} 54.28}} & \multicolumn{1}{c|}{\cellcolor[HTML]{FFFFFF}{\color[HTML]{000000} 51.02}} & \multicolumn{1}{c|}{\cellcolor[HTML]{FFFFFF}{\color[HTML]{000000} 53.45}} &\multicolumn{1}{c|}{\cellcolor[HTML]{FFFFFF}{\color[HTML]{000000} 57.39}} &\multicolumn{1}{c|}{\cellcolor[HTML]{FFFFFF}{\color[HTML]{000000} 54.21}} \\ \cline{2-14}
\multicolumn{1}{|c|}{\cellcolor[HTML]{EEFBED}{\color[HTML]{000000} }}& \multicolumn{1}{c|}{\cellcolor[HTML]{FFFFFF}{\color[HTML]{000000} }} & \multicolumn{1}{c|}{\cellcolor[HTML]{FFFFFF}{\color[HTML]{000000} Concat}} & \multicolumn{1}{c|}{\cellcolor[HTML]{FFFFFF}{\color[HTML]{000000} 54.86}} & \multicolumn{1}{c|}{\cellcolor[HTML]{FFFFFF}{\color[HTML]{000000} 56.98}} & \multicolumn{1}{c|}{\cellcolor[HTML]{FFFFFF}{\color[HTML]{000000} 52.38}} & \multicolumn{1}{c|}{\cellcolor[HTML]{FFFFFF}{\color[HTML]{000000} 54.02}} & \multicolumn{1}{c|}{\cellcolor[HTML]{FFFFFF}{\color[HTML]{000000} 56.87}} & \multicolumn{1}{c|}{\cellcolor[HTML]{FFFFFF}{\color[HTML]{000000} 53.23}} & \multicolumn{1}{c|}{\cellcolor[HTML]{FFFFFF}{\color[HTML]{000000} 54.62}} & \multicolumn{1}{c|}{\cellcolor[HTML]{FFFFFF}{\color[HTML]{000000} 52.24}} & \multicolumn{1}{c|}{\cellcolor[HTML]{FFFFFF}{\color[HTML]{000000} 54.89}} & \multicolumn{1}{c|}{\cellcolor[HTML]{FFFFFF}{\color[HTML]{000000} 57.26}} & \multicolumn{1}{c|}{\cellcolor[HTML]{FFFFFF}{\color[HTML]{000000} 55.07}} \\ \cline{3-14}
\multicolumn{1}{|c|}{\cellcolor[HTML]{EEFBED}{\color[HTML]{000000} }} & \multicolumn{1}{c|}{\cellcolor[HTML]{FFFFFF}{\color[HTML]{000000} }} & \multicolumn{1}{c|}{\cellcolor[HTML]{FFFFFF}{\color[HTML]{000000} SE-Concat}} & \multicolumn{1}{c|}{\cellcolor[HTML]{FFFFFF}{\color[HTML]{000000} 55.53}} & \multicolumn{1}{c|}{\cellcolor[HTML]{FFFFFF}{\color[HTML]{000000} 57.32}} & \multicolumn{1}{c|}{\cellcolor[HTML]{FFFFFF}{\color[HTML]{000000} 53.06}} & \multicolumn{1}{c|}{\cellcolor[HTML]{FFFFFF}{\color[HTML]{000000} 55.93}} & \multicolumn{1}{c|}{\cellcolor[HTML]{FFFFFF}{\color[HTML]{000000} 56.92}} & \multicolumn{1}{c|}{\cellcolor[HTML]{FFFFFF}{\color[HTML]{000000} 56.35}} & \multicolumn{1}{c|}{\cellcolor[HTML]{FFFFFF}{\color[HTML]{000000} 55.23}} & \multicolumn{1}{c|}{\cellcolor[HTML]{FFFFFF}{\color[HTML]{000000} 53.26}} &\multicolumn{1}{c|}{\cellcolor[HTML]{FFFFFF}{\color[HTML]{000000} 56.53}} &\multicolumn{1}{c|}{\cellcolor[HTML]{FFFFFF}{\color[HTML]{000000} 58.02}} &\multicolumn{1}{c|}{\cellcolor[HTML]{FFFFFF}{\color[HTML]{000000} 55.65}} \\ \cline{3-14} 
\multicolumn{1}{|c|}{\cellcolor[HTML]{EEFBED}{\color[HTML]{000000} }} & \multicolumn{1}{c|}{\cellcolor[HTML]{FFFFFF}{\color[HTML]{000000} }} & \multicolumn{1}{c|}{\cellcolor[HTML]{FFFFFF}{\color[HTML]{000000} Cross-Atten}} & \multicolumn{1}{c|}{\cellcolor[HTML]{FFFFFF}{\color[HTML]{000000} \underline{56.88}}} & \multicolumn{1}{c|}{\cellcolor[HTML]{FFFFFF}{\color[HTML]{000000} 59.63}} & \multicolumn{1}{c|}{\cellcolor[HTML]{FFFFFF}{\color[HTML]{000000} 53.66}} & \multicolumn{1}{c|}{\cellcolor[HTML]{FFFFFF}{\color[HTML]{000000} 55.89}} & \multicolumn{1}{c|}{\cellcolor[HTML]{FFFFFF}{\color[HTML]{000000} 57.98}} & \multicolumn{1}{c|}{\cellcolor[HTML]{FFFFFF}{\color[HTML]{000000} 57.26}} & \multicolumn{1}{c|}{\cellcolor[HTML]{FFFFFF}{\color[HTML]{000000} 57.01}} & \multicolumn{1}{c|}{\cellcolor[HTML]{FFFFFF}{\color[HTML]{000000} 53.82}} &\multicolumn{1}{c|}{\cellcolor[HTML]{FFFFFF}{\color[HTML]{000000} 56.87}} &\multicolumn{1}{c|}{\cellcolor[HTML]{FFFFFF}{\color[HTML]{000000}58.25}} & \multicolumn{1}{c|}{\cellcolor[HTML]{FFFFFF}{\color[HTML]{000000} 56.58}}  \\ \cline{3-14} 
\multicolumn{1}{|c|}{\cellcolor[HTML]{EEFBED}{\color[HTML]{000000} }} & \multicolumn{1}{c|}{\cellcolor[HTML]{FFFFFF}{\color[HTML]{000000} }} & \multicolumn{1}{c|}{\cellcolor[HTML]{FFFFFF}{\color[HTML]{000000} MLP-Mixer}} & \multicolumn{1}{c|}{\cellcolor[HTML]{FFFFFF}{\color[HTML]{000000} 56.08}} & \multicolumn{1}{c|}{\cellcolor[HTML]{FFFFFF}{\color[HTML]{000000} 61.02}} & \multicolumn{1}{c|}{\cellcolor[HTML]{FFFFFF}{\color[HTML]{000000} \textbf{55.24}}} & \multicolumn{1}{c|}{\cellcolor[HTML]{FFFFFF}{\color[HTML]{000000} \underline{57.46}}} & \multicolumn{1}{c|}{\cellcolor[HTML]{FFFFFF}{\color[HTML]{000000} 58.35}} & \multicolumn{1}{c|}{\cellcolor[HTML]{FFFFFF}{\color[HTML]{000000} \underline{58.35}}} & \multicolumn{1}{c|}{\cellcolor[HTML]{FFFFFF}{\color[HTML]{000000} 59.55}} &  \multicolumn{1}{c|}{\cellcolor[HTML]{FFFFFF}{\color[HTML]{000000} 54.38}} & \multicolumn{1}{c|}{\cellcolor[HTML]{FFFFFF}{\color[HTML]{000000} 57.05}} & \multicolumn{1}{c|}{\cellcolor[HTML]{FFFFFF}{\color[HTML]{000000} 59.36}} & \multicolumn{1}{c|}{\cellcolor[HTML]{FFFFFF}{\color[HTML]{000000} \underline{57.58}}} \\ \cline{3-14} 
\multicolumn{1}{|c|}{\cellcolor[HTML]{EEFBED}{\color[HTML]{000000} }} & \multicolumn{1}{c|}{\cellcolor[HTML]{FFFFFF}{\color[HTML]{000000} }} & \multicolumn{1}{c|}{\cellcolor[HTML]{FFFFFF}{\color[HTML]{000000} CLIP-Align}}    & \multicolumn{1}{c|}{\cellcolor[HTML]{FFFFFF}{\color[HTML]{000000}55.59}}  & \multicolumn{1}{c|}{\cellcolor[HTML]{FFFFFF}{\color[HTML]{000000}57.14}}       & \multicolumn{1}{c|}{\cellcolor[HTML]{FFFFFF}{\color[HTML]{000000}53.40}}  & \multicolumn{1}{c|}{\cellcolor[HTML]{FFFFFF}{\color[HTML]{000000}54.50}}  & \multicolumn{1}{c|}{\cellcolor[HTML]{FFFFFF}{\color[HTML]{000000}57.78}} & \multicolumn{1}{c|}{\cellcolor[HTML]{FFFFFF}{\color[HTML]{000000}53.58}} & \multicolumn{1}{c|}{\cellcolor[HTML]{FFFFFF}{\color[HTML]{000000}55.29}}& \multicolumn{1}{c|}{\cellcolor[HTML]{FFFFFF}{\color[HTML]{000000}52.46}} &\multicolumn{1}{c|}{\cellcolor[HTML]{FFFFFF}{\color[HTML]{000000}55.97}}&\multicolumn{1}{c|}{\cellcolor[HTML]{FFFFFF}{\color[HTML]{000000}57.80}}&\multicolumn{1}{c|}{\cellcolor[HTML]{FFFFFF}{\color[HTML]{000000}55.70}} \\ \cline{3-14}
\multicolumn{1}{|c|}{\multirow{-6}{*}{\cellcolor[HTML]{EEFBED}{\color[HTML]{000000} \begin{tabular}[c]{@{}c@{}}V-Face\\ +\\ V-AGA\\ +\\ A-Mel\end{tabular}}}} & \multicolumn{1}{c|}{\multirow{-5}{*}{\cellcolor[HTML]{FFFFFF}{\color[HTML]{000000} Feature}}} & \multicolumn{1}{c|}{\cellcolor[HTML]{FFFFFF}{\color[HTML]{000000} \textbf{Atten-Mixer}}} & \multicolumn{1}{c|}{\cellcolor[HTML]{FFFFFF}{\color[HTML]{000000} \textbf{57.23}}}       & \multicolumn{1}{c|}{\cellcolor[HTML]{FFFFFF}{\color[HTML]{000000} \textbf{63.93}}} & \multicolumn{1}{c|}{\cellcolor[HTML]{FFFFFF}{\color[HTML]{000000} \underline{55.05}}} & \multicolumn{1}{c|}{\cellcolor[HTML]{FFFFFF}{\color[HTML]{000000} \textbf{58.95}}} & \multicolumn{1}{c|}{\cellcolor[HTML]{FFFFFF}{\color[HTML]{000000} \textbf{59.65}}} & \multicolumn{1}{c|}{\cellcolor[HTML]{FFFFFF}{\color[HTML]{000000} \textbf{59.01}}} & \multicolumn{1}{c|}{\cellcolor[HTML]{FFFFFF}{\color[HTML]{000000} \textbf{61.14}}} & \multicolumn{1}{c|}{\cellcolor[HTML]{FFFFFF}{\color[HTML]{000000} 55.01}}  &\multicolumn{1}{c|}{\cellcolor[HTML]{FFFFFF}{\color[HTML]{000000} \textbf{59.96}}}&\multicolumn{1}{c|}{\cellcolor[HTML]{FFFFFF}{\color[HTML]{000000} \textbf{61.35}}}&\multicolumn{1}{c|}{\cellcolor[HTML]{FFFFFF}{\color[HTML]{000000} \textbf{58.63}}}  \\ \hline
\end{tabular}}
\vspace{1em}
\label{tab:multi2single_fusion_alter}
\end{table*}


\begin{table*}[!t]
\centering
\caption{Fusion results for multi-to-single cross-domain accuracy (\%) using domain-by-domain sampling.}
\resizebox{1.0\textwidth}{!}{
\begin{tabular}{|c
>{\columncolor[HTML]{FFFFFF}}c 
>{\columncolor[HTML]{FFFFFF}}c 
>{\columncolor[HTML]{FFFFFF}}c 
>{\columncolor[HTML]{FFFFFF}}c 
>{\columncolor[HTML]{FFFFFF}}c 
>{\columncolor[HTML]{FFFFFF}}c 
>{\columncolor[HTML]{FFFFFF}}c 
>{\columncolor[HTML]{FFFFFF}}c
>{\columncolor[HTML]{FFFFFF}}c 
>{\columncolor[HTML]{FFFFFF}}c
>{\columncolor[HTML]{FFFFFF}}c 
>{\columncolor[HTML]{FFFFFF}}c
>{\columncolor[HTML]{FFFFFF}}c|}
\hline
\multicolumn{14}{|c|}{\cellcolor[HTML]{EFEFEF}\textbf{Domain-by-Domain}}    \\ \hline  
\multicolumn{1}{|c|}{\cellcolor[HTML]{FFFFFF}{\color[HTML]{000000} \textbf{Input}}} & \multicolumn{1}{c|}{\cellcolor[HTML]{FFFFFF}{\color[HTML]{000000} \textbf{Type}}}         & \multicolumn{1}{c|}{\cellcolor[HTML]{FFFFFF}{\color[HTML]{000000} \textbf{Method}}}    & \multicolumn{1}{c|}{\cellcolor[HTML]{FFFFFF}{\color[HTML]{000000} \textbf{R\&M to B1}}} 
& \multicolumn{1}{c|}{\cellcolor[HTML]{FFFFFF}{\color[HTML]{000000} \textbf{R\&M to B2}}} 
& \multicolumn{1}{c|}{\cellcolor[HTML]{FFFFFF}{\color[HTML]{000000} \textbf{R\&M to D}}} 
& \multicolumn{1}{c|}{\cellcolor[HTML]{FFFFFF}{\color[HTML]{000000} \textbf{R\&M to E}}} 
& \multicolumn{1}{c|}{\cellcolor[HTML]{FFFFFF}{\color[HTML]{000000} \textbf{R\&B1 to B2}}} 
& \multicolumn{1}{c|}{\cellcolor[HTML]{FFFFFF}{\color[HTML]{000000} \textbf{R\&B1 to M}}} 
& \multicolumn{1}{c|}{\cellcolor[HTML]{FFFFFF}{\color[HTML]{000000} \textbf{R\&B1 to D}}} 
& \multicolumn{1}{c|}{\cellcolor[HTML]{FFFFFF}{\color[HTML]{000000} \textbf{R\&B1 to E}}} 
& \multicolumn{1}{c|}{\cellcolor[HTML]{FFFFFF}{\color[HTML]{000000} \textbf{B1\&M to R}}} 
& \multicolumn{1}{c|}{\cellcolor[HTML]{FFFFFF}{\color[HTML]{000000} \textbf{B1\&M to B2}}} 
& \multicolumn{1}{c|}{\cellcolor[HTML]{FFFFFF}{\color[HTML]{000000} \textbf{\shortstack{R\&B1\&M\\to B2}}}} \\  \hline
\multicolumn{1}{|c|}{\cellcolor[HTML]{ECF4FF}}   & \multicolumn{1}{c|}{\cellcolor[HTML]{FFFFFF}Score}   & \multicolumn{1}{c|}{\cellcolor[HTML]{FFFFFF}Avg}  & \multicolumn{1}{c|}{\cellcolor[HTML]{FFFFFF}54.77}  & \multicolumn{1}{c|}{\cellcolor[HTML]{FFFFFF}46.53} &\multicolumn{1}{c|}{\cellcolor[HTML]{FFFFFF}{\color[HTML]{000000}  50.73  }} &\multicolumn{1}{c|}{\cellcolor[HTML]{FFFFFF}{\color[HTML]{000000}   50.45 }}  & \multicolumn{1}{c|}{\cellcolor[HTML]{FFFFFF}59.41}   & \multicolumn{1}{c|}{\cellcolor[HTML]{FFFFFF}50.62} &\multicolumn{1}{c|}{\cellcolor[HTML]{FFFFFF}{\color[HTML]{000000}  55.68  }} &\multicolumn{1}{c|}{\cellcolor[HTML]{FFFFFF}{\color[HTML]{000000}  50.43 }} & \multicolumn{1}{c|}{\cellcolor[HTML]{FFFFFF}57.94}     & \multicolumn{1}{c|}{\cellcolor[HTML]{FFFFFF}59.41}  & \multicolumn{1}{c|}{\cellcolor[HTML]{FFFFFF}58.42}     \\ \cline{2-14} 
\multicolumn{1}{|c|}{\cellcolor[HTML]{ECF4FF}}   & \multicolumn{1}{c|}{\cellcolor[HTML]{FFFFFF}}  & \multicolumn{1}{c|}{\cellcolor[HTML]{FFFFFF}Concat}  & \multicolumn{1}{c|}{\cellcolor[HTML]{FFFFFF}53.85}      & \multicolumn{1}{c|}{\cellcolor[HTML]{FFFFFF}53.47}  &\multicolumn{1}{c|}{\cellcolor[HTML]{FFFFFF}{\color[HTML]{000000}  50.37  }} &\multicolumn{1}{c|}{\cellcolor[HTML]{FFFFFF}{\color[HTML]{000000}   51.02 }}  & \multicolumn{1}{c|}{\cellcolor[HTML]{FFFFFF}63.37}  & \multicolumn{1}{c|}{\cellcolor[HTML]{FFFFFF}51.88}   &\multicolumn{1}{c|}{\cellcolor[HTML]{FFFFFF}{\color[HTML]{000000}  56.32  }} &\multicolumn{1}{c|}{\cellcolor[HTML]{FFFFFF}{\color[HTML]{000000}   52.39 }}  & \multicolumn{1}{c|}{\cellcolor[HTML]{FFFFFF}57.94} & \multicolumn{1}{c|}{\cellcolor[HTML]{FFFFFF}\textbf{65.35}} & \multicolumn{1}{c|}{\cellcolor[HTML]{FFFFFF}58.42}  \\ \cline{3-14} 
\multicolumn{1}{|c|}{\cellcolor[HTML]{ECF4FF}} & \multicolumn{1}{c|}{\cellcolor[HTML]{FFFFFF}} & \multicolumn{1}{c|}{\cellcolor[HTML]{FFFFFF}SE-Concat}   & \multicolumn{1}{c|}{\cellcolor[HTML]{FFFFFF}56.62} & \multicolumn{1}{c|}{\cellcolor[HTML]{FFFFFF}44.55} &\multicolumn{1}{c|}{\cellcolor[HTML]{FFFFFF}{\color[HTML]{000000}  51.23  }} &\multicolumn{1}{c|}{\cellcolor[HTML]{FFFFFF}{\color[HTML]{000000}   52.39 }} & \multicolumn{1}{c|}{\cellcolor[HTML]{FFFFFF}63.37}     & \multicolumn{1}{c|}{\cellcolor[HTML]{FFFFFF}52.19} &\multicolumn{1}{c|}{\cellcolor[HTML]{FFFFFF}{\color[HTML]{000000}  56.56  }} &\multicolumn{1}{c|}{\cellcolor[HTML]{FFFFFF}{\color[HTML]{000000}   53.56 }} & \multicolumn{1}{c|}{\cellcolor[HTML]{FFFFFF}54.21}   & \multicolumn{1}{c|}{\cellcolor[HTML]{FFFFFF}58.42}  & \multicolumn{1}{c|}{\cellcolor[HTML]{FFFFFF}59.41}    \\ \cline{3-14} 
\multicolumn{1}{|c|}{\cellcolor[HTML]{ECF4FF}}  & \multicolumn{1}{c|}{\cellcolor[HTML]{FFFFFF}}  & \multicolumn{1}{c|}{\cellcolor[HTML]{FFFFFF}Cross-Atten}   & \multicolumn{1}{c|}{\cellcolor[HTML]{FFFFFF}54.15}  & \multicolumn{1}{c|}{\cellcolor[HTML]{FFFFFF}49.50} &\multicolumn{1}{c|}{\cellcolor[HTML]{FFFFFF}{\color[HTML]{000000}  52.46  }} &\multicolumn{1}{c|}{\cellcolor[HTML]{FFFFFF}{\color[HTML]{000000}   53.34 }} & \multicolumn{1}{c|}{\cellcolor[HTML]{FFFFFF}\underline{62.38}}  & \multicolumn{1}{c|}{\cellcolor[HTML]{FFFFFF}54.69} &\multicolumn{1}{c|}{\cellcolor[HTML]{FFFFFF}{\color[HTML]{000000}  56.98  }} &\multicolumn{1}{c|}{\cellcolor[HTML]{FFFFFF}{\color[HTML]{000000}   53.21 }} & \multicolumn{1}{c|}{\cellcolor[HTML]{FFFFFF}57.01}  & \multicolumn{1}{c|}{\cellcolor[HTML]{FFFFFF}60.40}  & \multicolumn{1}{c|}{\cellcolor[HTML]{FFFFFF}\underline{63.37}}  \\ \cline{3-14} 
\multicolumn{1}{|c|}{\cellcolor[HTML]{ECF4FF}}  & \multicolumn{1}{c|}{\cellcolor[HTML]{FFFFFF}} & \multicolumn{1}{c|}{\cellcolor[HTML]{FFFFFF}MLP-Mixer}  & \multicolumn{1}{c|}{\cellcolor[HTML]{FFFFFF}\textbf{57.54}} & \multicolumn{1}{c|}{\cellcolor[HTML]{FFFFFF}\underline{62.38}} &\multicolumn{1}{c|}{\cellcolor[HTML]{FFFFFF}{\color[HTML]{000000}  53.53  }} &\multicolumn{1}{c|}{\cellcolor[HTML]{FFFFFF}{\color[HTML]{000000}   53.45 }} & \multicolumn{1}{c|}{\cellcolor[HTML]{FFFFFF}\textbf{71.29}}  & \multicolumn{1}{c|}{\cellcolor[HTML]{FFFFFF}51.25} &\multicolumn{1}{c|}{\cellcolor[HTML]{FFFFFF}{\color[HTML]{000000}  57.59  }} &\multicolumn{1}{c|}{\cellcolor[HTML]{FFFFFF}{\color[HTML]{000000}   53.49 }} & \multicolumn{1}{c|}{\cellcolor[HTML]{FFFFFF}54.21}  & \multicolumn{1}{c|}{\cellcolor[HTML]{FFFFFF}62.38}       & \multicolumn{1}{c|}{\cellcolor[HTML]{FFFFFF}62.38}  \\ \cline{3-14} 
\multicolumn{1}{|c|}{\cellcolor[HTML]{ECF4FF}} & \multicolumn{1}{c|}{\cellcolor[HTML]{FFFFFF}}  & \multicolumn{1}{c|}{\cellcolor[HTML]{FFFFFF}CLIP-Align}  & \multicolumn{1}{c|}{\cellcolor[HTML]{FFFFFF}54.57}  & \multicolumn{1}{c|}{\cellcolor[HTML]{FFFFFF}53.66} &\multicolumn{1}{c|}{\cellcolor[HTML]{FFFFFF}{\color[HTML]{000000}51.41}} &\multicolumn{1}{c|}{\cellcolor[HTML]{FFFFFF}{\color[HTML]{000000}51.60}}  & \multicolumn{1}{c|}{\cellcolor[HTML]{FFFFFF}53.70}      & \multicolumn{1}{c|}{\cellcolor[HTML]{FFFFFF}52.75} &\multicolumn{1}{c|}{\cellcolor[HTML]{FFFFFF}{\color[HTML]{000000}56.78}} &\multicolumn{1}{c|}{\cellcolor[HTML]{FFFFFF}{\color[HTML]{000000}53.49}}   & \multicolumn{1}{c|}{\cellcolor[HTML]{FFFFFF}58.18}   & \multicolumn{1}{c|}{\cellcolor[HTML]{FFFFFF}66.26}  & \multicolumn{1}{c|}{\cellcolor[HTML]{FFFFFF}59.09}     \\ \cline{3-14} 
\multicolumn{1}{|c|}{\multirow{-6}{*}{\cellcolor[HTML]{ECF4FF}\begin{tabular}[c]{@{}c@{}}V-Face\\ +\\ V-AGA\end{tabular}}}  & \multicolumn{1}{c|}{\multirow{-5}{*}{\cellcolor[HTML]{FFFFFF}Feature}}  & \multicolumn{1}{c|}{\cellcolor[HTML]{FFFFFF}\textbf{Atten-Mixer}}  & \multicolumn{1}{c|}{\cellcolor[HTML]{FFFFFF}\textbf{57.54}}  & \multicolumn{1}{c|}{\cellcolor[HTML]{FFFFFF}58.42}&\multicolumn{1}{c|}{\cellcolor[HTML]{FFFFFF}{\color[HTML]{000000}  53.62 }} &\multicolumn{1}{c|}{\cellcolor[HTML]{FFFFFF}{\color[HTML]{000000}  54.32 }}  & \multicolumn{1}{c|}{\cellcolor[HTML]{FFFFFF}63.37}  & \multicolumn{1}{c|}{\cellcolor[HTML]{FFFFFF}52.19} &\multicolumn{1}{c|}{\cellcolor[HTML]{FFFFFF}{\color[HTML]{000000}  58.04  }} &\multicolumn{1}{c|}{\cellcolor[HTML]{FFFFFF}{\color[HTML]{000000}   53.78 }}  & \multicolumn{1}{c|}{\cellcolor[HTML]{FFFFFF}\textbf{69.16}} & \multicolumn{1}{c|}{\cellcolor[HTML]{FFFFFF}60.40} & \multicolumn{1}{c|}{\cellcolor[HTML]{FFFFFF}\textbf{64.36}}  \\ \hline
\multicolumn{1}{|c|}{\cellcolor[HTML]{FFFFEB}} & \multicolumn{1}{c|}{\cellcolor[HTML]{FFFFFF}Score}     & \multicolumn{1}{c|}{\cellcolor[HTML]{FFFFFF}Avg}   & \multicolumn{1}{c|}{\cellcolor[HTML]{FFFFFF}52.00}  & \multicolumn{1}{c|}{\cellcolor[HTML]{FFFFFF}50.50} &\multicolumn{1}{c|}{\cellcolor[HTML]{FFFFFF}{\color[HTML]{000000}  50.32 }} &\multicolumn{1}{c|}{\cellcolor[HTML]{FFFFFF}{\color[HTML]{000000}  50.87 }}  & \multicolumn{1}{c|}{\cellcolor[HTML]{FFFFFF}58.42}   & \multicolumn{1}{c|}{\cellcolor[HTML]{FFFFFF}51.56}  &\multicolumn{1}{c|}{\cellcolor[HTML]{FFFFFF}{\color[HTML]{000000}  56.98  }} &\multicolumn{1}{c|}{\cellcolor[HTML]{FFFFFF}{\color[HTML]{000000}   50.42 }} & \multicolumn{1}{c|}{\cellcolor[HTML]{FFFFFF}57.94}     & \multicolumn{1}{c|}{\cellcolor[HTML]{FFFFFF}61.39}   & \multicolumn{1}{c|}{\cellcolor[HTML]{FFFFFF}60.40}   \\ \cline{2-14}
\multicolumn{1}{|c|}{\cellcolor[HTML]{FFFFEB}} & \multicolumn{1}{c|}{\cellcolor[HTML]{FFFFFF}}  & \multicolumn{1}{c|}{\cellcolor[HTML]{FFFFFF}Concat}  & \multicolumn{1}{c|}{\cellcolor[HTML]{FFFFFF}50.46}     & \multicolumn{1}{c|}{\cellcolor[HTML]{FFFFFF}59.41}&\multicolumn{1}{c|}{\cellcolor[HTML]{FFFFFF}{\color[HTML]{000000}  51.24  }} &\multicolumn{1}{c|}{\cellcolor[HTML]{FFFFFF}{\color[HTML]{000000}   50.95 }}  & \multicolumn{1}{c|}{\cellcolor[HTML]{FFFFFF}62.38}     & \multicolumn{1}{c|}{\cellcolor[HTML]{FFFFFF}52.19} &\multicolumn{1}{c|}{\cellcolor[HTML]{FFFFFF}{\color[HTML]{000000}  57.25  }} &\multicolumn{1}{c|}{\cellcolor[HTML]{FFFFFF}{\color[HTML]{000000}   51.28 }}  & \multicolumn{1}{c|}{\cellcolor[HTML]{FFFFFF}60.75}    & \multicolumn{1}{c|}{\cellcolor[HTML]{FFFFFF}56.44}  & \multicolumn{1}{c|}{\cellcolor[HTML]{FFFFFF}57.43}    \\ \cline{3-14} 
\multicolumn{1}{|c|}{\cellcolor[HTML]{FFFFEB}}  & \multicolumn{1}{c|}{\cellcolor[HTML]{FFFFFF}}  & \multicolumn{1}{c|}{\cellcolor[HTML]{FFFFFF}SE-Concat}  & \multicolumn{1}{c|}{\cellcolor[HTML]{FFFFFF}55.38}   & \multicolumn{1}{c|}{\cellcolor[HTML]{FFFFFF}48.51} &\multicolumn{1}{c|}{\cellcolor[HTML]{FFFFFF}{\color[HTML]{000000}  51.35  }} &\multicolumn{1}{c|}{\cellcolor[HTML]{FFFFFF}{\color[HTML]{000000}   52.54 }}  & \multicolumn{1}{c|}{\cellcolor[HTML]{FFFFFF}58.42}    & \multicolumn{1}{c|}{\cellcolor[HTML]{FFFFFF}52.81} &\multicolumn{1}{c|}{\cellcolor[HTML]{FFFFFF}{\color[HTML]{000000}  57.83  }} &\multicolumn{1}{c|}{\cellcolor[HTML]{FFFFFF}{\color[HTML]{000000}   51.87 }}  & \multicolumn{1}{c|}{\cellcolor[HTML]{FFFFFF}62.62}    & \multicolumn{1}{c|}{\cellcolor[HTML]{FFFFFF}57.43}  & \multicolumn{1}{c|}{\cellcolor[HTML]{FFFFFF}59.41}    \\ \cline{3-14} 
\multicolumn{1}{|c|}{\cellcolor[HTML]{FFFFEB}} & \multicolumn{1}{c|}{\cellcolor[HTML]{FFFFFF}}  & \multicolumn{1}{c|}{\cellcolor[HTML]{FFFFFF}Cross-Atten}  & \multicolumn{1}{c|}{\cellcolor[HTML]{FFFFFF}52.31} & \multicolumn{1}{c|}{\cellcolor[HTML]{FFFFFF}53.47} &\multicolumn{1}{c|}{\cellcolor[HTML]{FFFFFF}{\color[HTML]{000000}  52.37  }} &\multicolumn{1}{c|}{\cellcolor[HTML]{FFFFFF}{\color[HTML]{000000}   54.68 }}  & \multicolumn{1}{c|}{\cellcolor[HTML]{FFFFFF}58.42} & \multicolumn{1}{c|}{\cellcolor[HTML]{FFFFFF}51.25} &\multicolumn{1}{c|}{\cellcolor[HTML]{FFFFFF}{\color[HTML]{000000}  57.39 }} &\multicolumn{1}{c|}{\cellcolor[HTML]{FFFFFF}{\color[HTML]{000000}   52.33 }}  & \multicolumn{1}{c|}{\cellcolor[HTML]{FFFFFF}64.49}& \multicolumn{1}{c|}{\cellcolor[HTML]{FFFFFF}58.42}  & \multicolumn{1}{c|}{\cellcolor[HTML]{FFFFFF}57.43}  \\ \cline{3-14} 
\multicolumn{1}{|c|}{\cellcolor[HTML]{FFFFEB}} & \multicolumn{1}{c|}{\cellcolor[HTML]{FFFFFF}}  & \multicolumn{1}{c|}{\cellcolor[HTML]{FFFFFF}MLP-Mixer}  & \multicolumn{1}{c|}{\cellcolor[HTML]{FFFFFF}53.23}   & \multicolumn{1}{c|}{\cellcolor[HTML]{FFFFFF}60.40} &\multicolumn{1}{c|}{\cellcolor[HTML]{FFFFFF}{\color[HTML]{000000}  52.45  }} &\multicolumn{1}{c|}{\cellcolor[HTML]{FFFFFF}{\color[HTML]{000000}   54.08 }}  & \multicolumn{1}{c|}{\cellcolor[HTML]{FFFFFF}62.38}   & \multicolumn{1}{c|}{\cellcolor[HTML]{FFFFFF}52.19} &\multicolumn{1}{c|}{\cellcolor[HTML]{FFFFFF}{\color[HTML]{000000}  57.89  }} &\multicolumn{1}{c|}{\cellcolor[HTML]{FFFFFF}{\color[HTML]{000000}   52.39}}  & \multicolumn{1}{c|}{\cellcolor[HTML]{FFFFFF}61.68}     & \multicolumn{1}{c|}{\cellcolor[HTML]{FFFFFF}58.42}  & \multicolumn{1}{c|}{\cellcolor[HTML]{FFFFFF}58.42}   \\ \cline{3-14} 
\multicolumn{1}{|c|}{\cellcolor[HTML]{FFFFEB}} & \multicolumn{1}{c|}{\cellcolor[HTML]{FFFFFF}}  & \multicolumn{1}{c|}{\cellcolor[HTML]{FFFFFF}CLIP-Align}  & \multicolumn{1}{c|}{\cellcolor[HTML]{FFFFFF}51.12}  & \multicolumn{1}{c|}{\cellcolor[HTML]{FFFFFF}60.23} &\multicolumn{1}{c|}{\cellcolor[HTML]{FFFFFF}{\color[HTML]{000000}51.89}} &\multicolumn{1}{c|}{\cellcolor[HTML]{FFFFFF}{\color[HTML]{000000} 51.46}}  & \multicolumn{1}{c|}{\cellcolor[HTML]{FFFFFF}63.17}      & \multicolumn{1}{c|}{\cellcolor[HTML]{FFFFFF}52.72} &\multicolumn{1}{c|}{\cellcolor[HTML]{FFFFFF}{\color[HTML]{000000}58.04}} &\multicolumn{1}{c|}{\cellcolor[HTML]{FFFFFF}{\color[HTML]{000000}51.89}}   & \multicolumn{1}{c|}{\cellcolor[HTML]{FFFFFF}61.53}   & \multicolumn{1}{c|}{\cellcolor[HTML]{FFFFFF}57.28}  & \multicolumn{1}{c|}{\cellcolor[HTML]{FFFFFF}58.08}     \\ \cline{3-14} 
\multicolumn{1}{|c|}{\multirow{-6}{*}{\cellcolor[HTML]{FFFFEB}\begin{tabular}[c]{@{}c@{}}V-Face\\ +\\ A-Mel\end{tabular}}}  & \multicolumn{1}{c|}{\multirow{-5}{*}{\cellcolor[HTML]{FFFFFF}Feature}}  & \multicolumn{1}{c|}{\cellcolor[HTML]{FFFFFF}\textbf{Atten-Mixer}}  & \multicolumn{1}{c|}{\cellcolor[HTML]{FFFFFF}53.85}  & \multicolumn{1}{c|}{\cellcolor[HTML]{FFFFFF}\textbf{66.34}} &\multicolumn{1}{c|}{\cellcolor[HTML]{FFFFFF}{\color[HTML]{000000}  54.32  }} &\multicolumn{1}{c|}{\cellcolor[HTML]{FFFFFF}{\color[HTML]{000000}   \underline{54.89} }}  & \multicolumn{1}{c|}{\cellcolor[HTML]{FFFFFF}63.37}  & \multicolumn{1}{c|}{\cellcolor[HTML]{FFFFFF}53.75} &\multicolumn{1}{c|}{\cellcolor[HTML]{FFFFFF}{\color[HTML]{000000}  57.96  }} &\multicolumn{1}{c|}{\cellcolor[HTML]{FFFFFF}{\color[HTML]{000000}   53.04 }}  & \multicolumn{1}{c|}{\cellcolor[HTML]{FFFFFF}58.88} & \multicolumn{1}{c|}{\cellcolor[HTML]{FFFFFF}\underline{64.36}} & \multicolumn{1}{c|}{\cellcolor[HTML]{FFFFFF}\underline{63.37}} \\ \hline
\multicolumn{1}{|c|}{\cellcolor[HTML]{FFF7F7}} & \multicolumn{1}{c|}{\cellcolor[HTML]{FFFFFF}Score}      & \multicolumn{1}{c|}{\cellcolor[HTML]{FFFFFF}Avg} & \multicolumn{1}{c|}{\cellcolor[HTML]{FFFFFF}48.62}   & \multicolumn{1}{c|}{\cellcolor[HTML]{FFFFFF}59.41}&\multicolumn{1}{c|}{\cellcolor[HTML]{FFFFFF}{\color[HTML]{000000}  51.28  }} &\multicolumn{1}{c|}{\cellcolor[HTML]{FFFFFF}{\color[HTML]{000000}   50.88 }}  & \multicolumn{1}{c|}{\cellcolor[HTML]{FFFFFF}51.49}     & \multicolumn{1}{c|}{\cellcolor[HTML]{FFFFFF}51.25} &\multicolumn{1}{c|}{\cellcolor[HTML]{FFFFFF}{\color[HTML]{000000}  56.83  }} &\multicolumn{1}{c|}{\cellcolor[HTML]{FFFFFF}{\color[HTML]{000000}   51.28 }} & \multicolumn{1}{c|}{\cellcolor[HTML]{FFFFFF}60.75}     & \multicolumn{1}{c|}{\cellcolor[HTML]{FFFFFF}62.38} & \multicolumn{1}{c|}{\cellcolor[HTML]{FFFFFF}\underline{63.37}}     \\ \cline{2-14} 
\multicolumn{1}{|c|}{\cellcolor[HTML]{FFF7F7}}& \multicolumn{1}{c|}{\cellcolor[HTML]{FFFFFF}}  & \multicolumn{1}{c|}{\cellcolor[HTML]{FFFFFF}Concat}  & \multicolumn{1}{c|}{\cellcolor[HTML]{FFFFFF}57.23}     & \multicolumn{1}{c|}{\cellcolor[HTML]{FFFFFF}60.40} &\multicolumn{1}{c|}{\cellcolor[HTML]{FFFFFF}{\color[HTML]{000000}  51.35  }} &\multicolumn{1}{c|}{\cellcolor[HTML]{FFFFFF}{\color[HTML]{000000}   50.39 }} & \multicolumn{1}{c|}{\cellcolor[HTML]{FFFFFF}56.44}      & \multicolumn{1}{c|}{\cellcolor[HTML]{FFFFFF}51.25} &\multicolumn{1}{c|}{\cellcolor[HTML]{FFFFFF}{\color[HTML]{000000}  57.38  }} &\multicolumn{1}{c|}{\cellcolor[HTML]{FFFFFF}{\color[HTML]{000000}   51.38 }} & \multicolumn{1}{c|}{\cellcolor[HTML]{FFFFFF}56.07}    & \multicolumn{1}{c|}{\cellcolor[HTML]{FFFFFF}57.43} & \multicolumn{1}{c|}{\cellcolor[HTML]{FFFFFF}56.44}      \\ \cline{3-14} 
\multicolumn{1}{|c|}{\cellcolor[HTML]{FFF7F7}} & \multicolumn{1}{c|}{\cellcolor[HTML]{FFFFFF}}                 & \multicolumn{1}{c|}{\cellcolor[HTML]{FFFFFF}SE-Concat}  & \multicolumn{1}{c|}{\cellcolor[HTML]{FFFFFF}55.08} & \multicolumn{1}{c|}{\cellcolor[HTML]{FFFFFF}60.40} &\multicolumn{1}{c|}{\cellcolor[HTML]{FFFFFF}{\color[HTML]{000000}  53.86  }} &\multicolumn{1}{c|}{\cellcolor[HTML]{FFFFFF}{\color[HTML]{000000}   52.83 }} & \multicolumn{1}{c|}{\cellcolor[HTML]{FFFFFF}59.41} & \multicolumn{1}{c|}{\cellcolor[HTML]{FFFFFF}54.06}  &\multicolumn{1}{c|}{\cellcolor[HTML]{FFFFFF}{\color[HTML]{000000}  57.92  }} &\multicolumn{1}{c|}{\cellcolor[HTML]{FFFFFF}{\color[HTML]{000000}   52.98 }} & \multicolumn{1}{c|}{\cellcolor[HTML]{FFFFFF}60.75} & \multicolumn{1}{c|}{\cellcolor[HTML]{FFFFFF}61.39} & \multicolumn{1}{c|}{\cellcolor[HTML]{FFFFFF}57.43}  \\ \cline{3-14} 
\multicolumn{1}{|c|}{\cellcolor[HTML]{FFF7F7}} & \multicolumn{1}{c|}{\cellcolor[HTML]{FFFFFF}}                 & \multicolumn{1}{c|}{\cellcolor[HTML]{FFFFFF}Cross-Atten}  & \multicolumn{1}{c|}{\cellcolor[HTML]{FFFFFF}50.15} & \multicolumn{1}{c|}{\cellcolor[HTML]{FFFFFF}56.44} &\multicolumn{1}{c|}{\cellcolor[HTML]{FFFFFF}{\color[HTML]{000000}  54.89  }} &\multicolumn{1}{c|}{\cellcolor[HTML]{FFFFFF}{\color[HTML]{000000}   53.58 }}  & \multicolumn{1}{c|}{\cellcolor[HTML]{FFFFFF}61.49} & \multicolumn{1}{c|}{\cellcolor[HTML]{FFFFFF}\textbf{57.50}} &\multicolumn{1}{c|}{\cellcolor[HTML]{FFFFFF}{\color[HTML]{000000}  57.85  }} &\multicolumn{1}{c|}{\cellcolor[HTML]{FFFFFF}{\color[HTML]{000000}   52.96 }}  & \multicolumn{1}{c|}{\cellcolor[HTML]{FFFFFF}59.81}  & \multicolumn{1}{c|}{\cellcolor[HTML]{FFFFFF}\underline{64.36}}  & \multicolumn{1}{c|}{\cellcolor[HTML]{FFFFFF}58.42}   \\ \cline{3-14} 
\multicolumn{1}{|c|}{\cellcolor[HTML]{FFF7F7}}  & \multicolumn{1}{c|}{\cellcolor[HTML]{FFFFFF}} & \multicolumn{1}{c|}{\cellcolor[HTML]{FFFFFF}MLP-Mixer}  & \multicolumn{1}{c|}{\cellcolor[HTML]{FFFFFF}52.62}   & \multicolumn{1}{c|}{\cellcolor[HTML]{FFFFFF}56.44} &\multicolumn{1}{c|}{\cellcolor[HTML]{FFFFFF}{\color[HTML]{000000}  54.78  }} &\multicolumn{1}{c|}{\cellcolor[HTML]{FFFFFF}{\color[HTML]{000000}   53.86 }}  & \multicolumn{1}{c|}{\cellcolor[HTML]{FFFFFF}58.42}    & \multicolumn{1}{c|}{\cellcolor[HTML]{FFFFFF}52.50} &\multicolumn{1}{c|}{\cellcolor[HTML]{FFFFFF}{\color[HTML]{000000}  \underline{58.96}  }} &\multicolumn{1}{c|}{\cellcolor[HTML]{FFFFFF}{\color[HTML]{000000}   \underline{53.87} }}   & \multicolumn{1}{c|}{\cellcolor[HTML]{FFFFFF}60.75}    & \multicolumn{1}{c|}{\cellcolor[HTML]{FFFFFF}61.39} & \multicolumn{1}{c|}{\cellcolor[HTML]{FFFFFF}58.42}      \\ \cline{3-14}
\multicolumn{1}{|c|}{\cellcolor[HTML]{FFF7F7}} & \multicolumn{1}{c|}{\cellcolor[HTML]{FFFFFF}}  & \multicolumn{1}{c|}{\cellcolor[HTML]{FFFFFF}CLIP-Align}  & \multicolumn{1}{c|}{\cellcolor[HTML]{FFFFFF}\underline{57.87}}  & \multicolumn{1}{c|}{\cellcolor[HTML]{FFFFFF}60.67} &\multicolumn{1}{c|}{\cellcolor[HTML]{FFFFFF}{\color[HTML]{000000}52.26}} &\multicolumn{1}{c|}{\cellcolor[HTML]{FFFFFF}{\color[HTML]{000000}51.42}}  & \multicolumn{1}{c|}{\cellcolor[HTML]{FFFFFF}56.62}      & \multicolumn{1}{c|}{\cellcolor[HTML]{FFFFFF}52.01} &\multicolumn{1}{c|}{\cellcolor[HTML]{FFFFFF}{\color[HTML]{000000}57.87}} &\multicolumn{1}{c|}{\cellcolor[HTML]{FFFFFF}{\color[HTML]{000000}52.46}}   & \multicolumn{1}{c|}{\cellcolor[HTML]{FFFFFF}56.42}   & \multicolumn{1}{c|}{\cellcolor[HTML]{FFFFFF}58.25}  & \multicolumn{1}{c|}{\cellcolor[HTML]{FFFFFF}57.00}     \\ \cline{3-14} 
\multicolumn{1}{|c|}{\multirow{-6}{*}{\cellcolor[HTML]{FFF7F7}\begin{tabular}[c]{@{}c@{}}V-AGA\\ +\\ A-Mel\end{tabular}}} & \multicolumn{1}{c|}{\multirow{-5}{*}{\cellcolor[HTML]{FFFFFF}Feature}}  & \multicolumn{1}{c|}{\cellcolor[HTML]{FFFFFF}\textbf{Atten-Mixer}} & \multicolumn{1}{c|}{\cellcolor[HTML]{FFFFFF}51.69}   & \multicolumn{1}{c|}{\cellcolor[HTML]{FFFFFF}56.44} &\multicolumn{1}{c|}{\cellcolor[HTML]{FFFFFF}{\color[HTML]{000000}  55.06  }} &\multicolumn{1}{c|}{\cellcolor[HTML]{FFFFFF}{\color[HTML]{000000}   54.87 }}   & \multicolumn{1}{c|}{\cellcolor[HTML]{FFFFFF}57.43} & \multicolumn{1}{c|}{\cellcolor[HTML]{FFFFFF}54.06} &\multicolumn{1}{c|}{\cellcolor[HTML]{FFFFFF}{\color[HTML]{000000}  58.91  }} &\multicolumn{1}{c|}{\cellcolor[HTML]{FFFFFF}{\color[HTML]{000000}   53.48 }}  & \multicolumn{1}{c|}{\cellcolor[HTML]{FFFFFF}{\color[HTML]{000000}\underline{66.36} }} & \multicolumn{1}{c|}{\cellcolor[HTML]{FFFFFF}\underline{64.36}} & \multicolumn{1}{c|}{\cellcolor[HTML]{FFFFFF}60.40}  \\ \hline
\multicolumn{1}{|c|}{\cellcolor[HTML]{EEFBED}} & \multicolumn{1}{c|}{\cellcolor[HTML]{FFFFFF}Score}      & \multicolumn{1}{c|}{\cellcolor[HTML]{FFFFFF}Avg}  & \multicolumn{1}{c|}{\cellcolor[HTML]{FFFFFF}53.85}  & \multicolumn{1}{c|}{\cellcolor[HTML]{FFFFFF}51.49} &\multicolumn{1}{c|}{\cellcolor[HTML]{FFFFFF}{\color[HTML]{000000}  54.76  }} &\multicolumn{1}{c|}{\cellcolor[HTML]{FFFFFF}{\color[HTML]{000000}   50.38 }}   & \multicolumn{1}{c|}{\cellcolor[HTML]{FFFFFF}65.35}   & \multicolumn{1}{c|}{\cellcolor[HTML]{FFFFFF}53.75} &\multicolumn{1}{c|}{\cellcolor[HTML]{FFFFFF}{\color[HTML]{000000}  55.64  }} &\multicolumn{1}{c|}{\cellcolor[HTML]{FFFFFF}{\color[HTML]{000000}   50.32}}   & \multicolumn{1}{c|}{\cellcolor[HTML]{FFFFFF}58.88}     & \multicolumn{1}{c|}{\cellcolor[HTML]{FFFFFF}57.43}  & \multicolumn{1}{c|}{\cellcolor[HTML]{FFFFFF}58.42}    \\ \cline{2-14}
\multicolumn{1}{|c|}{\cellcolor[HTML]{EEFBED}} & \multicolumn{1}{c|}{\cellcolor[HTML]{FFFFFF}}   & \multicolumn{1}{c|}{\cellcolor[HTML]{FFFFFF}Concat}& \multicolumn{1}{c|}{\cellcolor[HTML]{FFFFFF}52.92}  & \multicolumn{1}{c|}{\cellcolor[HTML]{FFFFFF}52.48} &\multicolumn{1}{c|}{\cellcolor[HTML]{FFFFFF}{\color[HTML]{000000}  55.93  }} &\multicolumn{1}{c|}{\cellcolor[HTML]{FFFFFF}{\color[HTML]{000000}   51.77 }}   & \multicolumn{1}{c|}{\cellcolor[HTML]{FFFFFF}55.45}  & \multicolumn{1}{c|}{\cellcolor[HTML]{FFFFFF}53.75}  &\multicolumn{1}{c|}{\cellcolor[HTML]{FFFFFF}{\color[HTML]{000000}  55.84  }} &\multicolumn{1}{c|}{\cellcolor[HTML]{FFFFFF}{\color[HTML]{000000}   51.28 }}   & \multicolumn{1}{c|}{\cellcolor[HTML]{FFFFFF}62.62}  & \multicolumn{1}{c|}{\cellcolor[HTML]{FFFFFF}62.38}  & \multicolumn{1}{c|}{\cellcolor[HTML]{FFFFFF}56.44}  \\ \cline{3-14} 
\multicolumn{1}{|c|}{\cellcolor[HTML]{EEFBED}} & \multicolumn{1}{c|}{\cellcolor[HTML]{FFFFFF}}  & \multicolumn{1}{c|}{\cellcolor[HTML]{FFFFFF}SE-Concat}  & \multicolumn{1}{c|}{\cellcolor[HTML]{FFFFFF}51.38}   & \multicolumn{1}{c|}{\cellcolor[HTML]{FFFFFF}60.40}&\multicolumn{1}{c|}{\cellcolor[HTML]{FFFFFF}{\color[HTML]{000000}  56.87  }} &\multicolumn{1}{c|}{\cellcolor[HTML]{FFFFFF}{\color[HTML]{000000}   52.36 }}  & \multicolumn{1}{c|}{\cellcolor[HTML]{FFFFFF}58.42}      & \multicolumn{1}{c|}{\cellcolor[HTML]{FFFFFF}51.56} &\multicolumn{1}{c|}{\cellcolor[HTML]{FFFFFF}{\color[HTML]{000000}  56.92 }} &\multicolumn{1}{c|}{\cellcolor[HTML]{FFFFFF}{\color[HTML]{000000}   51.98 }}   & \multicolumn{1}{c|}{\cellcolor[HTML]{FFFFFF}57.01}   & \multicolumn{1}{c|}{\cellcolor[HTML]{FFFFFF}62.38} & \multicolumn{1}{c|}{\cellcolor[HTML]{FFFFFF}59.41}     \\ \cline{3-14} 
\multicolumn{1}{|c|}{\cellcolor[HTML]{EEFBED}}  & \multicolumn{1}{c|}{\cellcolor[HTML]{FFFFFF}}  & \multicolumn{1}{c|}{\cellcolor[HTML]{FFFFFF}Cross-Atten}  & \multicolumn{1}{c|}{\cellcolor[HTML]{FFFFFF}51.80}  & \multicolumn{1}{c|}{\cellcolor[HTML]{FFFFFF}52.48} &\multicolumn{1}{c|}{\cellcolor[HTML]{FFFFFF}{\color[HTML]{000000}  56.82  }} &\multicolumn{1}{c|}{\cellcolor[HTML]{FFFFFF}{\color[HTML]{000000}   53.58 }}   & \multicolumn{1}{c|}{\cellcolor[HTML]{FFFFFF}62.38} & \multicolumn{1}{c|}{\cellcolor[HTML]{FFFFFF}53.12}&\multicolumn{1}{c|}{\cellcolor[HTML]{FFFFFF}{\color[HTML]{000000}  57.83  }} &\multicolumn{1}{c|}{\cellcolor[HTML]{FFFFFF}{\color[HTML]{000000}   52.38 }}    & \multicolumn{1}{c|}{\cellcolor[HTML]{FFFFFF}59.81}  & \multicolumn{1}{c|}{\cellcolor[HTML]{FFFFFF}\underline{64.36}}  & \multicolumn{1}{c|}{\cellcolor[HTML]{FFFFFF}60.40}  \\ \cline{3-14} 
\multicolumn{1}{|c|}{\cellcolor[HTML]{EEFBED}} & \multicolumn{1}{c|}{\cellcolor[HTML]{FFFFFF}}  & \multicolumn{1}{c|}{\cellcolor[HTML]{FFFFFF}MLP-Mixer}  & \multicolumn{1}{c|}{\cellcolor[HTML]{FFFFFF}53.85}  & \multicolumn{1}{c|}{\cellcolor[HTML]{FFFFFF}55.45} &\multicolumn{1}{c|}{\cellcolor[HTML]{FFFFFF}{\color[HTML]{000000}  \underline{57.82} }} &\multicolumn{1}{c|}{\cellcolor[HTML]{FFFFFF}{\color[HTML]{000000}   53.69 }}  & \multicolumn{1}{c|}{\cellcolor[HTML]{FFFFFF}60.40}      & \multicolumn{1}{c|}{\cellcolor[HTML]{FFFFFF}50.94} &\multicolumn{1}{c|}{\cellcolor[HTML]{FFFFFF}{\color[HTML]{000000}  58.92  }} &\multicolumn{1}{c|}{\cellcolor[HTML]{FFFFFF}{\color[HTML]{000000}   53.56 }}   & \multicolumn{1}{c|}{\cellcolor[HTML]{FFFFFF}59.81}   & \multicolumn{1}{c|}{\cellcolor[HTML]{FFFFFF}\underline{64.36}}  & \multicolumn{1}{c|}{\cellcolor[HTML]{FFFFFF}60.40}     \\ \cline{3-14} 
\multicolumn{1}{|c|}{\cellcolor[HTML]{EEFBED}} & \multicolumn{1}{c|}{\cellcolor[HTML]{FFFFFF}}  & \multicolumn{1}{c|}{\cellcolor[HTML]{FFFFFF}CLIP-Align}  & \multicolumn{1}{c|}{\cellcolor[HTML]{FFFFFF}53.15}  & \multicolumn{1}{c|}{\cellcolor[HTML]{FFFFFF}53.45} &\multicolumn{1}{c|}{\cellcolor[HTML]{FFFFFF}{\color[HTML]{000000}56.34}} &\multicolumn{1}{c|}{\cellcolor[HTML]{FFFFFF}{\color[HTML]{000000}52.45}}  & \multicolumn{1}{c|}{\cellcolor[HTML]{FFFFFF}56.55}      & \multicolumn{1}{c|}{\cellcolor[HTML]{FFFFFF}53.89} &\multicolumn{1}{c|}{\cellcolor[HTML]{FFFFFF}{\color[HTML]{000000}56.72}} &\multicolumn{1}{c|}{\cellcolor[HTML]{FFFFFF}{\color[HTML]{000000}51.87}}   & \multicolumn{1}{c|}{\cellcolor[HTML]{FFFFFF}62.93}   & \multicolumn{1}{c|}{\cellcolor[HTML]{FFFFFF}63.40}  & \multicolumn{1}{c|}{\cellcolor[HTML]{FFFFFF}57.17}     \\ \cline{3-14} 
\multicolumn{1}{|c|}{\multirow{-6}{*}{\cellcolor[HTML]{EEFBED}\begin{tabular}[c]{@{}c@{}}V-Face\\ +\\ V-AGA\\ +\\ A-Mel\end{tabular}}}   & \multicolumn{1}{c|}{\multirow{-5}{*}{\cellcolor[HTML]{FFFFFF}Feature}}  & \multicolumn{1}{c|}{\cellcolor[HTML]{FFFFFF}\textbf{Atten-Mixer}}  & \multicolumn{1}{c|}{\cellcolor[HTML]{FFFFFF}55.69} & \multicolumn{1}{c|}{\cellcolor[HTML]{FFFFFF}61.39} &\multicolumn{1}{c|}{\cellcolor[HTML]{FFFFFF}{\color[HTML]{000000}  \textbf{58.05}  }} &\multicolumn{1}{c|}{\cellcolor[HTML]{FFFFFF}{\color[HTML]{000000}   \textbf{55.72} }}   & \multicolumn{1}{c|}{\cellcolor[HTML]{FFFFFF}59.41}  & \multicolumn{1}{c|}{\cellcolor[HTML]{FFFFFF}\underline{55.00}}&\multicolumn{1}{c|}{\cellcolor[HTML]{FFFFFF}{\color[HTML]{000000}  \textbf{60.31}  }} &\multicolumn{1}{c|}{\cellcolor[HTML]{FFFFFF}{\color[HTML]{000000}   \textbf{54.34} }}    & \multicolumn{1}{c|}{\cellcolor[HTML]{FFFFFF}58.88}  & \multicolumn{1}{c|}{\cellcolor[HTML]{FFFFFF}59.41}   & \multicolumn{1}{c|}{\cellcolor[HTML]{FFFFFF}60.40}  \\ \hline

\multicolumn{1}{|c|}{\cellcolor[HTML]{FFFFFF}{\color[HTML]{000000} \textbf{Input}}} & \multicolumn{1}{c|}{\cellcolor[HTML]{FFFFFF}{\color[HTML]{000000} \textbf{Type}}}         & \multicolumn{1}{c|}{\cellcolor[HTML]{FFFFFF}{\color[HTML]{000000} \textbf{Method}}}      & \multicolumn{1}{c|}{\cellcolor[HTML]{FFFFFF}{\color[HTML]{000000} \textbf{D\&E to B1}}} & \multicolumn{1}{c|}{\cellcolor[HTML]{FFFFFF}{\color[HTML]{000000} \textbf{D\&E to B2}}} & \multicolumn{1}{c|}{\cellcolor[HTML]{FFFFFF}{\color[HTML]{000000} \textbf{D\&E to M}}} & \multicolumn{1}{c|}{\cellcolor[HTML]{FFFFFF}{\color[HTML]{000000} \textbf{D\&E to R}}} & \multicolumn{1}{c|}{\cellcolor[HTML]{FFFFFF}{\color[HTML]{000000} \textbf{\shortstack{R\&B1\&M\\to D}}}} & \multicolumn{1}{c|}{\cellcolor[HTML]{FFFFFF}{\color[HTML]{000000} \textbf{\shortstack{R\&B1\&M\\to E}}}} & \multicolumn{1}{c|}{\cellcolor[HTML]{FFFFFF}{\color[HTML]{000000} \textbf{\shortstack{R\&D\&E\\to B1}}}} & {\color[HTML]{000000} \textbf{\shortstack{R\&D\&E\\to M}}}         & {\color[HTML]{000000} \textbf{\shortstack{R\&M\&D\&E\\to B2}}}  & {\color[HTML]{000000} \textbf{\shortstack{R\&B1\&M\&\\D\&E to B2}}}     & {\color[HTML]{000000} \textbf{Avg}}       \\  \hline
\multicolumn{1}{|c|}{\cellcolor[HTML]{ECF4FF}{\color[HTML]{000000} }}   & \multicolumn{1}{c|}{\cellcolor[HTML]{FFFFFF}{\color[HTML]{000000} Score}}  & \multicolumn{1}{c|}{\cellcolor[HTML]{FFFFFF}{\color[HTML]{000000} Avg}}     & \multicolumn{1}{c|}{\cellcolor[HTML]{FFFFFF}{\color[HTML]{000000} 50.76}}   & \multicolumn{1}{c|}{\cellcolor[HTML]{FFFFFF}{\color[HTML]{000000} 53.62}}   & \multicolumn{1}{c|}{\cellcolor[HTML]{FFFFFF}{\color[HTML]{000000} 52.31}}   & \multicolumn{1}{c|}{\cellcolor[HTML]{FFFFFF}{\color[HTML]{000000} 51.22}}  & \multicolumn{1}{c|}{\cellcolor[HTML]{FFFFFF}{\color[HTML]{000000} 49.62}}  & \multicolumn{1}{c|}{\cellcolor[HTML]{FFFFFF}{\color[HTML]{000000} 50.38}}       & \multicolumn{1}{c|}{\cellcolor[HTML]{FFFFFF}{\color[HTML]{000000} 52.35}}  & \multicolumn{1}{c|}{\cellcolor[HTML]{FFFFFF}{\color[HTML]{000000} 51.15}} & \multicolumn{1}{c|}{\cellcolor[HTML]{FFFFFF}{\color[HTML]{000000} 54.37}} & \multicolumn{1}{c|}{\cellcolor[HTML]{FFFFFF}{\color[HTML]{000000} 52.12}} & \multicolumn{1}{c|}{\cellcolor[HTML]{FFFFFF}{\color[HTML]{000000} 52.98}} \\ \cline{2-14} 
\multicolumn{1}{|c|}{\cellcolor[HTML]{ECF4FF}{\color[HTML]{000000} }}  & \multicolumn{1}{c|}{\cellcolor[HTML]{FFFFFF}{\color[HTML]{000000} }}  & \multicolumn{1}{c|}{\cellcolor[HTML]{FFFFFF}{\color[HTML]{000000} Concat}}       & \multicolumn{1}{c|}{\cellcolor[HTML]{FFFFFF}{\color[HTML]{000000} 51.25}}  & \multicolumn{1}{c|}{\cellcolor[HTML]{FFFFFF}{\color[HTML]{000000} 54.21}}       & \multicolumn{1}{c|}{\cellcolor[HTML]{FFFFFF}{\color[HTML]{000000} 52.38}}  & \multicolumn{1}{c|}{\cellcolor[HTML]{FFFFFF}{\color[HTML]{000000} 51.42}}  & \multicolumn{1}{c|}{\cellcolor[HTML]{FFFFFF}{\color[HTML]{000000} 53.21}}      & \multicolumn{1}{c|}{\cellcolor[HTML]{FFFFFF}{\color[HTML]{000000} 52.46}}  & \multicolumn{1}{c|}{\cellcolor[HTML]{FFFFFF}{\color[HTML]{000000} 53.46}}  & \multicolumn{1}{c|}{\cellcolor[HTML]{FFFFFF}{\color[HTML]{000000} 52.31}} & \multicolumn{1}{c|}{\cellcolor[HTML]{FFFFFF}{\color[HTML]{000000} 54.88}} & \multicolumn{1}{c|}{\cellcolor[HTML]{FFFFFF}{\color[HTML]{000000} 54.38}} & \multicolumn{1}{c|}{\cellcolor[HTML]{FFFFFF}{\color[HTML]{000000} 54.49}}\\ \cline{3-14} 
\multicolumn{1}{|c|}{\cellcolor[HTML]{ECF4FF}{\color[HTML]{000000} }}  & \multicolumn{1}{c|}{\cellcolor[HTML]{FFFFFF}{\color[HTML]{000000} }}  & \multicolumn{1}{c|}{\cellcolor[HTML]{FFFFFF}{\color[HTML]{000000} SE-Concat}}   & \multicolumn{1}{c|}{\cellcolor[HTML]{FFFFFF}{\color[HTML]{000000} 51.36}}  & \multicolumn{1}{c|}{\cellcolor[HTML]{FFFFFF}{\color[HTML]{000000} 54.38}}  & \multicolumn{1}{c|}{\cellcolor[HTML]{FFFFFF}{\color[HTML]{000000} 52.36}}       & \multicolumn{1}{c|}{\cellcolor[HTML]{FFFFFF}{\color[HTML]{000000} 52.36}}  & \multicolumn{1}{c|}{\cellcolor[HTML]{FFFFFF}{\color[HTML]{000000} 54.87}} & \multicolumn{1}{c|}{\cellcolor[HTML]{FFFFFF}{\color[HTML]{000000} 53.49}}        & \multicolumn{1}{c|}{\cellcolor[HTML]{FFFFFF}{\color[HTML]{000000} 54.63}}  & \multicolumn{1}{c|}{\cellcolor[HTML]{FFFFFF}{\color[HTML]{000000} 52.33}} & \multicolumn{1}{c|}{\cellcolor[HTML]{FFFFFF}{\color[HTML]{000000} 55.62}}& \multicolumn{1}{c|}{\cellcolor[HTML]{FFFFFF}{\color[HTML]{000000} 55.72}} & \multicolumn{1}{c|}{\cellcolor[HTML]{FFFFFF}{\color[HTML]{000000} 54.27}}\\ \cline{3-14}
\multicolumn{1}{|c|}{\cellcolor[HTML]{ECF4FF}{\color[HTML]{000000} }} & \multicolumn{1}{c|}{\cellcolor[HTML]{FFFFFF}{\color[HTML]{000000} }}  & \multicolumn{1}{c|}{\cellcolor[HTML]{FFFFFF}{\color[HTML]{000000} Cross-Atten}}  & \multicolumn{1}{c|}{\cellcolor[HTML]{FFFFFF}{\color[HTML]{000000} 52.45}}  & \multicolumn{1}{c|}{\cellcolor[HTML]{FFFFFF}{\color[HTML]{000000} 55.32}}       & \multicolumn{1}{c|}{\cellcolor[HTML]{FFFFFF}{\color[HTML]{000000} 53.67}}  & \multicolumn{1}{c|}{\cellcolor[HTML]{FFFFFF}{\color[HTML]{000000} 53.34}}  & \multicolumn{1}{c|}{\cellcolor[HTML]{FFFFFF}{\color[HTML]{000000} 55.74}}       & \multicolumn{1}{c|}{\cellcolor[HTML]{FFFFFF}{\color[HTML]{000000} 54.42}}  & \multicolumn{1}{c|}{\cellcolor[HTML]{FFFFFF}{\color[HTML]{000000} 56.67}}  & \multicolumn{1}{c|}{\cellcolor[HTML]{FFFFFF}{\color[HTML]{000000} 53.87}} & \multicolumn{1}{c|}{\cellcolor[HTML]{FFFFFF}{\color[HTML]{000000} 55.71}}& \multicolumn{1}{c|}{\cellcolor[HTML]{FFFFFF}{\color[HTML]{000000} 55.84}} & \multicolumn{1}{c|}{\cellcolor[HTML]{FFFFFF}{\color[HTML]{000000} 55.45}} \\ \cline{3-14} 
\multicolumn{1}{|c|}{\cellcolor[HTML]{ECF4FF}{\color[HTML]{000000} }} & \multicolumn{1}{c|}{\cellcolor[HTML]{FFFFFF}{\color[HTML]{000000} }} & \multicolumn{1}{c|}{\cellcolor[HTML]{FFFFFF}{\color[HTML]{000000} MLP-Mixer}}    & \multicolumn{1}{c|}{\cellcolor[HTML]{FFFFFF}{\color[HTML]{000000} 52.66}}  & \multicolumn{1}{c|}{\cellcolor[HTML]{FFFFFF}{\color[HTML]{000000} 55.48}}       & \multicolumn{1}{c|}{\cellcolor[HTML]{FFFFFF}{\color[HTML]{000000} 53.48}}  & \multicolumn{1}{c|}{\cellcolor[HTML]{FFFFFF}{\color[HTML]{000000} 53.76}}  & \multicolumn{1}{c|}{\cellcolor[HTML]{FFFFFF}{\color[HTML]{000000} 56.72}} & \multicolumn{1}{c|}{\cellcolor[HTML]{FFFFFF}{\color[HTML]{000000} 56.73}} & \multicolumn{1}{c|}{\cellcolor[HTML]{FFFFFF}{\color[HTML]{000000} 56.72}}& \multicolumn{1}{c|}{\cellcolor[HTML]{FFFFFF}{\color[HTML]{000000} 54.83}}& \multicolumn{1}{c|}{\cellcolor[HTML]{FFFFFF}{\color[HTML]{000000} 56.82}}& \multicolumn{1}{c|}{\cellcolor[HTML]{FFFFFF}{\color[HTML]{000000} 55.71}}& \multicolumn{1}{c|}{\cellcolor[HTML]{FFFFFF}{\color[HTML]{000000} 56.78}} \\ \cline{3-14}
\multicolumn{1}{|c|}{\cellcolor[HTML]{ECF4FF}{\color[HTML]{000000} }} & \multicolumn{1}{c|}{\cellcolor[HTML]{FFFFFF}{\color[HTML]{000000} }} & \multicolumn{1}{c|}{\cellcolor[HTML]{FFFFFF}{\color[HTML]{000000} CLIP-Align}} & \multicolumn{1}{c|}{\cellcolor[HTML]{FFFFFF}{\color[HTML]{000000}51.98}} & \multicolumn{1}{c|}{\cellcolor[HTML]{FFFFFF}{\color[HTML]{000000}54.67}} & \multicolumn{1}{c|}{\cellcolor[HTML]{FFFFFF}{\color[HTML]{000000}53.42}} & \multicolumn{1}{c|}{\cellcolor[HTML]{FFFFFF}{\color[HTML]{000000}51.71}} & \multicolumn{1}{c|}{\cellcolor[HTML]{FFFFFF}{\color[HTML]{000000}54.09}} & \multicolumn{1}{c|}{\cellcolor[HTML]{FFFFFF}{\color[HTML]{000000}53.07}} & \multicolumn{1}{c|}{\cellcolor[HTML]{FFFFFF}{\color[HTML]{000000}53.63}} & \multicolumn{1}{c|}{\cellcolor[HTML]{FFFFFF}{\color[HTML]{000000}53.26}} & \multicolumn{1}{c|}{\cellcolor[HTML]{FFFFFF}{\color[HTML]{000000}55.27}} & \multicolumn{1}{c|}{\cellcolor[HTML]{FFFFFF}{\color[HTML]{000000}55.20}} & \multicolumn{1}{c|}{\cellcolor[HTML]{FFFFFF}{\color[HTML]{000000}54.66 }} \\ \cline{3-14} 
\multicolumn{1}{|c|}{\multirow{-6}{*}{\cellcolor[HTML]{ECF4FF}{\color[HTML]{000000} \begin{tabular}[c]{@{}c@{}}V-Face\\ +\\ V-AGA\end{tabular}}}}                               & \multicolumn{1}{c|}{\multirow{-5}{*}{\cellcolor[HTML]{FFFFFF}{\color[HTML]{000000} Feature}}} & \multicolumn{1}{c|}{\cellcolor[HTML]{FFFFFF}{\color[HTML]{000000} \textbf{Atten-Mixer}}} & \multicolumn{1}{c|}{\cellcolor[HTML]{FFFFFF}{\color[HTML]{000000} 53.86}}       & \multicolumn{1}{c|}{\cellcolor[HTML]{FFFFFF}{\color[HTML]{000000} 57.34}} & \multicolumn{1}{c|}{\cellcolor[HTML]{FFFFFF}{\color[HTML]{000000} 53.86}}  & \multicolumn{1}{c|}{\cellcolor[HTML]{FFFFFF}{\color[HTML]{000000} 54.11}}  & \multicolumn{1}{c|}{\cellcolor[HTML]{FFFFFF}{\color[HTML]{000000} 58.24}}  & \multicolumn{1}{c|}{\cellcolor[HTML]{FFFFFF}{\color[HTML]{000000} 57.02}}  & \multicolumn{1}{c|}{\cellcolor[HTML]{FFFFFF}{\color[HTML]{000000} 57.39}}  & \multicolumn{1}{c|}{\cellcolor[HTML]{FFFFFF}{\color[HTML]{000000} 54.87}}& \multicolumn{1}{c|}{\cellcolor[HTML]{FFFFFF}{\color[HTML]{000000} 58.78}}& \multicolumn{1}{c|}{\cellcolor[HTML]{FFFFFF}{\color[HTML]{000000} 58.72}} & \multicolumn{1}{c|}{\cellcolor[HTML]{FFFFFF}{\color[HTML]{000000} 57.59}}\\ \hline
\multicolumn{1}{|c|}{\cellcolor[HTML]{FFFFEB}{\color[HTML]{000000} }} & \multicolumn{1}{c|}{\cellcolor[HTML]{FFFFFF}{\color[HTML]{000000} Score}}  & \multicolumn{1}{c|}{\cellcolor[HTML]{FFFFFF}{\color[HTML]{000000} Avg}}     & \multicolumn{1}{c|}{\cellcolor[HTML]{FFFFFF}{\color[HTML]{000000} 50.38}} & \multicolumn{1}{c|}{\cellcolor[HTML]{FFFFFF}{\color[HTML]{000000} 53.25}} & \multicolumn{1}{c|}{\cellcolor[HTML]{FFFFFF}{\color[HTML]{000000} 52.33}}      & \multicolumn{1}{c|}{\cellcolor[HTML]{FFFFFF}{\color[HTML]{000000} 52.39}} & \multicolumn{1}{c|}{\cellcolor[HTML]{FFFFFF}{\color[HTML]{000000} 54.42}} & \multicolumn{1}{c|}{\cellcolor[HTML]{FFFFFF}{\color[HTML]{000000} 51.25}} & \multicolumn{1}{c|}{\cellcolor[HTML]{FFFFFF}{\color[HTML]{000000} 52.38}}  &
\multicolumn{1}{c|}{\cellcolor[HTML]{FFFFFF}{\color[HTML]{000000} 51.24}} &
\multicolumn{1}{c|}{\cellcolor[HTML]{FFFFFF}{\color[HTML]{000000} 54.31}} & \multicolumn{1}{c|}{\cellcolor[HTML]{FFFFFF}{\color[HTML]{000000} 53.43}} & \multicolumn{1}{c|}{\cellcolor[HTML]{FFFFFF}{\color[HTML]{000000} 53.63}} \\ \cline{2-14} 
\multicolumn{1}{|c|}{\cellcolor[HTML]{FFFFEB}{\color[HTML]{000000} }} & \multicolumn{1}{c|}{\cellcolor[HTML]{FFFFFF}{\color[HTML]{000000} }} & \multicolumn{1}{c|}{\cellcolor[HTML]{FFFFFF}{\color[HTML]{000000} Concat}} & \multicolumn{1}{c|}{\cellcolor[HTML]{FFFFFF}{\color[HTML]{000000} 51.49}} & \multicolumn{1}{c|}{\cellcolor[HTML]{FFFFFF}{\color[HTML]{000000} 53.48}}  & \multicolumn{1}{c|}{\cellcolor[HTML]{FFFFFF}{\color[HTML]{000000} 52.47}}      & \multicolumn{1}{c|}{\cellcolor[HTML]{FFFFFF}{\color[HTML]{000000} 53.18}}  & \multicolumn{1}{c|}{\cellcolor[HTML]{FFFFFF}{\color[HTML]{000000} 56.31}}  & \multicolumn{1}{c|}{\cellcolor[HTML]{FFFFFF}{\color[HTML]{000000} 53.32}}  & \multicolumn{1}{c|}{\cellcolor[HTML]{FFFFFF}{\color[HTML]{000000} 54.32}}  & \multicolumn{1}{c|}{\cellcolor[HTML]{FFFFFF}{\color[HTML]{000000} 52.33}} & \multicolumn{1}{c|}{\cellcolor[HTML]{FFFFFF}{\color[HTML]{000000} 54.33}}& \multicolumn{1}{c|}{\cellcolor[HTML]{FFFFFF}{\color[HTML]{000000} 56.72}} & \multicolumn{1}{c|}{\cellcolor[HTML]{FFFFFF}{\color[HTML]{000000} 54.68}} \\ \cline{3-14} 
\multicolumn{1}{|c|}{\cellcolor[HTML]{FFFFEB}{\color[HTML]{000000} }} & \multicolumn{1}{c|}{\cellcolor[HTML]{FFFFFF}{\color[HTML]{000000} }}  & \multicolumn{1}{c|}{\cellcolor[HTML]{FFFFFF}{\color[HTML]{000000} SE-Concat}}   & \multicolumn{1}{c|}{\cellcolor[HTML]{FFFFFF}{\color[HTML]{000000} 52.95}} & \multicolumn{1}{c|}{\cellcolor[HTML]{FFFFFF}{\color[HTML]{000000} 55.28}} & \multicolumn{1}{c|}{\cellcolor[HTML]{FFFFFF}{\color[HTML]{000000} 52.38}} & \multicolumn{1}{c|}{\cellcolor[HTML]{FFFFFF}{\color[HTML]{000000} 53.91}} & \multicolumn{1}{c|}{\cellcolor[HTML]{FFFFFF}{\color[HTML]{000000} 57.82}} & \multicolumn{1}{c|}{\cellcolor[HTML]{FFFFFF}{\color[HTML]{000000} 54.75}} & \multicolumn{1}{c|}{\cellcolor[HTML]{FFFFFF}{\color[HTML]{000000} 55.64}} & \multicolumn{1}{c|}{\cellcolor[HTML]{FFFFFF}{\color[HTML]{000000} 54.32}} & \multicolumn{1}{c|}{\cellcolor[HTML]{FFFFFF}{\color[HTML]{000000} 54.21}} & \multicolumn{1}{c|}{\cellcolor[HTML]{FFFFFF}{\color[HTML]{000000} 56.83}} & \multicolumn{1}{c|}{\cellcolor[HTML]{FFFFFF}{\color[HTML]{000000} 55.06}} \\ \cline{3-14} 
\multicolumn{1}{|c|}{\cellcolor[HTML]{FFFFEB}{\color[HTML]{000000} }} & \multicolumn{1}{c|}{\cellcolor[HTML]{FFFFFF}{\color[HTML]{000000} }} & \multicolumn{1}{c|}{\cellcolor[HTML]{FFFFFF}{\color[HTML]{000000} Cross-Atten}}& \multicolumn{1}{c|}{\cellcolor[HTML]{FFFFFF}{\color[HTML]{000000} 52.86}}& \multicolumn{1}{c|}{\cellcolor[HTML]{FFFFFF}{\color[HTML]{000000} 56.83}} & \multicolumn{1}{c|}{\cellcolor[HTML]{FFFFFF}{\color[HTML]{000000} 53.34}} & \multicolumn{1}{c|}{\cellcolor[HTML]{FFFFFF}{\color[HTML]{000000} 53.78}} & \multicolumn{1}{c|}{\cellcolor[HTML]{FFFFFF}{\color[HTML]{000000} 58.62}} & \multicolumn{1}{c|}{\cellcolor[HTML]{FFFFFF}{\color[HTML]{000000} 57.33}} & \multicolumn{1}{c|}{\cellcolor[HTML]{FFFFFF}{\color[HTML]{000000} 55.87}} & \multicolumn{1}{c|}{\cellcolor[HTML]{FFFFFF}{\color[HTML]{000000} 54.65}} & \multicolumn{1}{c|}{\cellcolor[HTML]{FFFFFF}{\color[HTML]{000000} 55.53}} & \multicolumn{1}{c|}{\cellcolor[HTML]{FFFFFF}{\color[HTML]{000000} 56.79}} & \multicolumn{1}{c|}{\cellcolor[HTML]{FFFFFF}{\color[HTML]{000000} 55.61}} \\ \cline{3-14}
\multicolumn{1}{|c|}{\cellcolor[HTML]{FFFFEB}{\color[HTML]{000000} }} & \multicolumn{1}{c|}{\cellcolor[HTML]{FFFFFF}{\color[HTML]{000000} }} & \multicolumn{1}{c|}{\cellcolor[HTML]{FFFFFF}{\color[HTML]{000000} MLP-Mixer}} & \multicolumn{1}{c|}{\cellcolor[HTML]{FFFFFF}{\color[HTML]{000000} 53.47}} & \multicolumn{1}{c|}{\cellcolor[HTML]{FFFFFF}{\color[HTML]{000000} 56.74}} & \multicolumn{1}{c|}{\cellcolor[HTML]{FFFFFF}{\color[HTML]{000000} 53.48}} & \multicolumn{1}{c|}{\cellcolor[HTML]{FFFFFF}{\color[HTML]{000000} 54.27}} & \multicolumn{1}{c|}{\cellcolor[HTML]{FFFFFF}{\color[HTML]{000000} 58.72}} & \multicolumn{1}{c|}{\cellcolor[HTML]{FFFFFF}{\color[HTML]{000000} 57.38}} & \multicolumn{1}{c|}{\cellcolor[HTML]{FFFFFF}{\color[HTML]{000000} 57.64}} & \multicolumn{1}{c|}{\cellcolor[HTML]{FFFFFF}{\color[HTML]{000000} 54.87}} & \multicolumn{1}{c|}{\cellcolor[HTML]{FFFFFF}{\color[HTML]{000000} 55.49}} & \multicolumn{1}{c|}{\cellcolor[HTML]{FFFFFF}{\color[HTML]{000000} 57.83}} & \multicolumn{1}{c|}{\cellcolor[HTML]{FFFFFF}{\color[HTML]{000000} 56.35}} \\ \cline{3-14}
\multicolumn{1}{|c|}{\cellcolor[HTML]{FFFFEB}{\color[HTML]{000000} }} & \multicolumn{1}{c|}{\cellcolor[HTML]{FFFFFF}{\color[HTML]{000000} }} & \multicolumn{1}{c|}{\cellcolor[HTML]{FFFFFF}{\color[HTML]{000000} CLIP-Align}} & \multicolumn{1}{c|}{\cellcolor[HTML]{FFFFFF}{\color[HTML]{000000}52.12}} & \multicolumn{1}{c|}{\cellcolor[HTML]{FFFFFF}{\color[HTML]{000000}54.52}} & \multicolumn{1}{c|}{\cellcolor[HTML]{FFFFFF}{\color[HTML]{000000}52.74}} & \multicolumn{1}{c|}{\cellcolor[HTML]{FFFFFF}{\color[HTML]{000000}54.06}} & \multicolumn{1}{c|}{\cellcolor[HTML]{FFFFFF}{\color[HTML]{000000}56.82}} & \multicolumn{1}{c|}{\cellcolor[HTML]{FFFFFF}{\color[HTML]{000000}53.48}} & \multicolumn{1}{c|}{\cellcolor[HTML]{FFFFFF}{\color[HTML]{000000}55.26}} & \multicolumn{1}{c|}{\cellcolor[HTML]{FFFFFF}{\color[HTML]{000000}53.06}} & \multicolumn{1}{c|}{\cellcolor[HTML]{FFFFFF}{\color[HTML]{000000}54.72}} & \multicolumn{1}{c|}{\cellcolor[HTML]{FFFFFF}{\color[HTML]{000000}57.08}} & \multicolumn{1}{c|}{\cellcolor[HTML]{FFFFFF}{\color[HTML]{000000}55.30}} \\ \cline{3-14} 
\multicolumn{1}{|c|}{\multirow{-6}{*}{\cellcolor[HTML]{FFFFEB}{\color[HTML]{000000} \begin{tabular}[c]{@{}c@{}}V-Face\\ +\\ A-Mel\end{tabular}}}}                               & \multicolumn{1}{c|}{\multirow{-5}{*}{\cellcolor[HTML]{FFFFFF}{\color[HTML]{000000} Feature}}} & \multicolumn{1}{c|}{\cellcolor[HTML]{FFFFFF}{\color[HTML]{000000} \textbf{Atten-Mixer}}} & \multicolumn{1}{c|}{\cellcolor[HTML]{FFFFFF}{\color[HTML]{000000} 54.87}} & \multicolumn{1}{c|}{\cellcolor[HTML]{FFFFFF}{\color[HTML]{000000} 58.69}} & \multicolumn{1}{c|}{\cellcolor[HTML]{FFFFFF}{\color[HTML]{000000} \underline{55.87}}} & \multicolumn{1}{c|}{\cellcolor[HTML]{FFFFFF}{\color[HTML]{000000} 54.86}} & \multicolumn{1}{c|}{\cellcolor[HTML]{FFFFFF}{\color[HTML]{000000} \underline{58.98}}}      & \multicolumn{1}{c|}{\cellcolor[HTML]{FFFFFF}{\color[HTML]{000000} 58.24}} & \multicolumn{1}{c|}{\cellcolor[HTML]{FFFFFF}{\color[HTML]{000000} 58.99}}  & \multicolumn{1}{c|}{\cellcolor[HTML]{FFFFFF}{\color[HTML]{000000} 55.68}} & \multicolumn{1}{c|}{\cellcolor[HTML]{FFFFFF}{\color[HTML]{000000} \textbf{61.39}}} & \multicolumn{1}{c|}{\cellcolor[HTML]{FFFFFF}{\color[HTML]{000000} 58.99}} & \multicolumn{1}{c|}{\cellcolor[HTML]{FFFFFF}{\color[HTML]{000000} \underline{58.13}}} \\ \hline
\multicolumn{1}{|c|}{\cellcolor[HTML]{FFF7F7}{\color[HTML]{000000} }} & \multicolumn{1}{c|}{\cellcolor[HTML]{FFFFFF}{\color[HTML]{000000} Score}} & \multicolumn{1}{c|}{\cellcolor[HTML]{FFFFFF}{\color[HTML]{000000} Avg}}     & \multicolumn{1}{c|}{\cellcolor[HTML]{FFFFFF}{\color[HTML]{000000} 52.86}} & \multicolumn{1}{c|}{\cellcolor[HTML]{FFFFFF}{\color[HTML]{000000} 56.31}} & \multicolumn{1}{c|}{\cellcolor[HTML]{FFFFFF}{\color[HTML]{000000} 51.85}} & \multicolumn{1}{c|}{\cellcolor[HTML]{FFFFFF}{\color[HTML]{000000} 52.38}} & \multicolumn{1}{c|}{\cellcolor[HTML]{FFFFFF}{\color[HTML]{000000} 53.21}} & \multicolumn{1}{c|}{\cellcolor[HTML]{FFFFFF}{\color[HTML]{000000} 51.27}} & \multicolumn{1}{c|}{\cellcolor[HTML]{FFFFFF}{\color[HTML]{000000} 57.67}} & \multicolumn{1}{c|}{\cellcolor[HTML]{FFFFFF}{\color[HTML]{000000} 51.23}} & \multicolumn{1}{c|}{\cellcolor[HTML]{FFFFFF}{\color[HTML]{000000} 54.06}} & \multicolumn{1}{c|}{\cellcolor[HTML]{FFFFFF}{\color[HTML]{000000} 54.72}} & \multicolumn{1}{c|}{\cellcolor[HTML]{FFFFFF}{\color[HTML]{000000} 54.43}} \\ \cline{2-14}
\multicolumn{1}{|c|}{\cellcolor[HTML]{FFF7F7}{\color[HTML]{000000} }} & \multicolumn{1}{c|}{\cellcolor[HTML]{FFFFFF}{\color[HTML]{000000} }} & \multicolumn{1}{c|}{\cellcolor[HTML]{FFFFFF}{\color[HTML]{000000} Concat}} & \multicolumn{1}{c|}{\cellcolor[HTML]{FFFFFF}{\color[HTML]{000000} 52.65}} & \multicolumn{1}{c|}{\cellcolor[HTML]{FFFFFF}{\color[HTML]{000000} 55.89}} & \multicolumn{1}{c|}{\cellcolor[HTML]{FFFFFF}{\color[HTML]{000000} 51.64}} & \multicolumn{1}{c|}{\cellcolor[HTML]{FFFFFF}{\color[HTML]{000000} 53.34}} & \multicolumn{1}{c|}{\cellcolor[HTML]{FFFFFF}{\color[HTML]{000000} 56.72}} & \multicolumn{1}{c|}{\cellcolor[HTML]{FFFFFF}{\color[HTML]{000000} 52.23}} & \multicolumn{1}{c|}{\cellcolor[HTML]{FFFFFF}{\color[HTML]{000000} 57.83}} & \multicolumn{1}{c|}{\cellcolor[HTML]{FFFFFF}{\color[HTML]{000000} 52.35}} & \multicolumn{1}{c|}{\cellcolor[HTML]{FFFFFF}{\color[HTML]{000000} 54.65}} & \multicolumn{1}{c|}{\cellcolor[HTML]{FFFFFF}{\color[HTML]{000000} 55.49}} & \multicolumn{1}{c|}{\cellcolor[HTML]{FFFFFF}{\color[HTML]{000000} 54.69}} \\ \cline{3-14}
\multicolumn{1}{|c|}{\cellcolor[HTML]{FFF7F7}{\color[HTML]{000000} }}& \multicolumn{1}{c|}{\cellcolor[HTML]{FFFFFF}{\color[HTML]{000000} }} & \multicolumn{1}{c|}{\cellcolor[HTML]{FFFFFF}{\color[HTML]{000000} SE-Concat}}  & \multicolumn{1}{c|}{\cellcolor[HTML]{FFFFFF}{\color[HTML]{000000} 53.49}} & \multicolumn{1}{c|}{\cellcolor[HTML]{FFFFFF}{\color[HTML]{000000} 57.28}} & \multicolumn{1}{c|}{\cellcolor[HTML]{FFFFFF}{\color[HTML]{000000} 52.38}} & \multicolumn{1}{c|}{\cellcolor[HTML]{FFFFFF}{\color[HTML]{000000} 54.96}} & \multicolumn{1}{c|}{\cellcolor[HTML]{FFFFFF}{\color[HTML]{000000} 57.46}} & \multicolumn{1}{c|}{\cellcolor[HTML]{FFFFFF}{\color[HTML]{000000} 56.93}} & \multicolumn{1}{c|}{\cellcolor[HTML]{FFFFFF}{\color[HTML]{000000} 58.72}} & \multicolumn{1}{c|}{\cellcolor[HTML]{FFFFFF}{\color[HTML]{000000} 52.39}} & \multicolumn{1}{c|}{\cellcolor[HTML]{FFFFFF}{\color[HTML]{000000} 55.68}} & \multicolumn{1}{c|}{\cellcolor[HTML]{FFFFFF}{\color[HTML]{000000} 57.89}} & \multicolumn{1}{c|}{\cellcolor[HTML]{FFFFFF}{\color[HTML]{000000} 56.35}} \\ \cline{3-14} 
\multicolumn{1}{|c|}{\cellcolor[HTML]{FFF7F7}{\color[HTML]{000000} }} & \multicolumn{1}{c|}{\cellcolor[HTML]{FFFFFF}{\color[HTML]{000000} }} & \multicolumn{1}{c|}{\cellcolor[HTML]{FFFFFF}{\color[HTML]{000000} Cross-Atten}} & \multicolumn{1}{c|}{\cellcolor[HTML]{FFFFFF}{\color[HTML]{000000} 53.64}} & \multicolumn{1}{c|}{\cellcolor[HTML]{FFFFFF}{\color[HTML]{000000} 58.72}} & \multicolumn{1}{c|}{\cellcolor[HTML]{FFFFFF}{\color[HTML]{000000} 53.03}} & \multicolumn{1}{c|}{\cellcolor[HTML]{FFFFFF}{\color[HTML]{000000} 54.73}} & \multicolumn{1}{c|}{\cellcolor[HTML]{FFFFFF}{\color[HTML]{000000} 57.68}} & \multicolumn{1}{c|}{\cellcolor[HTML]{FFFFFF}{\color[HTML]{000000} 56.73}} & \multicolumn{1}{c|}{\cellcolor[HTML]{FFFFFF}{\color[HTML]{000000} 58.31}} & \multicolumn{1}{c|}{\cellcolor[HTML]{FFFFFF}{\color[HTML]{000000} 53.48}} & \multicolumn{1}{c|}{\cellcolor[HTML]{FFFFFF}{\color[HTML]{000000} 57.83}} & \multicolumn{1}{c|}{\cellcolor[HTML]{FFFFFF}{\color[HTML]{000000} 59.83}} & \multicolumn{1}{c|}{\cellcolor[HTML]{FFFFFF}{\color[HTML]{000000} 56.73}} \\ \cline{3-14} 
\multicolumn{1}{|c|}{\cellcolor[HTML]{FFF7F7}{\color[HTML]{000000} }} & \multicolumn{1}{c|}{\cellcolor[HTML]{FFFFFF}{\color[HTML]{000000} }} & \multicolumn{1}{c|}{\cellcolor[HTML]{FFFFFF}{\color[HTML]{000000} MLP-Mixer}}  & \multicolumn{1}{c|}{\cellcolor[HTML]{FFFFFF}{\color[HTML]{000000} 54.67}} & \multicolumn{1}{c|}{\cellcolor[HTML]{FFFFFF}{\color[HTML]{000000} 59.81}}      & \multicolumn{1}{c|}{\cellcolor[HTML]{FFFFFF}{\color[HTML]{000000} 53.48}} & \multicolumn{1}{c|}{\cellcolor[HTML]{FFFFFF}{\color[HTML]{000000} 55.92}} & \multicolumn{1}{c|}{\cellcolor[HTML]{FFFFFF}{\color[HTML]{000000} 58.92}} & \multicolumn{1}{c|}{\cellcolor[HTML]{FFFFFF}{\color[HTML]{000000} 57.84}} & \multicolumn{1}{c|}{\cellcolor[HTML]{FFFFFF}{\color[HTML]{000000} 59.98}} & \multicolumn{1}{c|}{\cellcolor[HTML]{FFFFFF}{\color[HTML]{000000} 54.63}} & \multicolumn{1}{c|}{\cellcolor[HTML]{FFFFFF}{\color[HTML]{000000} 58.75}} & \multicolumn{1}{c|}{\cellcolor[HTML]{FFFFFF}{\color[HTML]{000000} 59.86}} & \multicolumn{1}{c|}{\cellcolor[HTML]{FFFFFF}{\color[HTML]{000000} 56.95}} \\ \cline{3-14}
\multicolumn{1}{|c|}{\cellcolor[HTML]{FFF7F7}{\color[HTML]{000000} }} & \multicolumn{1}{c|}{\cellcolor[HTML]{FFFFFF}{\color[HTML]{000000} }} & \multicolumn{1}{c|}{\cellcolor[HTML]{FFFFFF}{\color[HTML]{000000} CLIP-Align}} & \multicolumn{1}{c|}{\cellcolor[HTML]{FFFFFF}{\color[HTML]{000000}53.38}} & \multicolumn{1}{c|}{\cellcolor[HTML]{FFFFFF}{\color[HTML]{000000}56.31}} & \multicolumn{1}{c|}{\cellcolor[HTML]{FFFFFF}{\color[HTML]{000000}52.69}} & \multicolumn{1}{c|}{\cellcolor[HTML]{FFFFFF}{\color[HTML]{000000}53.52}} & \multicolumn{1}{c|}{\cellcolor[HTML]{FFFFFF}{\color[HTML]{000000}57.68}} & \multicolumn{1}{c|}{\cellcolor[HTML]{FFFFFF}{\color[HTML]{000000}52.80}} & \multicolumn{1}{c|}{\cellcolor[HTML]{FFFFFF}{\color[HTML]{000000}58.14}} & \multicolumn{1}{c|}{\cellcolor[HTML]{FFFFFF}{\color[HTML]{000000}53.19}} & \multicolumn{1}{c|}{\cellcolor[HTML]{FFFFFF}{\color[HTML]{000000}54.77}} & \multicolumn{1}{c|}{\cellcolor[HTML]{FFFFFF}{\color[HTML]{000000}56.17}} & \multicolumn{1}{c|}{\cellcolor[HTML]{FFFFFF}{\color[HTML]{000000}55.31}} \\ \cline{3-14} 
\multicolumn{1}{|c|}{\multirow{-6}{*}{\cellcolor[HTML]{FFF7F7}{\color[HTML]{000000} \begin{tabular}[c]{@{}c@{}}V-AGA\\ +\\ A-Mel\end{tabular}}}} & \multicolumn{1}{c|}{\multirow{-5}{*}{\cellcolor[HTML]{FFFFFF}{\color[HTML]{000000} Feature}}} & \multicolumn{1}{c|}{\cellcolor[HTML]{FFFFFF}{\color[HTML]{000000} \textbf{Atten-Mixer}}} & \multicolumn{1}{c|}{\cellcolor[HTML]{FFFFFF}{\color[HTML]{000000} 54.78}} & \multicolumn{1}{c|}{\cellcolor[HTML]{FFFFFF}{\color[HTML]{000000} \underline{60.45}}} & \multicolumn{1}{c|}{\cellcolor[HTML]{FFFFFF}{\color[HTML]{000000} 54.72}} & \multicolumn{1}{c|}{\cellcolor[HTML]{FFFFFF}{\color[HTML]{000000} 56.03}}& \multicolumn{1}{c|}{\cellcolor[HTML]{FFFFFF}{\color[HTML]{000000} 59.21}} & \multicolumn{1}{c|}{\cellcolor[HTML]{FFFFFF}{\color[HTML]{000000} 58.82}} & \multicolumn{1}{c|}{\cellcolor[HTML]{FFFFFF}{\color[HTML]{000000} \underline{60.12}}}& \multicolumn{1}{c|}{\cellcolor[HTML]{FFFFFF}{\color[HTML]{000000} 55.69}}  & \multicolumn{1}{c|}{\cellcolor[HTML]{FFFFFF}{\color[HTML]{000000} 60.33}} & \multicolumn{1}{c|}{\cellcolor[HTML]{FFFFFF}{\color[HTML]{000000} \underline{62.31}}} & \multicolumn{1}{c|}{\cellcolor[HTML]{FFFFFF}{\color[HTML]{000000} 57.89}} \\ \hline
\multicolumn{1}{|c|}{\cellcolor[HTML]{EEFBED}{\color[HTML]{000000} }} & \multicolumn{1}{c|}{\cellcolor[HTML]{FFFFFF}{\color[HTML]{000000} Score}} & \multicolumn{1}{c|}{\cellcolor[HTML]{FFFFFF}{\color[HTML]{000000} Avg}}    & \multicolumn{1}{c|}{\cellcolor[HTML]{FFFFFF}{\color[HTML]{000000} 52.38}} & \multicolumn{1}{c|}{\cellcolor[HTML]{FFFFFF}{\color[HTML]{000000} 57.33}} & \multicolumn{1}{c|}{\cellcolor[HTML]{FFFFFF}{\color[HTML]{000000} 52.88}}      & \multicolumn{1}{c|}{\cellcolor[HTML]{FFFFFF}{\color[HTML]{000000} 53.42}} & \multicolumn{1}{c|}{\cellcolor[HTML]{FFFFFF}{\color[HTML]{000000} 56.33}} & \multicolumn{1}{c|}{\cellcolor[HTML]{FFFFFF}{\color[HTML]{000000} 52.31}} & \multicolumn{1}{c|}{\cellcolor[HTML]{FFFFFF}{\color[HTML]{000000} 54.32}} & \multicolumn{1}{c|}{\cellcolor[HTML]{FFFFFF}{\color[HTML]{000000} 51.25}} & \multicolumn{1}{c|}{\cellcolor[HTML]{FFFFFF}{\color[HTML]{000000} 53.46}} & \multicolumn{1}{c|}{\cellcolor[HTML]{FFFFFF}{\color[HTML]{000000} 58.93}} & \multicolumn{1}{c|}{\cellcolor[HTML]{FFFFFF}{\color[HTML]{000000} 54.90}} \\ \cline{2-14}
\multicolumn{1}{|c|}{\cellcolor[HTML]{EEFBED}{\color[HTML]{000000} }}& \multicolumn{1}{c|}{\cellcolor[HTML]{FFFFFF}{\color[HTML]{000000} }} & \multicolumn{1}{c|}{\cellcolor[HTML]{FFFFFF}{\color[HTML]{000000} Concat}} & \multicolumn{1}{c|}{\cellcolor[HTML]{FFFFFF}{\color[HTML]{000000} 53.08}} & \multicolumn{1}{c|}{\cellcolor[HTML]{FFFFFF}{\color[HTML]{000000} 58.31}} & \multicolumn{1}{c|}{\cellcolor[HTML]{FFFFFF}{\color[HTML]{000000} 53.02}} & \multicolumn{1}{c|}{\cellcolor[HTML]{FFFFFF}{\color[HTML]{000000} 54.33}} & \multicolumn{1}{c|}{\cellcolor[HTML]{FFFFFF}{\color[HTML]{000000} 56.31}} & \multicolumn{1}{c|}{\cellcolor[HTML]{FFFFFF}{\color[HTML]{000000} 53.42}} & \multicolumn{1}{c|}{\cellcolor[HTML]{FFFFFF}{\color[HTML]{000000} 54.76}} & \multicolumn{1}{c|}{\cellcolor[HTML]{FFFFFF}{\color[HTML]{000000} 52.38}} & \multicolumn{1}{c|}{\cellcolor[HTML]{FFFFFF}{\color[HTML]{000000} 54.55}} & \multicolumn{1}{c|}{\cellcolor[HTML]{FFFFFF}{\color[HTML]{000000} 58.72}} & \multicolumn{1}{c|}{\cellcolor[HTML]{FFFFFF}{\color[HTML]{000000} 55.23}} \\ \cline{3-14}
\multicolumn{1}{|c|}{\cellcolor[HTML]{EEFBED}{\color[HTML]{000000} }} & \multicolumn{1}{c|}{\cellcolor[HTML]{FFFFFF}{\color[HTML]{000000} }} & \multicolumn{1}{c|}{\cellcolor[HTML]{FFFFFF}{\color[HTML]{000000} SE-Concat}} & \multicolumn{1}{c|}{\cellcolor[HTML]{FFFFFF}{\color[HTML]{000000} 54.89}} & \multicolumn{1}{c|}{\cellcolor[HTML]{FFFFFF}{\color[HTML]{000000} 58.39}} & \multicolumn{1}{c|}{\cellcolor[HTML]{FFFFFF}{\color[HTML]{000000} 53.76}} & \multicolumn{1}{c|}{\cellcolor[HTML]{FFFFFF}{\color[HTML]{000000} 55.41}} & \multicolumn{1}{c|}{\cellcolor[HTML]{FFFFFF}{\color[HTML]{000000} 57.35}} & \multicolumn{1}{c|}{\cellcolor[HTML]{FFFFFF}{\color[HTML]{000000} 56.65}} & \multicolumn{1}{c|}{\cellcolor[HTML]{FFFFFF}{\color[HTML]{000000} 55.65}} & \multicolumn{1}{c|}{\cellcolor[HTML]{FFFFFF}{\color[HTML]{000000} 54.83}} & \multicolumn{1}{c|}{\cellcolor[HTML]{FFFFFF}{\color[HTML]{000000} 56.73}} & \multicolumn{1}{c|}{\cellcolor[HTML]{FFFFFF}{\color[HTML]{000000} 58.79}} & \multicolumn{1}{c|}{\cellcolor[HTML]{FFFFFF}{\color[HTML]{000000} 56.24}} \\ \cline{3-14} 
\multicolumn{1}{|c|}{\cellcolor[HTML]{EEFBED}{\color[HTML]{000000} }} & \multicolumn{1}{c|}{\cellcolor[HTML]{FFFFFF}{\color[HTML]{000000} }} & \multicolumn{1}{c|}{\cellcolor[HTML]{FFFFFF}{\color[HTML]{000000} Cross-Atten}} & \multicolumn{1}{c|}{\cellcolor[HTML]{FFFFFF}{\color[HTML]{000000} 55.98}} & \multicolumn{1}{c|}{\cellcolor[HTML]{FFFFFF}{\color[HTML]{000000} 59.74}} & \multicolumn{1}{c|}{\cellcolor[HTML]{FFFFFF}{\color[HTML]{000000} 53.47}} & \multicolumn{1}{c|}{\cellcolor[HTML]{FFFFFF}{\color[HTML]{000000} 55.64}} & \multicolumn{1}{c|}{\cellcolor[HTML]{FFFFFF}{\color[HTML]{000000} 57.36}} & \multicolumn{1}{c|}{\cellcolor[HTML]{FFFFFF}{\color[HTML]{000000} 57.82}} & \multicolumn{1}{c|}{\cellcolor[HTML]{FFFFFF}{\color[HTML]{000000} 57.86}} &  \multicolumn{1}{c|}{\cellcolor[HTML]{FFFFFF}{\color[HTML]{000000} 55.73}} & \multicolumn{1}{c|}{\cellcolor[HTML]{FFFFFF}{\color[HTML]{000000} 56.79}} & \multicolumn{1}{c|}{\cellcolor[HTML]{FFFFFF}{\color[HTML]{000000} 58.22}} & \multicolumn{1}{c|}{\cellcolor[HTML]{FFFFFF}{\color[HTML]{000000} 56.84}}  \\ \cline{3-14} 
\multicolumn{1}{|c|}{\cellcolor[HTML]{EEFBED}{\color[HTML]{000000} }} & \multicolumn{1}{c|}{\cellcolor[HTML]{FFFFFF}{\color[HTML]{000000} }} & \multicolumn{1}{c|}{\cellcolor[HTML]{FFFFFF}{\color[HTML]{000000} MLP-Mixer}} & \multicolumn{1}{c|}{\cellcolor[HTML]{FFFFFF}{\color[HTML]{000000} \underline{56.73}}} & \multicolumn{1}{c|}{\cellcolor[HTML]{FFFFFF}{\color[HTML]{000000} 60.28}} & \multicolumn{1}{c|}{\cellcolor[HTML]{FFFFFF}{\color[HTML]{000000} 55.08}} & \multicolumn{1}{c|}{\cellcolor[HTML]{FFFFFF}{\color[HTML]{000000} \underline{58.02}}} & \multicolumn{1}{c|}{\cellcolor[HTML]{FFFFFF}{\color[HTML]{000000} 58.72}} & \multicolumn{1}{c|}{\cellcolor[HTML]{FFFFFF}{\color[HTML]{000000} \underline{58.93}}} & \multicolumn{1}{c|}{\cellcolor[HTML]{FFFFFF}{\color[HTML]{000000} 59.93}} & \multicolumn{1}{c|}{\cellcolor[HTML]{FFFFFF}{\color[HTML]{000000} \underline{55.76}}} & \multicolumn{1}{c|}{\cellcolor[HTML]{FFFFFF}{\color[HTML]{000000} 58.32}} & \multicolumn{1}{c|}{\cellcolor[HTML]{FFFFFF}{\color[HTML]{000000} 59.89}} & \multicolumn{1}{c|}{\cellcolor[HTML]{FFFFFF}{\color[HTML]{000000} 57.66}} \\ \cline{3-14} 
\multicolumn{1}{|c|}{\cellcolor[HTML]{EEFBED}{\color[HTML]{000000} }} & \multicolumn{1}{c|}{\cellcolor[HTML]{FFFFFF}{\color[HTML]{000000} }} & \multicolumn{1}{c|}{\cellcolor[HTML]{FFFFFF}{\color[HTML]{000000} CLIP-Align}} & \multicolumn{1}{c|}{\cellcolor[HTML]{FFFFFF}{\color[HTML]{000000}53.86}} & \multicolumn{1}{c|}{\cellcolor[HTML]{FFFFFF}{\color[HTML]{000000}59.04}} & \multicolumn{1}{c|}{\cellcolor[HTML]{FFFFFF}{\color[HTML]{000000}53.55}} & \multicolumn{1}{c|}{\cellcolor[HTML]{FFFFFF}{\color[HTML]{000000}55.26}} & \multicolumn{1}{c|}{\cellcolor[HTML]{FFFFFF}{\color[HTML]{000000}56.89}} & \multicolumn{1}{c|}{\cellcolor[HTML]{FFFFFF}{\color[HTML]{000000}54.47}} & \multicolumn{1}{c|}{\cellcolor[HTML]{FFFFFF}{\color[HTML]{000000}55.33}} & \multicolumn{1}{c|}{\cellcolor[HTML]{FFFFFF}{\color[HTML]{000000}53.18}} & \multicolumn{1}{c|}{\cellcolor[HTML]{FFFFFF}{\color[HTML]{000000}55.42}} & \multicolumn{1}{c|}{\cellcolor[HTML]{FFFFFF}{\color[HTML]{000000}59.19}} & \multicolumn{1}{c|}{\cellcolor[HTML]{FFFFFF}{\color[HTML]{000000}55.88}} \\ \cline{3-14} 
\multicolumn{1}{|c|}{\multirow{-6}{*}{\cellcolor[HTML]{EEFBED}{\color[HTML]{000000} \begin{tabular}[c]{@{}c@{}}V-Face\\ +\\ V-AGA\\ +\\ A-Mel\end{tabular}}}} & \multicolumn{1}{c|}{\multirow{-5}{*}{\cellcolor[HTML]{FFFFFF}{\color[HTML]{000000} Feature}}} & \multicolumn{1}{c|}{\cellcolor[HTML]{FFFFFF}{\color[HTML]{000000} \textbf{Atten-Mixer}}} & \multicolumn{1}{c|}{\cellcolor[HTML]{FFFFFF}{\color[HTML]{000000} \textbf{57.47}}}       & \multicolumn{1}{c|}{\cellcolor[HTML]{FFFFFF}{\color[HTML]{000000} \textbf{63.48}}} & \multicolumn{1}{c|}{\cellcolor[HTML]{FFFFFF}{\color[HTML]{000000} \textbf{56.71}}} & \multicolumn{1}{c|}{\cellcolor[HTML]{FFFFFF}{\color[HTML]{000000} \textbf{58.36}}} & \multicolumn{1}{c|}{\cellcolor[HTML]{FFFFFF}{\color[HTML]{000000} \textbf{59.55}}} & \multicolumn{1}{c|}{\cellcolor[HTML]{FFFFFF}{\color[HTML]{000000} \textbf{60.33}}} & \multicolumn{1}{c|}{\cellcolor[HTML]{FFFFFF}{\color[HTML]{000000} \textbf{61.35}}} & \multicolumn{1}{c|}{\cellcolor[HTML]{FFFFFF}{\color[HTML]{000000} \textbf{57.83}}} & \multicolumn{1}{c|}{\cellcolor[HTML]{FFFFFF}{\color[HTML]{000000} \underline{60.36}}} & \multicolumn{1}{c|}{\cellcolor[HTML]{FFFFFF}{\color[HTML]{000000} \textbf{62.35}}} & \multicolumn{1}{c|}{\cellcolor[HTML]{FFFFFF}{\color[HTML]{000000} \textbf{58.88}}}  \\ \hline
\end{tabular}}
\label{tab:multi2single_fusion_domainbydomain}
\vspace{-0.5em}
\end{table*}

\noindent\textbf{Single-to-Single Domain with Fusion.}\quad 
In this section, we conduct single-to-single cross-domain experiments with multimodal fusion. We consider two fusion levels: score-level fusion and feature-level fusion. For feature-level fusion, we compare several representative methods, including simple concatenation, SE-Concat~\cite{hu2018squeeze}, Cross-Atten~\cite{vaswani2017attention}, MLP-Mixer~\cite{tolstikhin2021mlp}, CLIP-style alignment (CLIP-ALign)~\cite{radford2021learning, oord2018representation}, and our proposed Atten-Mixer. Importantly, the benchmark includes stronger fusion baselines beyond simple concatenation. Cross-Atten serves as a Transformer-based cross-modal attention baseline, while MLP-Mixer provides a token/channel-mixing fusion baseline. Our Atten-Mixer further incorporates multi-head self-attention into the fusion module, enabling more effective cross-modal interaction. Specifically, simple concatenation directly combines the extracted modality features before classification. SE-Concat applies squeeze-and-excitation attention to the concatenated features. Cross-Atten models cross-modal interactions using the attention mechanism from the Transformer, while MLP-Mixer performs token- and channel-mixing over the extracted features. Instead of directly using a pretrained CLIP backbone, we employ a CLIP-style contrastive alignment strategy, since CLIP is pretrained for image-text semantic matching rather than direct fusion of heterogeneous modalities such as visual frames, behavioral cues, and audio. Moreover, deception detection depends on subtle task-specific cross-modal cues, which are not well aligned with CLIP's original pretraining objective. We therefore use modality-specific encoders to extract unimodal representations, project them into a shared normalized embedding space, and optimize them with a symmetric InfoNCE loss. The aligned embeddings are concatenated and fed into an MLP classifier, with joint supervision from unimodal classification losses, fused classification loss, and the multimodal alignment loss.

The modalities and input include three types, visual (face frames), visual (AU+gaze+affect), and audio (Mel spectrogram). As shown in Table~\ref{tab:single2singlefusion}, there are 25 domain transfer tasks, and the proposed Atten-Mixer consistently outperforms other methods across all modalities and fusion types, achieving the highest average accuracies. This demonstrates its strong ability to learn transferable representations. Feature fusion generally yields better performance than score-level fusion, highlighting the benefit of early interaction between modalities. Among the modalities, the best average performance across all the datasets is achieved when fused with Atten-Mixer. 

\vspace{0.3em}
\noindent\textbf{Multi-to-Single Domain with Fusion.}\quad Tables~\ref{tab:multi2single_fusion_simul}–\ref{tab:multi2single_fusion_domainbydomain} present multi-to-single domain generalization results across 21 tasks using four different modality combinations under three sampling strategies. Across all settings, the proposed Atten-Mixer consistently outperforms the baseline methods and achieves the highest average accuracies. Notably, feature-level fusion of Visual (AU+Gaze+Affect) + Audio using Atten-Mixer reaches the best overall average performance: 57.97\% under Domain-Simultaneous, 56.35\% under Domain-Alternating, and 58.88\% under Domain-by-Domain, showing strong cross-domain generalization ability. We also observe that the CLIP-style alignment variant, implemented by adding a contrastive loss to the concatenation baseline, improves over plain concatenation. This indicates that explicit cross-modal alignment is beneficial for audio-visual deception detection. However, its performance remains lower than Atten-Mixer, suggesting that alignment alone is insufficient and that deeper attention-based modality interaction is more effective. In addition, feature-level fusion generally outperforms score-level fusion, indicating the importance of learning deep modality interactions before prediction. Multimodal combinations, such as Visual + Audio, also consistently outperform unimodal inputs, further validating the effectiveness of jointly modeling visual and audio cues for deception detection. Overall, these results demonstrate the effectiveness of Atten-Mixer and underscore the advantage of modality fusion in audio-visual deception detection.

\begin{figure}[t]
\centering
\includegraphics[width=0.9\linewidth]{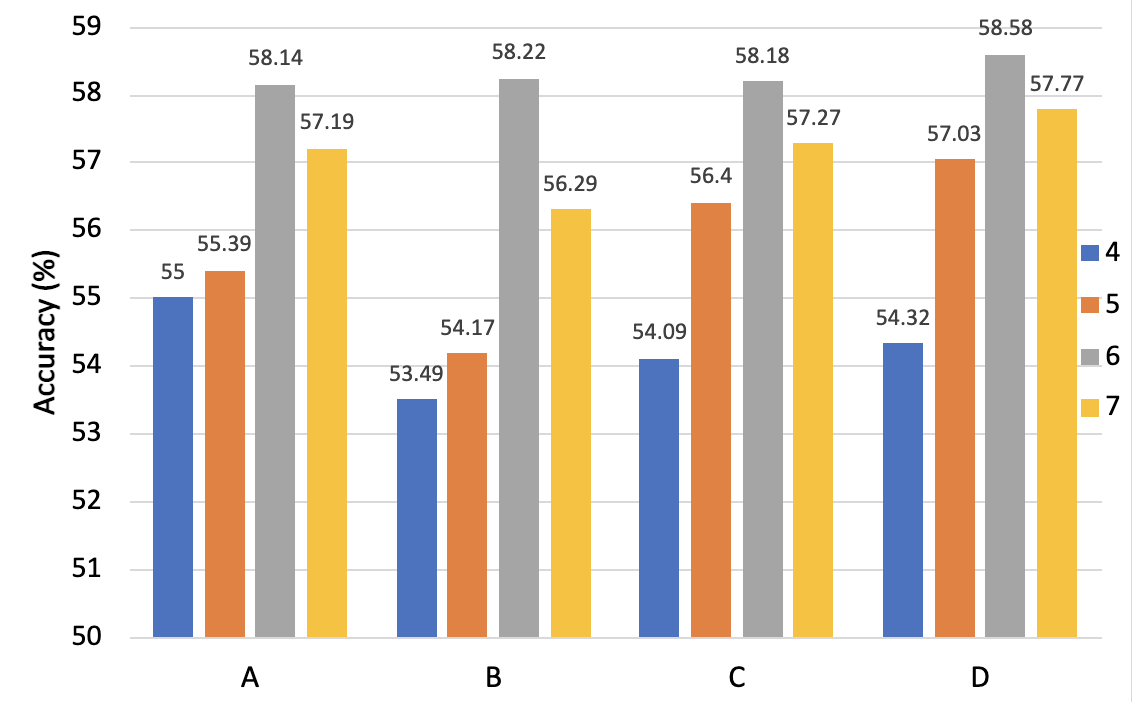}
\vspace{-1.5em}
  \caption{Ablation study for attention-mixer layers. The number of layers 4, 5, 6, and 7 are compared. The modality and inputs for A, B, C, and D are in line with those in Table~\ref{tab:single2singlefusion} from the top to the bottom.
  }
\label{fig:ablation}
\vspace{-1.5em}
\end{figure}

\noindent\textbf{Ablation Study for Attention-Mixer Fusion Module.}\quad
We conducted an ablation study for the proposed attention-mixer fusion module with the changes in the number of attention-mixer layers. The experiments were conducted on single-to-single domain testing, where the average accuracies were compared. As shown in Fig.~\ref{fig:ablation}, the number of attention-mixer layers was set to 4, 5, 6, and 7. The modalities and inputs in Table~\ref{tab:single2singlefusion} are compared, where ``A" had the inputs of Visual (Face frame) + Visual(AU+Gaze+Affect), ``B" had Visual (Face frames) + Audio(Mel spectrogram),  ``C" had Visual (AU+Gaze+Affect) + Audio (Mel spectrogram), and ``D" had Visual (Face frame) + Visual (AU+Gaze+Affect) + Audio (Mel spectrogram). The results showed that models with 6 attention-mixer layers achieved the best average accuracies, followed by 7 attention-mixer layers. The results showed that deeper fusion networks (such as 6 layers vs. 4 or 5 layers) improve generalization by enhancing cross-modal interactions, but overly deep architectures (e.g., 7 layers) may bring diminishing returns. Therefore, it is important to carefully select the fusion depth to balance representation capacity and the risk of overfitting.

\begin{table*}[!t]
\centering
\caption{The fusion results of multi-to-single cross-domain generalization accuracy (\%) for multimodal inter-domain gradient matching (MM-IDGM). The numbers in the brackets show the improved accuracies compared with Atten-Mixer.}
\resizebox{1.0\textwidth}{!}{
\begin{tabular}{|c
>{\columncolor[HTML]{FFFFFF}}c 
>{\columncolor[HTML]{FFFFFF}}c 
>{\columncolor[HTML]{FFFFFF}}c 
>{\columncolor[HTML]{FFFFFF}}c 
>{\columncolor[HTML]{FFFFFF}}c 
>{\columncolor[HTML]{FFFFFF}}c 
>{\columncolor[HTML]{FFFFFF}}c 
>{\columncolor[HTML]{FFFFFF}}c 
>{\columncolor[HTML]{FFFFFF}}c
>{\columncolor[HTML]{FFFFFF}}c 
>{\columncolor[HTML]{FFFFFF}}c 
>{\columncolor[HTML]{FFFFFF}}c|}
\hline
\multicolumn{1}{|c|}{\cellcolor[HTML]{FFFFFF}\textbf{Input}} & \multicolumn{1}{c|}{\cellcolor[HTML]{FFFFFF}\textbf{Model}} & \multicolumn{1}{c|}{\cellcolor[HTML]{FFFFFF}\textbf{R\&M to B1}} & \multicolumn{1}{c|}{\cellcolor[HTML]{FFFFFF}\textbf{R\&M to B2}} & \multicolumn{1}{c|}{\cellcolor[HTML]{FFFFFF}\textbf{R\&B1 to B2}} & \multicolumn{1}{c|}{\cellcolor[HTML]{FFFFFF}\textbf{R\&B1 to M}} & \multicolumn{1}{c|}{\cellcolor[HTML]{FFFFFF}\textbf{B1\&M to R}} & \multicolumn{1}{c|}{\cellcolor[HTML]{FFFFFF}\textbf{B1\&M to B2}} & \multicolumn{1}{c|}{\cellcolor[HTML]{FFFFFF}\textbf{R\&B1\&M to B2}} & \multicolumn{1}{c|}{\cellcolor[HTML]{FFFFFF}\textbf{R\&M to D}}  & \multicolumn{1}{c|}{\cellcolor[HTML]{FFFFFF}\textbf{R\&M to E}} & \multicolumn{1}{c|}{\cellcolor[HTML]{FFFFFF}\textbf{R\&B1 to D}} & \multicolumn{1}{c|}{\cellcolor[HTML]{FFFFFF}\textbf{R\&B1 to E}} \\ \hline

\multicolumn{1}{|c|}{\cellcolor[HTML]{ECF4FF}{\color[HTML]{000000} }}                                                                                                                                            & \multicolumn{1}{c|}{\cellcolor[HTML]{FFFFFF}{\color[HTML]{000000} Atten-Mixer}}             & \multicolumn{1}{c|}{\cellcolor[HTML]{FFFFFF}{\color[HTML]{000000} 53.23}}               & \multicolumn{1}{c|}{\cellcolor[HTML]{FFFFFF}{\color[HTML]{000000} 57.43}}               & \multicolumn{1}{c|}{\cellcolor[HTML]{FFFFFF}{\color[HTML]{000000} \textbf{63.37}}}       & \multicolumn{1}{c|}{\cellcolor[HTML]{FFFFFF}{\color[HTML]{000000} 52.19}}               & \multicolumn{1}{c|}{\cellcolor[HTML]{FFFFFF}{\color[HTML]{000000} 57.01}}               & \multicolumn{1}{c|}{\cellcolor[HTML]{FFFFFF}{\color[HTML]{000000} \textbf{63.37}}}       & \multicolumn{1}{c|}{\cellcolor[HTML]{FFFFFF}{\color[HTML]{000000} \textbf{62.38}}}                   & \multicolumn{1}{c|}{\cellcolor[HTML]{FFFFFF}{\color[HTML]{000000} 53.50}} &\multicolumn{1}{c|}{\cellcolor[HTML]{FFFFFF}{\color[HTML]{000000} 53.40}} & \multicolumn{1}{c|}{\cellcolor[HTML]{FFFFFF}{\color[HTML]{000000} 57.39}}&\multicolumn{1}{c|}{\cellcolor[HTML]{FFFFFF}{\color[HTML]{000000} 51.86}}                         \\ 
\multicolumn{1}{|c|}{\cellcolor[HTML]{ECF4FF}{\color[HTML]{000000} }}                                                                                                                                            & \multicolumn{1}{c|}{\cellcolor[HTML]{FFFFFF}{\color[HTML]{000000} IDGM}}                  & \multicolumn{1}{c|}{\cellcolor[HTML]{FFFFFF}{\color[HTML]{000000} 55.22}}               & \multicolumn{1}{c|}{\cellcolor[HTML]{FFFFFF}{\color[HTML]{000000} 57.45}}               & \multicolumn{1}{c|}{\cellcolor[HTML]{FFFFFF}{\color[HTML]{000000} 59.85}}                & \multicolumn{1}{c|}{\cellcolor[HTML]{FFFFFF}{\color[HTML]{000000} 51.41}}               & \multicolumn{1}{c|}{\cellcolor[HTML]{FFFFFF}{\color[HTML]{000000} \underline{61.34}}}               & \multicolumn{1}{c|}{\cellcolor[HTML]{FFFFFF}{\color[HTML]{000000} 59.41}}                & \multicolumn{1}{c|}{\cellcolor[HTML]{FFFFFF}{\color[HTML]{000000} 60.61}}                    & \multicolumn{1}{c|}{\cellcolor[HTML]{FFFFFF}{\color[HTML]{000000} 52.70}} &\multicolumn{1}{c|}{\cellcolor[HTML]{FFFFFF}{\color[HTML]{000000} 52.99}} & \multicolumn{1}{c|}{\cellcolor[HTML]{FFFFFF}{\color[HTML]{000000} 57.05}}&\multicolumn{1}{c|}{\cellcolor[HTML]{FFFFFF}{\color[HTML]{000000} 51.22}}                                      \\  
\multicolumn{1}{|c|}{\multirow{-3}{*}{\cellcolor[HTML]{ECF4FF}{\color[HTML]{000000} \begin{tabular}[c]{@{}c@{}}V-Face\\ +\\ V-AGA\end{tabular}}}}                       & \multicolumn{1}{c|}{\cellcolor[HTML]{FFFFFF}{\color[HTML]{000000} \textbf{MM-IDGM}}} & \multicolumn{1}{c|}{\cellcolor[HTML]{FFFFFF}{\color[HTML]{000000} \textbf{57.85}}}               & \multicolumn{1}{c|}{\cellcolor[HTML]{FFFFFF}{\color[HTML]{000000} \textbf{58.42}}}               & \multicolumn{1}{c|}{\cellcolor[HTML]{FFFFFF}{\color[HTML]{000000} 60.40}}       & \multicolumn{1}{c|}{\cellcolor[HTML]{FFFFFF}{\color[HTML]{000000} \textbf{52.50}}}               & \multicolumn{1}{c|}{\cellcolor[HTML]{FFFFFF}{\color[HTML]{000000} \textbf{65.22}}}               & \multicolumn{1}{c|}{\cellcolor[HTML]{FFFFFF}{\color[HTML]{000000} 62.05}}       & \multicolumn{1}{c|}{\cellcolor[HTML]{FFFFFF}{\color[HTML]{000000} \underline{61.42}}}                     & \multicolumn{1}{c|}{\cellcolor[HTML]{FFFFFF}{\color[HTML]{000000} 53.88}} &\multicolumn{1}{c|}{\cellcolor[HTML]{FFFFFF}{\color[HTML]{000000} 54.02}} & \multicolumn{1}{c|}{\cellcolor[HTML]{FFFFFF}{\color[HTML]{000000} 58.51}}&\multicolumn{1}{c|}{\cellcolor[HTML]{FFFFFF}{\color[HTML]{000000} 52.87}}                         \\ \hline

\multicolumn{1}{|c|}{\cellcolor[HTML]{FFFFEB}{\color[HTML]{000000} }}                                & \multicolumn{1}{c|}{\cellcolor[HTML]{FFFFFF}{\color[HTML]{000000} Atten-Mixer}}                &  \multicolumn{1}{c|}{\cellcolor[HTML]{FFFFFF}{\color[HTML]{000000} 55.69}}      & \multicolumn{1}{c|}{\cellcolor[HTML]{FFFFFF}{\color[HTML]{000000} \textbf{64.36}}}      & \multicolumn{1}{c|}{\cellcolor[HTML]{FFFFFF}{\color[HTML]{000000} 56.44}}                & \multicolumn{1}{c|}{\cellcolor[HTML]{FFFFFF}{\color[HTML]{000000} \textbf{53.75}}}               & \multicolumn{1}{c|}{\cellcolor[HTML]{FFFFFF}{\color[HTML]{000000} 58.88}}               & \multicolumn{1}{c|}{\cellcolor[HTML]{FFFFFF}{\color[HTML]{000000} 59.41}}                & \multicolumn{1}{c|}{\cellcolor[HTML]{FFFFFF}{\color[HTML]{000000} 58.42}}                    & \multicolumn{1}{c|}{\cellcolor[HTML]{FFFFFF}{\color[HTML]{000000} 53.97}} &\multicolumn{1}{c|}{\cellcolor[HTML]{FFFFFF}{\color[HTML]{000000} 53.43}} & \multicolumn{1}{c|}{\cellcolor[HTML]{FFFFFF}{\color[HTML]{000000} 57.82}}&\multicolumn{1}{c|}{\cellcolor[HTML]{FFFFFF}{\color[HTML]{000000} 52.35}}                          \\ 

\multicolumn{1}{|c|}{\cellcolor[HTML]{FFFFEB}{\color[HTML]{000000} }}                       & \multicolumn{1}{c|}{\cellcolor[HTML]{FFFFFF}{\color[HTML]{000000} IDGM}}                 & \multicolumn{1}{c|}{\cellcolor[HTML]{FFFFFF}{\color[HTML]{000000} 54.43}}               & \multicolumn{1}{c|}{\cellcolor[HTML]{FFFFFF}{\color[HTML]{000000} \underline{60.40}}}               & \multicolumn{1}{c|}{\cellcolor[HTML]{FFFFFF}{\color[HTML]{000000} 61.07}}                & \multicolumn{1}{c|}{\cellcolor[HTML]{FFFFFF}{\color[HTML]{000000} 51.40}}               & \multicolumn{1}{c|}{\cellcolor[HTML]{FFFFFF}{\color[HTML]{000000} \textbf{59.21}}}               & \multicolumn{1}{c|}{\cellcolor[HTML]{FFFFFF}{\color[HTML]{000000} 59.88}}                & \multicolumn{1}{c|}{\cellcolor[HTML]{FFFFFF}{\color[HTML]{000000} \textbf{59.75}}}                    & \multicolumn{1}{c|}{\cellcolor[HTML]{FFFFFF}{\color[HTML]{000000} 53.02}} &\multicolumn{1}{c|}{\cellcolor[HTML]{FFFFFF}{\color[HTML]{000000} 53.11}} & \multicolumn{1}{c|}{\cellcolor[HTML]{FFFFFF}{\color[HTML]{000000} 56.24}}&\multicolumn{1}{c|}{\cellcolor[HTML]{FFFFFF}{\color[HTML]{000000} 52.36}}                        \\ 

\multicolumn{1}{|c|}{\multirow{-3}{*}{\cellcolor[HTML]{FFFFEB}{\color[HTML]{000000} \begin{tabular}[c]{@{}c@{}}V-Face\\ +\\ A-Mel\end{tabular}}}}                        & \multicolumn{1}{c|}{\cellcolor[HTML]{FFFFFF}{\color[HTML]{000000} \textbf{MM-IDGM}}} & \multicolumn{1}{c|}{\cellcolor[HTML]{FFFFFF}{\color[HTML]{000000} 53.54}}               & \multicolumn{1}{c|}{\cellcolor[HTML]{FFFFFF}{\color[HTML]{000000} 59.41}}               & \multicolumn{1}{c|}{\cellcolor[HTML]{FFFFFF}{\color[HTML]{000000} \underline{62.38}}}                & \multicolumn{1}{c|}{\cellcolor[HTML]{FFFFFF}{\color[HTML]{000000} 51.56}}               & \multicolumn{1}{c|}{\cellcolor[HTML]{FFFFFF}{\color[HTML]{000000} 59.07}}               & \multicolumn{1}{c|}{\cellcolor[HTML]{FFFFFF}{\color[HTML]{000000} \textbf{62.38}}}                & \multicolumn{1}{c|}{\cellcolor[HTML]{FFFFFF}{\color[HTML]{000000} 59.50}}                    & \multicolumn{1}{c|}{\cellcolor[HTML]{FFFFFF}{\color[HTML]{000000} 54.26}} &\multicolumn{1}{c|}{\cellcolor[HTML]{FFFFFF}{\color[HTML]{000000} 54.29}} & \multicolumn{1}{c|}{\cellcolor[HTML]{FFFFFF}{\color[HTML]{000000} 58.33}}&\multicolumn{1}{c|}{\cellcolor[HTML]{FFFFFF}{\color[HTML]{000000} \textbf{53.46}}}                        \\ \hline

\multicolumn{1}{|c|}{\cellcolor[HTML]{FFF7F7}{\color[HTML]{000000} }}                  & \multicolumn{1}{c|}{\cellcolor[HTML]{FFFFFF}{\color[HTML]{000000} Atten-Mixer}}               & \multicolumn{1}{c|}{\cellcolor[HTML]{FFFFFF}{\color[HTML]{000000} 55.08}}               & \multicolumn{1}{c|}{\cellcolor[HTML]{FFFFFF}{\color[HTML]{000000} 55.45}}               & \multicolumn{1}{c|}{\cellcolor[HTML]{FFFFFF}{\color[HTML]{000000} 58.42}}                & \multicolumn{1}{c|}{\cellcolor[HTML]{FFFFFF}{\color[HTML]{000000} \underline{54.37}}}      & \multicolumn{1}{c|}{\cellcolor[HTML]{FFFFFF}{\color[HTML]{000000} 52.34}}               & \multicolumn{1}{c|}{\cellcolor[HTML]{FFFFFF}{\color[HTML]{000000} 61.39}}                & \multicolumn{1}{c|}{\cellcolor[HTML]{FFFFFF}{\color[HTML]{000000} 60.40}}                    & \multicolumn{1}{c|}{\cellcolor[HTML]{FFFFFF}{\color[HTML]{000000} 57.39}} &\multicolumn{1}{c|}{\cellcolor[HTML]{FFFFFF}{\color[HTML]{000000} 54.20}} & \multicolumn{1}{c|}{\cellcolor[HTML]{FFFFFF}{\color[HTML]{000000} 58.02}}&\multicolumn{1}{c|}{\cellcolor[HTML]{FFFFFF}{\color[HTML]{000000} 52.83}}                           \\

\multicolumn{1}{|c|}{\cellcolor[HTML]{FFF7F7}{\color[HTML]{000000} }}                   & \multicolumn{1}{c|}{\cellcolor[HTML]{FFFFFF}{\color[HTML]{000000} IDGM}}                 & \multicolumn{1}{c|}{\cellcolor[HTML]{FFFFFF}{\color[HTML]{000000} 55.49}}               & \multicolumn{1}{c|}{\cellcolor[HTML]{FFFFFF}{\color[HTML]{000000} 54.92}}               & \multicolumn{1}{c|}{\cellcolor[HTML]{FFFFFF}{\color[HTML]{000000} 58.36}}                & \multicolumn{1}{c|}{\cellcolor[HTML]{FFFFFF}{\color[HTML]{000000} 50.31}}               & \multicolumn{1}{c|}{\cellcolor[HTML]{FFFFFF}{\color[HTML]{000000} 53.15}}               & \multicolumn{1}{c|}{\cellcolor[HTML]{FFFFFF}{\color[HTML]{000000} 60.35}}                & \multicolumn{1}{c|}{\cellcolor[HTML]{FFFFFF}{\color[HTML]{000000} 59.21}}                    & \multicolumn{1}{c|}{\cellcolor[HTML]{FFFFFF}{\color[HTML]{000000} 56.87}} &\multicolumn{1}{c|}{\cellcolor[HTML]{FFFFFF}{\color[HTML]{000000} 53.40}} & \multicolumn{1}{c|}{\cellcolor[HTML]{FFFFFF}{\color[HTML]{000000} 57.54}}&\multicolumn{1}{c|}{\cellcolor[HTML]{FFFFFF}{\color[HTML]{000000} 52.33}}                            \\ 

\multicolumn{1}{|c|}{\multirow{-3}{*}{\cellcolor[HTML]{FFF7F7}{\color[HTML]{000000} \begin{tabular}[c]{@{}c@{}}V-AGA\\ +\\ A-Mel\end{tabular}}}}                     & \multicolumn{1}{c|}{\cellcolor[HTML]{FFFFFF}{\color[HTML]{000000} \textbf{MM-IDGM}}}    & \multicolumn{1}{c|}{\cellcolor[HTML]{FFFFFF}{\color[HTML]{000000} \underline{56.05}}}               & \multicolumn{1}{c|}{\cellcolor[HTML]{FFFFFF}{\color[HTML]{000000} 56.03}}               & \multicolumn{1}{c|}{\cellcolor[HTML]{FFFFFF}{\color[HTML]{000000} 59.37}}                & \multicolumn{1}{c|}{\cellcolor[HTML]{FFFFFF}{\color[HTML]{000000} 52.41}}               & \multicolumn{1}{c|}{\cellcolor[HTML]{FFFFFF}{\color[HTML]{000000} 53.40}}               & \multicolumn{1}{c|}{\cellcolor[HTML]{FFFFFF}{\color[HTML]{000000} 62.42}}                & \multicolumn{1}{c|}{\cellcolor[HTML]{FFFFFF}{\color[HTML]{000000} 60.98}}                   & \multicolumn{1}{c|}{\cellcolor[HTML]{FFFFFF}{\color[HTML]{000000} \underline{58.24}}} &\multicolumn{1}{c|}{\cellcolor[HTML]{FFFFFF}{\color[HTML]{000000} \underline{55.06}}} & \multicolumn{1}{c|}{\cellcolor[HTML]{FFFFFF}{\color[HTML]{000000} 58.72}}&\multicolumn{1}{c|}{\cellcolor[HTML]{FFFFFF}{\color[HTML]{000000} 51.24}}                            \\ \hline

\multicolumn{1}{|c|}{\cellcolor[HTML]{EEFBED}{\color[HTML]{000000} }}                & \multicolumn{1}{c|}{\cellcolor[HTML]{FFFFFF}{\color[HTML]{000000} Atten-Mixer}}               & \multicolumn{1}{c|}{\cellcolor[HTML]{FFFFFF}{\color[HTML]{000000} 52.31}}               & \multicolumn{1}{c|}{\cellcolor[HTML]{FFFFFF}{\color[HTML]{000000} 59.41}}              & \multicolumn{1}{c|}{\cellcolor[HTML]{FFFFFF}{\color[HTML]{000000} 58.42}}                & \multicolumn{1}{c|}{\cellcolor[HTML]{FFFFFF}{\color[HTML]{000000} 52.50}}               & \multicolumn{1}{c|}{\cellcolor[HTML]{FFFFFF}{\color[HTML]{000000} 58.88}}               & \multicolumn{1}{c|}{\cellcolor[HTML]{FFFFFF}{\color[HTML]{000000} 61.39}}                & \multicolumn{1}{c|}{\cellcolor[HTML]{FFFFFF}{\color[HTML]{000000} 60.40}}                    & \multicolumn{1}{c|}{\cellcolor[HTML]{FFFFFF}{\color[HTML]{000000} 57.30}} &\multicolumn{1}{c|}{\cellcolor[HTML]{FFFFFF}{\color[HTML]{000000} 54.57}} & \multicolumn{1}{c|}{\cellcolor[HTML]{FFFFFF}{\color[HTML]{000000} 59.41}}&\multicolumn{1}{c|}{\cellcolor[HTML]{FFFFFF}{\color[HTML]{000000} 52.84}}                             \\ 
\multicolumn{1}{|c|}{\cellcolor[HTML]{EEFBED}{\color[HTML]{000000} }}                  & \multicolumn{1}{c|}{\cellcolor[HTML]{FFFFFF}{\color[HTML]{000000} IDGM}}                   & \multicolumn{1}{c|}{\cellcolor[HTML]{FFFFFF}{\color[HTML]{000000} 52.45}}               & \multicolumn{1}{c|}{\cellcolor[HTML]{FFFFFF}{\color[HTML]{000000} 57.02}}               & \multicolumn{1}{c|}{\cellcolor[HTML]{FFFFFF}{\color[HTML]{000000} 60.42}}                & \multicolumn{1}{c|}{\cellcolor[HTML]{FFFFFF}{\color[HTML]{000000} 54.25}}               & \multicolumn{1}{c|}{\cellcolor[HTML]{FFFFFF}{\color[HTML]{000000} 58.42}}               & \multicolumn{1}{c|}{\cellcolor[HTML]{FFFFFF}{\color[HTML]{000000} 61.79}}                & \multicolumn{1}{c|}{\cellcolor[HTML]{FFFFFF}{\color[HTML]{000000} 60.03}}                     & \multicolumn{1}{c|}{\cellcolor[HTML]{FFFFFF}{\color[HTML]{000000} 58.02}} &\multicolumn{1}{c|}{\cellcolor[HTML]{FFFFFF}{\color[HTML]{000000} 54.21}} & \multicolumn{1}{c|}{\cellcolor[HTML]{FFFFFF}{\color[HTML]{000000} \underline{59.45}}}&\multicolumn{1}{c|}{\cellcolor[HTML]{FFFFFF}{\color[HTML]{000000} \underline{53.14}}}                                 \\

\multicolumn{1}{|c|}{\multirow{-3}{*}{\cellcolor[HTML]{EEFBED}{\color[HTML]{000000} \begin{tabular}[c]{@{}c@{}}V-Face+\\ V-AGA+\\ A-Mel\end{tabular}}}} & \multicolumn{1}{c|}{\cellcolor[HTML]{FFFFFF}{\color[HTML]{000000} \textbf{MM-IDGM}}}    & \multicolumn{1}{c|}{\cellcolor[HTML]{FFFFFF}{\color[HTML]{000000} 52.77}}             & \multicolumn{1}{c|}{\cellcolor[HTML]{FFFFFF}{\color[HTML]{000000} 58.43}}               & \multicolumn{1}{c|}{\cellcolor[HTML]{FFFFFF}{\color[HTML]{000000} 61.39}}                & \multicolumn{1}{c|}{\cellcolor[HTML]{FFFFFF}{\color[HTML]{000000} \textbf{55.57}}}               & \multicolumn{1}{c|}{\cellcolor[HTML]{FFFFFF}{\color[HTML]{000000} 59.46}}               & \multicolumn{1}{c|}{\cellcolor[HTML]{FFFFFF}{\color[HTML]{000000} \underline{62.58}}}                & \multicolumn{1}{c|}{\cellcolor[HTML]{FFFFFF}{\color[HTML]{000000} \textbf{60.60}}}                    & \multicolumn{1}{c|}{\cellcolor[HTML]{FFFFFF}{\color[HTML]{000000} \textbf{58.75}}} &\multicolumn{1}{c|}{\cellcolor[HTML]{FFFFFF}{\color[HTML]{000000} \textbf{55.87}}} & \multicolumn{1}{c|}{\cellcolor[HTML]{FFFFFF}{\color[HTML]{000000} \textbf{60.85}}}&\multicolumn{1}{c|}{\cellcolor[HTML]{FFFFFF}{\color[HTML]{000000} 53.01}}                               \\ \hline

\multicolumn{1}{|c|}{\cellcolor[HTML]{FFFFFF}\textbf{Input}} & \multicolumn{1}{c|}{\cellcolor[HTML]{FFFFFF}\textbf{Model}} & \multicolumn{1}{c|}{\cellcolor[HTML]{FFFFFF}\textbf{D\&E to B1}} & \multicolumn{1}{c|}{\cellcolor[HTML]{FFFFFF}\textbf{D\&E to B2}} & \multicolumn{1}{c|}{\cellcolor[HTML]{FFFFFF}\textbf{D\&E to M}} & \multicolumn{1}{c|}{\cellcolor[HTML]{FFFFFF}\textbf{D\&E to R}} & \multicolumn{1}{c|}{\cellcolor[HTML]{FFFFFF}\textbf{\shortstack{R\&B1\&M\\to D}}} & \multicolumn{1}{c|}{\cellcolor[HTML]{FFFFFF}\textbf{\shortstack{R\&B1\&M\\to E}}} & \multicolumn{1}{c|}{\cellcolor[HTML]{FFFFFF}\textbf{\shortstack{R\&D\&E\\to B1}}} & \multicolumn{1}{c|}{\cellcolor[HTML]{FFFFFF}\textbf{\shortstack{R\&D\&E\\to M}}}  & \multicolumn{1}{c|}{\cellcolor[HTML]{FFFFFF}\textbf{\shortstack{R\&M\&D\&E\\to B2}}} & \multicolumn{1}{c|}{\cellcolor[HTML]{FFFFFF}\textbf{\shortstack{R\&B1\&M\&\\D\&E to B2}}} & \multicolumn{1}{c|}{\cellcolor[HTML]{FFFFFF}\textbf{Avg}} \\ \hline
\multicolumn{1}{|c|}{\cellcolor[HTML]{ECF4FF}{\color[HTML]{000000} }}                                                                                                                                            & \multicolumn{1}{c|}{\cellcolor[HTML]{FFFFFF}{\color[HTML]{000000} Atten-Mixer}}             & \multicolumn{1}{c|}{\cellcolor[HTML]{FFFFFF}{\color[HTML]{000000} 53.47}}               & \multicolumn{1}{c|}{\cellcolor[HTML]{FFFFFF}{\color[HTML]{000000} 59.01}}               & \multicolumn{1}{c|}{\cellcolor[HTML]{FFFFFF}{\color[HTML]{000000} 53.62}}       & \multicolumn{1}{c|}{\cellcolor[HTML]{FFFFFF}{\color[HTML]{000000} 54.21}}               & \multicolumn{1}{c|}{\cellcolor[HTML]{FFFFFF}{\color[HTML]{000000} 58.43}}               & \multicolumn{1}{c|}{\cellcolor[HTML]{FFFFFF}{\color[HTML]{000000} 57.10}}       & \multicolumn{1}{c|}{\cellcolor[HTML]{FFFFFF}{\color[HTML]{000000} 57.54}}                   & \multicolumn{1}{c|}{\cellcolor[HTML]{FFFFFF}{\color[HTML]{000000} 53.75}} &\multicolumn{1}{c|}{\cellcolor[HTML]{FFFFFF}{\color[HTML]{000000} 60.40}} & \multicolumn{1}{c|}{\cellcolor[HTML]{FFFFFF}{\color[HTML]{000000} 60.99}}&\multicolumn{1}{c|}{\cellcolor[HTML]{FFFFFF}{\color[HTML]{000000} 56.84}}                         \\ 

\multicolumn{1}{|c|}{\cellcolor[HTML]{ECF4FF}{\color[HTML]{000000} }}  & \multicolumn{1}{c|}{\cellcolor[HTML]{FFFFFF}{\color[HTML]{000000} IDGM}}                  & \multicolumn{1}{c|}{\cellcolor[HTML]{FFFFFF}{\color[HTML]{000000} 54.12}}               & \multicolumn{1}{c|}{\cellcolor[HTML]{FFFFFF}{\color[HTML]{000000} 58.04}}               & \multicolumn{1}{c|}{\cellcolor[HTML]{FFFFFF}{\color[HTML]{000000} 54.36}}                & \multicolumn{1}{c|}{\cellcolor[HTML]{FFFFFF}{\color[HTML]{000000} 58.92}}               & \multicolumn{1}{c|}{\cellcolor[HTML]{FFFFFF}{\color[HTML]{000000} 59.03}}               & \multicolumn{1}{c|}{\cellcolor[HTML]{FFFFFF}{\color[HTML]{000000} 57.84}}                & \multicolumn{1}{c|}{\cellcolor[HTML]{FFFFFF}{\color[HTML]{000000} 57.93}}                    & \multicolumn{1}{c|}{\cellcolor[HTML]{FFFFFF}{\color[HTML]{000000} 54.46}} &\multicolumn{1}{c|}{\cellcolor[HTML]{FFFFFF}{\color[HTML]{000000} 60.24}} & \multicolumn{1}{c|}{\cellcolor[HTML]{FFFFFF}{\color[HTML]{000000} 60.54}}&\multicolumn{1}{c|}{\cellcolor[HTML]{FFFFFF}{\color[HTML]{000000} 56.89}}                                      \\  
\multicolumn{1}{|c|}{\multirow{-3}{*}{\cellcolor[HTML]{ECF4FF}{\color[HTML]{000000} \begin{tabular}[c]{@{}c@{}}V-Face\\ +\\ V-AGA\end{tabular}}}}                       & \multicolumn{1}{c|}{\cellcolor[HTML]{FFFFFF}{\color[HTML]{000000} \textbf{MM-IDGM}}} & \multicolumn{1}{c|}{\cellcolor[HTML]{FFFFFF}{\color[HTML]{000000} 54.88}}               & \multicolumn{1}{c|}{\cellcolor[HTML]{FFFFFF}{\color[HTML]{000000} 59.26}}               & \multicolumn{1}{c|}{\cellcolor[HTML]{FFFFFF}{\color[HTML]{000000} 54.38}}       & \multicolumn{1}{c|}{\cellcolor[HTML]{FFFFFF}{\color[HTML]{000000} \textbf{59.34}}}               & \multicolumn{1}{c|}{\cellcolor[HTML]{FFFFFF}{\color[HTML]{000000} 58.28}}               & \multicolumn{1}{c|}{\cellcolor[HTML]{FFFFFF}{\color[HTML]{000000} 58.28}}       & \multicolumn{1}{c|}{\cellcolor[HTML]{FFFFFF}{\color[HTML]{000000} 58.66}}                     & \multicolumn{1}{c|}{\cellcolor[HTML]{FFFFFF}{\color[HTML]{000000} 55.25}} &\multicolumn{1}{c|}{\cellcolor[HTML]{FFFFFF}{\color[HTML]{000000} 60.34}} & \multicolumn{1}{c|}{\cellcolor[HTML]{FFFFFF}{\color[HTML]{000000} 60.58}}&\multicolumn{1}{c|}{\cellcolor[HTML]{FFFFFF}{\color[HTML]{000000} \textbf{57.92 (+1.08)}}}                         \\ \hline

\multicolumn{1}{|c|}{\cellcolor[HTML]{FFFFEB}{\color[HTML]{000000} }}                                & \multicolumn{1}{c|}{\cellcolor[HTML]{FFFFFF}{\color[HTML]{000000} Atten-Mixer}}                &  \multicolumn{1}{c|}{\cellcolor[HTML]{FFFFFF}{\color[HTML]{000000} 54.24 }}      & \multicolumn{1}{c|}{\cellcolor[HTML]{FFFFFF}{\color[HTML]{000000} 59.41}}      & \multicolumn{1}{c|}{\cellcolor[HTML]{FFFFFF}{\color[HTML]{000000} 53.50}}                & \multicolumn{1}{c|}{\cellcolor[HTML]{FFFFFF}{\color[HTML]{000000} 55.14}}               & \multicolumn{1}{c|}{\cellcolor[HTML]{FFFFFF}{\color[HTML]{000000} 58.53}}               & \multicolumn{1}{c|}{\cellcolor[HTML]{FFFFFF}{\color[HTML]{000000} 58.65}}                & \multicolumn{1}{c|}{\cellcolor[HTML]{FFFFFF}{\color[HTML]{000000} 58.92}}                    & \multicolumn{1}{c|}{\cellcolor[HTML]{FFFFFF}{\color[HTML]{000000} 54.88}} &\multicolumn{1}{c|}{\cellcolor[HTML]{FFFFFF}{\color[HTML]{000000} 62.38}} & \multicolumn{1}{c|}{\cellcolor[HTML]{FFFFFF}{\color[HTML]{000000} 59.35}}&\multicolumn{1}{c|}{\cellcolor[HTML]{FFFFFF}{\color[HTML]{000000} 57.12}}                          \\ 

\multicolumn{1}{|c|}{\cellcolor[HTML]{FFFFEB}{\color[HTML]{000000} }}                       & \multicolumn{1}{c|}{\cellcolor[HTML]{FFFFFF}{\color[HTML]{000000} IDGM}}                 & \multicolumn{1}{c|}{\cellcolor[HTML]{FFFFFF}{\color[HTML]{000000} 53.29}}               & \multicolumn{1}{c|}{\cellcolor[HTML]{FFFFFF}{\color[HTML]{000000} 59.88}}               & \multicolumn{1}{c|}{\cellcolor[HTML]{FFFFFF}{\color[HTML]{000000} 54.01}}                & \multicolumn{1}{c|}{\cellcolor[HTML]{FFFFFF}{\color[HTML]{000000} 55.39}}               & \multicolumn{1}{c|}{\cellcolor[HTML]{FFFFFF}{\color[HTML]{000000} 58.98}}               & \multicolumn{1}{c|}{\cellcolor[HTML]{FFFFFF}{\color[HTML]{000000} \underline{59.35}}}                & \multicolumn{1}{c|}{\cellcolor[HTML]{FFFFFF}{\color[HTML]{000000} 58.43}}                    & \multicolumn{1}{c|}{\cellcolor[HTML]{FFFFFF}{\color[HTML]{000000} 55.34}} &\multicolumn{1}{c|}{\cellcolor[HTML]{FFFFFF}{\color[HTML]{000000} 62.57}} & \multicolumn{1}{c|}{\cellcolor[HTML]{FFFFFF}{\color[HTML]{000000} 60.42}}&\multicolumn{1}{c|}{\cellcolor[HTML]{FFFFFF}{\color[HTML]{000000} 57.07}}                        \\ 
\multicolumn{1}{|c|}{\multirow{-3}{*}{\cellcolor[HTML]{FFFFEB}{\color[HTML]{000000} \begin{tabular}[c]{@{}c@{}}V-Face\\ +\\ A-Mel\end{tabular}}}}                        & \multicolumn{1}{c|}{\cellcolor[HTML]{FFFFFF}{\color[HTML]{000000} \textbf{MM-IDGM}}} & \multicolumn{1}{c|}{\cellcolor[HTML]{FFFFFF}{\color[HTML]{000000} 53.56}}               & \multicolumn{1}{c|}{\cellcolor[HTML]{FFFFFF}{\color[HTML]{000000} 59.81}}               & \multicolumn{1}{c|}{\cellcolor[HTML]{FFFFFF}{\color[HTML]{000000} \textbf{55.36}}}                & \multicolumn{1}{c|}{\cellcolor[HTML]{FFFFFF}{\color[HTML]{000000} 56.03}}               & \multicolumn{1}{c|}{\cellcolor[HTML]{FFFFFF}{\color[HTML]{000000} 59.21}}               & \multicolumn{1}{c|}{\cellcolor[HTML]{FFFFFF}{\color[HTML]{000000} 59.32}}                & \multicolumn{1}{c|}{\cellcolor[HTML]{FFFFFF}{\color[HTML]{000000} 59.42}}                    & \multicolumn{1}{c|}{\cellcolor[HTML]{FFFFFF}{\color[HTML]{000000} 55.23}} &\multicolumn{1}{c|}{\cellcolor[HTML]{FFFFFF}{\color[HTML]{000000} 61.92}} & \multicolumn{1}{c|}{\cellcolor[HTML]{FFFFFF}{\color[HTML]{000000} \textbf{62.35}}}&\multicolumn{1}{c|}{\cellcolor[HTML]{FFFFFF}{\color[HTML]{000000} \textbf{57.64 (+0.52)}}}                        \\ \hline

\multicolumn{1}{|c|}{\cellcolor[HTML]{FFF7F7}{\color[HTML]{000000} }}                  & \multicolumn{1}{c|}{\cellcolor[HTML]{FFFFFF}{\color[HTML]{000000} Atten-Mixer}}               & \multicolumn{1}{c|}{\cellcolor[HTML]{FFFFFF}{\color[HTML]{000000} 55.11}}               & \multicolumn{1}{c|}{\cellcolor[HTML]{FFFFFF}{\color[HTML]{000000} 62.38}}               & \multicolumn{1}{c|}{\cellcolor[HTML]{FFFFFF}{\color[HTML]{000000} 54.02}}                & \multicolumn{1}{c|}{\cellcolor[HTML]{FFFFFF}{\color[HTML]{000000} 56.28}}      & \multicolumn{1}{c|}{\cellcolor[HTML]{FFFFFF}{\color[HTML]{000000} 58.48}}               & \multicolumn{1}{c|}{\cellcolor[HTML]{FFFFFF}{\color[HTML]{000000} 57.65}}                & \multicolumn{1}{c|}{\cellcolor[HTML]{FFFFFF}{\color[HTML]{000000} 60.19}}                    & \multicolumn{1}{c|}{\cellcolor[HTML]{FFFFFF}{\color[HTML]{000000} 54.64}} &\multicolumn{1}{c|}{\cellcolor[HTML]{FFFFFF}{\color[HTML]{000000} 60.50}} & \multicolumn{1}{c|}{\cellcolor[HTML]{FFFFFF}{\color[HTML]{000000} 60.13}}&\multicolumn{1}{c|}{\cellcolor[HTML]{FFFFFF}{\color[HTML]{000000} 57.11}}                           \\

\multicolumn{1}{|c|}{\cellcolor[HTML]{FFF7F7}{\color[HTML]{000000} }}                   & \multicolumn{1}{c|}{\cellcolor[HTML]{FFFFFF}{\color[HTML]{000000} IDGM}}                 & \multicolumn{1}{c|}{\cellcolor[HTML]{FFFFFF}{\color[HTML]{000000} 55.87}}               & \multicolumn{1}{c|}{\cellcolor[HTML]{FFFFFF}{\color[HTML]{000000} 61.24}}               & \multicolumn{1}{c|}{\cellcolor[HTML]{FFFFFF}{\color[HTML]{000000} 55.02}}                & \multicolumn{1}{c|}{\cellcolor[HTML]{FFFFFF}{\color[HTML]{000000} 56.21}}               & \multicolumn{1}{c|}{\cellcolor[HTML]{FFFFFF}{\color[HTML]{000000} 57.03}}               & \multicolumn{1}{c|}{\cellcolor[HTML]{FFFFFF}{\color[HTML]{000000} 57.95}}                & \multicolumn{1}{c|}{\cellcolor[HTML]{FFFFFF}{\color[HTML]{000000} 60.25}}                    & \multicolumn{1}{c|}{\cellcolor[HTML]{FFFFFF}{\color[HTML]{000000} 55.32}} &\multicolumn{1}{c|}{\cellcolor[HTML]{FFFFFF}{\color[HTML]{000000} 60.61}} & \multicolumn{1}{c|}{\cellcolor[HTML]{FFFFFF}{\color[HTML]{000000} 59.85}}&\multicolumn{1}{c|}{\cellcolor[HTML]{FFFFFF}{\color[HTML]{000000} 56.73}}                            \\ 

\multicolumn{1}{|c|}{\multirow{-3}{*}{\cellcolor[HTML]{FFF7F7}{\color[HTML]{000000} \begin{tabular}[c]{@{}c@{}}V-AGA\\ +\\ A-Mel\end{tabular}}}}                     & \multicolumn{1}{c|}{\cellcolor[HTML]{FFFFFF}{\color[HTML]{000000} \textbf{MM-IDGM}}}    & \multicolumn{1}{c|}{\cellcolor[HTML]{FFFFFF}{\color[HTML]{000000} 55.99}}               & \multicolumn{1}{c|}{\cellcolor[HTML]{FFFFFF}{\color[HTML]{000000} 62.68}}               & \multicolumn{1}{c|}{\cellcolor[HTML]{FFFFFF}{\color[HTML]{000000} \textbf{55.36}}}                & \multicolumn{1}{c|}{\cellcolor[HTML]{FFFFFF}{\color[HTML]{000000} 57.02}}               & \multicolumn{1}{c|}{\cellcolor[HTML]{FFFFFF}{\color[HTML]{000000} 59.31}}               & \multicolumn{1}{c|}{\cellcolor[HTML]{FFFFFF}{\color[HTML]{000000} 58.34}}                & \multicolumn{1}{c|}{\cellcolor[HTML]{FFFFFF}{\color[HTML]{000000} \underline{60.58}}}                   & \multicolumn{1}{c|}{\cellcolor[HTML]{FFFFFF}{\color[HTML]{000000} \underline{55.41}}} &\multicolumn{1}{c|}{\cellcolor[HTML]{FFFFFF}{\color[HTML]{000000} 60.89}} & \multicolumn{1}{c|}{\cellcolor[HTML]{FFFFFF}{\color[HTML]{000000} 61.78}}&\multicolumn{1}{c|}{\cellcolor[HTML]{FFFFFF}{\color[HTML]{000000} \textbf{57.68 (+0.57)}}}                            \\ \hline

\multicolumn{1}{|c|}{\cellcolor[HTML]{EEFBED}{\color[HTML]{000000} }}                & \multicolumn{1}{c|}{\cellcolor[HTML]{FFFFFF}{\color[HTML]{000000} Atten-Mixer}}               & \multicolumn{1}{c|}{\cellcolor[HTML]{FFFFFF}{\color[HTML]{000000} \underline{57.48}}}               & \multicolumn{1}{c|}{\cellcolor[HTML]{FFFFFF}{\color[HTML]{000000} \textbf{63.17}}}               & \multicolumn{1}{c|}{\cellcolor[HTML]{FFFFFF}{\color[HTML]{000000} 54.35}}                & \multicolumn{1}{c|}{\cellcolor[HTML]{FFFFFF}{\color[HTML]{000000} 58.88}}               & \multicolumn{1}{c|}{\cellcolor[HTML]{FFFFFF}{\color[HTML]{000000} 59.15}}               & \multicolumn{1}{c|}{\cellcolor[HTML]{FFFFFF}{\color[HTML]{000000} 58.96}}                & \multicolumn{1}{c|}{\cellcolor[HTML]{FFFFFF}{\color[HTML]{000000} 60.42}}                    & \multicolumn{1}{c|}{\cellcolor[HTML]{FFFFFF}{\color[HTML]{000000} 54.56}} &\multicolumn{1}{c|}{\cellcolor[HTML]{FFFFFF}{\color[HTML]{000000} \underline{62.58}}} & \multicolumn{1}{c|}{\cellcolor[HTML]{FFFFFF}{\color[HTML]{000000} 60.42}}&\multicolumn{1}{c|}{\cellcolor[HTML]{FFFFFF}{\color[HTML]{000000} 57.97}}                             \\ 
\multicolumn{1}{|c|}{\cellcolor[HTML]{EEFBED}{\color[HTML]{000000} }}                  & \multicolumn{1}{c|}{\cellcolor[HTML]{FFFFFF}{\color[HTML]{000000} IDGM}}                   & \multicolumn{1}{c|}{\cellcolor[HTML]{FFFFFF}{\color[HTML]{000000} 56.98}}               & \multicolumn{1}{c|}{\cellcolor[HTML]{FFFFFF}{\color[HTML]{000000} 62.11}}               & \multicolumn{1}{c|}{\cellcolor[HTML]{FFFFFF}{\color[HTML]{000000} 54.68}}                & \multicolumn{1}{c|}{\cellcolor[HTML]{FFFFFF}{\color[HTML]{000000} 58.25}}               & \multicolumn{1}{c|}{\cellcolor[HTML]{FFFFFF}{\color[HTML]{000000} \underline{59.64}}}               & \multicolumn{1}{c|}{\cellcolor[HTML]{FFFFFF}{\color[HTML]{000000} 59.31}}                & \multicolumn{1}{c|}{\cellcolor[HTML]{FFFFFF}{\color[HTML]{000000} 60.15}}                     & \multicolumn{1}{c|}{\cellcolor[HTML]{FFFFFF}{\color[HTML]{000000} 54.73}} &\multicolumn{1}{c|}{\cellcolor[HTML]{FFFFFF}{\color[HTML]{000000} 61.99}} & \multicolumn{1}{c|}{\cellcolor[HTML]{FFFFFF}{\color[HTML]{000000} 61.23}}&\multicolumn{1}{c|}{\cellcolor[HTML]{FFFFFF}{\color[HTML]{000000} 58.01}}                                 \\

\multicolumn{1}{|c|}{\multirow{-3}{*}{\cellcolor[HTML]{EEFBED}{\color[HTML]{000000} \begin{tabular}[c]{@{}c@{}}V-Face+\\ V-AGA+\\ A-Mel\end{tabular}}}} & \multicolumn{1}{c|}{\cellcolor[HTML]{FFFFFF}{\color[HTML]{000000} \textbf{MM-IDGM}}}    & \multicolumn{1}{c|}{\cellcolor[HTML]{FFFFFF}{\color[HTML]{000000} \textbf{58.13}}}               & \multicolumn{1}{c|}{\cellcolor[HTML]{FFFFFF}{\color[HTML]{000000} \underline{64.22}}}               & \multicolumn{1}{c|}{\cellcolor[HTML]{FFFFFF}{\color[HTML]{000000} \underline{55.05}}}                & \multicolumn{1}{c|}{\cellcolor[HTML]{FFFFFF}{\color[HTML]{000000} \underline{59.31}}}               & \multicolumn{1}{c|}{\cellcolor[HTML]{FFFFFF}{\color[HTML]{000000} \textbf{60.14}}}               & \multicolumn{1}{c|}{\cellcolor[HTML]{FFFFFF}{\color[HTML]{000000} \textbf{59.46}}}                & \multicolumn{1}{c|}{\cellcolor[HTML]{FFFFFF}{\color[HTML]{000000} \textbf{61.88}}}                    & \multicolumn{1}{c|}{\cellcolor[HTML]{FFFFFF}{\color[HTML]{000000} \textbf{56.93}}} &\multicolumn{1}{c|}{\cellcolor[HTML]{FFFFFF}{\color[HTML]{000000} \textbf{62.64}}} & \multicolumn{1}{c|}{\cellcolor[HTML]{FFFFFF}{\color[HTML]{000000} \underline{62.34}}}&\multicolumn{1}{c|}{\cellcolor[HTML]{FFFFFF}{\color[HTML]{000000} \textbf{59.02 (+1.05)}}}                               \\ \hline

\end{tabular}}
\label{tab:mm_idgm}
\end{table*}

\noindent\textbf{Performance of MM-IDGM}\quad
We proposed MM-IDGM to enhance multi-to-single domain generalization in deception detection. Table~\ref{tab:mm_idgm} presents the experimental results under the domain-simultaneous sampling strategy, comparing Atten-Mixer (baseline), IDGM, and the proposed MM-IDGM across various modality combinations. In all tested configurations, MM-IDGM consistently outperforms Atten-Mixer, achieving up to +1.08\% improvement in average accuracy, particularly in the case of Visual (Face Frames) + Visual (AU+Gaze+Affect). This demonstrates the effectiveness of aligning gradient directions across domains at the feature level in a multimodal setting.

While IDGM improves performance in many cases, it occasionally degrades accuracy. This may be due to conflicts from aligning gradients independently per modality. In contrast, MM-IDGM aligns all modalities jointly, yielding more stable and transferable features, and showing promising improvements in multimodal cross-domain generalization. However, its effectiveness can vary depending on the complexity of the target domain, data quality, and the severity of distribution shifts.

\vspace{-0.8em}
\subsection{Network Details}
Here, we present the detailed network architectures in Table~\ref{tab:full_networks}, including encoders, the fusion module, and classifiers. This network architecture is utilized for experiments on multi-to-single cross-domain generalization with fusion.

\begin{table}[h]
\centering
\caption{Architecture details for different modalities and fusion module.}
\label{tab:full_networks}
\resizebox{0.4\textwidth}{!}{
\begin{tabular}{|c|}
\hline
\textbf{Modality \& Inputs: Visual (Face Frames)} \\ 
Input Size: [B, 3, 32, 224, 224] \\ \hline
(For each frame) ResNet-18 [:-2] $\rightarrow$ [B, 32, 7, 7, 512] \\ Average Pooling  $\rightarrow$ [B, 32, 512]\\  (For sequence) GRU Layer $\rightarrow$ [B, 32, 512] \\ Fully connected layer $\rightarrow$ [B, 2]\\ 
 \hline\hline

\textbf{Modality \& Inputs: Visual (AU+Gaze+Affect)} \\ \hline
Input Size: [B, 50, 64] \\ \hline
 (On d-50) Linear(50, 64) + LN + ReLU \\Linear(64, 128) + LN + ReLU \\ Linear(128, 1) $\rightarrow$ [B*64, 1] \\ 
 Reshape to [B, 64, 1] \\
(On d-64) Linear(64, 32) + LN + ReLU \\ Linear(32, 16) + LN + ReLU \\ Linear(16, 2) $\rightarrow$ [B, 2]\\ \hline\hline

\textbf{Modality \& Inputs: Audio (Mel Spectrogram)} \\ 
Input Size: [B, 3, 224, 224] \\ \hline
 ResNet-18 [:12] $\rightarrow$ [B, 7, 7, 512]\\ 
 Average Pooling  $\rightarrow$ [B, 32, 512]\\ 
 Reshape to [B, 32*512]\\
 Linear(32*512, 12) $\rightarrow$ [B, 2] \\ \hline \hline

 \textbf{Atten-Mixer Fusion} \\ 
Input size: face frame [B, 32, 512], \\behavioral [B, 64, 1], \\ and audio  [B, 32, 512] \\ \hline
\textbf{Projection Layer}:  face frame $\rightarrow$[B, 64, 1], \\ behavioral$\rightarrow$ [B, 64, 1], \\and audio $\rightarrow$ [B, 64, 1]\\
 \textbf{Concatenation} $\rightarrow$ [B, 64, 3]\\ 
\textbf{Atten-Mixer Fusion}: number of patches $\rightarrow$ 64, \\channels $\rightarrow$ 3, \\and output dim $\rightarrow$ 128\\ 
\textbf{Output shape} $\rightarrow$ [B, 128]\\
\textbf{Classifier}: Linear(128, 64)+ReLU+Linear(64, 2) $\rightarrow$ [B, 2] \\ \hline

\end{tabular}}
\vspace{-1em}
\end{table}

\subsection{Discussion}
\label{sec:Discussion}

We observe that the general performance of cross-domain deception detection remains unsatisfactory, mainly because of the substantial domain gaps across datasets. The benchmark datasets differ in recording environments, visual resolution, illumination, audio quality, interaction format, and subject demographics, making it difficult for cues learned from one domain to transfer well to another. In particular, audio-based models are more sensitive to background noise, speech style, and recording conditions, while visual behavioral cues such as AUs, gaze, and affect show relatively stronger transferability in several settings. We also find that different domain sampling strategies work better for different audio-visual features, suggesting that cross-domain generalization depends on both the training strategy and the modality being used. Nevertheless, the results indicate that cross-domain deception detection is learnable. Multi-to-single training, multimodal fusion, and MM-IDGM all improve performance over weaker baselines, showing that transferable patterns do exist across domains. However, current paradigms are still limited in robustness, and stronger generalization methods are needed.

\vspace{0.3em}
\textbf{Ethical Consideration.}\quad 
Developing AI-based deception detection should emphasize privacy, fairness, and the prevention of psychological harm. If deployed without consent or proper safeguards, such systems may misuse sensitive personal data, including speech, facial expressions, and behavioral cues. Researchers should therefore ensure informed consent, responsible data use, and transparent model development.

\section{Conclusion} \label{sec:conclusion}
In this paper, we benchmark the cross-domain generalization performance for deception detection on publicly available datasets. We compare the single-to-single domain and multi-to-single domain generalization performances, where three strategies are used including domain-simultaneous, domain-alternating, and
domain-by-domain. We also investigate the effectiveness of the gradient reversal layer for domain-simultaneous strategy. Moreover, we propose the Attention-Mixer fusion method to alleviate the domain shift issue and boost the performance. Future works for deception detection are encouraged to propose better methods to improve the domain generalizability on audio-visual deception detection task.



\bibliographystyle{IEEEtran}
\bibliography{IEEEabrv,reference}

\end{document}